\documentclass[aps,amssymb,amsmath,amsfonts,pra,superscriptaddress]{book}
\usepackage{amssymb}
\usepackage{diagbox} %to use diagonal division in tables
\usepackage{amsmath}
\usepackage[english,catalan]{babel}
\usepackage[utf8]{inputenc}
\usepackage[usenames, dvipsnames]{color}
\usepackage{graphicx}
\usepackage{gensymb}
\usepackage{amsthm}
\usepackage{epstopdf}
\usepackage{xcolor}%para poner color en texto
\usepackage{hyperref}
\usepackage[T1]{fontenc}
\usepackage{graphics}
\usepackage{color}
\usepackage{float}
\usepackage{import}
\usepackage{bm}
\usepackage{enumerate}
\usepackage{appendix}
\usepackage{fancyhdr}

\newcommand{\one}{\mbox{$1 \hspace{-1.0mm}  {\bf l}$}}

\newcommand{\identity}{\one}
\newcommand{\trace}[2][]{\text{tr}_{#1}\left( #2 \right)}

\renewcommand{\vec}[1]{\bm{#1}}

\newcommand{\bra}[1]{\langle #1|}
\newcommand{\floor}[1]{\lfloor #1 \rfloor}
\newcommand{\ket}[1]{|#1\rangle}
\newcommand{\braket}[2]{\langle #1|#2\rangle}
\newcommand{\ketbra}[2]{| #1 \rangle \langle #2 |}
\newcommand{\expect}[1]{\langle #1\rangle}
\newcommand{\proj}[1]{\vert #1\rangle\!\langle#1 \vert}

\newcommand{\widebar}[1]{\overline{#1}}
\newcommand{\gras}[1]{\bold{#1}}
\newcommand\Tr{{\mathrm{Tr}}}
\newcommand{\tr}{\mathop{\mathrm{tr}}}
\newcommand\iu{{\mathrm{i}}}
\newcommand\e{{\mathrm{e}}}

\newcommand{\bed}{\[}
\newcommand{\eed}{\]}
\newcommand{\beq}{\begin{equation}}
\newcommand{\eeq}{\end{equation}}
\newcommand{\beqa}{\begin{eqnarray}}
\newcommand{\eeqa}{\end{eqnarray}}

\newcommand{\twopartdef}[4]
{
	\left\{
		\begin{array}{ll}
			#1 & \mbox{if } #2 \\
			#3 & \mbox{if } #4
		\end{array}
	\right.
}

\begin{document}

\frontmatter

\pagestyle{empty}
\begin{center}

\vfill
\centerline{\mbox{\includegraphics[width=60mm]{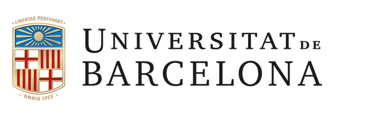}}}

\medskip
{\large Facultat de Física}

\vfill
{\bfseries\Large TESI DOCTORAL}

\vfill
\vspace{5mm}

{\LARGE Daniel Alsina Leal}

\vspace{15mm}

% Title in English according to the official assignment
{\LARGE\bfseries Multipartite entanglement and quantum algorithms}

\vfill

{\large Departament de Física Quàntica i Astrofísica}

\vfill

\begin{tabular}{rl}
{\large Director de tesi:} & {\large Dr. José Ignacio Latorre Sentís} \\   
\noalign{\vspace{2mm}}
{\large Tutor de tesi:} & {\large Dr. José Ignacio Latorre Sentís} \\   
\noalign{\vspace{2mm}}
{\large Programa de doctorat:} & {\large Física}\\
\end{tabular}

\vfill

{\large Barcelona, Abril de 2017}

\end{center}

\addcontentsline{toc}{chapter}{Agra\"iments}

\selectlanguage{catalan}

\chapter*{Agraïments}

He estat 5 anys treballant en aquesta tesi, i probablement no hauria trobat forces per arribar fins al final sense molta gent que, directa o indirectament, m'ha ajudat i animat a continuar.

Per començar, per suposat, el José Ignacio, el meu director de tesi, que des del principi m'ha fet confiança i m'ha acompanyat en aquest llarg viatge, amb unes sessions motivacionals a la seva pissarra que mai oblidaré. També el Sofyan i la Mari Carmen, que han exercit de directors en el seu moment, estant sempre disponibles a solucionar els meus dubtes.

Gràcies a l'Ignacio per la seva invitació a Munic, i al Karol i al Dardo per la meva estada a Cracòvia, i per treballar conjuntament amb mi en 2 articles. Gràcies als altres coautors d'Innsbruck, Barbara i Martin.

També els dec molt als companys de grup més experimentats que jo que em van ajudar molt al principi a adaptar-me a la feina de doctorand: Artur, Octavi, Arnau, Mauricio, Francesco. Espero haver complert amb els que han vingut després: Toni, Guillem, Salvo i Alba. Menció especial per l'Alba, amb qui hem compartit molt els meus dos últims anys de tesi: coautora, companya de grup, de despatx, proveïdora de Kinders i amiga, tot a la vegada.

Moltes gràcies a l'Enric i el Dani, amb qui hem tingut converses interminables de tot tipus en dinars, berenars i estones mortes variades durant tot aquest temps, i a l'Edu, amb qui ho fèiem al principi mentre encara estava fent la tesi. I a tots els del Lunchtime i de Físics pel món, que són massa i no els podria anomenar tots. Però es mereix una menció especial el Sergio, que m'ha ajudat diverses vegades amb els seus amplis coneixements de programació quan els meus ja no donaven per més.

De la vida pre-universitària vull agrair a l'Esquius, el Jon i el Jordi els treballs d'institut conjunts que van començar a marcar el meu camí i l'amistat que hem mantingut fins ara, i als professors Beth, Ferran i Toni la seva docència estimulant de física i matemàtiques, que em va ajudar a escollir la carrera. I de la vida universitària, a part de molts que ja he mencionat abans, l'Adrià i en Rimbi, amb qui vam compartir tantes i tantes aventures i desventures, i en Jordi novament, que va seguir sent un amic i company acadèmic inestimable.

En el món dels escacs, la meva carrera paral·lela a la física, també hi ha molta gent que m'ha donat suport en aquest temps, sense poder anomenar-los tots mencionaré els meus companys de l'anterior club, el Barcelona Uga, i els de l'actual, el Barberà, conjuntament amb els Wifis, amb qui hem passat grans estones últimament. Gràcies al Divis i Chess24 per permetre'm l'ús del seu tauler d'escacs a la portada. Segur que mai haurien esperat que es fes servir per posar un problema d'escacs quàntics en una portada de tesi.

I per suposat he d'agrair també el suport dels que han conviscut amb mi en els pisos compartits en què he passat aquests anys i han seguit dia a dia els meus èxits i fracassos: Cassius, Àlvar, Jordi i Marc.

Gràcies especials a amics que, simplement, sempre han estat allà, disponibles tant per una conversa entretinguda com per ajudar-me quan els necessitava: Pardo i Daniele.

I per últim, una gran abraçada als meus pares i al meu germà, que m'han estimat i donat suport des de molt abans que tots els altres.

\selectlanguage{english}

\addcontentsline{toc}{chapter}{List of publications}
\chapter*{List of publications}

This thesis is based on the following papers, sorted in chronological order.

\begin{itemize}

\item D. Alsina and J. I. Latorre, \emph{Tensor networks for frustrated systems: emergence of order from simplex entanglement}, arXiv:1312.0952 [quant-ph] (2013).

\item D. Goyeneche, D. Alsina, J. I. Latorre, A. Riera and K. \.{Z}yczkowski, \emph{Absolutely Maximally Entangled states, combinatorial designs and multi-unitary matrices}, Phys. Rev. A \textbf{92}, 032316 (2015). arXiv:1506.08857 [quant-ph].

\item D. Alsina and J. I. Latorre, \emph{Experimental test of Mermin inequalities on a 5-qubit quantum computer}, Phys. Rev. A \textbf{94}, 012314 (2016). arXiv:1605.04220 [quant-ph].

\item D. Alsina, A. Cervera, D. Goyeneche, J. I. Latorre and K. \.{Z}yczkowski, \emph{Operational approach to Bell inequalities: applications to qutrits}, Phys. Rev. A \textbf{94}, 032102 (2016). arXiv:1606.01991 [quant-ph].

\item M. Hebenstreit, D. Alsina, J. I. Latorre and B. Kraus, \emph{Compressed quantum computation using the IBM Quantum Experience}, Phys. Rev. A \textbf{95}, 052339 (2017). arXiv:1701.02970 [quant-ph].

\end{itemize}

\selectlanguage{catalan}

\addcontentsline{toc}{chapter}{Resum}

\chapter*{Resum}

\pagestyle{fancy}
\fancyhf{}
\fancyhead[LE]{\thepage}
\fancyhead[RE]{Resum}
\fancyhead[RO]{\thepage}

La informació quàntica ha crescut des d'un petit subcamp als anys setanta fins a esdevenir un dels camps més dinàmics de la física actualment, tant en aspectes fonamentals com en les seves aplicacions. En la secció teòrica, potser la propietat que ha atret més interès és la noció d'entrellaçament, la relació fantasmagòrica entre partícules que va deixar estupefacte Einstein i que ha suposat un enorme desafiament per a construir una interpretació coherent de la mecànica quàntica. Sense estar totalment solucionat, hem après prou per sentir-nos menys incòmodes amb aquest problema fonamental i el focus s'ha desplaçat a les seves aplicacions potencials. L'entrellaçament s'estudia avui en dia des de diferents perspectives com a recurs per realitzar tasques de processament de la informació.

\section*{Entrellaçament multipartit}

L'entrellaçament bipartit està ja molt ben comprès, però en el cas multipartit queden obertes moltes qüestions. La primera part d'aquesta tesi tracta l'entrellaçament multipartit en diferents contextos. Per començar el tractem en el marc de tot l'espai de Hilbert corresponent i ens centrem en com mesurar-lo i en determinar els estats màximament entrellaçats. Primerament estudiem el cas de 4 qubits, per ser un dels més simples on encara hi ha molt espai per investigar, i proposem l'hiperdeterminant com a mesura d'entrellaçament multipartit genuí, en contrast amb l'entropia d'entrellaçament, que sempre el mesura per biparticions. Trobem l'estat d'hiperdeterminant màxim per mètodes numèrics, i en donem algunes propietats interessants, i també trobem una certa correlació entre l'entropia i l'hiperdeterminant en el sentit que estats amb alt hiperdeterminant solen tenir alta entropia, però no a l'inrevés. Posteriorment estudiem els estats absolutament màximament entrellaçats (AME), que són aquells que tenen estats reduïts màximament mixtos en totes les possibles biparticions. Trobem una propietat matemàtica molt interessant de les matrius que els representen, la multiunitarietat, que implica que aquestes matrius mantenen la unitarietat sota diverses reordenacions dels seus coeficients. També els relacionem amb altres construccions matemàtiques conegudes, com els codis clàssics i els dissenys combinatoris.

Tot seguit entrem en el camp de les desigualtats de Bell, 
on escollim una manera operacional de tractar-les que ens permet de trobar els límits clàssics i quàntics fàcilment. Ens centrem en desigualtats de qutrits, motivats per l'intent de buscar una desigualtat màximament violada per l'estat AME de 4 qutrits. Malgrat no assolir aquest objectiu, trobem diverses desigualtats noves amb límits clàssics i quàntics que apunten a una estructura, i també trobem un mètode per obtenir desigualtats noves a partir d'estats màximament entrellaçats fent una identificació entre la base computacional de l'estat i els experiments de la desigualtat.

Després canviem el focus als hamiltonians que tenen estats entrellaçats com a estats fonamentals. Proposem l'espectre d'entrellaçament com a mesura de distància entre diferents teories, obtenint resultats numèrics coherents, com els pics de distància que apareixen al voltant dels punts crítics. I tanquem la primera part de la tesi estudiant el concepte de frustració geomètrica, que causa que les configuracions que minimitzen l'energia globalment no la minimitzin localment. Això genera complicació en els càlculs dels estats fonamentals d'hamiltonians que pateixen aquest efecte. Fem un estudi, clàssic i quàntic, de la frustració en sistemes petits amb diagonalització exacta, i després proposem un mètode basat en xarxes tensorials per poder computar més eficientment observables de sistemes amb frustració, utilitzant símplexs triangulars com a elements fonamentals enlloc dels parells màximament entrellaçats que utilitzen els PEPS.

\section*{Computació quàntica}

En l'apartat pràctic, el més prometedor avenç tecnològic del camp és l'adveniment dels ordinadors quàntics. Als anys 90 van aparèixer alguns algorismes quàntics que milloraven el rendiment de tots els algorismes clàssics per certs problemes, mentre que als anys 2000 es van començar a construir els primers ordinadors quàntics universals d'uns pocs àtoms, que van permetre implementacions d'aquests algorismes a petita escala. La computadora de D-Wave ja realitza recuita quàntica (\emph{quantum annealing}) en milers de qubits, encara que hi ha certa controvèrsia sobre si els processos interns de la màquina són vertaderament quàntics. Molts països del món estan destinant grans sumes de diners a aquest camp: el recent Flagship europeu i les inversions dels gegants informàtics dels Estats Units donen motius per a l'optimisme.

La segona part de la tesi tracta d'alguns aspectes de la computació quàntica. Comencem per inaugurar el camp de la computació quàntica al núvol, possible gràcies a l'aparició del primer ordinador disponible per a tots els públics a traves d'internet. L'hem usat i analitzat extensivament, demostrant que els seus processos són intrínsecament quàntics, a la vegada que encara generen molts errors, i hem proposat una manera d'adjudicar quantitativament incerteses a les mesures realitzades amb màquines a distància de les quals no coneixem tots els detalls.

També tractem la computació quàntica adiabàtica, un model alternatiu al convencional de circuits. Es basa en què s'identifica la solució del problema a resoldre amb l'estat fonamental d'un hamiltonià potencialment complicat. Es prepara un sistema en l'estat fonamental d'un altre hamiltonià simple del qual es coneix la solució, i es fa evolucionar el sistema lentament fins a l'hamiltonià problema, de manera que es manté sempre en l'estat fonamental de tots els hamiltonians instantanis, produint al final la solució. El problema d'aquest model és que si el gap es fa molt petit el temps de computació es fa molt gran, ja que s'ha d'evolucionar el sistema molt lentament per evitar que salti a un estat excitat. En el límit de gap zero, el temps es fa infinit. Hem intentat comprovar per un model d'Ising si aquest infinit es podia cancel·lar amb un altre factor, obtenint un resultat negatiu.

I finalment ens endinsem en el camp de la termodinàmica quàntica, per preguntar-nos si un sistema evolucionant sota un cert hamiltonià termalitza o no a temps prou grans. Si ens posem en el cas de subsistemes del sistema global i mirem la mitjana temporal dels observables a temps grans, trobem que els sistemes sempre tendeixen a un estat diagonal en la base d'energies, la col.lectivitat diagonal (\emph{diagonal ensemble}). Aleshores, esbrinar si un sistema termalitza es redueix a veure si la seva col.lectivitat diagonal coincideix amb l'estat tèrmic. Però calcular la col.lectivitat diagonal per sistemes grans és difícil computacionalment, i com que només esperem trobar termalització en sistemes grans, és necessari buscar un mètode eficient per calcular-la. Proposem un algorisme per fer-ho i demostrem que dóna els resultats correctes per sistemes petits, i ens queda pendent aplicar-lo a sistemes grans per poder complir amb l'objectiu inicial.

\selectlanguage{english}

\tableofcontents

\mainmatter

\addcontentsline{toc}{chapter}{Introduction}

\chapter*{Introduction}

\pagestyle{fancy}
\fancyhf{}
\fancyhead[LE]{\thepage}
\fancyhead[RE]{Introduction}
\fancyhead[RO]{\thepage}

Quantum information science has grown from being a very small subfield in the 70s until being one of the most dynamic fields in physics, both in fundamentals and applications. In the theoretical section, perhaps the feature that has attracted most interest is the notion of entanglement, the ghostly relation between particles that dazzled Einstein and has provided fabulous challenges to build a coherent interpretation of quantum mechanics. While not completely solved, we have today learned enough to feel less uneasy with this fundamental problem, and the focus has shifted towards its potential powerful applications. Entanglement is now being studied from different perspectives as a resource for performing information processing tasks.

With bipartite entanglement being largely understood nowadays, many questions remain unanswered in the multipartite case. The first part of this thesis deals with multipartite entanglement in different contexts. In the first chapters it is studied within the whole corresponding Hilbert space, and we investigate several entanglement measures searching for states that maximize them, including violations of Bell inequalities. Later, focus is shifted towards hamiltonians that have entangled ground states, and we investigate entanglement as a way to establish a distance between theories and we study frustration and methods to efficiently solve hamiltonians that exhibit it.

In the practical section, the most promised upcoming technological advance is the advent of quantum computers. In the 90s some quantum algorithms improving the performance of all known classical algorithms for certain problems started to appear, while in the 2000s the first universal computers of few atoms began to be built, allowing implementation of those algorithms in small scales. The D-Wave machine already performs quantum annealing in thousands of qubits, although some controversy over the true quantumness of its internal workings surrounds it. Many countries in the planet are devoting large amounts of money to this field, with the recent European flagship and the involvement of the largest US technological companies giving reasons for optimism.

The second part of this thesis deals with some aspects of quantum computation, starting with the creation of the field of cloud quantum computation with the appearance of the first computer available to the general public through internet, which we have used and analysed extensively. Also small incursions in quantum adiabatic computation and quantum thermodynamics are present in this second part.

\section*{Outline}

The first chapter of the thesis starts with a review of entanglement in general and entanglement measures in particular. Then the focus is put on entanglement of four-qubit states, where we have studied the hyperdeterminant as an entanglement measure and looked for the state that maximizes it.

In Chapter 2 we study absolutely maximally entangled states, those states that have the maximally mixed state in all its possible reductions, and discuss several mathematical properties and their relations to known mathematical objects like classical codes and combinatorial designs.

Chapter 3 deals with Bell inequalities studied from an operational approach. We investigate new ways of writing them and finding their classical and quantum bounds, focusing on Bell inequalities for qutrits, and we try to look for new ones.

The fourth chapter studies the concept of distance between theories, and we investigate the possible use of the entanglement spectrum as a distance.

Chapter 5 studies geometrical frustration, both in its classical and quantum version, establishes its relationship with difficulties in solving certain hamiltonians, and looks for methods to compute those cases more efficiently.

Chapter 6 introduces the field of cloud quantum computation, and we present the results obtained with the new quantum computer from IBM available to the general public, together with some considerations on how to evaluate the results and deal with uncertainties.

A shift towards adiabatic quantum computation is made in seventh chapter, where we try to check whether the computation time for the Ising hamiltonian really grows unboundedly when the gap goes to zero.

And in the last chapter we investigate on a new algorithm to compute the diagonal ensemble, with the motivation to elucidate whether a system thermalises or not for large evolution times. 

Finally we give some conclusions together with proposals on directions for future work.

\part{Multipartite entanglement}

\chapter{Entanglement measures of few parties} \label{ch:entanglement}

\pagestyle{fancy}
\fancyhf{}
\fancyhead[LE]{\thepage}
\fancyhead[RE]{MULTIPARTITE ENTANGLEMENT}
\fancyhead[LO]{CH.1 Entanglement measures of few parties}
\fancyhead[RO]{\thepage}

\section{Introduction to multipartite entanglement}

One of the most characteristic properties of quantum mechanics is the notion of entanglement, appearing for the first time in the discussion by Erwin Schr\"odinger of the famous EPR paper \cite{Einstein35,Schrodinger35,Schrodinger36}. Its modern definition says that a pure state of two or more components is said to be entangled if it cannot be written as a direct product of its parts. The most basic example is a maximally entangled state of two parties in the Hilbert space $\mathbb{C}^2 \otimes \mathbb{C}^2$
\begin{equation}
\ket{\psi}=\frac{1}{\sqrt{2}}\left(\ket{00}+\ket{11}\right).
\end{equation}
It has the striking yet now well established property that once you measure one of the system parts, the state of the other part changes instantaneously, no matter how far apart they are. This failure of local realism was first believed to be non-observable but it was turned into a falsifiable theory by Bell \cite{Bell64}. Numerous experiments have since confirmed with high confidence that quantum mechanics is not a local realistic theory \cite{Aspect82,Tittel98,Hensen15,Poh15}.
 
The recent explosion of quantum information science has brought the need to quantify entanglement and to classify states according to its measure, in order to be able to use them as a resource for certain information processing tasks, such as quantum teleportation \cite{Bennett93} or quantum key distribution \cite{Ekert91}. Nielsen and Chuang's book is a classical reference to the first developments of quantum information \cite{Nielsen00} and there is a broad amount of books and reviews of the theory of entanglement \cite{Plenio98,Schumacher00,Eisert01,Vedral02,Bengtsson06,Plenio07,Amico08,Horodecki09,
Eltschka14}.

This chapter will focus on entanglement measures for pure states of qubits. It will be useful to first define physically relevant transformations related to entanglement that can be applied to multipartite states.

\subsection{LU transformations} \label{sec:LU}

Entanglement is invariant under choices of local basis. It is then natural to introduce the concept of Local Unitary (LU) transformations among states.

An LU transformation between N-party states is defined as the product of N local unitary matrices such that
\begin{equation}
\ket{\Phi}= U_1 \otimes U_2 \otimes \ldots \otimes U_N \ket{\Psi} \, .
\end{equation}
$\ket{\Phi}$ and $\ket{\Psi}$ are called LU-equivalent and have thus the same amount of entanglement. One can therefore never create or destroy entanglement with LU transformations. Thus important effort has been given to the issue of finding in practice whether two states are interconvertible by LU transformations \cite{Kraus10a,Kraus10b}.

\subsection{LOCC transformations} \label{sec:locc}

Local Operations and Classical Communication (LOCC) transformations are defined as LU transformations plus some extra possibilities: measurements, operations involving additional local systems called \emph{ancillas} and classical communication between distant parties \cite{Bennett96a}. They are relevant in that they represent all possibilities available to process a state handled by various spatially separated parties. As measurements can only decrease the amount of entanglement, it never increases under LOCC. Two states $\ket{\Phi}$ and $\ket{\Psi}$ are said to be LOCC equivalent if they can be transformed into each other by LOCC transformations, and will have the same amount of entanglement. If $\ket{\Phi}$ can be transformed into $\ket{\Psi}$ but the opposite is not true, it is an indication that $\ket{\Phi}$ is more entangled than $\ket{\Psi}$. LOCC transformations thus introduce a natural ordering of states according to their amount of entanglement and give sense to the concept of entanglement as a resource for information processing tasks. Ref. \cite{Nielsen99} introduced a criterion to decide if one pure bipartite state can be transformed to another via LOCC.

As the mathematical characterization of LOCC classes is very difficult in the multipartite case, it is useful to introduce Stochastic LOCC (SLOCC) transformations \cite{Vidal00,Bennett00}. A SLOCC transformation is a LOCC transformation that will succeed at least with a certain probability. Diferent SLOCC equivalence classes have fundamentally different types of entanglement, although SLOCC do not allow to order entangled states according to their usefulness. A systematic classification of multipartite entanglement in terms of equivalence classes of states under SLOCC is presented in Ref.~\cite{Gour13}.

\subsection{Entanglement measures}

The following step is to define a function to measure entanglement that behaves as expected by the previously defined transformations. The following is a list of possible postulates for entanglement measures proposed by Ref. \cite{Plenio07}.

\begin{itemize}

\item An entanglement measure $E(\rho)$ is a mapping from density matrices into positive real numbers
\begin{equation}
\rho \rightarrow E(\rho) \in \mathbb{R^+}.
\end{equation}

\item $E(\rho) = 0$ if the state $\rho$ is separable.

\item $E$ does not increase on average under LOCC, i.e. 
\begin{equation}
E(\rho) \geq \sum_i p_i E (\frac{A_i \rho A_i^\dagger}{\tr A_i \rho A_i^\dagger}).
\end{equation}
where the $A_i$ are the Kraus operators describing
some LOCC protocol and the probability of obtaining
outcome $i$ is given by $p_i = \tr A_i \rho A_i^\dagger
$.
\end{itemize}

A measure satisfying these three conditions is called an \emph{entanglement monotone}. Other desirable properties, not always fulfilled by all measures, are

\begin{itemize}

\item Convexity

One common example for an additional property is convexity, which means that we require
\begin{equation}
E \left(\sum_i p_i \rho_i \right) \geq \sum p_i E (\rho_i),
\label{subadd}
\end{equation}
Requiring this is sometimes justified as capturing the notion of loss of information, i.e., describing the process of going from a set of identifiable states $\rho_i$ that appear with probabilities $p_i$ to a mixture of the form $\sum p_i \rho_i$.

\item Additivity of the tensor product

Additivity refers to asking the following requirement
\begin{equation}
E(\sigma^{\otimes n}) = n E(\sigma),
\end{equation}
to be satisfied for all integer n. A stronger requirement, that some measures fulfil, is full additivity, meaning
\begin{equation}
E(\sigma \otimes \rho) = E(\sigma) + E(\rho).
\end{equation}
\end{itemize}

\subsubsection{Von Neumann entropy of entanglement}

A relevant measure is the Von Neumann entropy of entanglement \cite{Bennett96b}. The entropy of subsystem A of a state living in the Hilbert space $\mathcal{H}_\mathcal{A} \otimes \mathcal{H}_\mathcal{B}$ is the Von Neumann entropy of its reduced density matrix $\rho_\mathcal{A}$, that is
\begin{equation}
S(A) = S(\rho_\mathcal{A}) \equiv -\rm{tr}(\rho_\mathcal{A} \log_2 \rho_\mathcal{A}).
\end{equation}
When defined in terms of a logarithm of base 2 as in this case, it is measured in \emph{ebits}.

We can also present it in terms of the Schmidt decomposition, which tells us that it is always possible to decompose a pure state in two parts and write it in the following way
\begin{equation}
\ket{\phi}=\sum_{i=1}^{N_\mathcal{A}} \sqrt{\lambda_i} \ket{\alpha_i^A} \ket{\beta_i^B}
\end{equation}
where $\ket{\alpha_i^A}$ and $\ket{\beta_i^B}$ are two orthonormal bases, $N_\mathcal{A}$ is the Hilbert dimension of subsystem A assuming $N_\mathcal{A} \leq N_\mathcal{B}$, and $\lambda_i$ are called the Schmidt coefficients. The entropy of entanglement can then be written as
\begin{equation}
S(A) = -\sum_{i=1}^{N_\mathcal{A}}{\lambda_i \log_2 \lambda_i} = S(B).
\label{eq:entropylambda}
\end{equation}
where we see that $S(A)=S(B)$ and thus it makes sense to talk about entropy as a measure of entanglement of the state. This expression also shows that the values of entanglement entropy can range from 0 for product states to $\log_2 N_\mathcal{A}$ ebits for maximally entangled states. In the two qubits case it can thus go from 0 to 1 ebits.

Besides all the properties of entanglement measures listed in the previous section, including full additivity, Von Neumann entanglement entropy has some additional interesting mathematical properties, many of them proved in Ref. \cite{Araki70}:

\begin{itemize}

\item Sub-additivity 
\begin{equation}
S(A,B) \leq S(A)+S(B)
\label{subadd}
\end{equation}
where $S(A,B)$ refers to the Von Neumann entropy of the joint state $\rho_\mathcal{AB}$, where $\rho_\mathcal{A} = Tr_\mathcal{B} \rho_\mathcal{AB}$. If $\rho_\mathcal{AB}$ is a pure state, $S(A,B)=0$.
\item Triangle inequality
\begin{equation}
S(A,B) \geq |S(A)-S(B)|
\label{triangle}
\end{equation}
\item Strong sub-additivity\\\\
If we consider three systems A,B and C and combine sub-additivity and triangle inequality, it is possible to deduce strong sub-additivity \cite{Lieb73}
\begin{equation}
S(A,B,C)+S(B) \leq S(A,B)+S(B,C).
\label{strongsubadd}
\end{equation}

The structure of states that saturate the previous inequality is presented in Ref. \cite{Hayden03}. An explicit bound for the difference between the terms was found in Ref. \cite{Carlen12}, whereas an operator extension was found in Ref. \cite{Kim12}.
\end{itemize}
 
We present now other measures of entanglement that we will need for the next section. It is possible to generalise Von Neumann entropy to a set of entropies depending on a parameter $\alpha$, called Renyi and Tsallis entropies.

\subsubsection{R\'enyi entropy}

R\'enyi introduced this entropy \cite{Renyi60} in order to find an information theoretic proof to the central limit theorem \cite{Harremoes06}. It has the following expression
\begin{equation}
E_R^{(\alpha)}(\lambda_i) \equiv \frac{1}{1-\alpha} \log_2\sum_{i=1}^{N_\mathcal{A}} \lambda_i^\alpha. \label{renyi} 
\end{equation}
If all the eigenvalues have the same value (and none of them is zero) we get the maximum possible R\'enyi entropy, which does not depend on $\alpha$: $E_R^{(\alpha)}(\lambda_i)= \log_2 N_\mathcal{A}$, where $N_\mathcal{A}$ is the dimension of Hilbert space of subsystem A. This coincides with the maximum Von Neumann entropy.\\
For $\alpha \rightarrow \infty$ then only the eigenvalue with greatest value counts and $E_R^{(\alpha)}\rightarrow 0$.\\
For $\alpha \rightarrow 0$ the weights of all eigenvalues become equal and $E_R^{(\alpha)}\rightarrow \log_2{M_\mathcal{A}}$, where $M_\mathcal{A}$ is the number of eigenvalues different from zero. If all eigenvalues are different from zero, then $M_\mathcal{A}$ = $N_\mathcal{A}$.\\

\subsubsection{Tsallis entropy}

Tsallis introduced this entropy \cite{Tsallis88} as a generalisation of the Boltzmann-Gibbs entropy. It has the following expression
\begin{equation}
E_T^{(\alpha)}(\lambda_i) \equiv \frac{1}{\alpha -1} \left(1-\sum_{i=1}^{N_\mathcal{A}} \lambda_i^\alpha\right). \label{tsallis} 
\end{equation}
If all the eigenvalues have the same value (and none of them is zero) we get the maximum possible Tsallis entropy for every $\alpha$: $E_T^{(\alpha)}(\lambda_i)=\frac{N_\mathcal{A}^{1-\alpha}-1}{1-\alpha}$. This coincides with maximum Von Neumann (and R\'enyi) entropy for $\alpha =1$.\\
For $\alpha \rightarrow \infty$ then only the eigenvalue with greatest value counts and $E_T^{(\alpha)}\rightarrow 0$.\\
For $\alpha \rightarrow 0$ the weights of all eigenvalues become equal and $E_T^{(\alpha)}\rightarrow M_\mathcal{A}-1$, where $M_\mathcal{A}$ is the number of eigenvalues different from zero. If all eigenvalues are different from zero, then $M_\mathcal{A}$ = $N_\mathcal{A}$.\\

Both entropies tend to Von Neumann entropy when $\alpha \rightarrow 1$

\begin{equation}
S(\lambda_i) = -\sum_{i=1}^{N_\mathcal{A}} \lambda_i\log_2\lambda_i. \label{shannon} 
\end{equation}

\subsubsection{Purity} \label{sec:purity}

The purity of a quantum state is a scalar that measures its degree of mixedness
\begin{equation}
\gamma \equiv \Tr(\rho^2).
\label{eq:purity}
\end{equation}
When applied to the reduced density matrix of a subsystem, $\gamma_\mathcal{A} = \Tr(\rho_\mathcal{A}^2)$, its degree of mixedness is directly related to the degree of entanglement of the whole state. Its complement, the linear entropy, defined as $S_\mathcal{A}^L \equiv 1-\gamma_\mathcal{A}$, is a first-order Taylor approximation to the Von Neumann entropy.

Purity ranges from $1/N_\mathcal{A}$ for a maximally mixed subsystem (and maximally entangled state) to 1 for a pure subsystem (and product state). Maximally mixed subsystems have always the following form proportional to identity
\begin{equation}
\rho_\mathcal{A} = \frac{1}{{N_\mathcal{A}}} \mathbb{I}.
\label{eq:maxmixed}
\end{equation}
The question of whether it is possible to minimize the purity of all bipartitions at the same time in a multipartite state will be a central one in the next chapter.

\subsubsection{Concurrence} \label{conc}

An alternative entanglement measure for the 2-qubits case is the \emph{concurrence} \cite{Hill97}. It is an entanglement monotone defined for a mixed state of 2 qubits as 
\begin{equation}
\mathcal{C}(\rho) \equiv \max (0,\lambda_1-\lambda_2-\lambda_3-\lambda_4)
\end{equation}
in which $\lambda_i$ are the eigenvalues in decreasing order of the hermitian matrix
\begin{equation}
R=\sqrt{\sqrt{\rho}\bar{\rho}\sqrt{\rho}}
\end{equation}
where $\bar{\rho} = (\sigma_y \otimes \sigma_y)\rho^*(\sigma_y \otimes \sigma_y)$ is the spin flipped state of $\rho$.

For pure states, it is equal to this much simpler expression,
\begin{equation}
\mathcal{C}(\ket{\psi}) = 2 \det \alpha  = 2 \left| \begin{array}{cc}
a_{00} & a_{01} \\
a_{10} & a_{11} \end{array} \right|, \label{eq:det} 
\end{equation}
where $\ket{\psi}=a_{00}\ket{00}+a_{01}\ket{01}+a_{10}\ket{10}+a_{11}\ket{11}$, and $\sum a^*_i a_i = 1$. $\alpha$ is the matrix formed by the coefficients of the state. It is normalized so that it goes from 0 to 1. In this case of pure states it behaves exactly as the Von Neumann entropy, so it doesn't add anything new, but it is convenient for us to introduce this determinant in order to generalise it to more qubits.

\subsubsection{Three-tangle} \label{tang}

The three-tangle is a measure for entanglement in systems of three parties $A, B$ and $C$ defined as 
\begin{equation}
\tau (A,B,C) \equiv \mathcal{C}^2(A,BC) - \mathcal{C}^2(A,B)- \mathcal{C}^2(A,C),
\label{monogamy3}
\end{equation}
where the quantities in the right hand side are squares of the concurrences between the corresponding subsystems. Its value is interpreted as a genuine three-partite entanglement, which entanglement between pairs cannot account for. For that reason it was called originally \emph{residual tangle} \cite{Coffman00}. It ranges from 0 to 1.

It can be defined in the spirit of the determinant of Eq. \ref{eq:det} in the 3-qubit case by making use of the hyperdeterminant, a generalisation of the determinant built in the following way: beginning from Eq. \ref{eq:det} we perform the substitution $a_{ij} \rightarrow (b_{ij0}+b_{ij1}x)$, where $ b_{ijk}$ are the coefficients of our 3-qubit state.
The discriminant $\Delta$ of the obtained polynomial $P_3(x)$ is the hyperdeterminant of this 3-tensor
\begin{eqnarray}
&&P_3(x) = {\rm det} \alpha /. a_{ij} \rightarrow (b_{ij0}+b_{ij1}x)\label{threetangle}\\ \nonumber 
&&{\rm Hdet_3}(\beta) = {\rm \Delta} [P_3(x)] = \frac{1}{4} \tau(A,B,C),
\end{eqnarray}
where /. means "substituting", and $\beta$ is the 3-tensor formed by the coefficients of the state. This hyperdeterminant is equal to the three-tangle with a factor of 4.

\begin{figure}[h!]
\centering
            \includegraphics[scale=1]{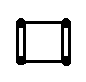}
             \caption{Graphical representation of the contractions building the three-tangle. The points represent a coefficient $b_{ijk}$, and each line represents an $\epsilon_{ij}$.}
\label{fig:grafictangle}     
\end{figure}

It can also be built by a tensor contraction
\begin{equation}
{\rm Hdet_3}(\beta) = \frac{1}{2} \left(\epsilon^{i_1 j_1}\epsilon^{i_2 j_2}b_{i_1 j_1 k_1}b_{i_2 j_2 k_2} \right) \left( \epsilon^{i_3 j_3}\epsilon^{i_4 j_4}b_{i_3 j_3 k_3}b_{i_4 j_4 k_4} \right) \epsilon^{k_1 k_3}\epsilon^{k_2 k_4},
\end{equation}
where $\epsilon$ is the Levi-Civita symbol. This can be represented pictorically as in Fig. \ref{fig:grafictangle}.
The extension to the 4-hyperdeterminant will be the main subject of the next section on 4-qubit states.

The next step is to decide which entanglement measure is the most appropriate given our aim and classify states accordingly. 

\subsection{2 qubits}

Choosing an entanglement measure for the 2-qubits case is an unambiguous task, as all possible measures behave equally as the entropy of entanglement \cite{Wooters98}. The maximally entangled states are therefore unambiguously characterized by the maximum entropy. One popular basis for those states is the Bell basis $\{\ket{\phi^+}, \ket{\phi^-}, \ket{\psi^+}, \ket{\psi^-}\}$ where
\begin{eqnarray}
&&\ket{\phi^\pm}=(\ket{00} \pm \ket{11}) / \sqrt{2}, \label{eq:bellstates} \\ \nonumber 
&&\ket{\psi^\pm}=(\ket{01} \pm \ket{10}) / \sqrt{2}.
\end{eqnarray}
These states can be transformed into each other by LU transformations. All of them have the maximum possible entropy of 1 ebit.

\subsection{3 qubits} \label{sec:3q}

In the 3-qubit case some difficulties already arise, although they are now considered completely solved, at least for pure states \cite{Dur00,Acin00}. Any state can be reduced to a canonical form where only six real parameters are kept, one of which is fixed by normalization
\begin{equation}
\ket{\psi} = \lambda_0 \ket{000} + \lambda_1 e^{i\phi} \ket{100} + \lambda_2 \ket{101} + \lambda_3 \ket{110} + \lambda_4 \ket{111}.
\end{equation}
There are therefore five independent entanglement invariants.  Convenient transformations can be chosen so that three of them are the three purities $Tr(\rho^2_i)$ with $i=\mathcal{A},\mathcal{B},\mathcal{C}$, another one the three-tangle from Eq. \ref{threetangle} and the last one the combination $Tr(\rho_\mathcal{A} \otimes \rho_\mathcal{B} \rho_\mathcal{AB})$.
Two fundamentally different types of entanglement exist, represented by two different SLOCC equivalence classes: one cannot go from a state of one class to one of the other by SLOCC transformations. The maximally entangled states of each class are
\begin{eqnarray}
&&\ket{GHZ}=(\ket{000} + \ket{111}) / \sqrt{2}, \label{eq:ghzw} \\ \nonumber 
&&\ket{W}=(\ket{001} + \ket{010} + \ket{100}) / \sqrt{3}.
\end{eqnarray}
$\ket{GHZ}$ state is maximally entangled in the sense that it maximizes the entropy of entanglement of all 3 possible bipartitions of the state at the same time, $S_\mathcal{A}=S_\mathcal{B}=S_\mathcal{C}=1$, and it is also the state which maximizes the three-tangle, $\rm{HDet_3}(\ket{GHZ}) =1$. Instead, $\ket{W}$ state is maximally entangled in the sense that it retains maximum bipartite entanglement when any one of the three qubits is traced out, whereas the three-tangle is 0 and the entropy of each partition is a non-maximum of 0.92 ebits.

Next section deals with the much richer 4-qubit case, where our research was focused.
 
\section{Entanglement of 4-qubit states}

The four-qubit case is much more complex. There is a natural growing of complexity with the growing number of parties, exemplified by the fact that one needs already 9 SLOCC inequivalent classes to completely classify all pure 4-qubit states \cite{Verstraete02}. But, more interestingly, it is the first instance where it is not possible to build a state that has maximally mixed reduced states in all its balanced bipartitions at the same time, as was proven analytically by Ref. \cite{Higuchi00}. We call states fulfilling this requirement \emph{absolutely maximally entangled states} (AMEs) \cite{Helwig12,Helwig13}, and study them extensively in next chapter. The impossibility of maximum entropy in all bipartitions comes from the incompatibility of the associated system of equations. This phenomenon has been called \emph{frustration}, and we will study a variant of it in Chapter 5. 

The proposal from Ref. \cite{Higuchi00} for a maximally entangled state is a state with the maximum allowed average entropy, and the same entropy in each bipartition, which we will call the Higuchi-Sudbery state $\ket{HS}$. Other sources have proposed other states as candidates for being maximally entangled in some sense. The following subsection is a selection of those proposals

\subsection{Candidates for maximally entangled states}

\begin{itemize}

\item Ref. \cite{Higuchi00} proposed $|HS\rangle$ state as the one with maximum average Von Neumann entropy in its bipartitions of 2 qubits.
\begin{equation}
|HS\rangle=\frac{1}{\sqrt{6}}\left(|0011\rangle + |1100\rangle + w \left(|0101\rangle + |1010\rangle \right) + w^2 \left(|0110\rangle + |1001\rangle \right)\right),
\label{higuchi}
\end{equation}
where $w=\rm{exp}(\frac{2i\pi}{3}).$\\
It has an entropy on each bipartition of 1.79 ebits. The theoretical maximum, according to Eq. \ref{eq:entropylambda}, would be $\log_2 N_\mathcal{A} = \log_2 4 = 2$ ebits, but as we said in the introduction to this chapter there is no state reaching that maximum.

\item Ref. \cite{Yeo06} presented the state $|YC\rangle$, with which they were able to show that it was possible to
perform a faithful teleportation of an arbitrary two-qubit entangled state, and they called it a genuine four-partite 
entangled state.
\begin{equation}
   \begin{split}
   |YC\rangle= &\frac{1}{\sqrt{8}}\left(|0000\rangle - |0011\rangle - |0101\rangle + |0110\rangle + |1001\rangle +
   |1010\rangle + |1100\rangle\right. \\
   & \left. + |1111\rangle\right).\\
   \end{split}
\end{equation}
\item The cluster states $|C_1\rangle$, $|C_2\rangle$ and $|C_3\rangle$, introduced in Ref. \cite{Briegel01}, are identified by Ref. \cite{Gour10} as the only states
that have the property that for 2, out of the 3 bipartite cuts, the Von Neumann entropy is 2 ebits and for the last bipartite cut it is 1 ebit. Their average of 1.66 ebits is therefore lower than $\ket{HS}$ state. In addition, they maximize the Renyi $\alpha$ entropy for $\alpha\geq 2$.
\begin{eqnarray}
&&|C_1\rangle=\frac{1}{2}\left(|0000\rangle + |0011\rangle + |1100\rangle - |1111\rangle\right),\\ \nonumber
&&|C_2\rangle=\frac{1}{2}\left(|0000\rangle + |0110\rangle + |1001\rangle - |1111\rangle\right),\\ \nonumber
&&|C_3\rangle=\frac{1}{2}\left(|0000\rangle + |0101\rangle + |1010\rangle - |1111\rangle\right).
\end{eqnarray}
\item Ref. \cite{Gour10} found $|L\rangle$ and $|M\rangle$ states while searching, respectively, for the states 
that maximize the average Tsallis $\alpha$ entropy of entanglement for $\alpha >2$ and for $0<\alpha <2$.
\begin{equation}
   \begin{split}
   |L\rangle =&\frac{1}{\sqrt{12}}\left(\left((1+w)(|0000\rangle + |1111\rangle\right)+(1-w)\left(|0011\rangle + 
   |1100\rangle\right)\right.\\
   &\left.+w^2\left(|0101\rangle + |0110\rangle + |1001\rangle + |1010\rangle\right)\right).\\
   \end{split}
   \label{statel}
\end{equation} 
where $w=\rm{exp}(\frac{2i\pi}{3}).$\\
\begin{equation}
   \begin{split}
   |M\rangle=&\frac{1}{\sqrt{2}}\left(\left(\frac{i}{2}+\frac{1}{\sqrt{12}}\right)\left(|0000\rangle + |1111\rangle\right) + \left(\frac{i}{2}-\frac{1}{\sqrt{12}}\right)\left(|0011\rangle + |1100\rangle\right)\right. + \\
&\left.\frac{1}{\sqrt{3}}\left(|0101\rangle + |1010\rangle\right)\right).
  \end{split}
\end{equation} 
\item In our work we found state $\ket{HD}$ as the one with maximum hyperdeterminant, as we will discuss in Sect. \ref{sec:maxhd}.
\begin{equation}
|HD\rangle=\frac{1}{\sqrt{6}}(|1000\rangle + |0100\rangle + |0010\rangle + |0001\rangle +\sqrt{2}|1111\rangle ).
\label{eq:hd}
\end{equation}
\end{itemize}

\subsection{The hyperdeterminant as an entanglement measure}

An issue with the entropy as an entanglement measure is that it explicitly depends on bipartitions. The same problem applies to the three-tangle for n>3, although of course it is also an interesting measure \cite{Osterloh16}. One could think that for the multipartite scenario it would be more interesting, at least for some tasks, to have an entanglement measure that is genuinely multipartite. Ref.\cite{Wong01} generalizes the 3-tangle to the n-tangle, and therefore to the 4-tangle, but it admits that, although being an entanglement monotone, it cannot be fully considered a multipartite entanglement measure. Monogamy inequalities that replicate Eq. \ref{monogamy3} for the four-qubit case have been found \cite{Regula16}, but the resulting residual tangle is not a SLOCC invariant and is thus not optimal as an entanglement measure.  One natural candidate for such a measure is the hyperdeterminant, a mathematical generalisation for higher-dimensional matrices of the determinant of a matrix \cite{Cayley45,Gelfand94}.
The hyperdeterminant as a possibly relevant quantity to measure 4-qubit entanglement was introduced in Ref. \cite{Miyake02}, and as explained in Sect. \ref{tang} it is a generalization of the determinant and the three-tangle, entanglement measures for two and three qubits respectively. A deep analysis of different polynomial invariants, with the hyperdeterminant as one of its combinations, and an application of these invariants to the classification in SLOCC classes was done in Ref. \cite{Luque03}.
 
We wanted to find the four-qubit state that maximizes the absolute value of hyperdeterminant. We believe that this state could accomplish certain information processing tasks that need genuinely multipartite entanglement that other maximally entangled states such as $\ket{HS}$ state from Eq. \ref{higuchi} state could not accomplish.
The hyperdeterminant of a 4-tensor A (${\rm Hdet_4(A)}$) can be 
constructed in three ways: by using the relationship between the concepts of determinant and discriminant of a polynomial, 
by extension of the concept of determinant to higher dimensions using tensor contractions, and by writing the state in a 
suitable basis, constructing some polynomial invariants and working the hyperdeterminant out of them. We describe the three procedures in detail now. \\

\subsubsection{Discriminant method}

The first process is described in Ref. \cite{Miyake02}, and is the same as the one to find the three-tangle, with an extra step. We begin by writing the determinant of an arbitrary 2x2 matrix $\alpha$ and identify it with the 2-hyperdeterminant
\begin{equation} 
\det (\alpha) = {\rm Hdet_2}(\alpha) = a_{00}a_{11}-a_{10}a_{01}.
\end{equation}
Next, we perform the substitution $a_{ij} \rightarrow (b_{ij0}+b_{ij1}x)$, where $ b_{ijk}$ is an arbitrary 3-tensor.
The discriminant $\Delta$ of the obtained polynomial $P_3(x)$ is the hyperdeterminant of this 3-tensor
\begin{eqnarray}
&&P_3(x) = {\rm Hdet_2} /. a_{ij} \rightarrow (b_{ij0}+b_{ij1}x),\\ \nonumber
&&{\rm Hdet_3}(\beta) = {\rm \Delta} [P_3(x)],
\end{eqnarray}
where /. means `substituting'.
Finally we extend the process to the fourth dimension, now using $c_{ijkl}$, the coefficients of the 4-tensor $\gamma$ from which we want to calculate the hyperdeterminant. We insert a $\frac{1}{256}$ factor to make the result coincide with the tensor contraction process from next section,
\begin{eqnarray} 
&&P_4(x) = {\rm Hdet_3} /. b_{ijk} \rightarrow (c_{ijk0}+c_{ijk1}x),\label{p4}\\ \nonumber
&&{\rm Hdet_4}(\gamma) = \frac{1}{256} {\rm \Delta} [P_4(x)].
\end{eqnarray}
From now on, to simplify the notation we identify the hyperdeterminant of the tensor ${\rm Hdet_4}(\gamma)$ with the hyperdeterminant of the state itself ${\rm Hdet_4}(\ket{\psi})$ or just ${\rm Hdet}(\ket{\psi})$.
\subsubsection{Tensor contraction method}

\begin{figure}[h!]
\centering
            \includegraphics[scale=0.3]{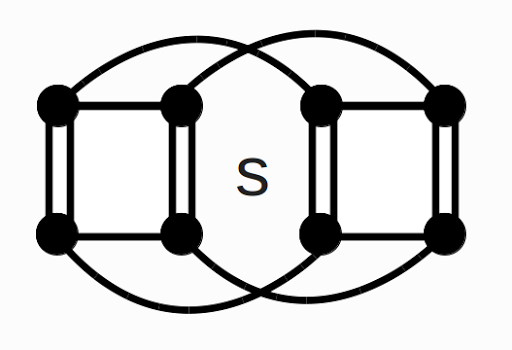}
             \includegraphics[scale=0.3]{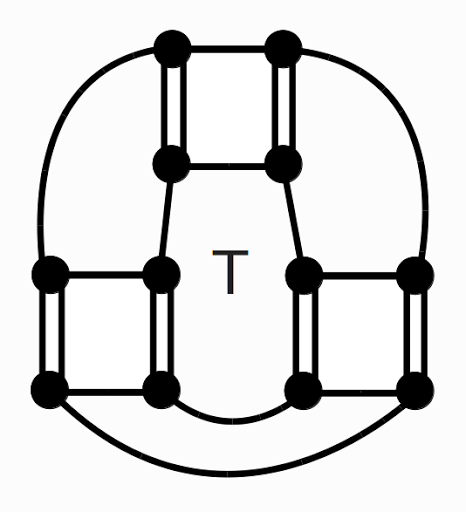}            
             \caption{Graphical representation of the contractions building S and T invariants. Each square represents a coefficient $c_{ijkl}$, and each line going from a point of a square to a point of another square represents an $\epsilon_{ij}$. Lines connecting points inside the same square corresponding to previous contractions to build the three-tangle of Fig. \ref{fig:grafictangle}.}
\label{fig:stgrafic}     
\end{figure}

The second method consists on finding the $\rm{Hdet_4}$ from tensor contractions. We obtain in that way two independent polynomial invariants: S and T. Its expressions in terms of some coefficients $b_{ijkl}$ to be defined later are the following
\begin{eqnarray}
&&S=\frac{1}{2}\epsilon^{i_1j_1}\epsilon^{i_2j_2}\epsilon^{i_3j_3}\epsilon^{i_4j_4}b_{i_1i_2i_3i_4}b_{j_1j_2j_3j_4},\label{eq:st} \\ \nonumber
&&T=\frac{1}{6}\epsilon^{i_3j_1}\epsilon^{i_4j_2}\epsilon^{j_3k_1}\epsilon^{j_4k_2}\epsilon^{k_3i_1}\epsilon^{k_4i_2}
b_{i_1i_2i_3i_4}b_{j_1j_2j_3j_4}b_{k_1k_2k_3k_4}.
\end{eqnarray}
Figure \ref{fig:stgrafic} gives a graphical representation of these contractions.

Now we perform the substitution $$b_{ijkl} \rightarrow b_{n}; \hspace{10mm}i,j,k,l = 0,1 ; \hspace{10mm} n = 0,1,2,3,4.$$
where n is the number of "1" in the set $\{ijkl\}$. The result is
\begin{eqnarray}
&&S=3{b_2}^2 - 4b_1b_3 + b_0b_4,\label{invs}\\ 
&&T=-{b_2}^3 + 2b_1b_2b_3 - b_0{b_3}^2 - {b_1}^2b_4 + b_0b_2b_4. \label{invt}
\end{eqnarray}
Now, if we write $P_4(x)$ (\ref{p4}) in the form
\begin{equation}
P_4(x) = x^4b_0 + 4x^3b_1 + 6x^2b_2 +4xb_3 +b_4,
\end{equation}
we can obtain the values $b_i$ and insert them into the expressions for S and T. The hyperdeterminant is obtained by the following combination
\begin{equation}
\rm{Hdet_4(\ket{\psi})} = S^3-27T^2,
\label{eq:st}
\end{equation}
which is the one that coincides with the discriminant method.

This way of computing the hyperdeterminant brings connections to the theory of elliptic curves \cite{Gibbs10}. There is a function of a complex variable $\tau$ called \emph{modular discriminant} which is defined as $\Delta(\tau)=g_2^3(\tau)-27g_3^2(\tau)$ with $g_2$ and $g_3$ being invariants of the theory, which is an obvious parallelism of Eq. \ref{eq:st}. The modular discriminant is related to Dedekind eta function which is a modular form $\rm{\Delta}(\tau) = (\eta (\tau))^{24}$. Since the hyperdeterminant is a polynomial of degree 24, this links its degree to the presence of number 24 in elliptic curves, which in turn is related to the number of dimensions in bosonic string theory \cite{Baez08}.

\subsubsection{Bell basis method}

The third way is based on the classification of four-qubit states explained by Ref. \cite{Verstraete02}. It divides all states in nine different classes, within which any state can be transformed into each other by SLOCC operations. Ref. \cite{Luque03} proves that eight of these nine classes have zero hyperdeterminant, so we don't need to worry about them. The remaining class, called the \emph{generic} class, is presented in the following compact way in Ref. \cite{Gour10}
\begin{equation}
\mathcal{A} \equiv \lbrace z_0u_0+z_1u_1+z_2u_2+z_3u_3 \mid z_0,z_1,z_2,z_3 \in \mathbb{C} \rbrace ,
\end{equation}
where $$ u_0 \equiv |\phi^+\rangle|\phi^+\rangle , \hspace{2mm} u_1 \equiv |\phi^-\rangle|\phi^-\rangle, $$
         \vspace {-4mm}
      $$ u_2 \equiv |\psi^+\rangle|\psi^+\rangle , \hspace{2mm} u_3 \equiv |\psi^-\rangle|\psi^-\rangle. $$
and $|\phi^\pm\rangle|$ and $|\psi^\pm\rangle|$ 
are the Bell states defined in Eq. \ref{eq:bellstates}.\\

The method to transform a state from the computational basis into the Bell basis (which means obtaining the values of the 
$z_i$) is explained in detail in Ref. \cite{Verstraete02}. With the state in this Bell basis, Ref. \cite{Luque03} presents a very simple way of obtaining the hyperdeterminant
\begin{equation}
{\rm Hdet_4}(\ket{\psi}) = \frac{1}{256}V(z_0^2,z_1^2,z_2^2,z_3^2)^2, 
\end{equation}
where V denotes the Vandermonde determinant. We can write it explicitly as
\begin{equation}
{\rm Hdet_4}(\ket{\psi}) = \frac{1}{256}\prod_{0\leq i<j \leq 3}(z_j^2-z_i^2)^2. 
\end{equation}

Before presenting our results on the hyperdeterminant, we devote next section to a reduction of the Hilbert space in order to simplify numerical searches.

\subsection{Canonical form of 4-qubit states}

Ref. \cite{Acin00} provided a reduction of the Hilbert space of states of 3 qubits to only 5 real coefficients out of the total of 16, and called it the canonical form. We want to do the same with the 4-qubit space as it simplifies searches and allows us to have a reference form out of all possible equivalent forms for writing a state.
The standard form of 4-qubit states has 16 complex coefficients (32 real). Using the various possible changes of basis (LU transformations, see Sect. \ref{sec:LU}) to eliminate the maximum number of coefficients, we can reduce them to 19 real numbers
different from zero, one of which is fixed by normalization, so we have 18 degrees of freedom. We describe now the procedure.

We first parametrize our state as a tensor with 4 coefficients
\begin{equation}
\ket{\psi} = \sum_{i,j,k,l=0}^1 t_{ijkl} |ijkl\rangle.
\end{equation}
Now we create two 3-tensors, one named $T_0$ with the coefficients with i=0 and the other named $T_1$ with the coefficients with i=1
\begin{equation}
(T_i)_{jkl} = t_{ijkl}.
\end{equation}
We now make a unitary transformation on the first qubit
\begin{equation}
T'_i=\sum_{j}u_{ij}T_j,
\end{equation}
such that the three-tangle of $T'_0$ equals 0
\begin{equation}
{\rm Hdet_3}\hspace{1mm} T'_0=0. \label{tangle0}
\end{equation}
Now we define four matrices in the same way as we did before
\begin{equation}
(T_{ij})_{kl} = t_{ijkl},
\end{equation}
and with a unitary transformation on the second qubit we impose
\begin{equation}
{\det} \hspace{1mm} T'_{00}=0. \label{det0}
\end{equation}
Now we diagonalize $T'_{00}$, using our last two available unitary transformations. Due to Eq. \ref{det0} the resulting matrix will have at least three zeros. At the same time, due to  Eq. \ref{tangle0} $T''_{01}$ (the matrix resulting from the transformation of $T'_{01}$) will have at least one zero. Thus, we have reduced the initial 16 coefficients to only 12.\\

We can further simplify our form by absorbing four phases into the "0" of each of the four qubits, and then eliminating the phase of $|0000\rangle$ using global phase invariance. Thus only seven relevant phases remain and our canonical form reads:

\begin{equation}
	\begin{split}
		|\phi\rangle &=c_0|0000\rangle +c_1e^{i\theta_1}|0100\rangle +c_2e^{i\theta_2}|0101\rangle +c_3e^{i\theta_3}
		|0110\rangle \\
		&+c_4e^{i\theta_4}|1000\rangle +c_5e^{i\theta_5}|1001\rangle +c_6e^{i\theta_6}|1010\rangle +c_7|1011\rangle \\
		&+c_8e^{i\theta_8}|1100\rangle +c_9|1101\rangle +c_{10}|1110\rangle +c_{11}|1111\rangle,\\
		&c_i \geq 0 , \hspace{3mm} 0 \leq \theta \leq 2\pi , \hspace{3mm} \mu_{i} \equiv c_i^2 , \hspace{3mm} \sum_i \mu_i = 1.
	\end{split}
\end{equation}
Ref. \cite{Higuchi00} proposed a different canonical form where they imposed
$t_{1000}=t_{0100}=t_{0010}=t_{0001}=0$.\\

\subsection{Results for maximum hyperdeterminant} \label{sec:maxhd}

Our search produced the state from Eq. \ref{eq:hd} as the one with maximum hyperdeterminant. It was found by maximizing the Hdet numerically in an exhaustive search algorithm. Afterwards a genetic
algorithm found the same result.  The variables were limited to the canonical forms, to speed up the process. As a safe
check, a genetic algorithm was run on a generic state with all 16 coefficients, also returning the same results. We also 
searched the maximum Hdet with a genetic algorithm varying the 4 coefficients of the Bell basis, producing once again the 
same result. Ultimately we found that the state had already been discovered previously in Ref. \cite{Osterloh05}, and after our work an analytical proof of it being the maximum hyperdeterminant state appeared \cite{Gour14,Chen13}.

It is written here in the canonical form, and can be seen to have a very simple expression. It is fully symmetric in all its parts, which means that all bipartite entanglement measures like the various entropies will be the same for all partitions.

It has the interesting property that it can be written as a simple linear combination of maximum spin states
\begin{equation}
|HD\rangle = \frac{\sqrt{6}}{2}\left(\sqrt{2} ~ |2,2\rangle + |2,-1\rangle \right),
\label{hdspin}
\end{equation}
where we have used the notation $|j,m\rangle$.

Remarkably, state $|L\rangle$ from Eq. \ref{statel}, found by Ref. \cite{Gour10} to be the state with maximum average Tsallis entropy for $\alpha>2$, has the same Hdet, while state $\ket{HD}$ has the same Tsallis entropy as $\ket{L}$. In fact when writing state $\ket{L}$ in the canonical form it turns out that their real coefficients are the same as $\ket{HD}$, while their phases differ. These states also maximize the invariant T from Eq.\ref{invt}, while the invariant S from Eq. \ref{invs} equals zero.  We have checked that inserting arbitrary relative phases in the state $|HD\rangle$ does not produce changes in the result of the Hdet, but the opposite happens if we insert them in the state $|L\rangle$. So in $|HD\rangle$ only absolute values matter, which is in accord to the fact that $|L\rangle$ has the same modules but different phases as $|HD\rangle$ in the canonical form, while $|L\rangle$ appears to have finely-tuned phases to produce the maximum Hdet. In the Bell basis, they have exactly the same expression.

The invariant S is maximized by a variety of states, for example the $|GHZ\rangle$ of 4 parties. But all of them happen to 
have the precise value of T to have zero Hdet.

The following table summarizes the results for the states that we have already presented. $\ket{GHZ}$ and $\ket{W}$ refer to the 4-party version of states from Eq. \ref{eq:ghzw}. If the states have complex coefficients, the exact
results of the hyperdeterminant and the various approximations are given in absolute values.\\\\

\begin{tabular}{ | c | c  c  c  c  c  c|}
    \hline
      State & S & S($\cdot10^{3}$) & T & T($\cdot10^{5}$)& Hdet & Hdet($\cdot10^{7}$) \\ \hline
    $\ket{HD}$ & 0 & 0 & $-\frac{1}{2^4\cdot3^6}$ & $-8.57$
& $\frac{1}{2^8\cdot3^9}$ & $1.98$ \\
    $\ket{L}$ & 0 & 0 & $-\frac{1}{2^4\cdot3^6}$ & $-8.57$
& $\frac{1}{2^8\cdot3^9}$ & $1.98$ \\
    $\ket{GHZ}$ & $\frac{1}{2^6\cdot3}$ & $5.21$ 
& $-\frac{1}{2^9\cdot3^3}$ & $-7.23$ & 0 & 0 \\
    $\ket{C_i}$ & $\frac{1}{2^6\cdot3}$ & $5.21$ 
& $-\frac{1}{2^9\cdot3^3}$ & $-7.23$ & 0 & 0 \\
    $\ket{YC}$ & $\frac{1}{2^6\cdot3}$ & $5.21$ 
& $\frac{1}{2^9\cdot3^3}$ & $7.23$ & 0 & 0\\
    $\ket{W}$ & 0 & 0 & 0 & 0 & 0 & 0\\
    $\ket{HS}$ & 0 & 0 & 0 & 0 & 0 & 0 \\
    $\ket{M}$ & 0 & 0 & 0 & 0 & 0 & 0 \\ \hline 

\end{tabular}
\begin{center}\textbf {Table 1:} Exact and approximated results of the hyperdeterminant of our special
states. The various numbers are the two invariants S and T and the total hyperdeterminant. \end{center}

\subsection{Results for random states}

\begin{figure}[h!]
\centering
            \includegraphics[scale=0.5]{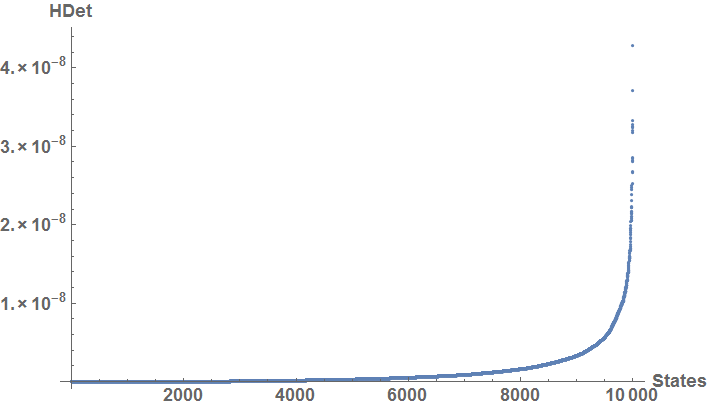}
            \caption{Value of the hyperdeterminant for 10 000 random states, ordered here
according to their Hdet. We can see that more than half of them have a value not greater than $10^{-9}$, and around 5\% of 
them have values of order $10^{-8}$. The mean value is $1.32 \hspace{1mm} 10^{-9}$. A similar study was done in Ref. \cite{Enriquez15}}.
\label{fig:grafichd}      
\end{figure}

\begin{figure}[h!]
\centering
            \includegraphics[scale=0.5]{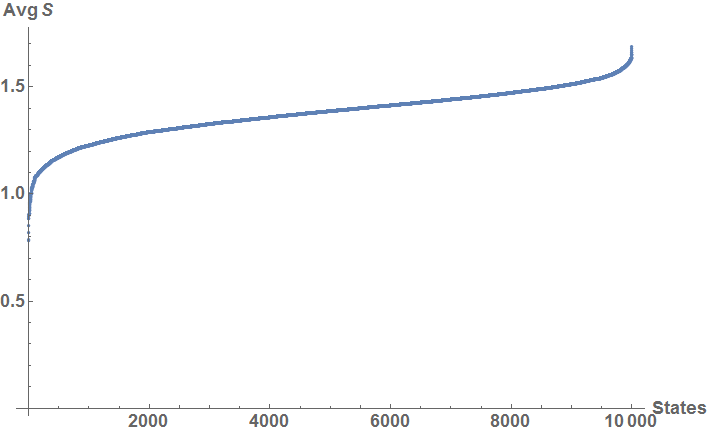}
             \caption{Value of the average Von Neumann entropy for 10 000 random states, ordered here
according to their entropy. We can see that around 70\% of them have a value between 1.3 and 1.5 ebits.  The mean value is
1.38 ebits.}
\label{fig:graficent}     
\end{figure}

\begin{figure}[h!]
\centering
            \includegraphics[scale=0.5]{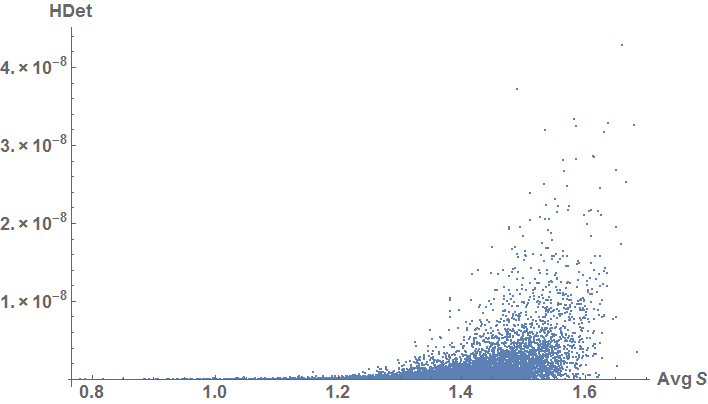}                        
            \caption{The X-axis shows the value of the average Von Neumann entropy and the Y-axis the 
value of the hyperdeterminant of the previous 10 000 random states. We can see that there is a correlation up to a certain 
point in one direction: all the states with high hyperdeterminant are also states with high entropy, but the converse appears not to be true.}
\label{fig:corrhdetent}
\end{figure}

We have tried to obtain a picture of the typical entanglement obtained when looking at random states. We have generated 10,000 random states and have calculated their hyperdeterminant and their average Von Neumann entropy, to see which are their typical values and if there is some correlation between them. Figs. \ref{fig:grafichd}, \ref{fig:graficent} and \ref{fig:corrhdetent} show the distribution of HDet and entropy values of the random states.

Fig. \ref{fig:grafichd} shows us that most states group themselves around a hyperdeterminant two orders of magnitude below our maximum $\ket{HD}$ state, which has a value of $1.98 \cdot 10^{-7}$. Having a value of the Hdet close to the maximum is very rare. This is not the case for the entropy: Fig. \ref{fig:graficent} tells us that most states concentrate on a region of relatively high entanglement between 1.3 and 1.5 ebits, which is in the same order of magnitude as the maximum of 1.79 ebits. So the traditional assertion that random states are highly entangled appears to be true for entropy but not for Hdet.
Fig. \ref{fig:corrhdetent} shows us a certain correlation in the sense that all states with high hyperdeterminant are also states with high entropy. For example, our state with maximum Hdet, the $|HD\rangle$ state, with its average entropy of 1.58 ebits, would be among the highest 3\% within the 10,000 random states. However, the converse appears not to be true: there are many states with high entropy and low hyperdeterminant, as we can see both in the graphic and in our special 
states: the cluster states and $|HS\rangle$ and $|M\rangle$ states have high entropy but zero hyperdeterminant. The statistical correlation between hyperdeterminant and entropy is slightly below 0.5.

\begin{figure}[h!]
\centering
            \includegraphics[scale=0.5]{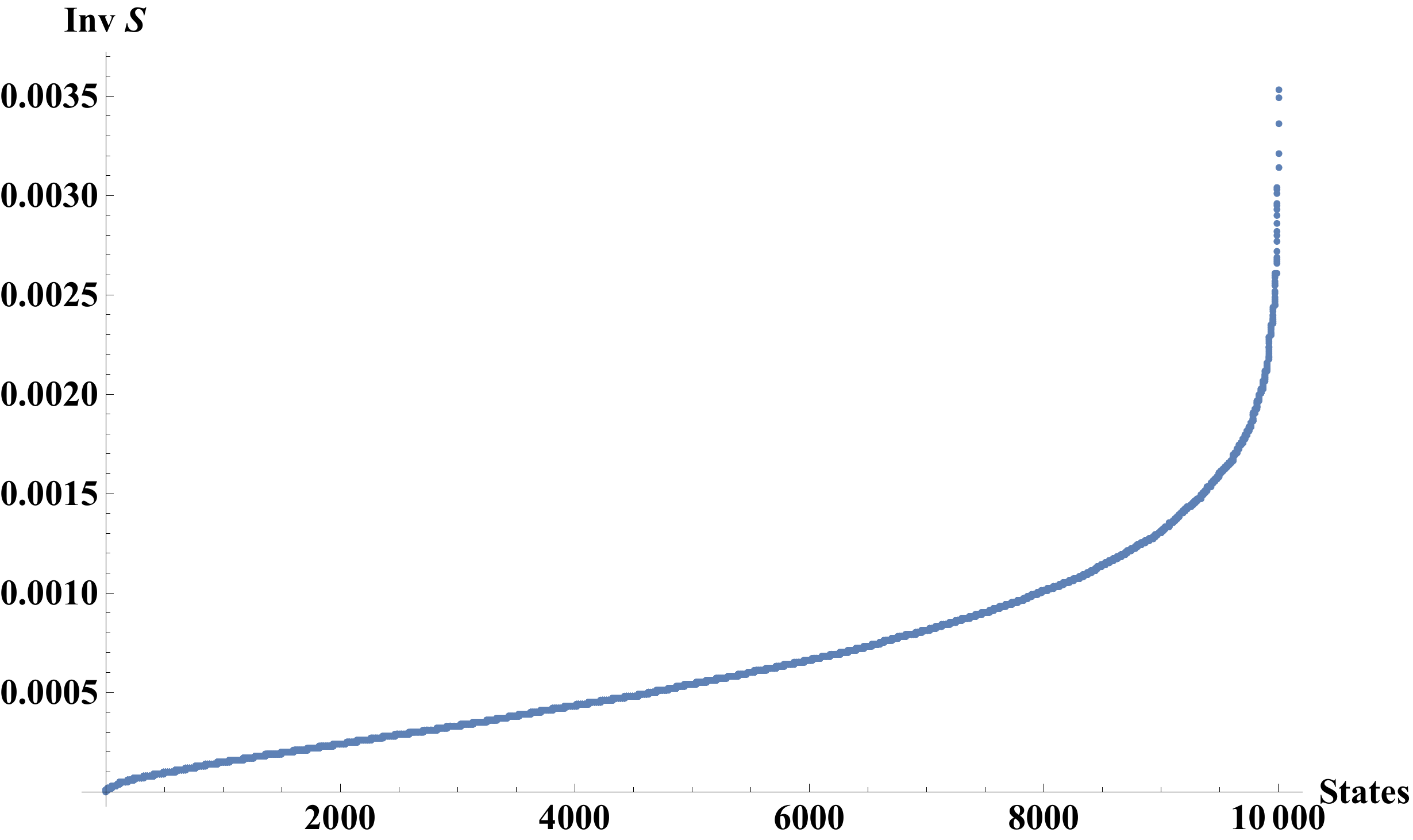}
            \caption{Value of the S invariant for 10 000 random states, ordered here
according to their S. We can see that more than half of them have a value not greater than $10^{-3}$, and none of them has a value higher than those of our special states. The mean value is $6.5~10^{-4}$.}
\label{fig:randoms}      
\end{figure}

\begin{figure}[h!]
\centering
            \includegraphics[scale=0.5]{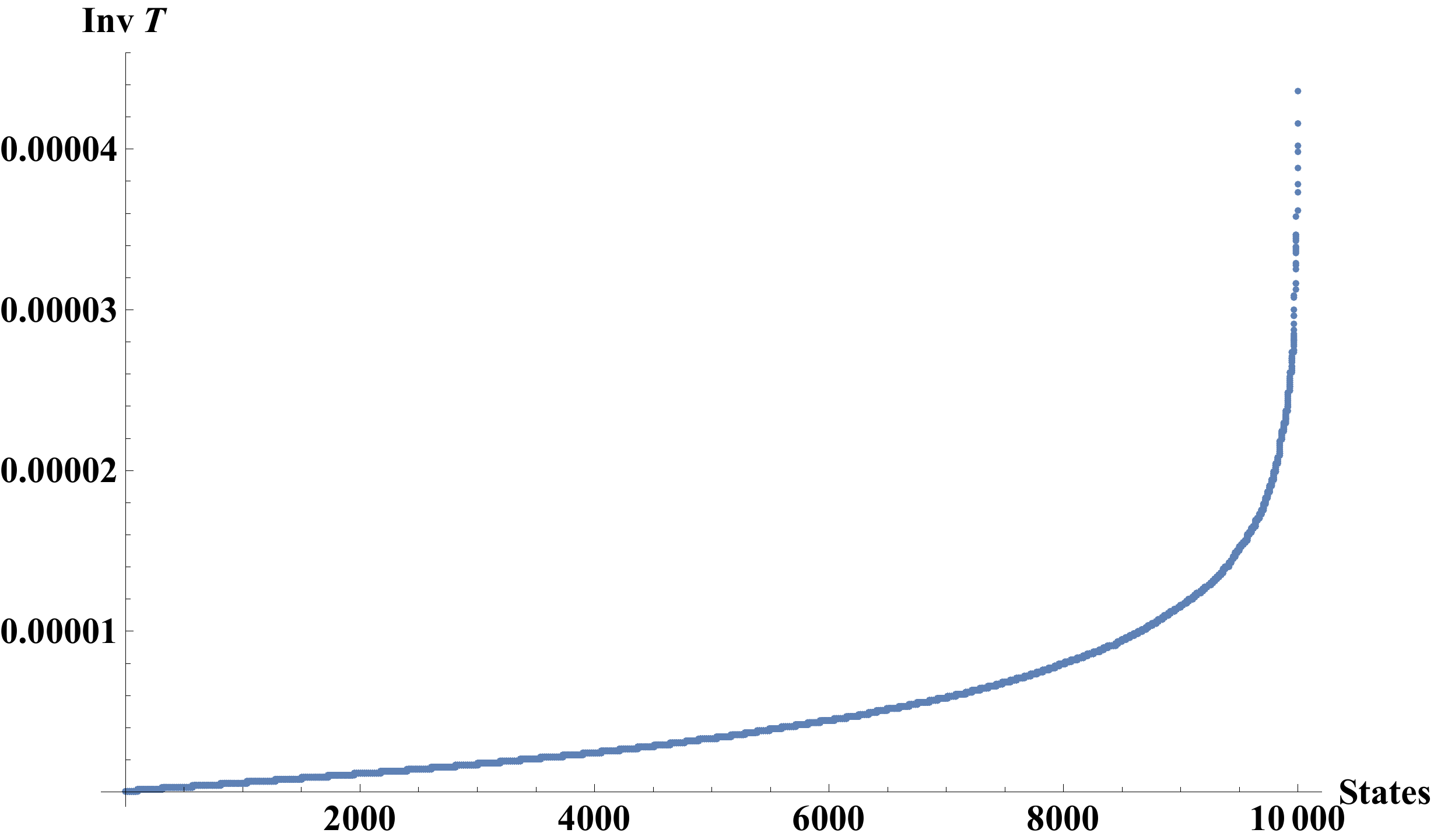}
            \caption{Value of the T invariant for 10 000 random states, ordered here
according to their T. We can see that more than 80\% of them have a value not greater than $10^{-5}$, and none of them has a value higher than those of our special states. The mean value is $5.0~10^{-6}$.}
\label{fig:randomt}      
\end{figure}

\begin{figure}[h!]
\centering
            \includegraphics[scale=0.5]{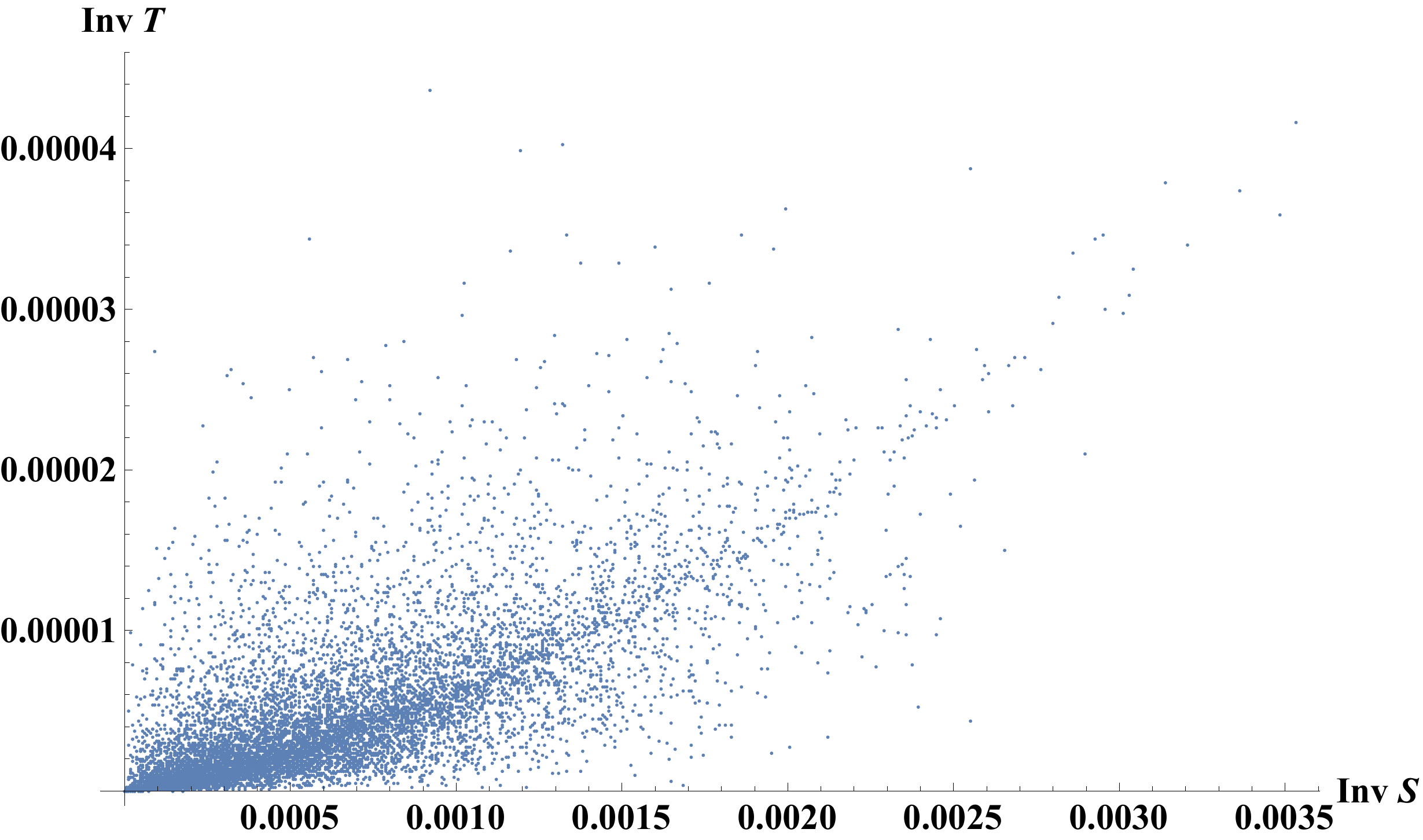}
            \caption{A quite high correlation between values of the S and T invariants, in accord with the fact that most states have close to zero hyperdeterminant, with the invariants cancelling out.}
\label{fig:corrst}      
\end{figure}

We have also plotted results for the S and T invariants in Figs. \ref{fig:randoms} and \ref{fig:randomt}. One can see that all of our special states have values for those invariants higher than any of the random states, except for those that have a value of zero. So, for example, $\ket{GHZ}$ has a very high value of both S and T, while $\ket{HD}$ has an even higher value of T. Fig. \ref{fig:corrst} shows an important correlation between the values of S and T, of 0.7, which is an accord to the fact that most states have a very small hyperdeterminant, as S and T tend to cancel out. Naturally the $\ket{HD}$ state is an exception.

\section{Summary}

We have presented the topic of multipartite entanglement and have showed that different entanglement measures give different maximally entangled states, and that one has to decide which measure to use according to the particular situation. 

The focus has been put on the states of four qubits, as it is the simplest of the yet not so well understood situations, and a specially interesting case because it is the first one where it is not possible to attain the maximum theoretical entropy in all its bipartitions at the same time.

The hyperdeterminant is a good candidate for a measure of genuine multipartite entanglement for various reasons: it is a mathematical extension of the three-tangle, it captures in a single real number information from the whole system and it is built from the two only global polynomial invariants S and T, see Eq. \ref{eq:st}.

We have found the state that maximizes the hyperdeterminant, which is remarkably simple and symmetric, see Eq. \ref{eq:hd}. It could be interesting to find a parent hamiltonian for this state and a Bell inequality maximally violated by it, to better understand in which tasks it could provide an advantage over others.

We found that most random states have an hyperdeterminant close to 0, and that there is a certain correlation between hyperdeterminant and average Von Neumann entropy in that states with high Hdet tend to also have high entropy, although the opposite is not true.

A canonical form for states of four qubits has also been presented, using all possible local transformations to reduce the number of real coefficients from 32 to 18, which gives the true number of entanglement parameters, two of them being the global invariants S and T, from which the hyperdeterminant can be readily obtained.

\chapter{Absolutely maximally entangled states}

\pagestyle{fancy}
\fancyhf{}
\fancyhead[LE]{\thepage}
\fancyhead[RE]{MULTIPARTITE ENTANGLEMENT}
\fancyhead[LO]{CH.2 Absolutely maximally entangled states}
\fancyhead[RO]{\thepage}

After introducing multipartite entanglement and studying the interesting case of four qubits in the first chapter, here we turn our attention to a special kind of states that have been recently called \emph{absolutely maximally entangled states} (AMEs) in an attempt to establish them as unambiguously maximally entangled. We consider now general \emph{qudit} d-dimensional systems and do not focus exclusively on qubits. We define AMEs and present their relationships with different fields: holography, classical codes, combinatorial designs and a new concept called \emph{multiunitary matrices}. 

\section{AME States: Definition and basic properties and examples}

\subsection{Definition}
%AME
A lot of attention was recently paid to identify entangled states of $N$--party systems,
such that after tracing out arbitrary $N-k$  subsystems, the remaining $k$ subsystems are maximally mixed \cite{Gisin98,Higuchi00,Brown05,Facchi08,Facchi10b,Arnaud13}.
Such states are often called $k$--uniform \cite{Scott04,Goyeneche14}
and by construction the integer number $k$ cannot exceed $N/2$.
In this chapter, we shall focus on the extremal case, $k=\floor{N/2}$ (we put the floor function $\floor{}$ to include cases of N even and odd) which defines an AME (see Refs. \cite{Helwig12,Helwig13}). Such states had been previously known as perfect maximally multipartite entangled states \cite{Facchi09}.

The definition of AME states corresponds to those quantum states that carry maximum entropy in {\sl all} their bipartitions.
 It is a remarkable fact that the existence of  
such states is not at all trivial and deepens into several branches of mathematics. Let us be more precise and  define an AME($N$,$d$) state
$|\psi\rangle \in {\cal H}$, made with $N$ qudits of local dimension $d$, 
 ${\cal H}=(C^d)^{\otimes N}$ as a state such that all its reduced density matrices in any subspace
${\cal A}=(C^d)^{\otimes \frac{N}{2}}$, ${\cal H}={\cal A} \otimes \bar{\cal A}$,
carry maximal entropy
\begin{equation}
  S(\rho_{\cal A})=\frac{N}{2} \log d \qquad \forall {\cal A}  \ .
\end{equation}
This is tantamount to asking the reduced density matrices to $k$ qudits to be
proportional to the identity
\begin{equation}
\rho_k=\frac{1}{d^k} \mathbb{I}_{d^k} \qquad \forall k\leq \frac{N}{2}\ .
\end{equation}
which matches the maximally mixed expression from Eq. \ref{eq:maxmixed}.
Let us note the fact that a $k$--uniform state is also $k^{\prime}$--uniform for any $0<k^{\prime}<k$.

%monogamy
There is an obstruction for a state to reach maximal entanglement in all partitions
due to the concept of monogamy of entanglement \cite{Coffman00,Terhal03}. Every local degree of freedom
that tries to get maximally entangled with another one is, then, forced to disentangle
from any third party. Therefore, entanglement can be seen
as a resource to be shared with other parties. If two local degrees of freedom get largely entangled among themselves, 
then they are less able to be entangled with the rest of the system. We have already seen in chapter 1 that this obstruction prevents the existence of an AME state for the case of four qubits. But this rule is not always fulfilled. There are
cases where the values of the local dimensions  $d$ and the total number of
qudits $N$ are such that AME states exist. 
For a given $N$, there is always a large enough $d$ 
for which there exists an AME state \cite{Helwig12}. 
However, the lowest value of $d$ such that an AME state exists is not known in general.

% relation AME to multipartite protocols
Let us mention that AME states are 
 useful and necessary to accomplish 
certain classes of multipartite protocols. In particular, in Ref. \cite{Helwig12},
it was shown that AME states are needed to implement two different categories of 
protocols. First, they are needed to achieve perfect multipartite
teleportation. Second, they provide the
resource needed for quantum secret sharing. 
These connections hint at further
relations between AME states and different branches of Mathematics.
For instance, AME states are related to Reed-Solomon codes \cite{Reed60}; 
Also, AME states (and $k$--uniform states in general) are deeply linked to error correction codes \cite{Scott04}. We develop these relations in the following sections.

%AME and holography
There is yet another surprising connection between AME states and holography \cite{Latorre15,Pastawski15}. It can be seen that AME states provide the basis for a tensor network structure that distributes entanglement in a most efficient and isotropic way. This tensor network can be proved to deliver holographic codes, that may be useful as quantum memories and as microscopic models for quantum gravity. These new developments are related to the
property of multiunitarity that will be explored in Sect. \ref{sec:multiunitarity}.

\subsection{Local Unitary equivalence}

As we said in chapter 1, entanglement is invariant under choices of local basis. It is then
natural to introduce the concept of Local Unitary (LU) equivalence among AME states, in the spirit of Sect. \ref{sec:LU}.
If $\ket{\Psi}$ is an AME, any other state 
LU-equivalent to it is also an AME. %Thus, if in a Hilbert space ${\mathcal H}(N,d)=(\mathbb{C}^d)^{\otimes N}$ there exists an AME state,
%this means that there are an infinite number of them.
In this respect, we will define AME($N,d$) as the set of all AME states in the
Hilbert space ${\mathcal H}(N,d)$  and denote their elements by a Greek letter,
e.~g.\ $\ket{\Omega_{4,3}}\in \textrm{AME}(4,3)$ is an AME of four qutrits.

The LU transformations introduce equivalence classes of states. 
A question arises naturally about which state should be chosen as the representative of the class,
that will be denoted as \emph{canonical form} of an AME state. 
It is possible to 
argue in two different directions. On one hand, we may consider that a
natural representative may carry all the elements of the computational
basis. It would then be necessary to establish theorems and a criterion
to fix the coefficients. On the other hand, an alternative possibility is to choose the element of the class with a minimal support on the computational basis. Results in both directions are presented in Sect. \ref{sec:maximalsup} and Sect. \ref{sec:codes}.

It is not known in general how many different LU classes there are in the set AME($N,d$) for every $N$ and $d$. This question can be tackled by the construction of LU-invariants. 
A few examples are at hand for few qubits. For three qubits, it is known how to obtain
a canonical form of any state using LU and that all
states are classified by 5 invariants as we showed in Sect. \ref{sec:3q}, only one of them
is genuinely multipartite, the tangle. For four qubits we presented a full derivation of the canonical form and its invariants in the previous chapter.
Yet, it is unknown how to proceed to larger local dimensions and number
of parties. It is arguable that the subset of AME states is
characterized by several LU-invariants, probably related to distinct
physical tasks. In such a case, there would be different AME states not
related to each other by LU.  

\subsection{AME and holography}

Quantum holography amounts to the fact that the information content of a quantum system is that of its boundary. It follows that the information present in the system is far less than the maximum allowed. Degrees of freedom in the bulk will not carry maximal correlations, neither the
von Neumann entropy of any sub-part of the system will scale
as its volume.

To gain insight in quantum holography, it is natural to investigate the bulk/boundary correspondence of the operator content of the theory. On the other
hand, Quantum Information brings a new point of view on this issue, since
it focuses on the properties of states rather than on the dynamics
that generates it. In this novel context, we may ask what is
the structure of quantum states that display holographic properties.
That is, we aim at finding which
is the detailed entanglement scaffolding that guarantees that
information flows from
the boundary to the bulk of a system in a perfect way.

A concept separate from holography turns out to be very useful to address the
analysis of holography from this new Quantum Information perspective, that is Tensor Networks of the kind of
matrix product states (MPS), projected entangled pair states (PEPS) 
and multi-scale entanglement renormalization ansatz (MERA). 
 Indeed, Tensor Networks
provide a frame to analyze how correlations get
 distributed in quantum states, and thus to understand holography
at the level of quantum states. Each connection among ancillary indices
quantifies the amount of entanglement which links parts of the system.
Holography must necessarily relay on some very peculiar entanglement
structure. We talk about tensor networks in Chapter 5.

A first attempt to understand the basic property behind holography
of quantum states was presented in Ref. \cite{Latorre15}. There, it was proposed to create a quantum state on a triangular lattice based on
a tensor network that uses as ancillary states an absolutely maximally entangled state. To be precise, the state $\ket{\Omega} \in AME(4,3)$ (see Eq. \ref{eq:ame43}) was defined on tetrahedrons, in such a
way that the vertices in its basis connect the tensor network and the tip
of the tetrahedron corresponds to a physical index.

In a subsequent work \cite{Pastawski15}, another construction was based on
the 5-qubit $\ket{\Upsilon_{5,2}} \in AME(5,2)$ state, see Eq.  \ref{eq:ame52}. Again, the fact that the internal construction
of the state is based on isometries is at the heart of the holographic property.

There are two obvious observations on the surprising relation between
AME states and holography. The first is related to the natural link between
AME states and error correction codes, that we will develop in Sect. \ref{sec:codes}. It is arguable, then, that
the essence of holography is error correction, which limits the amount
of information in the system. The second is that the very property
responsible for holography is multiunitarity, which is analyzed in depth in section \ref{sec:multiunitarity}. It is further arguable
that multiunitarity is the building block of symmetries, since the
sense of direction is lost and can be defined at will. Those ideas deserve
a much deeper analysis.

\subsection{Related definitions}

\subsubsection{Maximally Entangled sets}

As we said throughout Chapter 1, multipartite entanglement is significantly different from the bipartite one.
While in the bipartite case, there is a single maximally entangled state (up to local unitaries) that can be transformed into any other state by LOCC (see Sect. \ref{sec:locc}) and cannot be obtained from any other,
in the multipartite scenario this is no longer true.
In Ref.~\cite{Vicente13}, the notion of the Maximally Entangled (ME) set of $N$-partite states is introduced as the set of states from which any state outside of it can be obtained via LOCC from one of the states within the set and no state in the set can be obtained from any other state via LOCC. 
Note that this notion of maximal entanglement is strictly weaker than the AME, in the sense that
most of (or all) states in the ME set will not be an AME state,
 but any AME state will be in its corresponding ME set. 
In Refs.~\cite{Vicente13,Spee16}, the ME set is characterized for the cases of three and four qubits, and in Ref. \cite{Hebenstreit16} it is characterized for generic states of three qutrits.
It is interesting to point out that, unlike the 3-qubit case, deterministic LOCC transformations are almost never possible among fully entangled 4-partite states. 
As a consequence of this,
while the ME set is of measure zero for 3-qubit states, almost all states are in the 4-qubit ME set, even when allowing finitely many rounds of communication \cite{Spee17}.
This suggests the following picture; given a fixed local dimension and for an increasing number of parties, 
the AME states become more an more rare at the same time that more and more states need to be included
in the ME set. In other words, while maximally entangled states defined from an operational point of view become typical when the number of parties increases, AME states are exotic.

\subsubsection{Maximally Multipartite Entangled states}

In Ref.~\cite{Facchi08}, \emph{Maximally Multipartite Entangled} (MME) states are
introduced as those states that maximize the average entanglement (measured in terms of purity, see Sect. \ref{sec:purity}) 
where the average is taken over all the balanced bipartitions i.~e.~ $|\mathcal{A}|=\floor{N/2}$.
More specifically, the MME states are defined as the minimizers of the \emph{potential of multipartite entanglement}
\begin{equation}
\pi_{ME}=\binom{N}{\floor{N/2}} \sum_{|\mathcal{A}|=\floor{N/2}}\pi_\mathcal{A}\, ,
\end{equation}
where $\pi_\mathcal{A}=\Tr(\rho_\mathcal{A}^2)$ is the purity of the partition $\mathcal{A}$.
Note that the above potential is bounded by $1/d^{\floor{N/2}}\le \pi_{ME}\le 1$ and 
its lower bound is only saturated by AME states.

By minimizing the multipartite entanglement potential, explicit examples of AME states of 5 and 6 qubits are presented in Ref.~\cite{Facchi08}. It is remarkable that even for a relatively
small number of qubits ($N\ge 7$), such minimization problem
has a landscape of
the parameter space with a large number
of local minima, what implies a very slow convergence.
The reason for that is frustration. The condition that purity saturates its minimum
can be satisfied for some but not for all the bipartitions (see \cite{Facchi10b} for details).
In this respect, in Ref. \cite{Facchi10a}, the minimization of the multipartite entanglement potential is mapped
into a classical statistical mechanics problem.
The multipartite entanglement potential is seen there as a Hamiltonian which is minimized by
simulated annealing techniques.

\subsection{Qubit AMEs}

Let us consider states made out of qubits.
The simplest cases of AME states are any of the Bell states from Eq. \ref{eq:bellstates}. There is a unique partition of two qubits and it is possible to entangle both parties maximally. 
As we already said in that space there is a unique quantity that describes the amount of entanglement in the system. All two particle states, whatever their local dimension, can be entangled maximally. Those states are of no interest to our present discussion which is genuinely centered in multipartite entanglement.

In the case of 3 qubits the GHZ state, see Eq. \ref{eq:ghzw}, is an AME, but as we already said there is no 4-qubit AME \cite{Higuchi00}. The amount of degrees of freedom in the definition of the state is insufficient to fulfill all the constraints coming from the requirement of maximum entanglement.

For 5 and 6 qubits, there are AME states. In particular, a 5-qubit state $\ket{\Upsilon_{5,2}} \in AME(5,2)$ can be defined by the coefficients of the superposition of basis states that form it:
\begin{equation}
\ket{\Upsilon_{5,2}} = \frac{1}{2^{5/2}}\sum_{i=0}^{2^5-1} c^{(\Upsilon)}_i |i\rangle,
\end{equation}
where we used the usual shorthand notation for the elements in the computational basis and the coefficients have the same modulus and signs given by \cite{Facchi08}
\begin{eqnarray}
c^{(\Upsilon)}&=&\{1, 1, 1, 1, 1, -1,-1, 1, 1, -1,-1,\nonumber\\ 
&&1, 1, 1, 1,1,1, 1, -1, -1,1, -1,1,\nonumber\\ 
&&-1, -1, 1, -1, 1, -1, -1, 1, 1\} .
\label{eq:ame52}
\end{eqnarray}
A state AME(5,2) found in Ref. \cite{Brown05} proved to be useful for a number of multipartite tasks in Ref.
\cite{Muralidharan08}. It can also be found as the superposition
of a perfect error correcting code as presented in Ref. \cite{Laflamme96} and also from orthogonal arrays \cite{Goyeneche14}.

For the sake of completeness, let us also provide an absolutely maximally entangled 6-qubit state
\begin{equation}
\ket{\Xi_{6,2}} = \frac{1}{\sqrt 2^6}\sum_{i=0}^{2^6-1} c^{(\Xi)}_i |i\rangle,
\end{equation}
with 
\begin{eqnarray}
c^{(\Xi)}&=&\{
 -1,- 1,- 1,+ 1,- 1, 1, 1, 1, \nonumber\\ 
&&-1,- 1,- 1, 1,1,- 1,- 1,- 1, \nonumber\\ 
&&-1, -1, 1, -1,- 1, 1, -1,- 1, \nonumber\\ 
&&1, 1, -1, 1,- 1, 1, -1,- 1, \nonumber\\ 
&&-1, 1, -1,- 1,- 1,- 1, 1,- 1, \nonumber\\ 
&&1, -1, 1,1,- 1,- 1, 1,- 1, \nonumber\\ 
&&1, -1,- 1,-1, 1, 1, 1,- 1, \nonumber\\ 
&&1, -1,- 1,-1,- 1,- 1,- 1, 1 \} .
\label{eq:ame62}
\end{eqnarray}

The case of 7 qubits was recently solved: it was proved that no 7-qubit AME exists \cite{Huber16}. For 8 qubits or more, there are no AME states \cite{Scott04}.

\subsection{A central example: AME(4,3)}

The first non-trivial example of an AME for larger local dimensions corresponds to
a state made of 4 qutrits, AME(4,3). Its explicit construction is
\begin{equation}\label{eq:ame43}
\ket{\Omega_{4,3}}=
\frac{1}{3}\sum_{i,j,=0,1,2} |i\rangle|j\rangle |i+j\rangle |i+2 j\rangle
\ ,
\end{equation}
where all qutrit indices are computed mod(3). It is easy to verify that
all the reduced density matrices to two qutrits are equal to $\rho=\frac{1}{9} \mathbb{I}_9$,
so that
this state carries entropy $S=2 \log 3$ for every one of its bipartitions.

The state $\ket{\Omega_{4,3}}$ can be viewed as a map of a two-qutrit product basis into a second one. That is
\begin{equation}
  \ket{\Omega_{4,3}} = \sum_{i=0,8} |u_i\rangle |v_i\rangle
  = \sum_{i,j=0,8} |u_i\rangle U_{ij} |u_j\rangle
\end{equation}
where $\{|u_i\rangle\}$ and $\{|v_i\rangle\}$ are product basis for two qutrits, and  $|v_i\rangle = U_{ij} |u_j\rangle$.
The matrix $U_{ij}$ is not only unitary  (as it must as a consequence of multiunitarity) but
also a bijective map between the sets of words of length $2$ over an alphabet $\mathbb{Z}_3=\{0,1,2\}$.
In other words, the entries of $U_{ij}$ are $0$ with a single $1$ per row/column.
This property remains the same whatever partition is analysed, though the unitary will vary. 

A second feature of the state $\ket{\Omega_{4,3}}$ is that the \emph{Hamming distance} between any pair of elements in the state is three, $D_H=3$, where the Hamming distance between two
codewords is defined as the number of positions in which they differ, e.~g.\ $D_H(00010,10000)=2$. As all the sequences in $\ket{\Omega_{4,3}}$ differ in 3 elements, any single qutrit error can be corrected. This hints at the relationship between AMEs and error correction codes. Both of these properties are related to the fact that $|\Omega_{4,3}\rangle$ is an AME of minimal support, a concept we are going to explore in Sect. \ref{sec:codes}.

It is also natural to expect AME states to accommodate easily to some magic-square like relations. A simple example goes as follows. Write the coefficients of $\ket{\Omega_{4,3}}$ as a matrix of row $i$ and column $j$, giving the composed value $a$ of the remaining two qutrits from 0 to 8, that is $a=3(i+j) {\rm mod}(3)+(i+2j) {\rm mod}(3)$. The square reads
\[
\begin{array}{ccc}
    $0$&  $5$& $7$\\
    $4$ & $6$&  $2$\\
    $8$ & $1$&  $3$
\end{array}
\]
where all rows and columns add up to 12. The same properties are maintained if we interchange the indices in the state. These kind of combinatorial designs are going to be explored in Sect. \ref{sec:combinatorial}.

From an experimental point of view, $\ket{\Omega_{4,3}}$ can be created defining
a quantum circuit that generates it.
Such a circuit makes use of the following
two gates:
\begin{eqnarray}
  \label{eq:gates}
\nonumber
   {\rm Fourier} &\quad& F_3 |0\rangle=\frac{1}{\sqrt 3}\left(|0\rangle+
  |1\rangle+|2\rangle\right),
  \\
   {\rm C_3-adder}&\quad &U_{C-adder} |i\rangle|j\rangle=|i\rangle|(i+j) {\rm mod} 3\rangle.\nonumber\\
\end{eqnarray}
This gate $C_3$ generalizes the CNOT gate for qubits and is represented in the circuit using the usual symbol for CNOT with the subscript 3.
\begin{figure}
\begin{center}
%\scalebox{.3}{\includegraphics{circuit2.eps}} WARNING. if EPS will be considered you need to change H -> F_3 and add a 3 to the CNOT
\scalebox{.3}{\includegraphics{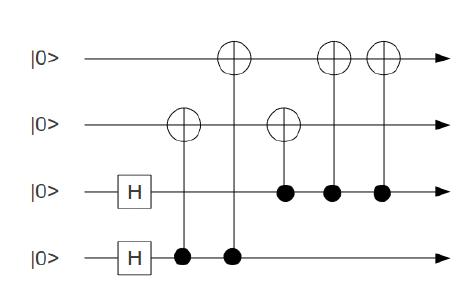}}
\end{center}
\caption{Quantum circuit required to generate the state AME $\Omega_{4,3}$ (4 qutrits) based on the Fourier gate $F_3$ and control-adders gate $C_3-adder$ .}
\label{fig:circuit2}
\end{figure}
The $\ket{\Omega_{4,3}}$ state  can be constructed as a
sequence of a Fourier $F_3$ and $C_3$-adders following the
circuit  depicted in Fig.\ref{fig:circuit2} acting
on the initial state $|0000\rangle$ of 4 qutrits.

\subsection{Support of AME states}

From the explicit examples we have presented so far,
different AME states appear to need different numbers of elements to be written.
For instance, AME(4,3) is made of the superposition of 9 states, all
weighted with the same coefficient. Yet, AME(6,2) is written using
the 64 basis states with coefficients either 1 or -1. 

Let us define the \emph{support} of a state $\ket{\psi}$ as 
the number of non-zero coefficients when $\ket{\psi}$ is written in the computational basis.
The \emph{support of a class} 
is defined as the support of the state inside the class with minimal support.
Note that the support of a class defines in turn another equivalence class.
Two states are \emph{support} equivalent if they belong to a LU class
with equal support.

In this sense, it is interesting
to point out that the state AME(6,2), defined in Eq.~\ref{eq:ame62}
with the maximal support of $2^6$ states, 
is LU equivalent to a state of support 16 by applying some Hadamard gates on its basis. 
It can be proved that minimal support for k-uniform states of qudits of d levels is $d^k$, therefore minimal support for an AME of 6 qubits would be 8 instead of 16. However it is known that AME(6,2) cannot have support 8.
Also in Ref. \cite{Brown05} a state AME(5,2) is built using 8 elements, while the theoretical minimum would be 4. 
We may wonder why the naive minimum possible number of $2^{\floor{N/2}}$ elements is not always attained. The answer to this question is given in Ref. \cite{Goyeneche14}, where a one-to-one relationship between $k$--uniform states having minimal support and a kind of combinatorial arrangements known as orthogonal arrays is given. Therefore, the non-existence of such states having minimal support is due to the non-existence of some classes of orthogonal arrays (those having strength one). Next section deals in detail with states with minimal support, but first we deal in the next subsection with an attempt to have a general expression for an AME state with maximal support.

\subsection{General expression for AME states} \label{sec:maximalsup}

It is interesting to ask whether it is possible to have always the same type of expression for an AME state, as a general reference form. As we already know that not all Hilbert spaces with AMEs have a minimal support AME, one could think on maximal support forms for this aim. Ref. \cite{Facchi09} partially solved the question for qubits. It can be shown that a state expressed as
\begin{equation}
|\Psi\rangle= \sum_{k=0}^{2^{n}-1}  z_k |k\rangle , \hspace{5mm} z_k = r_k \xi_k ,
\end{equation}
where the $\ket{k}$ span the whole computational basis and $r_k$ and $\xi_k$ are respectively the modulus and the phase of the complex coefficients $z_k$, is an AME if it satisfies the following equations:

\begin{equation}
\label{eqphases}
\sum_{m}  r_{l \bigoplus m} r_{l' \bigoplus m} \xi_{l \bigoplus m} \xi_{l' \bigoplus m} = 0,
\end{equation}
where $l,l'$ and $m$ stand for both parts of a certain balanced bipartition (of $\floor{N/2}$ parties).

The general form of the squared modulus of the coefficients will be:

\begin{equation}
\label{eq:modulus}
r_k^2 = 2^{-N} + \sum_{\frac{N}{2} < n \leq N} \sum_{j \in S^n} c_j^{(n)} \prod_{1 \leq {h \leq n}} (2k_{h}^{(j,n)} -1),
\end{equation}
where $S^n$ stands for the set of bipartitions of the system into groups of $n$ and $N-n$ particles and $j$ is an index for the bipartitions of this set. $h$ is an index for particles contained in the bipartition $j$, and $k_{h}^{(j,n)}$ stands for the value (0 or 1) of particle $h$ in ket $k$, in the case of a certain bipartition $j$ of the set $S^n$. The real coefficients $c_j^{(n)}$ are free as long as they satisfy Eq.~(\ref{eqphases}) and the normalization condition.

Ref. \cite{Facchi09} classifies all forms of AMEs with a small number of qubits using these equations. 

It is worth noting that AME states with maximal support and uniform amplitudes $r_k = 1 / \sqrt{d^N} \hspace{2mm} \forall k$ are possible to define for all numbers of qubits where there is an AME at all, by just setting all $c_j$ to 0 in Eq.~(\ref{eq:modulus})). Indeed we know of AMEs of this form for 2,3, 5 and 6 qubits. We call AME states having this property \emph{uniform} AMEs, and they are good candidates for general expressions, but it is not clear if they always exist for other local dimensions. It is an interesting open question whether one can build uniform AMEs in all the Hilbert spaces of any local dimension where AMEs exist, as it would provide a way to greatly simplify their discovery and classification. We conjecture that it is true and give further evidence in favour of it in section \ref{sec:AMEconstruct}.

\section{AME states of minimal support and classical codes} \label{sec:codes}

In Ref.~\cite{Helwig13}, a subclass of AME$(N,d)$ states 
is shown to be constructed by means of classical maximum-distance
separable (MDS) codes. In this section we show that such a subclass corresponds to the set of AME states of minimal support and exploit these ideas to get conditions for their existence.

\subsection{Equivalence between AME states of minimal support and MDS codes}

A state in AME$(N,d)$ belongs to the class of minimal support iff
it is LU-equivalent to a state $\ket{\Psi}$ with support $d^{\floor{N/2}}$ i.~e.
\begin{equation}
\label{eq:tuples-state}
\ket{\Psi}= \frac{1}{d^{\floor{N/2}}}\sum_{k=1}^{d^{\floor{N/2}}} 
r_k\e^{\iu \theta_k}\ket{x_k}\, ,
\end{equation}
where $x_k\in \mathbb{Z}_d^N$ are words of length $N$
over the alphabet $\mathbb{Z}_d=\{0,\ldots, d-1\}$, and $r_k>0$ and $\theta_k\in [0,2\pi)$ are their modulus and phases respectively.

Given a bipartition $A=\{a_1,\ldots,a_{\floor{n}} \}$, it will be useful to introduce the subword $x_k[A]$ of the word $x_k$
for the partition $A$ as the concatenation of the $a_1$-th, $a_2$-th, $\ldots$ $a_n$-th letters of $x_k$, that is, 
$x_k[A]=x_k[a_1]x_k[a_2]\ldots x_k[a_n]$.
Let us also denote by $X_\Psi = \{x_k, k=1,\ldots,d^{\floor{N/2}}\}$ the set of words which $\ket{\Psi}$ has support on, and 
by $X_\Psi[A]=\{x_k[A], k=1,\ldots,d^{\floor{N/2}}\}$ the set of all subwords $x_k[A]$ corresponding to the bipartition $A$.
With this notation, the reduced density matrix of a partition $A$ can be written as
\begin{equation}
\rho_A=\sum_{k'}\bra{x_{k'}[\bar A]}\proj{\Psi} \ket{x_{k'}[\bar A]},
\end{equation}
where
\begin{equation}
\ket{\Psi}= \frac{1}{d^{\floor{N/2}}}\sum_{k=1}^{d^{\floor{N/2}}} r_k\e^{\iu \theta_k}\ket{x_k[A]}\ket{x_k[\bar A]}\, .
\end{equation}
In order for $\ket{\Psi}$ to be an AME, the reduced density matrix of 
any bipartition $A=\{a_1,\ldots,a_{\floor{N/2}} \}$ needs to be the completely mixed state.
It is easy to see that this has the following implications:
\begin{enumerate}
\item The modulus are all the same, $r_k=1$ for all $k$.
\item The phases $\theta_k$ are arbitrary.
\item For any balanced bipartition $A$, with $|A|=\floor{N/2}$, 
two words $x_i,x_j\in X_\Psi$ have subwords $x_i[A]=x_j[A]$ if and only if $i=j$.
Equivalently, \ $X_\Psi[A]=\mathbb{Z}_d^{\floor{N/2}}$.
\end{enumerate}

Condition 3
implies that any pair of different words $x_i,x_j\in X_\Psi$ have Hamming
distance
\begin{equation}
d_H(x_i,x_j)\ge \floor{N/2}+1.
\end{equation}
To see this, note that otherwise there would exist
a balanced bipartition $A'$ for which $x_i[A']=x_j[A']$ for $i\neq j$ and
consequently the set $X_\Psi[A']$ would not contain all the possible words of length
$\floor{N/2}$, that is, $X_\Psi[A']\subset \mathbb{Z}_d^{\floor{N/2}}$.

A set of $M$ words of length $N$ over an alphabet of size $d$ that differ pairwise by at least a 
Hamming distance $\delta$ is called a \emph{classical code}. 
What we have shown above is that the existence of AME($N,d$) states of minimal support
imply the existence of classical codes of $M=|X_\Psi|=d^{\floor{N/2}}$ words with Hamming distance 
$\delta=\floor{N/2}+1$.
The codes produced by AME states are special in the sense that they saturate the \emph{Singleton bound} \cite{Singleton64},
\begin{equation}
 M\le d^ {N-\delta+1}\, .
\end{equation}
This type of codes that saturate the Singleton bound are called maximum-distance separable (MDS) codes.

The converse statement is also true. That is, the existence of a MDS code also implies
the existence of an AME state with minimal support \cite{Helwig13}.
The argument is the following: a code of $M=d^{\floor{N/2}}$ words of length $N$
 and Hamming distance $\delta=\floor{N/2}+1$
has all its subwords associated to any balanced bipartition $A$ of size $|A|=\floor{N/2}$ different, which implies the condition 3 above.
 Thus, \emph{AME states of minimal support are equivalent to
 classical MDS codes}.

This equivalence can be exploited to see that
a necessary condition for AME($N,d$) states with minimal support
(and equivalently of MDS codes) to exist is that the local dimension $d$ and the number of parties $N$ fulfill
\begin{equation}
d \ge \floor{N/2} +1 \, .
\label{eq:dn21}
\end{equation}

We can prove it as follows: let us try to construct an AME state by building an MDS code.
Due to the relabeling freedom, the first $d+1$ words of the code can be chosen as
\begin{equation}
{\footnotesize
\begin{array}{c}
d+1\\
\textrm{code}\\
\textrm{words}
\end{array}}
\left\{
\begin{array}{rrr}
0&\ldots \ldots \ldots 00 & 0\ldots \ldots  \ldots 0 , \\ 
0&\ldots\ldots \ldots  01 & 1\ldots \ldots  \ldots 1 ,\\ 
\vdots & \vdots &   \vdots \\
0&\ldots 0 (d-1)& (d-1)\ldots (d-1),\\
0&\ldots \ldots\ldots 10 &  x_d[\floor{N/2}+1] \ldots x_d[N],
\end{array}
\right.\nonumber
\end{equation}
where the letters $x_d[i]$, for $\floor{N/2}+1 \le i \le N$, are still unknown
and every word is written in two subwords of lengths $\floor{N/2}$
and $\floor{(N+1)/2}$ respectively.

Note that:
({\it i}) none of the unknown letters $x_d[i]$ can be $0$ 
in order for the word $x_d$ to have Hamming distance $\floor{N/2} +1$ with the first word $0\ldots0 0\ldots0$.
({\it ii}) None of the unknown letters $x_d[i]$ can be repeated in order for $x_d$ to
have Hamming distance $\floor{N/2}+1$ with the other $d-1$ words.
Therefore, if $\floor{(N+1)/2}$ variables $x_d[i]$,  must take $\floor{(N+1)/2}$ different values, 
and all of them must be different from 0,
it is necessary to extend the alphabet, forcing that $d\ge \floor{N/2}+1$.

Interestingly, Eq. \ref{eq:dn21} forbids the existence of AME(N,2) states having minimal support for $N>3$. However, this does not represent a proof that an AME(4,2) does not exist, just that it couldn't be of minimal support. It does prove that existing states AME(5,2) and AME(6,2) cannot be of minimal support. Also, this inequality is saturated for the existing states AME(4,3) and AME(6,4). An AME(7,5) of minimal support was recently shown not to exist in Ref. \cite{Bernal16}, thus proving that Eq. \ref{eq:dn21} is a necessary condition but not sufficient. The existence of the cases AME(8,5) and AME(8,6) is still open whereas the states AME(8,7) (see Sect. \ref{sec:constructioname}) and AME(8,8) are known.

\subsection{A less trivial example: AME(6,4)}
An application of the above connection between AME states of minimal support and MDS codes is the construction
of AME states by exploring the set of all words and selecting those which differ in at least
$\floor{N/2}+1$ elements.
For the case of the AME(6,4), such a search gives a state with an equal superposition of the following entries:
\begin{eqnarray}
\label{eq:ame64}
&\{&000000, 001111, 002222, 003333, 010123, 011032, \nonumber \\
&&012301, 013210, 020231, 021320, 022013, 023102, \nonumber \\
&&030312, 031203, 032130, 033021, 100132, 101023, \nonumber \\
&&102310, 103201, 110011, 111100, 112233, 113322, \nonumber \\
&&120303, 121212, 122121, 123030, 130220, 131331, \nonumber \\
&&132002, 133113, 200213, 201302, 202031, 203120, \nonumber \\
&&210330, 211221, 212112, 213003, 220022, 221133, \nonumber \\
&&222200, 223311, 230101, 231010, 232323, 233232, \nonumber \\
&&300321, 301230, 302103, 303012, 310202, 311313, \nonumber \\
&&312020, 313131, 320110, 321001, 322332, 323223, \nonumber \\
&&330033, 331122, 332211, 333300\}\, .
\end{eqnarray}
This state has $64 = 4^{6/2}$ terms and is thus an AME of minimal support.

\subsection{Construction of AMEs with minimal support} \label{sec:constructioname}

Finding AME states by exploring the set of all words is highly inefficient
and becomes in practice unfeasible from a relatively small number of parties.
In this context, the Reed-Solomon codes \cite{Reed60} can be a useful tool to produce
systematic construction of MDS codes and equivalently AME states.

Let us review here the particular case of $d$ prime and $N=d+1$.
Let us refer to the elements of the superposition
in the quantum states as words $x_i$ and the word of a half-partition
as $u_i$. The code words are obtained using
the action of a generator $G$, $x_i=u_i \cdot G$. The problem is then reduced to
fixing $G$. It can be shown that a family of valid generators is given by
\begin{equation}
  G=\left( 
  \begin{array}{ccccc}
  1&1&\ldots & 1& 0\\ 
  g_0 & g_1 &\ldots & g_{d-1}&0\\
  \ldots & \ldots & \ldots & \ldots & \ldots \\
  g_0^k & g_1^k &\ldots & g_{d-1}^k &1\\  
    \end{array}
  \right),
\end{equation}   
where $d\in \textrm{Prime}$, $N=d+1$, and $k=N/2$.

The case of AME(4,3) can be re-obtained using 
\begin{equation}
  G=\left( 
  \begin{array}{cccc}
  1&1 & 1& 0\\ 
  0 & 1 & 2 &1
   \end{array}
  \right).
\end{equation}
 Another concrete example corresponds to $g_0=1$, $g_1=1$,..., $g_6=6$ that
corresponds to AME(8,7), with $N=8$ $d=7$-dits. Then a total of $7^4$ codewords
are obtained that differ by a minimum Hamming distance $d_H=5$. An alternative way to build MDS codes and thus AME states of $N=d+1$ with d prime with the use of stabilizer formalism was developed very recently in Ref. \cite{Raissi17}. These systematic constructions prove that primality of $d$ and $N=d+1$ together are sufficient conditions to have an AME state of minimal support. For example, for $d=2$ we have the GHZ state (support 2), and for $d=3$ we obtain the AME(4,3) defined in Eq. \ref{eq:ame43}, which has support 9.

In order to try to use the above constructions for different cases, it is interesting to address the question of whether given some AME($N,d$) it is possible
to construct another AME($N',d'$). In this context, the following lemma can be useful which
allows to construct an AME($N',d$) for  any $N'<N$:
if there exists an AME state with minimal support in ${\mathcal H}(N,d)$
then there exist other AME states with minimal support in the Hilbert spaces 
${\mathcal H}(N', d)$ for any $N'\le N$.

To prove this last statement, 
we will assume N is even and will consider separately the transitions $N\to N-1$ and $N-1\to N-2$. 

Transition $N\to N-1$:
The existence of an AME state with minimal support implies the existence of a
code of $d^{\floor{N/2}}$ words of length $N$ with Hamming distance $d_H\ge \floor{N/2}+1$. 
Let us order the words in the code in increasing order and take the subset of the first $d^{\floor{N/2}-1}$ words which start with $0$. 
Note that by suppressing such a $0$, we get a code of $d^{\floor{N/2}-1}$ words of length $N-1$ with Hamming distance
$\floor{N/2}+1$, forming an AME.

Transition $N-1 \to N-2$:
From the previous steps, we are left with a code of $d^{\floor{N/2}-1}$ words of length $N-1$ and
Hamming distance $\floor{N/2}+1$. Note that by suppressing an arbitrary letter from all the words of the code,
one is left with a set of $d^{\floor{N/2}-1} = d^{\floor{(N-2)/2}} $ words of length $N-2$ with Hamming distance $\floor{N/2}+1-1=\floor{N/2}=\floor{(N-2)/2}+1$,
which is an MDS code. 

By iterating the previous procedure, we obtain MDS codes (and AME states of minimal support) for any $2\le N'\le N$. Thus the sufficient conditions for existence of AMEs of minimal support can be relaxed to both primality of $d$ and 
\begin{equation}
N\leq d+1.
\label{eq:nd1}
\end{equation}
which is the combination of both results of this section. Let us note that our central example $\ket{\Omega_{4,3}}$ (\ref{eq:ame43}) is the only state that saturates both the necessary (\ref{eq:dn21}) and the sufficient (\ref{eq:nd1}) conditions for existence of AMEs of minimal support.

\subsection{Non-minimal support AMEs and perfect quantum error correcting codes}

AME states are also deeply related to quantum error correction codes and compression \cite{Laflamme96,Rains99,Scott04}.
This is somewhat intuitive since maximal entropy is related to maximally mixed subsets.
The measure of any local degree of freedom delivers an output which is completely random. This is, in turn, the basic element to correct errors. Hence, a relation between the elements superposed to form an AME state and error correction codes is expected.

Le us illustrate the connection of an AME(5,2) state with the
well-known 5-qubit code \cite{Laflamme96}. It is easy to see that by applying some Hadamard gates on local qubits, the AME(5,2) state, defined through
the 32 coefficient given in Eq. \ref{eq:ame52}, is LU-equivalent to 
a state with fewer non-zero coefficients. Actually, a representative of
the same AME(5,2) class of only 8 coefficients can be found. That state corresponds to a superposition of the two
logical states in the error correcting codes found in Ref. \cite{Laflamme96}
\begin{equation}
|\Omega_{5,2}\rangle = \frac{1}{\sqrt {2}} \left( |0\rangle_L + |1\rangle_L\right) ,
\end{equation}
where the logical qubits are defined as
\begin{eqnarray}
 |0\rangle_L &=& \frac{1}{2} \left( | 00000\rangle +|00011\rangle + |01100\rangle -|01111\rangle\right), \nonumber\\ 
 |1\rangle_L&=& \frac{1}{2} \left( | 11010\rangle +|11001\rangle+|10110\rangle -|10101\rangle \right).\nonumber
\end{eqnarray}
Note the fact that the coefficients carry both plus and minus signs
as is the case in the non-minimal support AMEs.

\section{AME states and combinatorial designs}\label{sec:combinatorial}

Combinatorial designs are arrangements of elements satisfying some specific so called \emph{balanced} properties \cite{Stinson04}. Such elements are restricted to a finite set, typically considered as subsets of integer numbers. Some remarkable examples are block designs, t-designs, orthogonal Latin squares and orthogonal arrays (see Ref. \cite{Hedayat99} and references
therein). Combinatorial designs have important applications in quantum physics \cite{Zauner99}. Indeed, a connection between genuinely multipartite maximally entangled states and orthogonal arrays has recently been found \cite{Goyeneche14}. Furthermore, they are a fundamental tool in optimal design of experiments \cite{Raghvarao88,Atkinson07}. The existence of some combinatorial designs
can be extremely difficult to prove. For example the existence of Hadamard matrices in every dimension multiple of four (i.e., the \emph{Hadamard conjecture}) is a question raised in 1893 \cite{Hadamard93} that represents one of the most important open problems in combinatorics.

\subsection{Relation to mutually orthogonal Latin squares}

Let us consider the explicit expression of the AME(4,3) state defined in  Eq. \ref{eq:ame43}:
\begin{eqnarray}
&\ket{\Omega_{4,3}} &=\frac{1}{3}\left( |0000\rangle+|0112\rangle+|0221\rangle\right. \nonumber\\
&&+|1011\rangle+|1120\rangle+|1202\rangle\nonumber\\
&&\left. +|2022\rangle+|2101\rangle+|2210\rangle\right).
\label{eq:pop33}
\end{eqnarray}
Here, the third and fourth symbols appearing in every term of the state can be arranged into 2 Greco-Latin squares of size three:
\begin{equation}
\label{eq:greco}
\begin{array}{ccc}
A\alpha&B\gamma&C\beta\\
B\beta&C\alpha&A\gamma\\
C\gamma&A\beta&B\alpha
\end{array} 
\ = \ 
\begin{array}{ccc}
{ A\spadesuit}  & K\clubsuit& {Q\diamondsuit}\\
{ K\diamondsuit} & Q\spadesuit & { A\clubsuit}\\
{Q\clubsuit}    & A\diamondsuit  &{K\spadesuit}
\end{array},
\end{equation}
where every symbol of the sets $\{A,\alpha\}$, $\{B,\beta\}$ and $\{C,\gamma\}$, is associated to $0,1$ and $2$, respectively.
Hence a pair of symbols may represent a card of a given rank and suit.
Note that the first two digits in Eq. \ref{eq:pop33} may be interpreted as addresses determining the position of a symbol in the square. Furthermore, by considering the following four Mutually Orthogonal Latin Squares (MOLS) \cite{Colbourn96} of size five
\begin{equation}
\begin{array}{cccccc}
0000&4321&3142&2413&1234\\
1111&0432&4203&3024&2340\\
2222&1043&0314&4130&3401\\
3333&2104&1420&0241&4012\\
4444&3210&2031&1302&0123&,
\end{array}
\end{equation}
we define a 2-uniform state of 6 subsystems with five levels each:
\begin{eqnarray}
|\Phi_5\rangle&=\frac{1}{5}&(|000000\rangle+|104321\rangle+|203142\rangle+|302413\rangle+|401234\rangle+\nonumber\\
&&|011111\rangle+|110432\rangle+|214203\rangle+|313024\rangle+|412340\rangle+\nonumber\\
&&|022222\rangle+|121043\rangle+|220314\rangle+|324130\rangle+|423401\rangle+\nonumber\\
&&|033333\rangle+|132104\rangle+|231420\rangle+|330241\rangle+|434012\rangle+\nonumber\\
&&|044444\rangle+|143210\rangle+|242031\rangle+|341302\rangle+|440123\rangle).
\end{eqnarray}
A state locally equivalent to $|\Phi_5\rangle$ has been previously found in Ref. \cite{Goyeneche14} but its connection to MOLS is first given here.
By considering the standard construction of maximal sets of $d-1$ MOLS of prime size $d$ we can generalize the above construction for quantum states of a prime number of levels $d$ and $N=d+1$ parties as follows
\begin{equation}
|\Phi_{d+1,d}\rangle=\frac{1}{d}\sum_{i,j=0}^{d-1}|i,j\rangle\bigotimes_{m=1}^{d-1}|i+jm\rangle.
\label{eq:popgen}
\end{equation}

It is well known that a maximal set of $d-1$ MOLS of size $d$ exist for every prime power $d=p^m$ \cite{Stinson04}. This means that the above general expression can be extended to the case of prime power level systems. For instance, the maximal set of 3 MOLS of order $d=4$ can be represented by a color figure
\begin{equation}
\begin{array}{ccccc}
{\color{blue}A\spadesuit}&{\color{green}K\diamondsuit}&{\color{orange}
Q\heartsuit}&{\color{red}J\clubsuit}\\
{\color{orange}J\diamondsuit}&{\color{red}Q\spadesuit}&{\color{blue}
K\clubsuit}&{\color{green}A\heartsuit}\\
{\color{green}Q\clubsuit}&{\color{blue}J\heartsuit}&{\color{red}
A\diamondsuit}&{\color{orange}K\spadesuit}\\
{\color{red}K\heartsuit}&{\color{orange}A\clubsuit}&{\color{green}
J\spadesuit}&{\color{blue}Q\diamondsuit}&.
\end{array}
\end{equation}
This design determines an AME(5,4) given by
\begin{eqnarray}\label{eq:ame54}
|\Omega_{5,4}\rangle&=&|00000\rangle+|10312\rangle+|20231\rangle+|30123\rangle+\nonumber\\
&&|01111\rangle+|11203\rangle+|21320\rangle+|31032\rangle+\nonumber\\
&&|02222\rangle+|12130\rangle+|22013\rangle+|32301\rangle+\nonumber\\
&&|03333\rangle+|13021\rangle+|23102\rangle+|33210\rangle\nonumber,
\end{eqnarray}
where every symbol of the sets $\{A,\spadesuit,blue\}$, $\{J,\diamondsuit,orange\}$, $\{Q,\clubsuit,green\}$ and $\{K,\heartsuit,red\}$, is associated to $0,1,2$ and $3$, respectively,
while the first two digits of every term label the position of a symbol in the pattern. In the above expression a normalization factor is required. Note that this state, or a state equivalent with respect to local unitary transformations, 
arises from the Reed-Solomon code of length five \cite{Reed60}. 

Furthermore, the construction can be extended to any dimension $d$ in the following way:
\begin{equation}
|\Phi_{d+1,d}\rangle=\frac{1}{d}\sum_{i,j=0}^{d-1}|i,j\rangle\bigotimes_{m=1}^{\mathcal{N}(d)}|\lambda_m[i,j]\rangle,
\label{eq:ame43d}
\end{equation}
where $\mathcal{N}(d)$ denotes the maximal number of MOLS of size $d$.
Here $\lambda_m[ij]$ denotes the entries of the $m$-th Latin square,
so the above expression can be considered as a direct generalization of Eq. \ref{eq:popgen}.
It is worth adding that for dimensions $d\ge 12$ not equal to a prime power number 
only lower bounds for the function  $\mathcal{N}(d)$ are known \cite{Colbourn96}.
The problem is solved only for smaller dimensions, as
$\mathcal{N}(6)=1$ in agreement with unsolvability of the famous Euler problem of $36$
officers, while $\mathcal{N}(10)=2$ (see \cite{Stinson04}), where an explicit form
of a pair of MOLS of size ten is derived.
Thus for $d=10$ expression (\ref{eq:ame43d}) describes a $2$-uniform state of
$4$ subsystems with $10$ levels each.

In general, the problem of constructing $N-2$ MOLS of size $d$ is equivalent to building a 2-uniform state of $N$ qudits of $d$ levels having $d^2$ positive terms. Note that $d^2$ is the minimal number of terms that a 2-uniform state of qudits of $d$ level systems can have.

\subsection{AME states and hypercubes}
In the previous section we considered maximal sets of MOLS to construct 2-uniform states of qudits.
 However, this construction is not useful to find AME states for $d>4$. The aim of this section is to consider combinatorial arrangements for constructing AME states in such cases. The main result is inspired in a generalization of the AME(4,3) state $\ket{\Omega}$ given in Eq. \ref{eq:ame43}
and the AME state of 6 ququarts presented in Eq.~\ref{eq:ame64}.
\begin{figure}[!h]
\centering % centering figure
{\includegraphics[width=7.7cm]{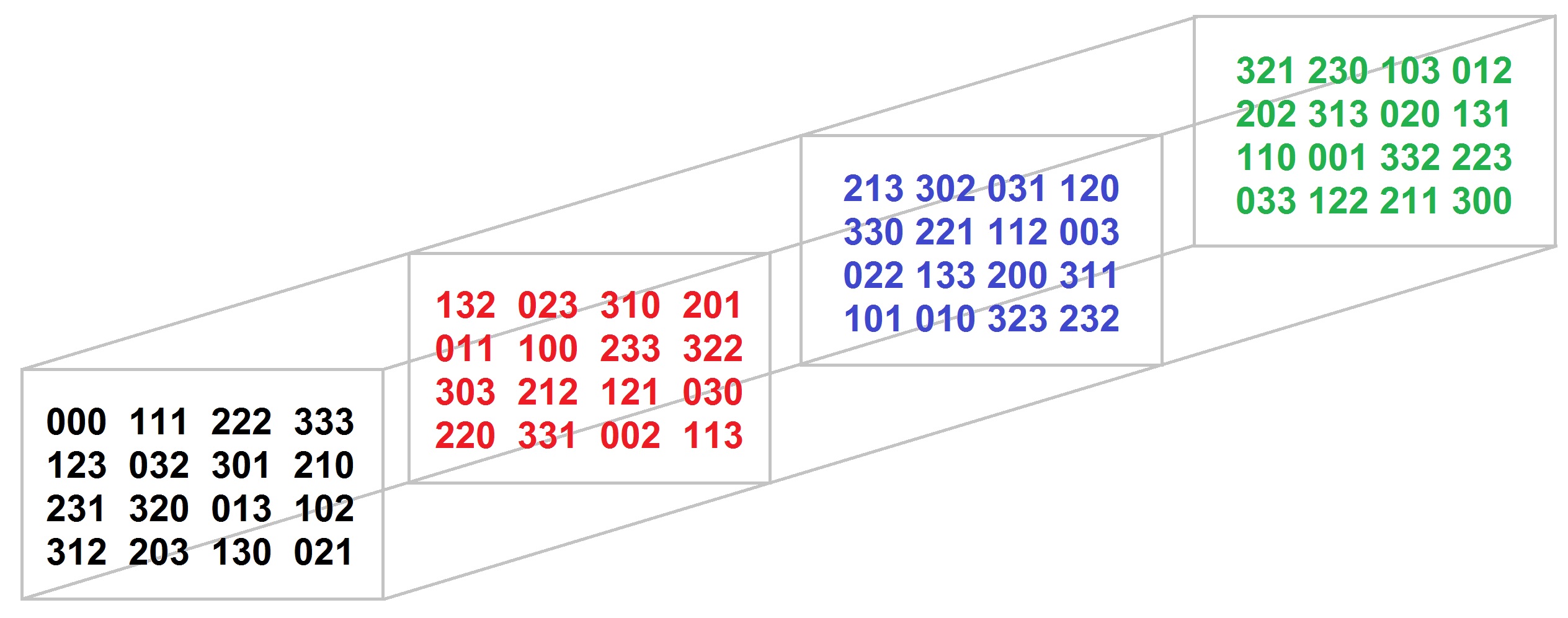}} % importing figure
\caption{Three mutually orthogonal Latin cubes of dimension $3$ and size $4$. 
This arrangement allows us to generate a state AME(6,4) of 6 ququarts.
 Each of the 12 planes (4 horizontal, 4 vertical and 4 oblique) contains a set of 3 MOLS of size 4.}
\label{Fig1}
\end{figure}
In Ref. \cite{Goyeneche14} it was shown that this state can be derived from the irredundant orthogonal array IrOA$(64,6,4,3)$. Furthermore, this orthogonal array can be interpreted as a set of three mutually orthogonal Latin cubes (see Fig.\ref{Fig1}). Also note that the AME(4,3) state $\ket{\Omega}$ arises from an IrOA(9,4,3,2) (See Eq.(B1) in \cite {Goyeneche14}. Thus, if $k$ mutually orthogonal hypercubes of dimension $k$ having $k+1$ symbols exist they are one to one connected with an IrOA($(k+1)^k,2k,k+1,k$) and, therefore, it would produce an AME(2k,k+1) state. This family of states would saturate the Singleton bound $d\geq\frac{N}{2}+1$. Note that for $k=1$ one obtains the  standard Bell state, $|\Phi^+\rangle=(|01\rangle+|10\rangle)/\sqrt{2}$ (1 Latin square of size 2 with 2 symbols). Taking $k=2$ and using 2 MOLS of size 3 and 3 symbols we arrive to the AME(4,3) state of $\ket{\Omega}$ (\ref{eq:ame43}) which, from this point of view, can be considered as a generalization of the Bell state. Furthermore, the AME(6,4) state (\ref{eq:ame64}) corresponding to $k=3$ 
also belongs to this family and it is associated to three mutually orthogonal Latin 
cubes of dimension $3$ and size $4$. It is interesting to check
whether there exist other states with $k\ge 4$ belonging to this family.

\section{AME states and multiunitarity} \label{sec:multiunitarity}

\subsection{Unitary matrices and bipartite systems}

Let us illustrate the connection between unitary matrices and AME states for the simplest case of 2 qubits. Let us assume that the state of the system is given by
\begin{equation}
|\phi\rangle=\frac{1}{\sqrt{2}}(U_{0,0}|00\rangle+U_{0,1}|01\rangle+U_{1,0}|10\rangle+U_{1,1}|11\rangle),
\end{equation}
where $\rho_A=\frac{1}{2}UU^{\dag}$, 
$\rho_B=\frac{1}{2}(U^T)(U^T)^{\dag}$  and $T$ denotes transposition. Any unitary matrix $U$ of size $2$ represents a Bell--like state.
Furthermore, the Pauli set of four unitary matrices $\mathcal{U}=\{\mathbb{I},\sigma_x,\sigma_y,\sigma_z\}$,
orthogonal in the sense of the Hilbert-Schmidt product, 
defines the maximally entangled Bell basis in ${\cal H}_2 \otimes {\cal H}_2$.

\subsection{Multiunitarity for AME(4,3)}

We shall consider again the AME(4,3) state of four qutrits $\ket{\Omega_{4,3}}$, and 
represent its coefficients by a four--index tensor 
\begin{equation}
\ket{\Omega_{4,3}}=\sum_{i,j,k,l=0,1,2} t_{ijkl} \; |ijkl\rangle .
\label{eq:tensorame43}
\end{equation}
where the entries of the tensor $t$ can be expressed as product of the Kronecker delta functions
\begin{equation}
 t_{ijkl}=\frac{1}{9}\delta_{k,i+j}\delta_{l,i+2j}.
\label{eq:tensorcoef}
\end{equation}
Here, the addition operations are modulo 3. As discussed in the previous section, all non-zero coefficients are equal.
The tensor $t_{ijkl}$ consists of $3^4=81$ elements which can be reshaped
to form a square matrix of order $9$. Note that there exist altogether
$\binom{4}{2}=6$ different ways of choosing a bipartition of the indices and forming a matrix $U_{\mu,\nu}$. That is,
\begin{equation}\label{munu}
(\mu,\nu) = \left\{
\begin{array}{c}
(i+2j,k+2l),\,(k+2l,i+2j)\\
(i+2k,j+2l),\,(j+2l,i+2k)\\
(i+2l,j+2k),\,(j+2k,i+2l)\\
\end{array}.
\right.
\end{equation}
The non-trivial property of an AME(4,3) tensor is that these six matrices are unitary. As transposition of a unitary matrix 
remains unitary, it is sufficient, in this case, to check
unitarity for the three cases appearing in the first column of the right side of Eq. \ref{munu}. That is, taking combined indices in the original tensor,
\begin{equation}
  t_{ijkl}=U^{(1)}_{(ij)(kl)} = U^{(2)}_{(ik)(jl)} 
   = U^{(3)}_{(ij)(kl)} .
\label{multiunitarityame4}
\end{equation}
absolute maximal entanglement is achieved if the matrices  $U^{(1)}$, $U^{(2)}$ and $U^{(3)}$ correspond to different changes of bases, that is unitary matrices. We refer to such particular kind of unitary matrices as {\sl multiunitary}. %\textcolor{red}{Appendix B?}%
 
\subsection{General multiunitarity }
\label{sec:general-multi-unitaritity}
 
Let us consider a more 
general case of pure states of $N$ subsystems with $d$ levels each. That is,
\begin{equation}
|\phi\rangle=\sum_{s_0,\dots,s_{N-1}=0}^{d-1}
  t_{s_0,\dots,s_{N-1}}|s_0,\dots,s_{N-1}\rangle .
\end{equation}
Let us assume here that the number of subsystems is even, $N=2k$,
so there exist $M=\binom{2k}{k}$ possible splittings of the system
into two parts of the same size.

A necessary condition for $|\phi\rangle$ to be an AME state 
is that the tensor $t$ with $2k$ indices, reshaped into a square matrix of
size $d^{k}$, forms a unitary matrix $U$. 
This is so, as the reduction associated to the first $k$ qudits,
given by $\rho_{k}=UU^{\dag}$, should be proportional to the identity.
To arrive at an AME(2k,d) state, similar conditions have to hold
for all $M$ different square matrices obtained from the tensor $t$
by all possible ways of reshaping its entries into a square matrix. This observation provides a clear motivation to introduce 
the notion of {\sl multiunitarity}:
A square matrix $A$ of order $d^k$ ($k\geq2$), acting on a composed Hilbert space ${\cal H}_d^{\otimes k}$ and represented in a product basis by 
$A_{\stackrel{\scriptstyle n_1, \dots , n_k}{\nu_1, \dots,  \nu_k}}:=
\langle n_1, \dots , n_k |A| \nu_1, \dots,  \nu_k\rangle$
is $k$--unitary if it is unitary for $M=\binom{2k}{k}$ reorderings of its entries corresponding to all possible choices of $k$ indices out of $2k$.

In this way, we can establish the following one-to-one connections:
\begin{equation*}
AME(2,d) \equiv \mbox{ unitary of order }d,
\end{equation*}
for bipartite systems of $2$ qudits having $d$ levels each and, in general:
\begin{equation*}\label{eq:multiunitarityame43}
AME(2k,d) \equiv \mbox{ $k$-unitary of order }d^k,
\end{equation*}
for multipartite systems of $N=2k$ qudits having $d$ levels each. By construction, $1$--unitarity reduces to standard unitarity.
Any  $k$--unitary matrix with $k>1$ is called {\sl multiunitary}. It is well-known that unitarity of matrices is invariant under multiplication. Multiunitarity imposes more restrictions on a given matrix $U$ than unitarity. Therefore, the product of two
multiunitary matrices in general is not multiunitary. For instance, the matrix $O_8$ (see Eq. \ref{eq:O8} below) is hermitian and 
$3$-unitary, but $O_8^2=\mathbb{I}$ is only $1$-unitary.

Similarly, the case of the AME(4,3) state $\ket{\Omega}$ reduces to analyzing the properties of the tensor $t_{ijkl}$ in Eq. \ref{eq:tensorame43} and verifying the multiunitarity of $U$. Indeed, for this state we have $U=Perm(0,5,7,4,6,2,8,1,3)$, $U^{T_2}=Perm(0,5,7,1,3,8,2,4,6)$ and $U^R=Perm(0,2,1,4,3,5,8,7,6)$, where $T_2$ and $R$ mean partial transposition and reshuffling %(see appendix \ref{appendixB} for further details).
Here, $Perm$ denotes a permutation matrix.% A mini-catalogue of all the multi-unitary matrices defined in this work can be found in Appendix \ref{appendixD}.}

\subsection{AME and Hadamard matrices} \label{sec:hadamard}

For six qubits, the AME(6,2) state $\ket{\Xi_{6,2}}$ of Eq. \ref{eq:ame62} with maximal support arises from graph states \cite{Hein04}. Let us write it explicitly
\begin{eqnarray*}
\ket{\Xi_{6,2}}=\frac{1}{8}\left(\right. 
&&-|000000\rangle -|000001\rangle -|000010\rangle +|000011\rangle -|000100\rangle +|000101\rangle \\
&& +|000110\rangle+ |000111\rangle-|001000\rangle -|001001\rangle -|001010\rangle +|001011\rangle \\
&& +|001100\rangle -|001101\rangle -|001110\rangle -|001111\rangle -|010000\rangle -|010001\rangle \\
&&+ |010010\rangle -|010011\rangle -|010100\rangle +|010101\rangle -|010110\rangle -|010111\rangle\\
&& +|011000\rangle+ |011001\rangle -|011010\rangle+ |011011\rangle -|011100\rangle+ |011101\rangle \\
&&-|011110\rangle -|011111\rangle -|100000\rangle+ |100001\rangle -|100010\rangle -|100011\rangle\\
&& -|100100\rangle -|100101\rangle +|100110\rangle -|100111\rangle + |101000\rangle -|101001\rangle \\
&&+|101010\rangle +|101011\rangle -|101100\rangle -|101101\rangle +|101110\rangle -|101111\rangle\\ 
&&+|110000\rangle -|110001\rangle -|110010\rangle -|110011\rangle +|110100\rangle +|110101\rangle \\
&&+|110110\rangle -|110111\rangle +|111000\rangle -|111001\rangle -|111010\rangle -|111011\rangle \\
&& \left. -|111100\rangle -|111101\rangle -|111110\rangle +|111111\rangle
\right).
\end{eqnarray*}
This state leads to the following Hadamard matrix of order $D=2^3=8$
which is $3$--unitary:
\begin{equation}\label{eq:O8}
O_8=\frac{1}{\sqrt{8}}\left(\begin{array}{rrrrrrrr}
-1& -1& -1& 1& -1& 1& 1& 1\\
-1& -1& -1& 1& 1& -1& -1& -1\\
-1& -1&  1& -1& -1& 1& -1& -1\\
1& 1& -1& 1& -1& 1& -1& -1\\
-1& 1& -1& -1& -1& -1& 1& -1\\ 
1& -1& 1& 1& -1& -1& 1& -1\\
1& -1& -1& -1& 1& 1& 1& -1\\
1& -1& -1& -1& -1& -1& -1& 1
\end{array}\right).
\end{equation}
Note that the entries of $\ket{\Xi_{6,2}}$ are given by the concatenation of the rows of $O_8$, up to normalization. This matrix is symmetric and equivalent up to enphasing and permutations \cite{Tadej06}
to the symmetric Sylvester Hadamard matrix $H_8=H_2^{\otimes3}$.

 Note that $O_8$ is 3-unitary but $H_2^{\otimes3}$ is not,
so permutation or enphasing  of a unitary matrix 
can spoil its multiunitarity.
Moreover, from the concatenation of the rows of $H_2^{\otimes3}$ 
we only generate a 1-uniform state, which means that $H_2^{\otimes3}$ is only 1-unitary (i.e., unitary). 
% This means that AME states are not invariant under the consideration 
% of equivalent Hadamard matrices. 

\subsection{Further constructions of AME states}\label{sec:AMEconstruct}

It is not possible to bring the state AME(4,3) $\ket{\Omega_{4,3}}$ into a real uniform state with local unitaries, e.g. having all its entries of the form $\pm 3^{-2}$. This is a consequence of the fact that a real Hadamard matrix of size 9 does not exist. However, the state $\ket{\Omega_{4,3}}$ is equivalent under local unitary operations to the complex uniform state (See Eq.(3) in Ref. \cite{Gaeta15})
\begin{equation}
\ket{\Omega'_{4,3}}=\frac{1}{9}\sum_{i,j,k,l=0}^2\omega^{j(i-k)+l(i+k)}|i,j,k,l\rangle,
\end{equation}
where $\omega=e^{2\pi i/3}$. This state is associated to the following $2$-unitary complex Hadamard matrix:
\begin{equation}
\label{eq:uame43}
U_{P}=\frac{1}{3}\left(\begin{array}{ccccccccc}
1&1&1&1&w&w^2&1&w^2&w\\
1&1&1&w^2&1&w&w&1&w^2\\
1&1&1&w&w^2&1&w^2&w&1\\
1&w&w^2&1&w^2&w&1&1&1\\
w&w^2&1&1&w^2&w&w^2&w^2&w^2\\ 
w^2&1&w&1&w^2&w&w&w&w\\
1&w^2&w&1&1&1&1&w&w^2\\
w^2&w&1&w&w&w&1&w&w^2\\
w&1&w^2&w^2&w^2&w^2&1&w&w^2
\end{array}\right).
\end{equation}
Interestingly, every integer power $(U_P)^m$ is a complex Hadamard matrix for $m\neq 4\,(\mathrm{Mod}\,4)$ and $(U_P)^8=\mathbb{I}$. 
Moreover, $U_P$ is equivalent to the tensor product of Fourier matrices $F_3\otimes F_3^{\dag}$, that is
\begin{equation}\label{equiv}
U_P=\mathcal{D}F_3\otimes F_3^{\dag}\mathcal{PD},
\end{equation}
where 
$\mathcal{D}=\mathrm{Diag}(1, 1, 1, 1,\omega,\omega^2, 1,\omega^2,\omega)$ is a diagonal unitary matrix,
while $\mathcal{P}$ is a permutation matrix which changes the order of the 
columns from $\{1,\dots,9\}$ to $\{1,4,7,2,5,8,3,6,9\}$.
In order to construct a $2$-unitary matrix one has to take a unitary $U$ such that its partially transpose $U^{T_2}$ and the 
reshuffled matrix $U^R$ are unitary %-- see Appendix \ref{appendixB}}. 
In the case of a matrix $U$ of size $D=3^2$ this implies that the set of nine $3\times3$ unitary matrices appearing in the $3\times3$ blocks of Eq. \ref{eq:uame43} define an orthogonal basis for the Hilbert-Schmidt product.
% which can be easily verified. 
It is thus possible to obtain AME states by considering orthogonal bases of unitary operators.
For instance, one can construct  the $\ket{\Omega}$ state from the following matrix:
\begin{equation}\label{U_displa}
U^{\prime}_P=\frac{1}{\sqrt{3}}\left(\begin{array}{ccc|ccc|ccc}
1&0&0&1&0&0&1&0&0\\
0&1&0&0&\omega&0&0&\omega^2&0\\
0&0&1&0&0&\omega^2&0&0&\omega\\\hline
0&0&1&0&0&\omega&0&0&\omega^2\\
1&0&0&\omega^2&0&0&\omega&0&0\\
0&1&0&0&1&0&0&1&0\\\hline
0&1&0&0&\omega^2&0&0&\omega&0\\
0&0&1&0&0&1&0&0&1\\
1&0&0&\omega&0&0&\omega^2&0&0
\end{array}\right).
\end{equation}
We applied  here the orthogonal basis defined by the displacement operators of size $d=3$:
\begin{equation}\label{Displa}
D_{p_1,p_2}=\tau^{p_1p_2} X^{p_1}Z^{p_2},
\end{equation}
where $p=(p_1,p_2)\in\mathbb{Z}_d^2$, $\tau=-e^{\pi i/d}$, $\omega=e^{2\pi i/d}$, $X|k\rangle=|k+1\rangle$ and $Z|k \rangle=\omega^k|k\rangle$. These operators define the discrete Weyl-Heisenberg group. This approach can be easily generalized to any prime $d>2$. Indeed, for $d$ prime every reordering of indices lead us to the same matrix up to permutation of columns and rows, and therefore it remains unitary.
This shows a construction of AME($4$,$d$) working for any prime number of levels $d$.  Moreover, it is likely that this construction can be generalized for $d$ being a power of a prime by considering the theory of Galois fields. 
Observe that the above construction is essentially different from the construction of AME states used in coding theory. 
Indeed, the tensor products of $N$ displacement operators bases of size $d$, i.e. the set $\{D_{p^1_1,p^1_2}\otimes \dots\otimes D_{p^N_1,p^N_2}\}$, produce codes and states AME($N$, $d$) \cite{Scott04}.

We conjecture that
for any AME state one can choose suitable local unitary operations such that its associated matrix transforms into a multiunitary complex Hadamard matrix. 
That matrix represents a maximally entangled state with the maximum number of terms having all entries of the same amplitude, what we called a uniform state in section \ref{sec:maximalsup}, and this conjecture is equivalent to the one we formulated there. 

\section{Summary}

In this chapter we have analysed some new properties of Absolutely Maximally Entangled (AME)
states in multipartite systems. First of all, we have reviewed and extended
several ways of constructing them. Then, we have explored their relation to the
field of combinatorial designs. For instance, 
a state AME(4,3) consisting of four maximally entangled qutrits
is linked to the set of two mutually orthogonal
Latin squares of order $3$, while a state AME(6,4) made of 6 ququarts is related to the set
of three mutually orthogonal Latin cubes of order $4$.

A deep relation between AME states and matrices that display the property of multiunitarity has been found: AME states made out of an even number $N$ of degrees of freedom are equivalent to multiunitary matrices (i.e., matrices being unitary after $M=\binom{N}{N/2}$ rearrangements of its entries). This remarkable property may be at the core of the use of AME states in holography \cite{Latorre15,Pastawski15}. 

Furthermore, making use of a state AME($2k$,$d$) consisting of $N=2k$ parties, one can construct a $(k-1)$-uniform state state of $2k-1$ parties by removing a single subsystem.

We have proved the existence of states AME(4,$d$) for every prime $d>2$ with the help of the concept of multiunitarity. Note that this also provides the existence of states AME(3,$d$) for every prime $d>2$. Our proof is in accord with the known result from coding theory that there are AMEs of minimal support for every prime $d>2$ and $n\leq d+1$.

AME states remain largely unexplored. Let us bring a number of open problems
that deserve to be solved.
\begin{itemize}

\item Classification of AME and LU invariants.\\
It is unclear whether there is a clear cut classification of AME, which is related to
LU invariants. The fact that some AME states carry different minimal support or that
some AME are right away related to Reed-Solomon codes hints at a some unknown structure
among AME states.

\item Non-minimal support.\\
Examples of AME states with non-minimal support are 
only explicitly known for 5 and 6 qubits. It would
be natural to find examples for higher local dimensions.

\item Computation of invariants.\\
LU invariants grow exponentially with the size of the number of parties.
An example of them is the hyperdeterminant, which has only been computed up to
4 qubits. There are no computations of hyperdeterminants of 4 qutrits.
Do AME states carry maximum values for some LU invariant? From the
general theory of hyperdeterminants, we know that the rank of the tensor
defining a four-qutrit state is 1269, which seems out of reach for 
any practical computation.

\item Bell inequalities.

Bell inequalities maximally violated by AME(4,3) are unknown, but it is possible to build a Bell inequality out of the state, as will be shown in Sect. \ref{sec:bellmapping}. 

\end{itemize}

\pagestyle{fancy}
\fancyhf{}
\fancyhead[LE]{\thepage}
\fancyhead[RE]{MULTIPARTITE ENTANGLEMENT}
\fancyhead[LO]{CH.3 Operational approach to Bell inequalities}
\fancyhead[RO]{\thepage}

\chapter{Operational approach to Bell inequalities: application to qutrits}

\section{Introduction}
As we commented in the introduction to the first chapter, in 1964 Bell introduced an inequality that provided a tool to discern between quantum non-locality and any local theory of hidden variables \cite{Bell64}. Ultimately this answers the question of whether entanglement is a fundamental property of nature, for what it needs quantum mechanics to be nonlocal.

A new Bell inequality was proposed in the 1969 CHSH paper \cite{Clauser69}, which was simpler and easier to test experimentally.
It placed constraints on expected values of measurements of correlations of two outcomes with two settings per observer. Experimenters quickly began to test the inequality, and by 1982 there was already 
a strong evidence that local hidden variable theories were being ruled out \cite{Aspect82}. The experiment kept being repeated for larger distances or different components \cite{Tittel98}. However, the question of loopholes remained alive: hypotheses on the experimental setting that were taken for granted while computing the expectation values and that were not necessarily true in strict analysis. 
Recent experiments \cite{Hensen15,Poh15} claim to have closed all ``closable" loopholes.

There have been numerous attempts to go beyond the CHSH inequalities.
 Mermin introduced a set of inequalities for an arbitrary number of qubits that were maximally violated
 by the GHZ state \cite{Mermin90,Greenberger90}. A systematic mathematical treatment of these inequalities was carried
 out a decade later \cite{Werner01,Cereceda01,Zukowski02}. It was also at that time that an inequality for two parties, each performing quantum measurements with
$d$ outcomes was discovered \cite{Collins02} and with it came the first realization that Bell inequalities are not always maximally violated by a maximally entangled state \cite{Acin02}, which showed that entanglement is not in a one-to-one correspondence with nonlocality. Progress in generalization to a larger number of d-dimensional particles has been more modest \cite{Acin04}. For a general recent review of Bell nonlocality and a large list of references, see Ref. \cite{Brunner14}.

This chapter constructs new Bell inequalities for systems composed of several subsystems of more than two levels each. In particular, we focus our attention on quantum systems consisting on qutrits. Inequalities for three outcomes have been written more often in terms of probabilities but they can also be treated with expectation values \cite{Chen02,Arnault12}. We have extended this formalism in order to build new inequalities for three outcomes and a different number of parties and find its classical and quantum bounds for qutrits in a semi-systematic way. We have found some regular patterns for the coefficients of the inequalities and for the settings and states that maximally violate these inequalities. This mechanism is potentially generalizable to other dimensions.

This chapter is organized as follows. In Sect. \ref{sec:bellqubits}, a review of the CHSH and Mermin inequalities for two outcomes and several parties is presented. We focus on an interesting pattern involving commutators, which we use to write n-particle inequalities and classical and quantum bounds in a simple way. 
In Sect. \ref{sec:bellqutrits}, the work done for qutrits is reviewed and we present our formalism and methods to construct new inequalities and find their classical and quantum bounds. In Sect. \ref{sec:bellmapping}, a new strategy is presented to find Bell inequalities from the expressions of maximally entangled states.
%Some further issues, including the multiplets of optimal settings (MOS) and potential generalization of the results obtained for higher dimensions are discussed in the Appendix.

\section{Bell inequalities for two outcomes}\label{sec:bellqubits}

\subsection{Two parties} \label{sec:2parties}

In the case of two parties the only relevant Bell inequality is the one of 
Clauser, Horne, Shimony and  Holt \cite{Clauser69}. It is obtained out of the following Bell polynomial
\begin{equation}
B_{CHSH} = ab + ab' +a'b -a'b'.
\label{eq:chshn}
\end{equation}
Here, $a,a'=\pm 1$ and $b,b'=\pm 1$ are the possible outcomes detected by observers Alice and Bob, respectively. Note that Eq. \ref{eq:chshn} can be factorized as
\begin{equation}
B_{CHSH} = a(b+b')+a'(b-b'),
\label{chshbrackets}
\end{equation}
so one of the terms is $\pm 2$, while the other one is equal to zero, 
which means that the maximum value that can be obtained with a local realistic theory is $\langle B_{CHSH}\rangle_{LR}=2$. In a more general case, this \textit{classical} bound can be obtained by computing the value of the Bell polynomial with all possible outcomes for $a,a',b$ and $b'$ and selecting its maximum.

In quantum mechanics, the variables $a,a'$ and $b,b'$ are represented by Hermitian operators acting on the Hilbert spaces $\mathcal{H}_{a}$ and $\mathcal{H}_{b}$, respectively. For dichotomic variables the operators satisfy $a^{2}=a'^{2}=b^{2}=b'^{2}=\mathbb{I},$ because the measurement operators $a,a',b$ and $b'$ have eigenvalues $\pm 1$. The quantum Bell operator reads then
\begin{equation}\label{eq:CHSH2}
B_{CHSH} = a \otimes b + a \otimes b' +a' \otimes b -a' \otimes b',
\end{equation}
where $\otimes$ denotes the Kronecker product. The \emph{quantum} bound $\langle B_{CHSH}\rangle_{QM}$ corresponds to the maximal eigenvalue of all possible Bell operators (\ref{eq:CHSH2}) satisfying the previously stated conditions. A Bell operator $B$ defines a Bell inequality if $\langle B \rangle_{LR}<\langle B\rangle_{QM}.$
In the case of CHSH, it was proven by Tsirelson \cite{Tsirelson80} that the maximum quantum value is $\langle B_{CHSH}\rangle_{QM}=2\sqrt{2}$.
An enlightening proof of this quantum value is given in Ref. \cite{Landau87} and is reproduced now. The square of the Bell operator shown in Eq. \ref{eq:CHSH2} is
\begin{equation}
B^{2}_{CHSH}= 4\mathbb{I}_a\otimes \mathbb{I}_b-[\hat{a},\hat{a'}]\otimes[\hat{b},\hat{b'}]\ .
\label{chsh2o}
\end{equation}
For a local hidden variable theory all observables commute, so the classical value is determined by $\langle B_{CHSH}\rangle_{LR}=\sqrt{\langle B^2_{CHSH}\rangle_{LR}}=\sqrt{4}=2.$
On the other hand, the largest absolute value of all the possible eigenvalues for commutators of hermitian operators is 2 and it is achieved by considering the Pauli matrices, as they have the property
$[\sigma_j,\sigma_k]=2i \epsilon_{jkl} \sigma_l$ and $\sigma_l$ has eigenvalues $\pm 1$. Here, $\epsilon_{jkl}$ is the antisymmetric Levy-Civita tensor. Therefore, the quantum value of the square Bell operator \eqref{chsh2o} is given by $
\langle B_{CHSH}\rangle_{QM}=\sqrt{\langle B^2_{CHSH}\rangle_{QM}}=\sqrt{8}=2\sqrt{2}.$
In Sect. \ref{sec:mermin} we give a more formal treatment of this technique.

It is interesting to study the ratio associated to a Bell polynomial
\begin{equation}
R(B)=\frac{\langle B\rangle_{QM}}{\langle B\rangle_{LR}},
\end{equation}
as it quantifies the strength of the inequality generated by the Bell operator $B$.
Note that a Bell inequality is characterised by the ratio $R(B)>1$. 
For example, for the CHSH inequality we have $R(B_{CHSH})=\sqrt{2}$.

Quantum states producing $R(B)>1$ are non-local in the sense that those ratios cannot be reproduced by considering a local hidden variable theory. As consequence, non-local quantum states cannot be fully separable. However, entanglement and non-locality are different concepts. Indeed, some entangled states do not violate any Bell inequality. Furthermore, states producing the maximal ratio are typically highly entangled \cite{Kar95}, but not necessarily maximally entangled \cite{Acin02}.

This chapter focuses on the study of this ratio, although more elaborated measures can be studied, like the p-value \cite{Hensen15} or the Kullback-Leibler relative entropy \cite{Vandam05}. 

\subsection{Three parties}\label{TP}

In the case of three qubits the most general symmetric Bell operator can be written as
\begin{eqnarray}\label{B3}
B_{3}&=& z_{0} (a\otimes b\otimes c)+z_{3} (a'\otimes b'\otimes c')+\nonumber \\ &&z_{1} (a\otimes b\otimes c'+a\otimes b'\otimes c+a'\otimes b\otimes c)+ \nonumber \\
 &&z_{2} (a\otimes b'\otimes c'+a'\otimes b\otimes c'+a'\otimes b'\otimes c),
\end{eqnarray}
where $z_0,\dots,z_3\in\mathbb{R}$. The following values for $z_i$ \cite{Moradi09}
\begin{equation}
z_{i}^M=\lbrace z_0,z_1,z_2,z_3 \rbrace^M = \lbrace 0,1,0,-1 \rbrace,
\end{equation}
lead us to the 3-qubit Mermin operator
\begin{eqnarray}
M_{3}&=& (a\otimes b\otimes c'+a\otimes b'\otimes c+a'\otimes b\otimes c)-\nonumber\\ &&(a'\otimes b'\otimes c'),
\label{mermin3}
\end{eqnarray}
having a square
\begin{eqnarray}\label{eq:C3a}
M_{3}^{2}&=&4\mathbb{I}_{ABC}-\bigl([a,a']\otimes [b,b']\otimes\mathbb{I}_C+\\
&&[a,a']\otimes\mathbb{I}_B\otimes [c,c']+\mathbb{I}_A\otimes [b,b']\otimes [c,c']\bigr).\nonumber
\end{eqnarray}
For brevity the symbols of the Kronecker product and identities are suppressed in every subsequent equation.
Eq. \ref{eq:C3a} allows us to obtain the classical value $\langle M_3\rangle_{LR}=2$ and the quantum value $\langle M_3\rangle_{QM}=4$, since each commutator can achieve a maximum absolute value of 2.

A different set of coefficients $z_{i}^S=\lbrace 1,1,-1,1\rbrace$ was proposed by Svetlichny \cite{Svetlichny87}. This choice leads to the form
\begin{eqnarray}\label{S_3}
S_{3}&=& (abc)+ (abc'+ab'c+a'bc) \nonumber \\
 && -(ab'c'+a'bc'+a'b'c) - (a'b'c'),
\end{eqnarray}
having the square form
\begin{eqnarray}
S_{3}^{2}&=&8-2\left([a,a'][b,b']+[a,a'][c,c']+[b,b'][c,c']\right)-\nonumber\\
&&\lbrace a,a'\rbrace \lbrace b,b'\rbrace \lbrace c,c'\rbrace.
\label{eq:C3b}
\end{eqnarray}
Note that this squared operator includes both commutators and anticommutators. For Pauli matrices $\lbrace\sigma_{i},\sigma_{j}\rbrace=2\delta_{ij}$, so a maximal value for the commutator implies a minimum value for the anticommutator, and vice versa. The commutators vanish and the anticommutators are maximum
while estimating the classical value and $\langle S_3\rangle_{LR}=4$. For the quantum value the optimal case occurs when the commutators take the maximum amplitude $\pm2$ and the anticommutators vanish, so that $\langle S_{3}\rangle_{QM}=4\sqrt{2}$. The ratios for the Bell operators of Eqs.  \ref{mermin3} and \ref{S_3} are given by
$R(M_3)=2\,\mbox{ and }\, R(S_3)=\sqrt{2}.$ The interesting point about inequality $S_3$ is not that it produces a maximum violation ratio, which it doesn't, but that it can detect genuine multipartite entanglement.
It is known that Mermin inequality generated by the Bell operator (\ref{mermin3}) can be violated by biseparable states, whereas Svetlichny inequality defined by the operator (\ref{S_3}) cannot. Bell inequalities generated by operators like $S_3$ are called \emph{multipartite Bell inequalities}. This topic is analysed in detail by Collins et al. \cite{Collins02b}.

These inequalities are already well tested experimentally. Violation of inequalities  $M_3$ and $S_3$ have been reported in Ref. \cite{Pan00} and \cite{Lavoie09}, respectively.

\subsection{Mermin polynomials}
\label{sec:mermin}

There exists an entire family of $n$-qubit inequalities first discovered by Mermin \cite{Mermin90,Werner01}. Here, we construct Mermin operators as in Ref. \cite{Collins02b}. Let us change the notation of observables $\{a,b,c...\} \equiv \{a_1,a_2,a_3...\}$, which is more convenient to treat the multipartite case. Defining $M_1 \equiv a_1$, the Mermin polynomials are obtained recursively as
\begin{equation}
M_n=\frac{1}{2} M_{n-1}(a_n+a'_n) + \frac{1}{2}M'_{n-1}(a_n-a'_n),
\label{generalmermin}
\end{equation}
where $M'_k$ is obtained from $M_k$ by interchanging primed and nonprimed observables $a_n$. In particular, $M_2$ and $M_3$ correspond to the operators \ref{eq:CHSH2} and \ref{mermin3}, respectively, up to a constant factor. It was proved in Ref. \cite{Cereceda01} that all Mermin operators have a square form composed by  the identity and commutators, as operators \ref{eq:CHSH2} and \ref{mermin3}. Let us now proceed with our version of the proof. The square of Mermin operators gives an expression containing commutators $[\cdot,\cdot]$ and anticommutators $\{\cdot,\cdot\}$
\begin{eqnarray}
M^2_n  = &&\frac{1}{4} \bigl(M^2_{n-1}(2+\{a_n,a'_n\})+M'^2_{n-1}(2-\{a_n,a'_n\}) \nonumber \\ 
&&-[M_{n-1},M'_{n-1}][a_n,a'_n]\bigr), 
\label{msquaredrec}
\end{eqnarray}
\begin{eqnarray}
M'^2_n = &&\frac{1}{4} \bigl(M'^2_{n-1}(2+\{a_n,a'_n\})+M^2_{n-1}(2-\{a_n,a'_n\}) \nonumber \\
&&-[M_{n-1},M'_{n-1}][a_n,a'_n]\bigr).
\end{eqnarray}
It is easy to see from here that, if $M^2_{n-1}=M'^2_{n-1}$, then $M^2_{n}=M'^2_{n}$. As this is true for $M^2_1=M'^2_1=1$, by induction it is true for every $n$. Therefore, Eq. \ref{msquaredrec} can be simplified to
\begin{equation}
M^2_n = M^2_{n-1} - \frac{1}{4} [M_{n-1},M'_{n-1}] [a_n,a'_n],
\end{equation}
where
\begin{equation*}
[M_{n-1},M'_{n-1}] = [M_{n-2},M'_{n-2}] + M^2_{n-2} [a_{n-1},a'_{n-1}].
\end{equation*}
Given that $[M_1,M'_1]=[a_1,a'_1]$, every operator $M^2_n$ can be expressed as a sum of products of an even number of commutators. Thus the operator $M^2_n$ reads,
\begin{equation}
M^2_n = 1 + \sum_{s=1}^{[n/2]} \frac{(-1)^s}{2^{2s}} \sum_{i_j \in D} \prod_{j=1}^{2s} [a_{i_j},a'_{i_j}],
\label{msquared}
\end{equation}
where $D$ is the set of $n$ operators taken in groups of $2s$ elements. This result is implicitly presented in Ref. \cite{Werner01}. The classical and quantum values arise immediately. On one hand, $\langle M_n\rangle_{LR}=1$, as the second term in Eq. \ref{msquared} is always zero due to the presence of commutators. On the other hand, for the quantum value every commutator takes $\pm2$, the sign conveniently chosen to maximize it. Thus,
\begin{equation}
\langle M^2_n\rangle_{QM} = 1 + {{n}\choose{2}} + {{n}\choose{4}}+... = 2^{n-1}.
\label{msquaredquant}
\end{equation}
The quantum value for $M_n$ is, therefore, $\langle M_n\rangle_{QM}=\sqrt{\langle M^2_n\rangle_{QM}}=2^{\frac{n-1}{2}}$,
which matches the rate computed by Werner and Wolf \cite{Werner01}.
Let us note that when computing this last step it is assumed that the maximum eigenvalue of a sum of matrices is equal to the sum of the maximum eigenvalues, a fact that is not true in general but is true in this case.

The optimal states for the Mermin inequalities are the GHZ states \cite{Mermin90,Werner01}. 
For $n=2$ and $n=3$ these states can be considered as maximally entangled.
However, for $n\ge 4$ it is not the case, as we saw throughout Chapter \ref{ch:entanglement}. Therefore, the Mermin inequalities 
provide an example for which the maximal violation does not correspond to maximally entangled states. Let us mention that the experimental violation of Mermin inequalities has been verified up to 14 qubits with ion traps \cite{Lanyon14}. Recently, we implemented $M_3, M_4$ and $M_5$ cases on a 5 superconducting qubits quantum computer designed by IBM, a project that will be described in Chapter \ref{ch:ibm}.
\section{Bell inequalities for three outcomes} 
\label{sec:bellqutrits}
In this section we study Bell inequalities for three outcomes and their maximum violations in the cases of hermitian and unitary setting operators. We remark that all the maximal violations presented for Bell inequalities and having three outcomes have been found for qutrit states. Therefore, they are lower bounds for the maximal possible quantum value which, in principle, could be attained for qudits with more than three number of levels each.

\subsection{Two parties with hermitian operators}
A Bell inequality for two parties, two settings and $d$ outcomes was proposed by Collins et al. \cite{Collins02} and it is known as CGLMP inequality. The violation of some of these inequalities has been verified experimentally \cite{Vaziri02}. In the case of three outcomes the inequality is given by
\begin{eqnarray}
&p(a=b)+p(b=a'+1)+p(a'=b')+& \nonumber \\
&p(b'=a)-p(a=b-1)-p(b=a')-\nonumber \\
&p(a'=b'-1)-p(b'=a-1) \leq 2,&
\label{CGLMP}
\end{eqnarray}
where the possible outcomes are $\{0,1,2\}$ and the sum inside probabilities is modulo $d=3$. This Bell inequality can be associated with the following Bell operator 
\begin{eqnarray}
C_{223}&=&2-3(a^2+b'^2)+\frac{3}{4}(ab+a^2b-a'b-a'^2b-ab^2+\nonumber\\
&&a'b^2+ab'-a^2b'+a'b'+a'^2b'+ab'^2-a'b'^2)+\nonumber\\
&&\frac{9}{4}(a^2b^2-a'^2b^2+a^2b'^2+a'^2b'^2),
\label{cglmpreal}
\end{eqnarray}
where the notation $C_{nsd}$ stands for $n$ parties, $s$ settings and $d$ outcomes. The optimal settings can be obtained by choosing one arbitrary setting and obtaining the other one with a phase transformation followed by the Fourier transform, as discussed extensively in Ref. \cite{Collins02}. The quantum value is given by $\langle C_{223}\rangle_{QM}=2(5-\gamma^2)/3\approx2.9149$ for the optimal state $|\psi\rangle= (|00\rangle+\gamma|11\rangle+|22\rangle)/\sqrt{(2+\gamma^2)}$ where $\gamma =(\sqrt{11}-\sqrt{3})/2 \approx 0.7923$ \cite{Acin02}. The violation rate for this quasi GHZ state reads $R_{2t}=(5-\gamma^2)/3 \approx 1.4547$. In Ref. \cite{Acin02} the ratios for CGLMP inequalities are found up to $d=8$ levels. The optimal settings can be conveniently expressed in terms of the eight Gell-Mann matrices $\lambda_i$, 
the traceless generators of SU(3) \cite{Gellmann64}.
The optimal settings for the Bell inequality generated by the operator \ref{cglmpreal} are
\begin{eqnarray}
A=B&=&\lambda_3, \nonumber \\
A'=B'&=&\frac{2}{3} (\lambda_1+\lambda_6) + \frac{1}{6} (\lambda_3+\sqrt{3}\lambda_8) = \frac{2\sqrt{2}}{3} J_1 + \frac{1}{3} J_3.
\label{gellmann2qt}
\end{eqnarray}
where $J_1$ and $J_3$ are two elements of the representation of SU(2) in three dimensions. 

The Bell operator in Eq. \ref{cglmpreal} has a rather long and unenlightening form. In the next subsection we will show how the consideration of unitary setting operators instead of hermitian operators simplifies the study of these Bell inequalities.

\subsection{Two parties with unitary operators}

A more convenient way to represent Bell inequalities for three outcomes is by considering complex outcomes associated to the third roots of unity \cite{Zukowski97,Chen02,Arnault12}. In this way, settings turn from hermitian to unitary operators with eigenvalues $\{1,w,w^2\}$, where $w=\exp(2\pi i/ 3)$. Note that for qubits the Pauli matrices are both hermitian and unitary, while for qutrits a choice between one of these properties has to be made. Note that any operator that can be expressed as a linear combination 
(with real or complex coefficients) of rank one projectors forming a POVM allows for a physical interpretation. Note also that sum of unitary operators is in general, not a normal operator. 
A complex operator $M$ is normal if $[M,M^{\dag}]=0$. However, any operator 
can be decomposed into its \emph{hermitian} 
and \emph{anti-hermitian} part, $B=[B]_H +i [B]_A$, where
$[B]_H:= \frac{1}{2}(B+B^\dagger)$ and
$[B]_A:= \frac{1}{2i}(B-B^\dagger)$ are hermitian operators and, therefore, they have real eigenvalues.

The Bell operator \eqref{cglmpreal} can be written as
the anti-hermitian part of a non-hermitian operator,
\begin{equation}
C_{223}=  \left[ a(w b-b')+a'(wb'-b) \right]_A.
\label{cglmpim}
\end{equation}
This form appears to be a direct generalization of the CHSH operator \eqref{chshbrackets}, with different signs and relative phases added. If one of the terms reaches the maximum value $\sqrt{3}$ then the other one is forced to be zero. The classical and quantum values for this operator are $\langle C_{223}\rangle_{LR} = \sqrt{3}\approx 1.73$ and $\langle C_{223}\rangle_{QM}  = (1/2)(\sqrt{3}+\sqrt{11})\approx 2.52$, and the ratio is given by $R(C_{223}) = (1/3)(5-\gamma^2)\approx 1.45$. The violation rate is therefore the same as for CGLMP inequality \eqref{CGLMP} as expected, because it is the same inequality albeit written in a different language. Let us now find the optimal settings for the operator (\ref{cglmpim}). The convenient representation for unitary operators are the generalized unitary Pauli matrices which form the Weyl-Heisenberg group. The generators of the group are
\begin{equation}\label{XZ}
X = \left( \begin{array}{ccc}
0 & 0 & 1 \\
1 & 0 & 0 \\
0 & 1 & 0 \end{array} \right)\hspace{0.5cm}\mbox{and}\hspace{0.5cm}
Z = \left( \begin{array}{ccc}
1 & 0 & 0 \\
0 & w & 0 \\
0 & 0 & w^2 \end{array} \right),
\end{equation}
where $\omega=e^{2\pi i/3}$. An orthonormal basis is given by the nine elements
\begin{equation}
X^k Z^j = \sum_{m=0}^2 |m+k\rangle w^{jm} \langle m|\, ,
\label{genpauli}
\end{equation}
which are proportional to the elements of the Weyl-Heisenberg group. By numerical optimization it is possible to show that the optimal settings for the operators \eqref{cglmpim} are
\begin{eqnarray}
&A=B=X,&\nonumber \\
&A'=B'=\frac{1}{3} (- X + 2w XZ + 2w^2 XZ^2).&
\label{complexap}
\end{eqnarray}
In matrix notation, $A'$ has a simple structure
\[ A' = \left( \begin{array}{ccc}
0 & 0 & 1 \\
-1 & 0 & 0 \\
0 & -1 & 0 \end{array} \right).\] 
The optimal settings for all the complex CGLMP inequalities, in this case ($\{X,A'\}$), are called \emph{multiplets of optimal settings} (MOS). In Appendix \ref{mubs} some properties of MOS are discussed.

Let us investigate the square of the operator $C_{223}$ introduced in Eq. \ref{cglmpreal}. Making use of this identity for the hermitian and antihermitian
 parts of an operator $C$
\begin{equation}
%(\textrm{Im}(C))^2 = \frac{1}{4}(CC^\dagger + C^\dagger C) - \frac{1}{2} \textrm{Re}(C^2) \, ,
 (C_A)^2 = \frac{1}{4}(CC^\dagger + C^\dagger C) - \frac{1}{2} (C^2)_H \, ,
\label{cglmpim2}
\end{equation}
it is easy to show that $C_{223}C_{223}^\dagger$ has an interesting structure
\begin{equation}
C_{223}C_{223}^\dagger = 3 + (1+\{\{a,a'\}\})(1+\{\{b,b'\}\}) .
\label{ccdagger}
\end{equation}
Here  $\{\{a,a'\}\}$ is called the \emph{complex anticommutator} $\{\{a,a'\}\}= aa'^\dagger + a'a^\dagger$. The complex anticommutator attains its maximum value 2 both for MOS and MUB (see appendix \ref{mubs} for a definition of these pairs of matrices). 
However, its classical value can also be equal to 2 by using $a=a'=1$. Thus the form in Eq. \ref{ccdagger} does not allow us to distinguish between classical and quantum values.

\subsection{Three parties}

A three parties Bell inequality was proposed by Ac\'in et al. in Ref. \cite{Acin04}. 
In the probability formalism it reads 
\begin{eqnarray}
&p(a+b+c=0) + p(a+b'+c'=1) +&\nonumber \\
&p(a'+b+c'=1)+p(a'+b'+c=1)-&\nonumber \\
&2p(a'+b'+c'=0) - p(a'+b+c=2) -& \nonumber \\
& p(a+b'+c=2)- p(a+b+c'=2) \leq 3.&
\label{3qtprob}
\end{eqnarray}
The analysis here is very similar to the CGLMP case: the maximal violation is given by a quasi maximally entangled state $|\psi\rangle= (|000\rangle+\gamma |111\rangle+|222\rangle)/\sqrt{2+\gamma^2}$ where now $\gamma\approx1.186$. The quantum value is $4.37$ and the violation rate is $R=(5-\gamma^2)/3 \approx 1.4574$, as for 2 qutrits. The corresponding hermitian Bell operator has a rather long form, so we will not reproduce it here. The optimal settings can be expressed in terms of the Gell-Mann matrices as
\begin{eqnarray}
&&A=B=C=\lambda_3,\nonumber \\
&&A'=B'=C'=\frac{1}{\sqrt{3}} (\lambda_2+\lambda_4+\lambda_6) \, .
\end{eqnarray}
Let us now consider the case of unitary settings having complex eigenvalues. The Bell operator associated to inequality \ref{3qtprob} can be expressed this hermitian part of an operator
\begin{eqnarray}
C_{333}&=&\mathbb{I}+\frac{2}{3}  \bigl[abc +2a'b'c' +w(a'b'c+a'b c'+ab'c') \nonumber \\
&&-w^2(a'b c+a b'c +a b c')\bigr]_H.
\end{eqnarray}
One can also drop the additive and multiplicative terms and study the simplified operator
\begin{eqnarray}
C'_{333}&=&  \bigl[abc +2a'b'c' +w(a'b'c+a'b c'+ab'c') \nonumber \\
&&-w^2(a'b c+a b'c +a b c')\bigr]_H.
\label{3qutrits}
\end{eqnarray}
Here, the classical value is $\langle C'_{333}\rangle_{LR}=3$ and the quantum value is $\langle C'_{333}\rangle_{QM}=(3/4)(1+\sqrt{33})\approx5.058$, which yields to the ratio $R(C'_{333}) = (1/4)(1+\sqrt{33})\approx1.686$. The optimal settings are given by
\begin{eqnarray}
&A=B=C=X,& \nonumber \\
&A'=B'=C'=Z.&
\end{eqnarray}
Note that the settings are mutually unbiased (see Appendix \ref{mubs}). Now the violation rate is greater because the additive constant term has been eliminated. This appears somewhat arbitrary but it is more convenient to compare  inequalities  for different number of parties without additive terms. In this way, it is expected that the rate of violation increases with the number of parties, as it happens for qubits. Intriguingly, the 3-qutrit operator \eqref{3qutrits} can be derived from the 2-qutrit CGLMP operator \eqref{cglmpim} by adding a third party such that the resulting 3-qutrit operator is symmetric, as shown in Appendix \ref{2to3}.

\subsection{Larger number of parties}
In the case of four parties, two settings and three outcomes we have found the following symmetric Bell operator
% expressed here by anti-hermitian part of an operator 
%
\begin{eqnarray}
&&C_{423}=  \bigl[2(abcd) + (a'bcd+ab'cd+abc'd+abcd')   \nonumber \\
&&+w(a'b'cd+a'bc'd+a'bcd'+ab'c'd+ab'cd'+abc'd') \nonumber \\
&&+(a'b'c'd+a'bc'd'+a'b'cd'+ab'c'd') +2 (a'b'c'd')\bigl]_A,\nonumber \\
\label{4qt}
\end{eqnarray}
which produces $\langle C_{423}\rangle_{LR}=3\sqrt{3}\approx5.19$, $\langle C_{423}\rangle_{QM}\approx 9.766$ and $R(C_{423})\approx 1.879$ for the optimal settings
\begin{eqnarray}
&&A=B=C=D=X, \nonumber \\
&&A'=B'=C'=D'=Z,
\end{eqnarray}
which are again mutually unbiased settings. In this case the optimal state has entanglement properties equivalent to those of the exact GHZ of four parties and three settings $|GHZ_{4,3}\rangle=(|0000\rangle+|1111\rangle+|2222\rangle)\sqrt{3}$.

For 6 parties we have also found a symmetric Bell operator. To simplify the notation, the polynomials having terms with the same number of primes are denoted by its number of primes in parenthesis, for example:
$(1') \equiv a'bcdef+ab'cdef+abc'def+abcd'ef+abcde'f+abcdef'$. In this notation, the 6 parties operator reads
\begin{equation}
C_{623}=-w(0')+(1')-(2')+w(3')-(4')+(5')-w(6').
\label{primenotation}
\end{equation}

For this inequality, $\langle C_{623}\rangle_{LR}=9\sqrt{3}\approx15.589$, $\langle C_{623}\rangle_{QM}\approx32.817$ and $R(C_{623})\approx2.105$, with MOS optimal settings. The maximal violation is a given by a \emph{quasi} GHZ state, as for the case of 2 and 3 qutrits. \\

Let us summarize the results for the symmetric Bell operators for $n$-qutrit systems studied in this section. Unfortunately, we could not find a 5-qutrit inequality that follows all the patterns. The inequalities considered are those determined by the coefficients of Table \ref{tabcoefficients}, and the results are summarized in Table \ref{tabresults}.

\begin{table}[h!]
\centering
\begin{tabular}{c | c | c | c | c | c}
\hline
\backslashbox{Terms}{Parties} & 2 & 3 & 4 & 5  & 6 \\
\hline
(0') & $\omega$ & $1$    & $2$   & $\omega^2$  & $-\omega$ \\
(1') & $1$ & $-\omega^2$ & $1$   & $-\omega^2$ & $1$ \\
(2') & $\omega$ & $\omega$    & $\omega$   & $-\omega^2$ & $-1$ \\
(3') &     & $2$    & $1$   & $-\omega^2$ & $\omega$ \\ 
(4') &     &        & $2$   & $\omega^2$  & $-1$ \\  
(5') &     &        &       & $\omega^2$  & $1$ \\
(6') &     &        &       &        & $-\omega$ \\
\hline
\end{tabular}
\caption{Coefficients for symmetric Bell inequalities  from two to six parties and two settings and three outcomes, where $\omega=e^{2\pi i/3}$. The primed notation $(k')$ identifies all terms having $k$ primed settings, as defined before in Eq. \ref{primenotation}.}
\label{tabcoefficients}
\end{table}

\begin{table}[h!]
\begin{tabular}{c | c | c | c | c | c}
\hline
\backslashbox{}{\hspace{2mm}Qutrits} & 2 & 3 & 4 & 5 & 6 \\
 \hline
$\langle [B]_A\rangle_{LR}$ & \boldmath{$\sqrt{3}$} & $3\sqrt{3}$ & \boldmath{$3\sqrt{3}$} & $9\sqrt{3}$&  \boldmath{$9\sqrt{3}$}\\
$\langle [B]_A\rangle_{LR}^{(-)}$ & $-2\sqrt{3}$ & $-3\sqrt{3}$ & $-6\sqrt{3}$ & $-9\sqrt{3}$ & $-18\sqrt{3}$\\
$\langle [B]_H\rangle_{LR}$ & $3$ & \boldmath{$3$} & $9$ & \boldmath{$9$} & $27$\\
$\langle [B]_H\rangle_{LR}^{(-)}$ & $-3$ & $-6$ & $-9$ & $-18$ & $-27$ \\
$\langle [B]_x\rangle_{QM}$ & $ 2.524 $ & $5.058$ & $9.766$ & $\mathit{15.575}$ & $32.817$ \\
$\mathrm{R}$ & $1.457$ & $1.686$ & $1.879$ & $\mathit{1.731}$ & $2.105$ \\
$Settings$ & MOS & MUB & MUB & \it{Num.} & MOS \\
P & $0.347$ & $0.342$ & $1/3$ & $\mathit{0.351}$ & $0.334$\\ 
\hline
\end{tabular}
\caption{Main results for inequalities from 2 to 6 qutrits, where it can be seen that the classical patterns match perfectly, while the 5-qutrit inequality appears not to follow the quantum pattern. Here, $\langle B\rangle_{LR}$ and $\langle B\rangle_{LR}^{(-)}$ denote the maximum and minimum classical value for optimizations of anti-hermitian or hermitian part of the operator, respectively. The quantity that we take as the extremal classical bound is marked in bold, and
$\langle [B]_x\rangle_{QM}$ stands for its corresponding quantum value, where $x=A$ for an even number of qutrits and $x=H$ for an odd number of qutrits. $R=\langle B\rangle_{QM}/\langle B\rangle_{LR}$ and \emph{Settings} denotes the optimal settings. $P$ denotes the purity of the $\lfloor n/2 \rfloor$ party reductions of the optimal state and \emph{Num.} means numerical approximate solution, and italic font in the 5-qutrits case is written to note that this case does not follow the same patterns of the others. We remark that optimal values appearing in this table have been achieved by optimizing over qutrit systems.}
\label{tabresults}
\end{table}

The main patterns that can be seen in Table \ref{tabresults} are
\begin{itemize}
\item[\emph{(i)}] 
For an even number of qutrits 
the classical values $\langle B\rangle_{LR}$ arise from the anti-hermitian part of an operator 
while for odd number of qutrits one takes its hermitian part.
%and with the real part for an odd number of qutrits (bold).
The following relation between the minimal and the maximal classical values
holds, $\langle B^-\rangle_{LR}=-2\langle B\rangle_{LR}$.

\item[\emph{(ii)}] There is a factor of $\sqrt{3}$ between the maximum value of the 
hermitian and anti-hermitian parts, 
and also a factor of $\sqrt{3}$ between the maximal value of two consecutive numbers of qutrits. The maximal value of the hermitian parts are the same for $n$ and $n+1$ qutrits if $n$ is even. Also, the maximal value of the anti-hermitian parts are the same for $n$ and $n+1$ if $n$ is odd.

\item[\emph{(iii)}] The quantum value $\langle B\rangle_{QM}$ of a non-hermitian operator $B$ is computed as the maximum over quantum values of the hermitian and anti-hermitian parts, i.e.,  $\langle B\rangle_{QM}=\mathrm{Max}\{\langle B_H\rangle_{QM},\langle B_A\rangle_{QM}\}$. The rate of violation increases with the number of qutrits except for the 5-qutrit case, which do not follow the patterns.

\item[\emph{(iv)}] The optimal settings are either MUB or MOS, with the exception of the 5-qutrit inequality.

\item[\emph{(v)}] The optimal states have entanglement properties close to a GHZ or exactly those of a GHZ state in the case of four qutrits. In Table \ref{tabresults} the closeness to the GHZ state is measured by the purity $P$ of the reduced matrix $\sigma$ over $\lfloor n/2 \rfloor$ particles. The GHZ state of $n$ qutrits has reductions to two parties with $P={\rm Tr} \sigma^2=1/3$, 
whereas the absolutely maximally entangled state has $P=1/3^{[n/2]}$ 

\end{itemize}

\section{Mapping states to Bell operators}\label{sec:bellmapping}

Let us now present a novel idea to generate Bell inequalities based on a mapping from maximally entangled states to Bell operators. We shall illustrate the construction through an example and, then, generalize it to different cases. 

The two-qubit state
\begin{equation}
|\psi \rangle = (|+\rangle\otimes\,|0\rangle +|-\rangle\otimes\,|1\rangle)/\sqrt{2},
\end{equation}
where $|\pm\rangle=\sqrt{1/2}\bigl(|0\rangle\pm|1\rangle\bigr)$, can be expanded to match the form
\begin{equation}\label{mes}
| \psi \rangle = \frac{1}{2} \left( |0_A0_B\rangle + |0_A1_B\rangle + |1_A0_B\rangle - |1_A1_B\rangle \right).
\end{equation}
This state belongs to the set of maximally entangled  Bell states.
The CHSH Bell operator can be obtained from this state by identifying first and second party with observables for Alice and Bob, respectively. We identify symbol $0$ with non-primed settings and symbol $1$ with primed settings, as in Table \ref{legend}.
\begin{table}[h!]
\centering
\begin{tabular}{c  c  c}

   $| \psi \rangle$ & $\rightarrow$ & $\mathcal{B}$ \\
\hline
   $| 0_A \rangle$ & $\rightarrow$ & $a$ \\
   
   $| 1_A \rangle$ & $\rightarrow$ & $a'$ \\
   
   $| 0_B \rangle$ & $\rightarrow$ & $b$ \\
   
   $| 1_B \rangle$ & $\rightarrow$ & $b'$ \\

\end{tabular}
\caption{Substitution legend for mapping states to Bell operators for the CHSH case.}
\label{legend}
\end{table}

By removing the normalization term the CHSH operator arises
\begin{equation}
\mathcal{B}_{CHSH} = ab + ab' + a'b - a'b' \, .
\end{equation}
Furthermore, the maximally entangled state (\ref{mes}) is the optimal state for a suitable choice of the measurement settings. This remarkable fact motivates us to study new multipartite Bell inequalities generated from multipartite states.

\subsection{Bell inequalities from entangled states}\label{BIFES}

The general strategy is to construct Bell inequalities associated 
to some distinguished maximally entangled states.
Starting from the Bell state for two qutrits,
$|\psi_3^+\rangle=(|00\rangle+|11\rangle+|22\rangle)/\sqrt{3}$
 and applying the Fourier transform to the second party we obtain 
\begin{equation}\label{IF_GHZ}
|\phi\rangle=\mathbb{I}\otimes F_3|\psi_3^+\rangle.
\end{equation}
From this state, using legend from table \ref{legend} and adding the case $|2_A \rangle \rightarrow a''$ and analogously for party B, a new Bell operator for 2 qutrits and 3 settings arises,
\begin{equation}
C_{233} = [\,\vec{a}\cdot F_3\vec{b}\,]_H,
\end{equation}
where $\vec{a}=(a,a',a'')$, $\vec{b}=(b,b',b'')$ and $F_3$ is the Fourier matrix of order three, $(F_3)_{jk}=e^{2\pi ijk/3}$. This operator has a classical value $\langle C_{233}\rangle_{LR}=9/2$ and it is maximally violated by a state with the same entanglement properties of the GHZ with a violation ratio $R(C_{233})=2/\sqrt{3} \cos (\pi/18)\approx1.137$ for the optimal MUB settings
\begin{eqnarray}
&&A=B=X , \nonumber \\
&&A'=B'=Z , \nonumber \\
&&A''=B''=X^2 Z^2 ,
\end{eqnarray}
where $X$ and $Z$ are given in Eq. \ref{XZ}. An equivalent inequality with the same properties was found in Refs. \cite{Ji08, Liang09}.

We can apply the same strategy for four qutrits 
starting with the GHZ state 
$|GHZ_4^3\rangle=(|0000\rangle + |1111\rangle +|2222\rangle)/\sqrt{3}$. 
Acting with Fourier transform $F_3$ on three parties we obtain a locally
equivalent state
\begin{equation}\label{IFFF_GHZ}
|GHZ_4^{3'}\rangle=\mathbb{I}\otimes F_3\otimes F_3\otimes F_3|GHZ_4^3\rangle,
\end{equation}
which leads to the Bell operator
\begin{equation}
C'_{433} = [\,\vec{a}\cdot F_3\vec{b}\cdot F_3\vec{c}\cdot F_3\vec{d}\,]_H,
\end{equation}
where $\vec{a}=(a,a',a'')$, $\vec{b}=(b,b',b'')$, and analogously for other parties. The generalized inner product of four vectors is defined as $w\cdot x\cdot y\cdot z=\sum_{j=0}^{2} w_j x_j y_j z_j$. The optimal state has the entanglement properties of the GHZ,
but with a larger violation ratio than for the operator \ref{4qt}. 

\subsection{Bell inequalities from AME state}

Let us now try the strategy above described for the AME of 4 qutrits studied in previous chapter
\begin{equation}\label{AME43}
AME(4,3)=\frac{1}{9}\sum_{i,j,k,l=0}^2 w^{j(i-k)+l(i+k)}|ijkl\rangle.
\end{equation}

% for a detailed discussion about AME states. 
The recipe to construct the Bell operator consists in taking representation (\ref{AME43})
which contains $3^4=81$ terms with coefficients of the form $\{1,w,w^2\}$.
In the next step one uses the same legend from previous subsection. This procedure 
leads us to a Bell operator for four parties, three settings and three outcomes, 
which can be written in a compact way as
\begin{equation}
C_{433} = \sum_{i,j,k,l=0}^2 w^{j(i-k)+l(i+k)}a_{i}b_{j}c_{k}d_{l},
\label{c433}
\end{equation}
where $a_0 = a, a_1 = a', a_2 = a''$, and the same for the rest of the observables.

After transformations $d' \rightarrow wd'$ and $d' \leftrightarrow d''$, 
numerical optimization produces the following configuration of optimal settings 
\begin{eqnarray}
&&A=B=C=D=X \, , \nonumber \\
&&A'=C'=D'=X^2Z^2 \hspace{5mm} B'=X, \nonumber \\
&&A''=C''=D''=Z \hspace{9mm} B''=N,
\end{eqnarray}
where $N$ is a certain matrix of size three obtained numerically. 
The optimal settings are not symmetric because the AME state 
is not symmetric under interchange of particles.

Numerical optimization suggests that the optimal state is not AME. 
Surprisingly, it has almost the same entanglement properties as the GHZ state, namely its purity is $P=1/3$ for the density matrices of reductions to 2 parties, and $P=1/3$ for three of the possible reductions to one party, while the fourth one (party B) has $P=1$, indicating that party B is in a product state with the other three. The same violation ratio as for four qutrits with two settings is obtained, see Eq. \ref{4qt}. 
This result, and the fact that the optimal settings include $B=B'$ 
suggests that the third setting is not adding anything new 
and that this inequality is essentially the same as in the case of two settings.

Table \ref{tab3settings} summarizes the results for the 3-settings qutrit inequalities arising 
from entangled states.

\begin{table}[h!]
\begin{tabular}{c | c | c | c }
\hline
\backslashbox{}{\hspace{2mm}Qutrits} & 2 & 4 (GHZ) & 4 (AME)\\
\hline
$\langle [B]_A\rangle_{LR}$ & $3\sqrt{3}$ & $9\sqrt{3}$ & $9\sqrt{3}$ \\
$\langle [B]_A\rangle_{LR}^{(-)}$ & $-3\sqrt{3}$ & $-9\sqrt{3}$ & $-9\sqrt{3}$\\
$\langle [B]_H\rangle_{LR}$ & \boldmath{$4.5$} & \boldmath{$13.5$} & \boldmath{$13.5$} \\
$\langle [B]_H\rangle_{LR}^{(-)}$ & $-4.5$ & $-27$ & $-27$ \\
$\langle [B]_H\rangle_{QM}$ & $ 5.117 $ & $26.025$ & $25.372$ \\
R & $1.137$ & $1.928$ & $1.879$\\
$Settings$ & MUB & Num. & MUB and Num. \\
P & 1/3 & 1/3 & 1/3 \\
\hline
\end{tabular}
\caption{Characterization of Bell inequalities for 2 and 4 parties, 3 settings and 3 outcomes. There is one 4-qutrit inequality built from the GHZ state and another one built from the AME state. For all the cases the optimal states are states with the same entanglement properties as the GHZ. Abbreviations and symbols are considered as in Table \ref{tabresults}, although in this case the quantum bound is computed always with the hermitian part.}
\label{tab3settings}
\end{table}

\section{Concluding remarks}

We have used the formalism of unitary matrices with complex roots of unity as eigenvalues to construct Bell inequalities of multipartite systems, 2 settings and 3 outcomes (see Section \ref{sec:bellqutrits}). We have shown that the 2-party and 3-party inequalities from Ref. \cite{Collins02} and Ref. \cite{Acin04} are closely related. Furthermore, we have extended these cases to 4 and 6 parties and, less convincingly, to 5 parties. We obtained regular patterns for this set of inequalities, as shown in Table \ref{tabresults}. Two of the most striking patterns are: \emph{a}) the structure of the classical bounds and their simple arithmetic progression with the number of particles, and \emph{b}) the fact that the inequalities tend to have a maximal quantum bound for settings that are either MUBs or multiplets of optimal settings (MOS) (see Appendix \ref{mubs}).

We also introduced a mapping from entangled states to Bell operators that allows us to define Bell inequalities for multipartite systems (see Section \ref{BIFES}). In particular, we have constructed new inequalities for two and four parties with three settings, which are maximally violated by states with the same entanglement properties as the GHZ state.  We also demonstrated that a Bell inequality generated by a given quantum state is not necessarily maximally violated by the same state. For example, the inequality from Eq. \ref{c433} is generated by the absolutely maximally entangled state of 4 qutrits, but maximally violated by a GHZ-like state. This novel formalism has the potential to generate a wide range of Bell inequalities for an arbitrarily large number of parties, settings and outcomes.

Let us also mention here some important questions, which remain open. Concerning the approach to Bell inequalities from squares of operators represented by commutators it would be interesting to find a procedure to determine whether a given Bell operator allows such a form. Analyzing the mapping between states and Bell operators one can raise the question of whether a maximally entangled state is necessary to produce a tight Bell inequality in the case of 2 outcomes (e.g. it holds for the CHSH and all Mermin inequalities). On the other hand, the mathematical characterization of the entire set of MOS for the CGLMP inequalities defined in Appendix \ref{mubs} is a pending task. Finally, it would be interesting to have a generating polynomial for Bell inequalities with 3 outcomes, in the same way that we have the Mermin polynomials for Bell inequalities with 2 outcomes (see Eq \ref{generalmermin}).

\chapter{Distance between theories from entanglement spectra}

\pagestyle{fancy}
\fancyhf{}
\fancyhead[LE]{\thepage}
\fancyhead[RE]{MULTIPARTITE ENTANGLEMENT}
\fancyhead[LO]{CH.4 Distance between theories from entanglement spectra}
\fancyhead[RO]{\thepage}

\section{Introduction}

So far we have considered entanglement of conveniently defined states, without considering how these states might be generated. Instead, in the last two chapters of the first part we are going to study entanglement of ground states of certain hamiltonians of interest. 

In this chapter we are going to analyse quantitatively the difference between quantum states. It can be associated with a notion of distance between their state vectors or density matrices. There has been a huge number of proposals of distance measures in order to study quantum distinguishability of states, since a seminal paper by Wooters \cite{Wooters81}. Applying the ideas of Fisher information metric from classical probability distributions, the Fubini-Study metric and the Bures metric have been widely used for bipartite pure and mixed states respectively. The parameters of the Bloch sphere as a distance element were used  and the metric was built thereafter.

In multipartite systems this procedure is more complicated, as the number of parameters needed to describe the state grows exponentially with the size of the system and therefore a notion of distance based on them is less and less useful. But an idea was suggested in Ref.  \cite{Zanardi06} to change the focus from the state to the Hamiltonian having it as the ground state. Then we can work with just one or very few parameters, and we can use them as a distance element to compare states with a huge number of particles. Only those states that are ground states of simple Hamiltonians will be considered, but these are the physically most relevant so we really have a gain. In that paper, they proposed the overlap $\braket {\phi (\lambda)} {\phi (\lambda + d\lambda)}$ as the distance function. In the world of mixed states, this would correspond to the quantum fidelity $\Tr \left( \sqrt {\sqrt{\rho} \sigma \sqrt{\rho} } \right)$. Other distance measures such as an adaptation of the Bures metric for this case \cite{Zanardi07} or a Chernoff-bound based distance \cite{Calsamiglia08} have appeared since.

We are going to propose an alternative using entanglement. We are going to use as a distance function the entanglement spectrum (a presentation of the Schmidt decomposition analogous to a set of 'energy levels'), which has already been used to detect topological order \cite{Li08} between states and quantum phase transitions.

In section \ref{sect:RG} we build the theoretical ground for using the entanglement spectrum as a distance measure. Then we use it to compute distances between various models: free fermions in section \ref{sect:FF}, Ising model converted to free fermions by a Jordan-Wigner transformation in section \ref{sect:PT}, Ising and Heisenberg models in finite spin chains in section \ref{sect:FSC}, its infinite counterparts using the available analytical solutions in section \ref{sect:ISC} and finally an application to a 2D system in section \ref{sect:2D}.

\section{Equivalence and distance}\label{sect:RG}
%%%%%%%%%%%%%%%%%%%%%%%%%%

Let us first introduce some concepts that will be needed for the theoretical discussion.

A \emph{matrix product state} (MPS) \cite{Perez07} is a representation of a pure quantum state written in this form
\beq
\ket{\psi} = \sum_{s} \Tr [A_1^{(s_1)} A_2^{(s_2)} ... A_n^{(s_n)}] \ket{s_1 s_2 ... s_n},
\eeq
where $A_i^{(s_i)}$ are complex square matrices of order $\chi$, the bond dimension. It is the simplest of the so called \emph{tensor network} states, with the general aim to reduce drastically the amount of information needed to describe a state.  The parameter $\chi$ controls the precision of the approximation. We will investigate a bit more tensor networks in the next chapter, but for now this is what we need for the present discussion.

The \emph{Wightman axioms} \cite{Streater64} are an attempt at a rigorous mathematical formulation of quantum field theory. \emph{Wightman fields} are operator-valued distributions satisfying the Wightman axioms, and \emph{Wightman functions} are the correlator functions of Wightman fields.
  
\begin{figure}[h]
\begin{center}
\includegraphics[width=100mm,height=30mm]{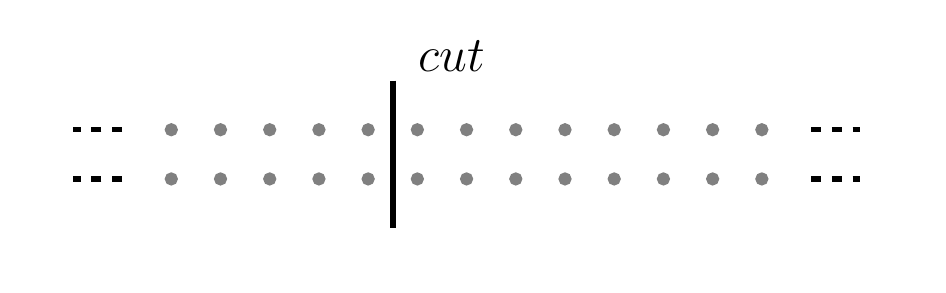} 
\caption{Two spin chains with the same Schmidt spectrum for any bipartition.}\label{FIG:BIPARTITION}
\end{center}
\end{figure}

Let us consider two matrix product states $\ket{\Psi},\ket{\Psi'}$ defined on an infinite one-dimensional lattice. We will assume these states are translationally invariant. These two hypotheses are actually not necessary to our discussion, but they will simplify it. Besides, the local Hilbert spaces of $\ket{\Psi}$ and $\ket{\Psi'}$ need not match. We wish to discuss the meaning of the situation where  $\ket{\Psi}$ and $\ket{\Psi'}$ have the same Schmidt spectrum for any bipartition of the chain (see Fig.\ref{FIG:BIPARTITION}), that is
\begin{equation}\label{eq:state-equivalence}
\ket{\Psi}=\sum_{\alpha=1}^{\chi} \lambda_{\alpha}^{1/2} \ket{\ell_\alpha} \ket{r_\alpha}, \hspace{0.7cm}
\ket{\Psi'}=\sum_{\alpha=1}^{\chi} \lambda_{\alpha}^{1/2} \ket{\ell'_\alpha} \ket{r'_\alpha}.
\end{equation}

We are going to show that there is a fixed local unitary correspondence between the Wightman functions of $\Psi$ and those of $\Psi'$, whereas if two states $\Psi$ and $\Psi^{\#}$ have different entanglement spectra $\Lambda$ and $\Lambda^{\#}$, the difference between their Wightman functions is dictated by the distance between $\Lambda$ and $\Lambda^{\#}$.

We will argue that, when looked at sufficiently large scale, $\ket{\Psi}$ and $\ket{\Psi'}$ can be regarded as essentially equivalent quantum states; in good approximation ${\Psi}$ and ${\Psi'}$ only differ by local changes. Barring degenerate cases, all states with the same Schmidt spectrum can be gathered into an equivalence class, represented by a particularly simple element. 

Let us first focus on the first of these two states
\[
\ket{\Psi}= \lim_{n \to \infty} \sum_{s_1 \ldots s_n} \tr[A(s_1) \ldots A(s_n)] \ket{s_1 \ldots s_n},
\]
and let us consider a renormalisation group (RG) operation that gathers neighbouring pairs of particles in the following way \cite{Verstraete05}:
\begin{equation}\label{eq:RG}
A_{\alpha \mu}(s) A_{\mu \beta}(t)= \sum_{\lambda=1}^\chi U_{(st),\lambda} (\Sigma V^*)_{\lambda,(\alpha \beta)} \to (\Sigma V^*)_{\lambda,(\alpha \beta)} \equiv \widetilde{A}_{\alpha \beta}(\lambda).
\end{equation}
The rationale behind this operation is to eliminate local degrees of freedom in \emph{real} space. The transfer matrix of $\Psi$, defined as 
\[
E_{(\alpha,\beta),(\alpha',\beta')}= \sum_{s=1}^d A_{\alpha,\alpha'}(s) \; \widebar{A}_{\beta,\beta'}(s)
\]
transforms very simply under this RG operation: 
\[
E \to \widetilde{E}=E^2.
\]
$E$ is not hermitian but it can be diagonalised:
\begin{equation}
E=\sum_{i} \nu_i \ket{\varphi^R_i} \bra{\varphi^L_i},
\end{equation}
The left and right eigenvectors of $E$ need not match but they satisfy $\braket{\varphi^L_i}{\varphi^R_j} \propto \delta_{ij}$.

It is well-known that for any invertible $Y$, $\Psi$ is left invariant by the transformation 
\beq
A_{\alpha,\beta}(s) \to \sum_{\mu,\nu=1}^\chi Y_{\alpha,\mu} \; A_{\mu,\nu}(s) \; Y^{-1}_{\nu,\beta}.
\eeq

\begin{figure}[h]
\begin{center}
\includegraphics[width=95mm,height=45mm]{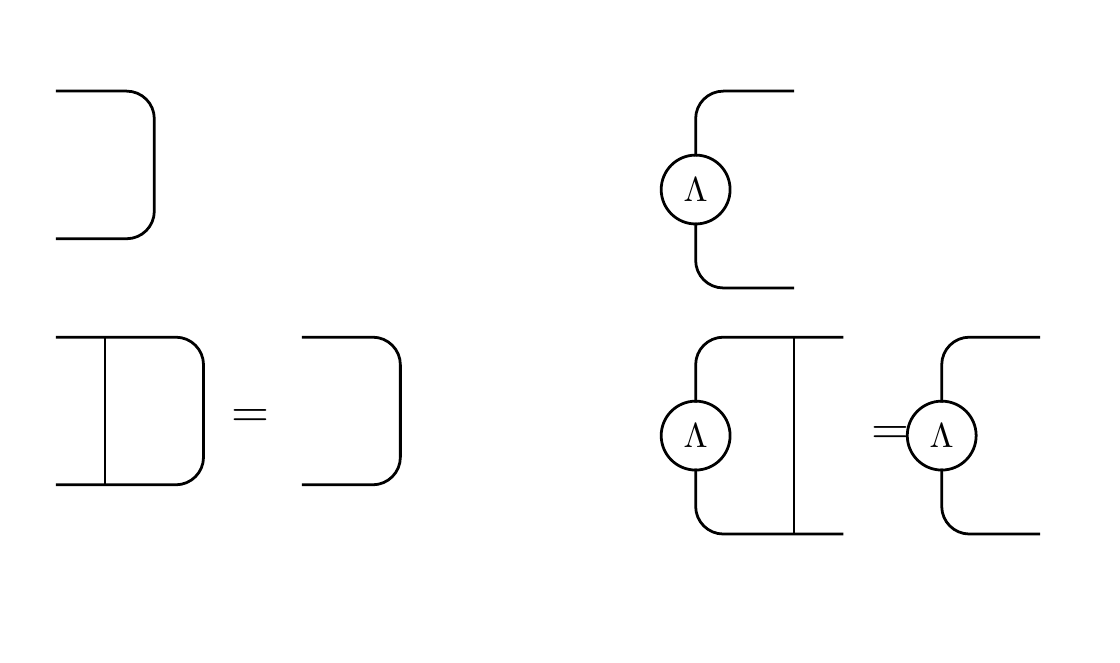} 
\caption{Graphical representation of the identities satisfied by an open boundary conditions (OBC) MPS in canonical form. Upper left: right eigenvector $\ket{\Phi^R}=\sum_{\alpha} \ket{\alpha,\alpha}$. Upper right: left eigenvector $\bra{\Phi^L}= \sum_{\alpha=1}^\chi \lambda_{\alpha} \bra{\alpha,\alpha}$. Lower left: $E^\star \ket{\Phi^R}=\ket{\Phi^R}$. Lower right: $\bra{\Phi^L} E^\star=\bra{\Phi^L}$.
}\label{FIG:CANONICAL-IDENTITIES}
\end{center}
\end{figure}

Also, it is natural to regard two states differing by local unitaries as essentially equivalent:
\beq
A_{\alpha,\beta}(s) \sim
A'_{\alpha,\beta}(s)=\sum_{t=1}^d U_{st} \; A_{\alpha,\beta}(t).
\eeq
Such two states only differ by a local change of basis performed on the physical degrees of freedom. These two symmetries can be used to impose that the largest eigenvalue of $E$ has unit magnitude. Of course,
\bed
E^m=\sum_{i} \nu_i^m \ket{\varphi^R_i} \bra{\varphi^L_i}.
\eed
Thus, the fixed point of the RG flow $E^\star=\lim_{m \to \infty} E^m$ has only two sorts of eigenvectors: those corresponding to unit magnitude eigenvalues, and those corresponding to zero eigenvalue. To simplify matters, we will focus on the (mathematically generic) case where $E^\star$ admits only one non-zero eigenvalue:
\bed
E^\star=e^{i \theta} \ket{\Phi^R} \bra{\Phi^L}.
\eed
Since the phase $e^{i \theta}$ can be absorbed on the definition of $\Phi^R$ or the definition of $\Phi^L$, we can assume that the only non-trivial eigenvalue of $E^\star$ is actually equal to $1$. Actually, it is shown in Ref. \cite{Verstraete05} that there is always a transformation that leaves the state $\Psi$ invariant and such that 
\bed
\ket{\Phi^R}= \sum_{\alpha=1}^\chi \ket{\alpha,\alpha}, \hspace{0.5cm}
\bra{\Phi^L}= \sum_{\alpha=1}^\chi \lambda_{\alpha} \bra{\alpha,\alpha}.
\eed
That these states should be the eigenvectors of $E^\star$ is not surprising if one thinks of the identities satisfied by the canonical form, see Fig.\ref{FIG:CANONICAL-IDENTITIES}.

\begin{figure}[h]
\begin{center}
\includegraphics[width=60mm,height=20mm]{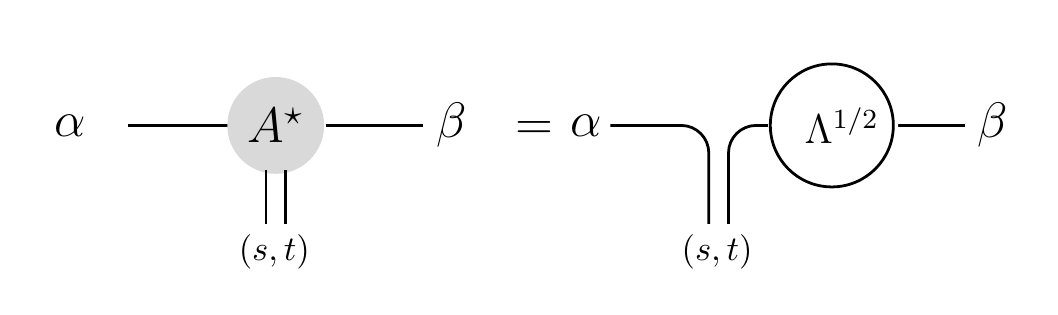} 
\caption{Graphical representation of the MPS described by Eq.(\ref{eq:REPRESENTATIVE}).
}\label{FIG:REPRESENTATIVE}
\end{center}
\end{figure}

In summary, all MPS that have the same Schmidt spectrum $\Lambda=\{\lambda_1,\ldots,\lambda_\chi\}$ form a class characterised by the same left and right eigenvectors for the transfer matrix of their RG fixed point. A simple representative of this class is the MPS $\ket{\Psi^\star}$ characterised by the matrices
\beq\label{eq:REPRESENTATIVE}
A_{\alpha,\beta}^\star(s t)=\sqrt{\lambda_\beta} \delta_{\alpha,s} \delta_{\beta,t}.
\eeq
$A^\star$ is represented on Fig.\ref{FIG:REPRESENTATIVE}. %In other words, the RG fixed point of an MPS is essentially governed by its Schmidt spectrum. This is natural: when two neighbouring spins are blocked, we actually eliminate local degrees of freedom. We thus expect the entanglement properties between the new supersite and the rest of the chain to be unaffected. 

We will now demonstrate physical equivalence between quantum states that have the same Schmidt spectrum. For that, we consider again Eq. \ref{eq:RG}. Let $\gras{v}^{(\alpha \beta)}$ denote the vector with components: $v^{\alpha \beta}_\lambda=(\Sigma V^*)_{\lambda,(\alpha \beta)}$ and let
\bed
\mathcal{V}=\textrm{span} \{\gras{v}^{(\alpha \beta)}, \alpha,\beta=1\ldots \chi \}.
\eed
$\mathcal{V}$ is certainly a vector space, which dimension is at most equal to $\chi$. The matrix $U$ appearing in the singular decomposition (\ref{eq:RG}) is an isometry when $d^2 > \chi$:
\bed
U:  \mathcal{V} \to \mathbb{C}^d \otimes \mathbb{C}^d ,
\hspace{0.5cm}
\sum_{s,t=1}^d U_{(s t),\lambda} \widebar{U}_{(s t),\lambda'}=\delta_{\lambda \lambda'}.
\eed
We observe that it can be lifted to a unitary as
\bed
\widetilde{U}= U \oplus \gras{1}_{\mathcal{V}^\perp}: \mathbb{C}^d \otimes \mathbb{C}^d \to \mathbb{C}^d \otimes \mathbb{C}^d,
\eed
and clearly, the operation of glueing two neighbouring sites can be (formally) expressed in terms of $\widetilde{U}$ instead of $U$:
\bed
\sum_{\nu=1}^\chi A_{\alpha \nu}(s_1) A_{\nu \beta}(s_2)=\sum_{\lambda=1}^\chi \widetilde{U}_{(s_1 s_2),\lambda} (\Sigma V^*)_{\lambda, (\alpha \beta)} 
\eed
\beq
=\sum_{\lambda=1}^\chi \widetilde{U}_{(s_1 s_2),\lambda} \; \widetilde{A}_{\alpha \beta}(\lambda) \to \widetilde{A}_{\alpha \beta}(\lambda).
\eeq
The change $U \to \widetilde{U}$ will be useful in determining how the RG flow transforms operators. 

Let us consider $n$ contiguous regions $\mathcal{R}_1,\mathcal{R}_2, \ldots,\mathcal{R}_n$, each containing $2^m$ microscopic spins, and let  $X_1(\mathcal{R}_1),X_2(\mathcal{R}_2),\ldots,X_n(\mathcal{R}_n)$ denote $n$ operators with support on each of these regions. We are interested in the mean value of the product of these operators ("Wightman functions"):
\bed
\bra{\Psi} X_1(\mathcal{R}_1) \otimes \ldots \otimes X_n(\mathcal{R}_n) \ket{\Psi}.
\eed
The first step of the RG flow allows to re-express this mean value as
\beq\label{eq:RG-1}
\bra{\Psi} X_1(\mathcal{R}_1) \otimes \ldots \otimes X_n(\mathcal{R}_n) \ket{\Psi}
=
\bra{\widetilde{\Psi}} \widetilde{X}_1(\widetilde{\mathcal{R}}_1) \otimes \ldots \otimes \widetilde{X}_n(\widetilde{\mathcal{R}}_n) \ket{\widetilde{\Psi}},
\eeq
where $\ket{\widetilde{\Psi}}$ is the state resulting from the RG $A \to \widetilde{A}$. For the operators, the RG transformation $X(\mathcal{R}) \to \widetilde{X}(\widetilde{\mathcal{R}})$ explicitly reads
\bed
\mathcal{L}_1: X \to \widetilde{X}= \bigotimes_{j=1}^{2^{m-1}} \widetilde{U}^{(j,j+1) \dagger} X \bigotimes_{j=1}^{2^{m-1}} \widetilde{U}^{(j,j+1)}.
\eed
It has two features, which are central to our purposes: it is \emph{unitary} and it acts on regions of finite size, that is quasi-locally. Similarly, we define the superoperators $\mathcal{L}_2, \mathcal{L}_3, \ldots$ %Its action has support on a region of size $2^m$.

Let $X^{(m)} \equiv \big( \mathcal{L}_m \circ \ldots \circ \mathcal{L}_1 \big) (X)$, and let $\Psi^{(m)}$ denote the state obtained after $m$ renormalization steps. Repeated applications of the identity (\ref{eq:RG-1}) show that:
\bed
\bra{\Psi} X_1(\mathcal{R}_1) \otimes X_2(\mathcal{R}_2) \otimes \ldots \otimes X_n(\mathcal{R}_n) \ket{\Psi}=
\eed
\beq
\bra{\Psi^{(m)}} \; X^{(m)}_1(1) \otimes X^{(m)}_2(2) \otimes \ldots \otimes X^{(m)}_n(n) \; \ket{\Psi^{(m)}}.
\eeq
Let $\ket{\Psi'^{(m)}}$ denote the states obtained after $m$ renormalisation steps performed on $\ket{\Psi'}$, and let $A^{(m)}$ and $A'^{(m)}$ denote the corresponding matrices. We assume $m$ is large enough that both $\Psi^{(m)}$ and $\Psi'^{(m)}$ can be confused with their respective RG fixed points in good approximation. Then, according to Ref.  \cite{Verstraete05}, there exist invertible matrices $Z^{(1)},Z^{(2)}$ and unitary matrices $W^{(1)},W^{(2)}$ that relate these two fixed points to the representative of their class. That is,
\bed
\sum_{\mu,\nu=1}^\chi \sum_{s',t'=1}^d Z^{(1)}_{\alpha,\mu} \; A^{(m)}_{\mu,\nu}(s',t') \; \big( Z^{(1)} \big )^{-1}_{\nu,\beta} \; W^{(1)}_{(s',t'),(s,t)}= 
\eed
\beq
\sum_{\mu,\nu=1}^\chi \sum_{s',t'=1}^d Z^{(2)}_{\alpha,\mu} \; A'^{(m)}_{\mu,\nu}(s',t') \; \big( Z^{(2)} \big )^{-1}_{\nu,\beta} \; W^{(2)}_{(s',t'),(s,t)}= 
A_{\alpha,\beta}^\star(s,t).
\eeq
(The unitaries $W_1,W_2$ act on renormalised spins, that is on regions of size $2^m$ of the original lattice.) Of course such an action is long range if compared to the scale of a microscopic spin of the original lattice, but it can safely be considered \emph{local}, when compared to the scale of the whole system. Let us define $W=W_2 W_1^{-1}$. We see that 
\bed
\bra{\Psi^{(m)}} X^{(m)}_1(1) \otimes \ldots \otimes X^{(m)}_n(n) \ket{\Psi^{(m)}}=
\eed
\beq
\bra{\Psi'^{(m)}} W X^{(m)}_1(1) W^{\dagger} \otimes \ldots \otimes 
W X^{(m)}_n(n) W^{\dagger} \ket{\Psi'^{(m)}}.
\eeq
This last identity is the main result of this section: it shows that, with the proviso that the highest eigenvalue of the transfer operator $E$ is not degenerate, \emph{at sufficiently large scale, the states $\Psi$ and $\Psi'$ can be regarded as locally unitarily equivalent. Any observation made on one state can be made on the other, modulo a local change of basis, independent of the precise operators $X_1, \ldots,X_n$ we are interested in. There is a fixed local unitary correspondence between all Wightman functions for $\Psi$ and $\Psi'$.} 

Moreover, entanglement spectra naturally lead to a notion of distance between states that is invariant under local changes of bases. It is easy  to see that the mean value of any string of operators can be expressed in terms of operators acting on the virtual degrees of freedom as
\beq
\bra{\Psi} X_1(\mathcal{R}_1) \otimes \ldots \otimes X_n(\mathcal{R}_n) \ket{\Psi}=
\bra{\Phi_L} \Upsilon_1(1) \ldots \Upsilon_n(n) \ket{\Phi_R},
\eeq
where the $\Upsilon_k$ are $\chi^2 \times \chi^2$ matrices acting on virtual degrees of freedom. The proof of this identity is trivial when expressing the l.h.s. diagrammatically.
A similar expression holds for any other state $\Psi^{\#}$ with a different entanglement spectrum $\Lambda^{\#}$. So,
\bed
|\bra{\Phi_L} \Upsilon_1(1) \ldots \Upsilon_n(n) \ket{\Phi_R}-\bra{\Phi_L} \Upsilon_1(1) \ldots \Upsilon_n(n) \ket{\Phi^{\#}_R}|
\eed
\beq\label{eq:dist-ent-sp}
\leq 
|| \bra{\Phi_L} || \times ||\Upsilon_1(1) \ldots \Upsilon_n(n) || \times ||\Lambda-\Lambda^{\#}||.
\eeq
\emph{The difference between entanglement spectra sets bounds on the difference between observable quantities. The inequality \ref{eq:dist-ent-sp} justifies that we use the distance between entanglement spectra as a distance between theories}. We loosely use the word 'theory' to refer to all states that only differ by local changes of bases.

%=====

%After $m$ steps of renormalisation, we find that 

%=====

%We now prove equivalence between the two states of Eq.(\ref{eq:state-equivalence}). Let $\ket{\Psi^{(m)}}$ and $\ket{\Psi'^{(m)}}$ denote the states obtained after $m$ renormalisation steps, and let $A^{(m)}$ and $A'^{(m)}$ denote the corresponding matrices. We assume $m$ is large enough that both $\Psi^{(m)}$ and $\Psi'^{(m)}$ can be confused with their respective RG fixed points in good approximation. According to \cite{Wolf}, there exist invertible matrices $X_1,X_2$ and unitary matrices $W_1,W_2$ such that the identities depicted on Fig.*** hold. Thus the diagram *** holds. The unitaries $W_1,W_2$ act on one renormalised spin, that is on regions of size $2^m$ of the original lattice. Let us define $W=W_2 W_1^{-1}$ and let us consider $n$ contiguous regions $\mathcal{R}_1,\mathcal{R}_2, \ldots,\mathcal{R}_n$, each containing to $2^m$ microscopic spins, and let  $X_1(\mathcal{R}_1),X_2(\mathcal{R}_2),\ldots,X_n(\mathcal{R}_n)$ denote $n$ operators with support on these regions. We see that
%\bed
%\bra{\Psi} X_1(\mathcal{R}_1) \otimes X_2(\mathcal{R}_2) \otimes \ldots \otimes X_n(\mathcal{R}_n) \ket{\Psi}
%\eed
%\beq
%=\bra{\Psi'} W X_1(\mathcal{R}_1) W^{\dagger} \otimes W X_2(\mathcal{R}_2) W^{\dagger} \otimes \ldots \otimes 
%W X_n(\mathcal{R}_n) W^{\dagger} \ket{\Psi'}.
%\eeq

%We will see shortly that such a class is actually an \emph{equivalence} class.

\section{Free Fermions}\label{sect:FF}
%%%%%%%%%%%%%%%%%%%%%%

Consider a system where a fermionic mode is associated with each site $k$, and let $c_k^{\dagger}$ and $c_k$ denote the corresponding creation and annihilation operators. With each such fermionic mode, one can associate two Majorana fermion operators
\beq
\check{a}_{2k-1}=c_k+c^\dagger_k, \hspace{0.3cm} \check{a}_{2k}=\frac{1}{i}(c_k-c_k^\dagger).
\eeq
We are interested in Hamiltonians of the form:
\beq\label{eq:Majornana-form}
H=\frac{i}{4}\sum_{k,l} A_{k,l} \check{a}_{k} \check{a}_{l}.
\eeq
The techniques that can be used to analyse the entanglement properties of part of this system are well known. The reduced density matrix $\rho_A$ of a part $A$ of the whole system corresponding to $n$ sites is completely characterised by a set of $n$ symplectic values 
\bed
\text{Sp}_{\text{symp}}(A)=\{\nu_1,\ldots,\nu_n\}. 
\eed
The entanglement spectrum of $\rho_A$ is the set
\beq
\lambda(\nu_1,\ldots,\nu_n)=\prod_{k=1}^n \frac{1+(-)^{x_k} \nu_k}{2}, \; x_k \in \{0,1\}.
\eeq

Two free fermionic entanglement spectra $\Lambda$ and $\Lambda'$ can be easily compared if their presentation in decreasing order match, i.e. the $\alpha$-th eigenvalue for both $\Lambda$ and $\Lambda'$ correspond to the same bit string $(x_1,\ldots,x_n)$ for all $\alpha \in \{1,\ldots,2^n\}$. Then it is easy to see that the (square) distance decomposes into modes:
\beq
||\Lambda-\Lambda'||^2_2
=\prod_{k=1}^n \frac{1+\nu^2_k}{4} 
+\prod_{k=1}^n \frac{1+\nu'^2_k}{4} 
-2 \prod_{k=1}^n \frac{1+\nu_k \nu'_k}{4}.
\eeq
and can therefore be computed efficiently.

Since the entanglement spectrum is directly related to the symplectic spectrum, it makes sense to compare the ground state of two quadratic Hamiltonians by looking at the difference 
\beq
||\Lambda-\Lambda'||_{\text{symp}}=||\text{Sp}_{\text{symp}}(A)-\text{Sp}_{\text{symp}}(A')||.
\eeq

Finally, motivated by the abundant literature on entanglement entropy, it is natural to try and compare entanglement spectra using logarithmic functions. A possibility for that is to use the (symmetrised!) Kullback-Leibler (KL) pseudo-distance between two spectra:
\beq
D(\Lambda || \Lambda')+D(\Lambda' || \Lambda)
=\sum_{\alpha} \lambda_\alpha \log\frac{\lambda_\alpha}{\lambda'_\alpha}
+\sum_{\alpha} \lambda'_\alpha \log\frac{\lambda'_\alpha}{\lambda_\alpha}.
\label{defkullback}
\eeq
When the presentation of $\Lambda$ and $\Lambda'$ match, this KL distance also decomposes into modes and can be computed efficiently:
\bed
||\Lambda-\Lambda'||_{\text{KL}}=D(\Lambda || \Lambda')+D(\Lambda' || \Lambda)
\eed
\beq
=\sum_{k=1}^n D_2(\frac{1+\nu}{2} ||\frac{1+\nu'}{2})
+\sum_{k=1}^n D_2(\frac{1+\nu'}{2} ||\frac{1+\nu}{2}),
\eeq
where $D_2(p || q)= p \log \frac{p}{q}+(1-p) \log \frac{1-p}{1-q}$. 

In what follows we will use these three measures of distance and show that they lead to equivalent results.

\section{Phase transition in the Ising model}\label{sect:PT}
%%%%%%%%%%%%%%%%%%%%%%%%

We will consider the Ising model on a finite chain with a transverse field and open boundary conditions.
\bed
H=-\sum_{i=1}^{n-1} \sigma_i^x \sigma^x_{i+1}-h \sum_{i=1}^n \sigma^z_i.
\label{ising}
\eed
where $\sigma$ are the Pauli matrices, $h$ is the value of the transverse field, and n is the chain  length.  A Jordan-Wigner transformation allows to express this Hamiltonian in the form (\ref{eq:Majornana-form}).

Although the analysis presented in the previous section was concerned with the case of an infinite lattice, it is not difficult to see that it also holds true for a large finite system. When the system is gapped, the area law guarantees that the ground state can be faithfully represented by a matrix product state which bond dimension grows gently with the system size. Actually, even at criticality, we know that the bond dimension need only grow polynomially with the system size in order to obtain a faithful description.

We have found that Entanglement Spectrum (ES) distance can be used to construct an indicator sensitive to the phase transition at $h=1$:
\beq\label{eq:susc-1}
E(h)=\lim_{\delta h \to 0} \frac{||\Lambda(h+\delta h)-\Lambda(h)||}{\delta h}.
\eeq

\begin{figure}[h]
\begin{center}
\includegraphics[width=80mm,height=80mm]{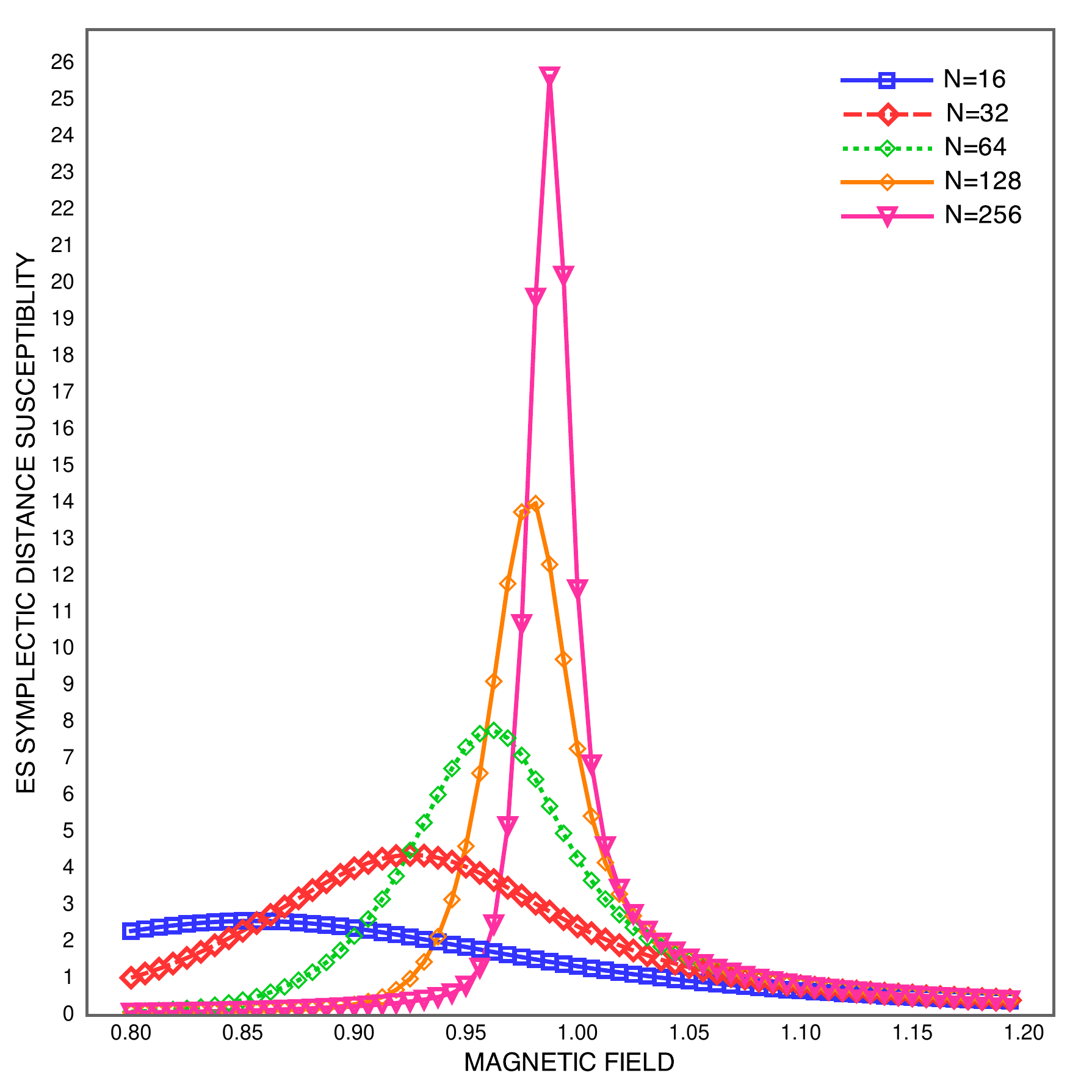} 
\caption{Entanglement distance susceptibility in the Ising model, as a function of the magnetic field, see Eq. \ref{eq:susc-1} for finite systems of various sizes (open boundary conditions). We clearly see that the peak becomes more pronounced and shifts towards the critical point $h^*=1$ as the system size is increased. The Euclidean distance between symplectic spectra has been used.
}\label{FIG:Ising-symp-dist}
\end{center}
\end{figure}

\begin{figure}[h]
\begin{center}
\includegraphics[width=80mm,height=80mm]{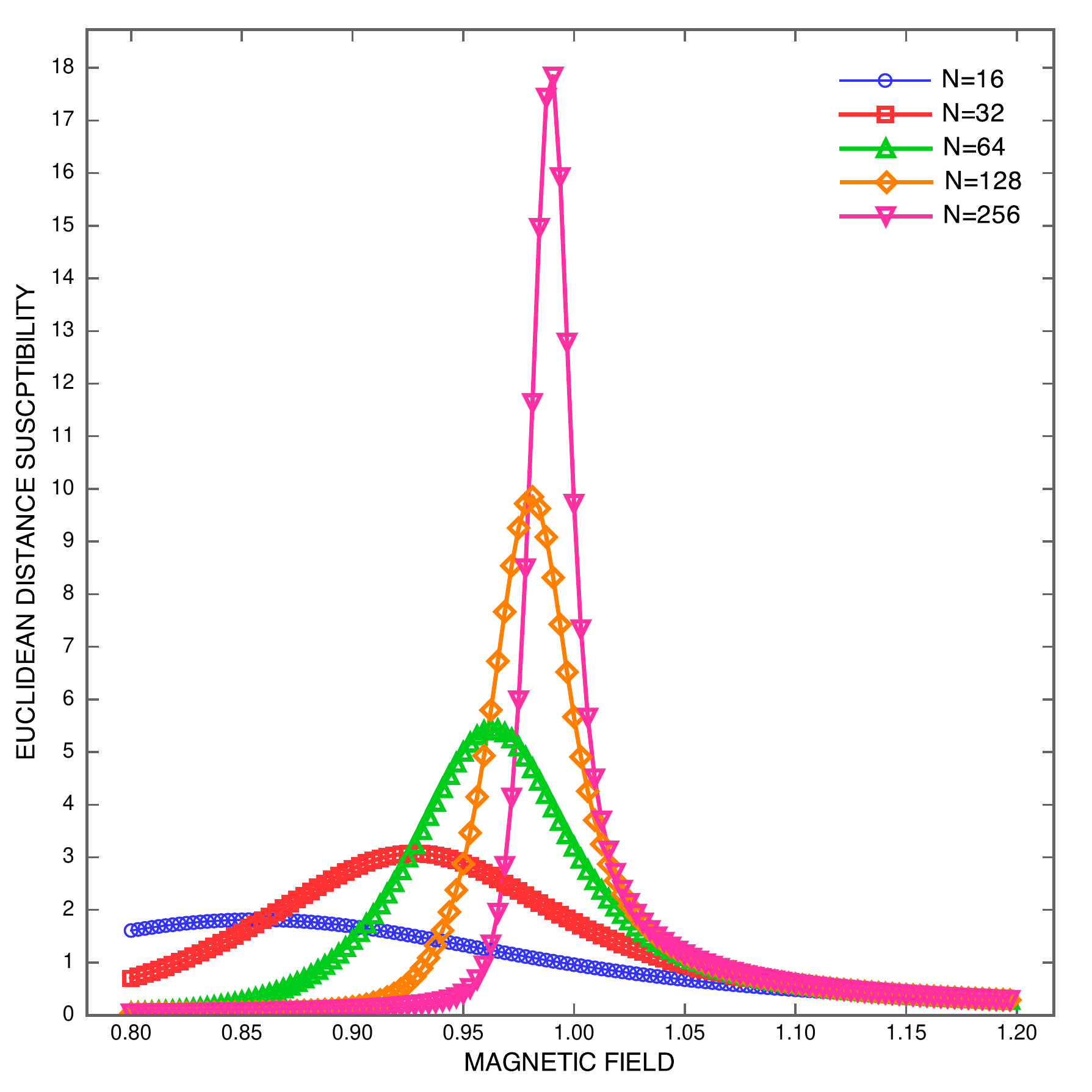} 
\caption{Entanglement distance susceptibility in the Ising model, as a function of the magnetic field, see Eq. \ref{eq:susc-1} for finite systems of various sizes (open boundary conditions). We clearly see that the peak becomes more pronounced and shifts towards the critical point $h^*=1$ as the system size is increased.
The Euclidean distance between full entanglement spectra has been used.
}\label{FIG:Ising-full-dist}
\end{center}
\end{figure}

\begin{figure}[h]
\begin{center}
\includegraphics[width=80mm,height=80mm]{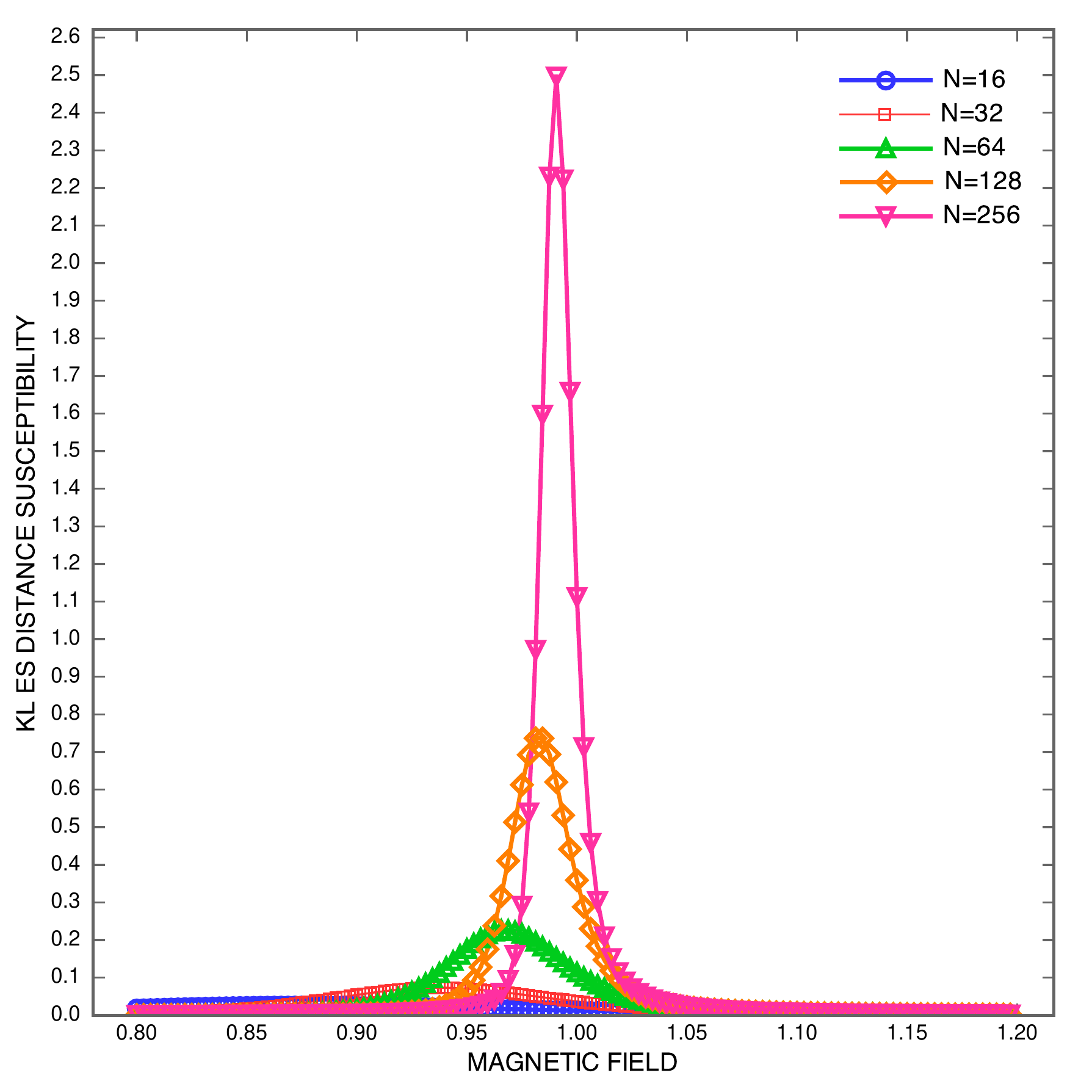} 
\caption{Entanglement distance susceptibility in the Ising model, as a function of the magnetic field, see Eq. \ref{eq:susc-1} for finite systems of various sizes (open boundary conditions). We clearly see that the peak becomes more pronounced and shifts towards the critical point $h^*=1$ as the system size is increased.
The Kullback-Leibler distance between full entanglement spectra has been used.
}\label{FIG:Ising-KL-dist}
\end{center}
\end{figure}

The plot of this quantity is shown in the Figures. \ref{FIG:Ising-symp-dist}, \ref{FIG:Ising-full-dist}, \ref{FIG:Ising-KL-dist}.

%\begin{center}
%\begin{figure}[ht]
%\begin{minipage}[b]{0.5\linewidth}
%\centering
%\includegraphics[width=30mm,height=30mm]{1D-ISING-SYMP-SUSCEP.PDF} 
%\includegraphics[width=30mm,height=30mm]{ES-ISING-FULL-DIST-SUSC.PDF} 
%\includegraphics[width=30mm,height=30mm]{ES-ISING-FULL-DIST-SUSC.PDF} 
%\end{minipage}
%\hspace{-1cm}
%\begin{minipage}[b]{0.6\linewidth}
%\centering
%\includegraphics[width=40mm,height=40mm]{KULLBACK-OBC-ISING-CHAIN.PDF} 
%\end{minipage}
%\begin{minipage}[b]{0.6\linewidth}
%\centering
%\includegraphics[width=20mm,height=40mm]{KULLBACK-OBC-ISING-CHAIN.PDF} 
%\end{minipage}
%\label{sixknots}
%\caption{}
%\end{figure}
%\end{center}

%\section{Phase transitions in the XXZ model}
%%%%%%%%%%%%%%%%%%%

\section{Finite spin chains}\label{sect:FSC}

We now study the distance between different instances of the Ising and Heisenberg chain models with different number of spins, using the Kullback-Leibler (KL) distance defined above \eqref{defkullback}, divided by two.

First we compute the distance between Ising models with different values of the transverse field. We will use the same expression as before (\ref{ising}) for the Ising model, but now without any transformation and with periodic boundary conditions,

\begin{equation}
H = - \sum_{i} \left( \sigma^x_i \sigma^x_{i+1} + h \sigma^z_i \right).
\end{equation}

\begin{figure}[h!]
      \begin{center}
            \includegraphics[scale=0.35]{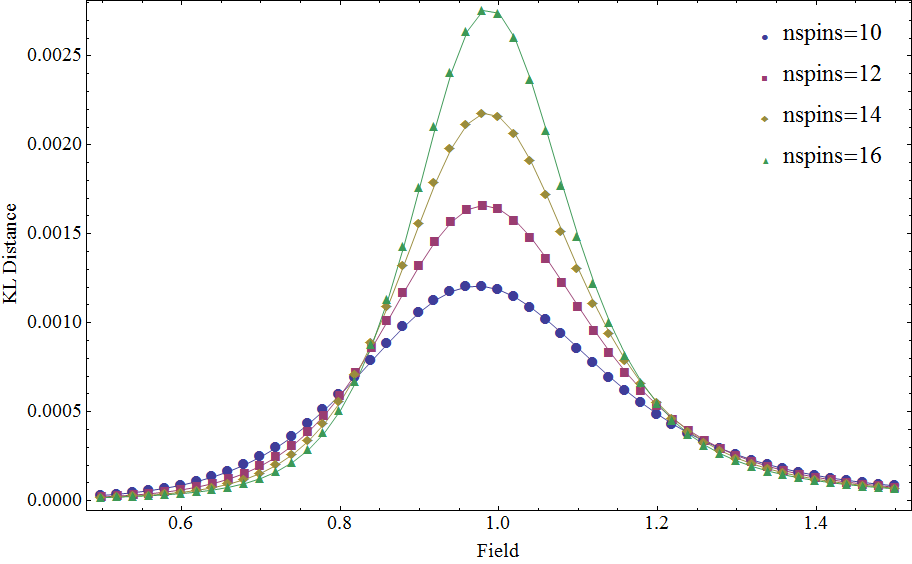}
            \caption{\label{fig:ising}Average Kullback-Leibler distances between Ising models with fields $h$ and $h+dh$ (with $dh=0.02$) for finite spin chains of various sizes. Periodic boundary conditions are assumed. We clearly see a peak around $h=1$ (critical point), that gets more pronounced by increasing the size of the chain.
             }
      \end{center}
     
\end{figure}

We can see the results in Fig.\ref{fig:ising}.

Next we show the distance between the Ising model with different values of the transverse field and the isotropic Heisenberg model, defined as

\begin{equation}
H = \sum_{i} \sigma^x_i \sigma^x_{i+1} + \sigma^y_i \sigma^y_{i+1} + \sigma^z_i \sigma^z_{i+1}.
\end{equation}

\begin{figure}[h!]
      \begin{center}
            \includegraphics[scale=0.35]{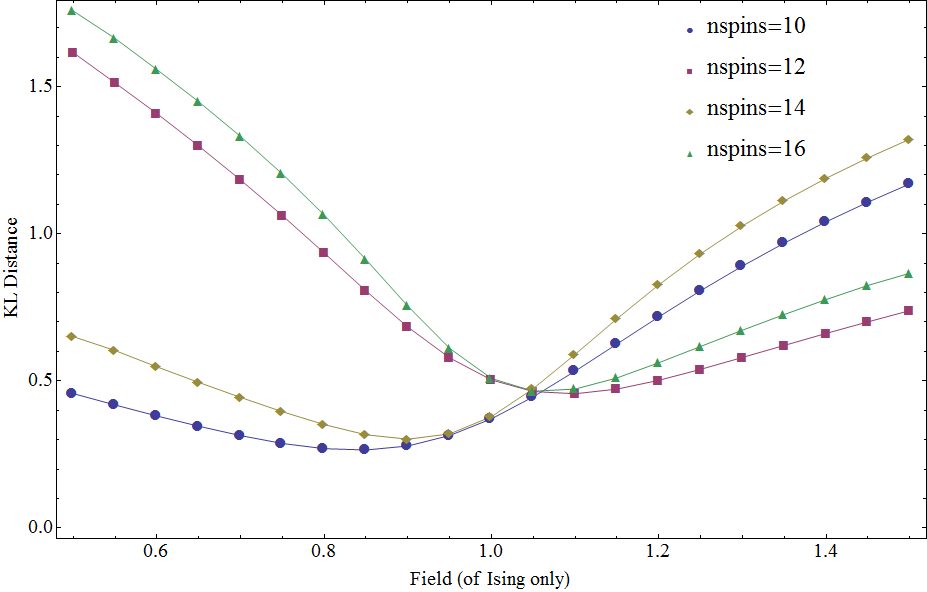}
            \caption{\label{fig:isingheis}Average Kullback-Leibler distances between Ising model with different transverse fields ($dh=0.05$) and the isotropic Heisenberg model for finite spin chains of various sizes. Periodic boundary conditions are assumed. We see a minimum around $h=1$, the critical point for both models, and a different pattern for N=10, 14 and N=12, 16, probably related to the finite size effects of the different parity of their half-chains (odd and even respectively).
             }
      \end{center}
     
\end{figure}

The resulting plot is shown in Fig.\ref{fig:isingheis}.

\begin{figure}[h!]
      \begin{center}
            \includegraphics[scale=0.35]{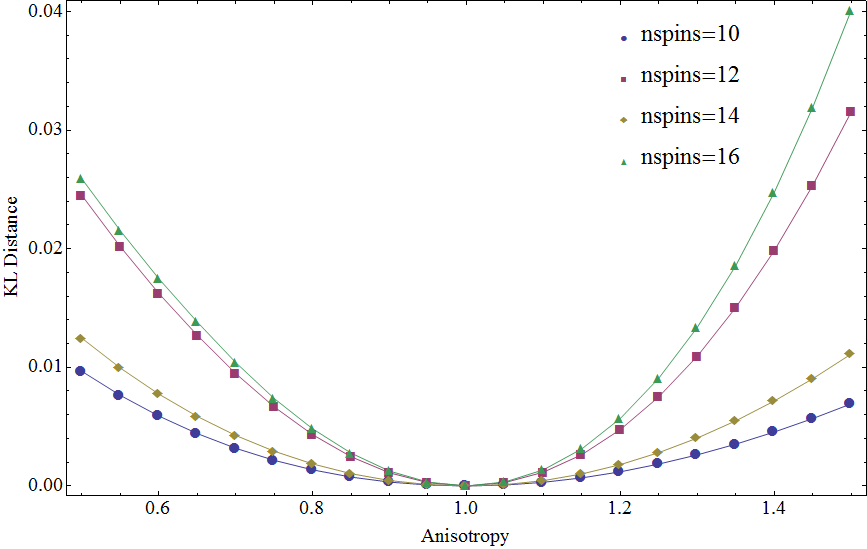}
            \caption{\label{fig:anis}Average Kullback-Leibler distances between Heisenberg models with a certain anisotropy and the isotropic one for different sizes of the spin chain. Periodic boundary conditions are assumed. Of course the difference goes to zero at $\Delta=1$ as it compares the model to itself.
             }
      \end{center}
      
\end{figure}

Finally we show the distance between the Heisenberg model with different values of the anisotropy parameter and the isotropic Heisenberg model. We now define the anisotropic model

\begin{equation}
H = \sum_{i} \sigma^x_i \sigma^x_{i+1} + \sigma^y_i \sigma^y_{i+1} + \Delta (\sigma^z_i \sigma^z_{i+1}),
\end{equation}

with $\Delta$ as the value of the anisotropy. The results are plotted in Fig.\ref{fig:anis}.

We see a very simple pattern in figure \ref{fig:ising}: the maximum distance is always around $h = 1$ (the critical point) and increases with the number of spins. In figure \ref{fig:isingheis} we see that now the distance is minimal between Ising models and the Heisenberg model around $h = 1$, probably caused by the underlying similarity between different models in their critical point. We also see quite a different shape between cases of 10,14 spins and 12,16 spins. We assume it has to do with their half-chains having odd and even number of spins respectively and conjecture that it is a finite-size effect that will disappear in bigger sizes. In figure \ref{fig:anis} we see that the KL distance of the anisotropic models with the isotropic one uniformly decreases until $\Delta = 1$ and then uniformly increases. Again we see a different pattern for 10,14 and 12,16 cases.
     
\section{Infinite spin chains}\label{sect:ISC}

We now show the same figures as in the previous section, but with infinite chains. We use the exact results from Ref. \cite{Orus06} to obtain the expression for the density matrix of half of the chain. For the Heisenberg model:

\begin{equation}
\rho_{\Delta} (n_0,n_1,...,n_{\infty}) = \frac{1}{Z_{\Delta}}e^{-\sum_{k=0}^{\infty} n_k \epsilon_k},
\end{equation}
where 
\begin{equation}
 \epsilon_k = 2k\arccos{\Delta},
\end{equation}
and
\begin{equation}
 \hspace{5mm} Z_{\Delta}=\prod_{k=0}^{\infty} (1+e^{-\epsilon_{k}}).
\end{equation}

For the Ising model it's slightly more complicated:

\beq
\rho_{h} (n_0,n_1,...,n_{\infty}) = \frac{1}{Z_{h}}e^{-\sum_{k=0}^{\infty} n_k \epsilon_k}, 
\eeq
where
\beq
 Z_{h}=\prod_{k=0}^{\infty} (1+e^{-\epsilon_{k}}),
\eeq
and
\beq
 \epsilon_k = \twopartdef { 2k\epsilon } {h < 1} {2(k+1)\epsilon} {h > 1}.
\eeq
Furthermore
\beq
 \epsilon = \pi \frac{I(\sqrt{1-x^2})}{I(x)},
\eeq
where $I(x)$ is the elliptic integral of the second kind:
\beq
I(x) = \int^{\pi/2}_{0} \frac {d\theta}{\sqrt{1-x^2 sin^2(\theta)}},
\eeq
where
\beq
x = \twopartdef { h } {h < 1} {1/h} {h > 1}.
\eeq

\begin{figure}[h!]
      \begin{center}
            \includegraphics[scale=0.30]{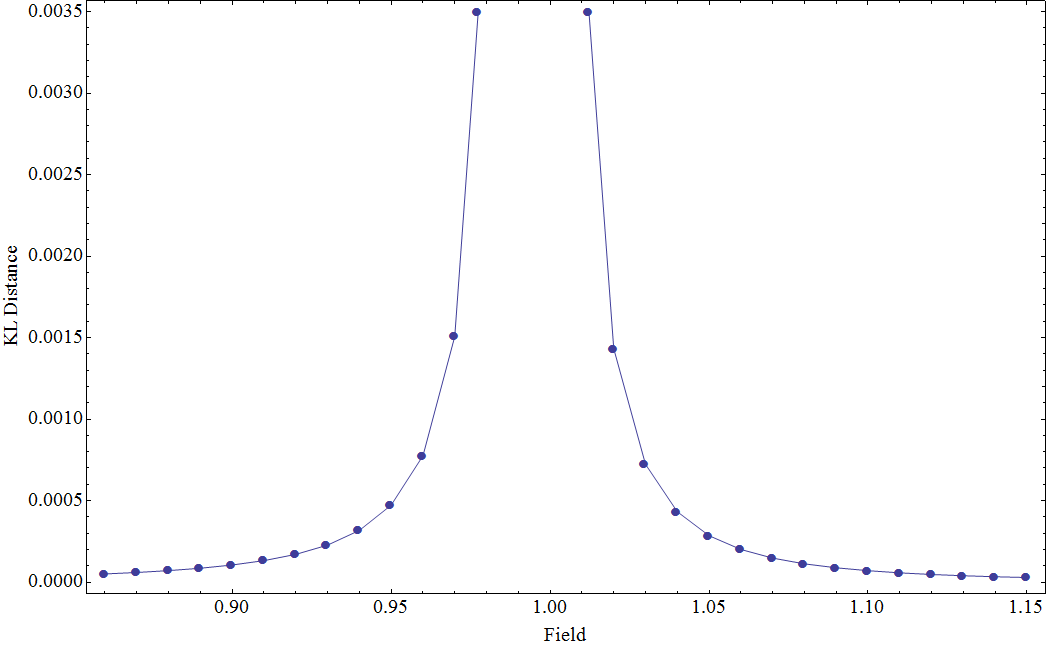}
            \caption{\label{fig:isinginf}Average Kullback-Leibler distances between infinite Ising models with fields $\lambda$ and $\lambda+d\lambda$ (with $d\lambda=0.01$). We see a very narrow peak around the critical point $h=1$.
             }
      \end{center}
     
\end{figure}

\begin{figure}[h!]
      \begin{center}
            \includegraphics[scale=0.30]{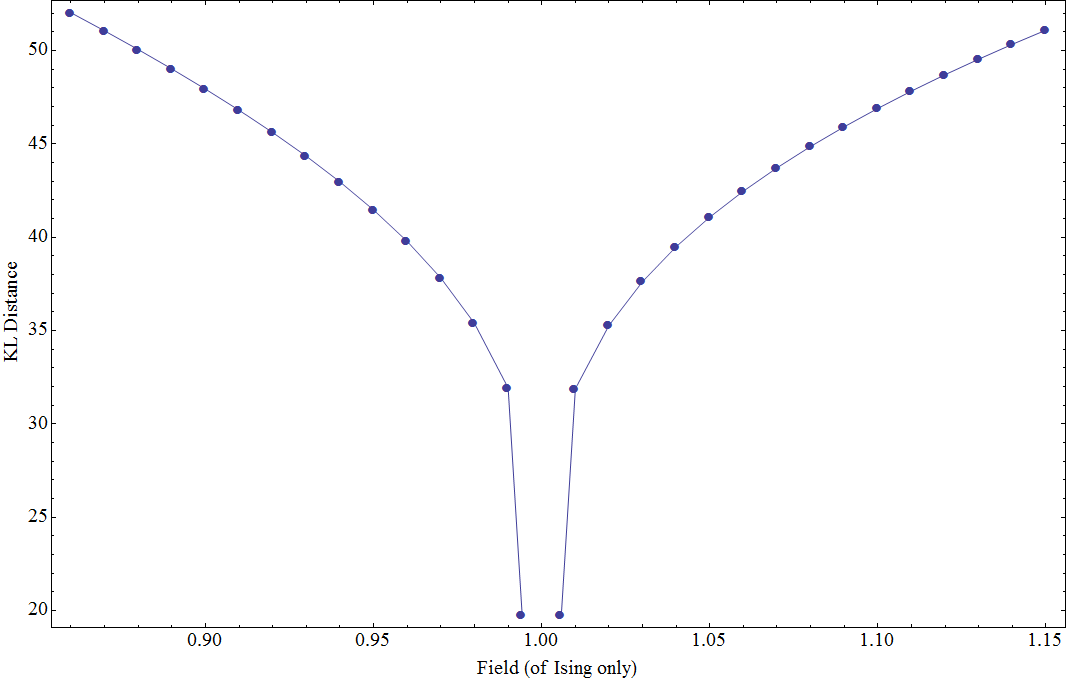}
            \caption{\label{fig:isingheisinf}Average Kullback-Leibler distances between infinite Ising model with different transverse fields ($dh=0.01$) and the infinite Heisenberg model. We see a very narrow minimum in the critical point of the Ising model $h=1$.
             }
      \end{center}
     
\end{figure}

\begin{figure}[h!]
      \begin{center}
            \includegraphics[scale=0.30]{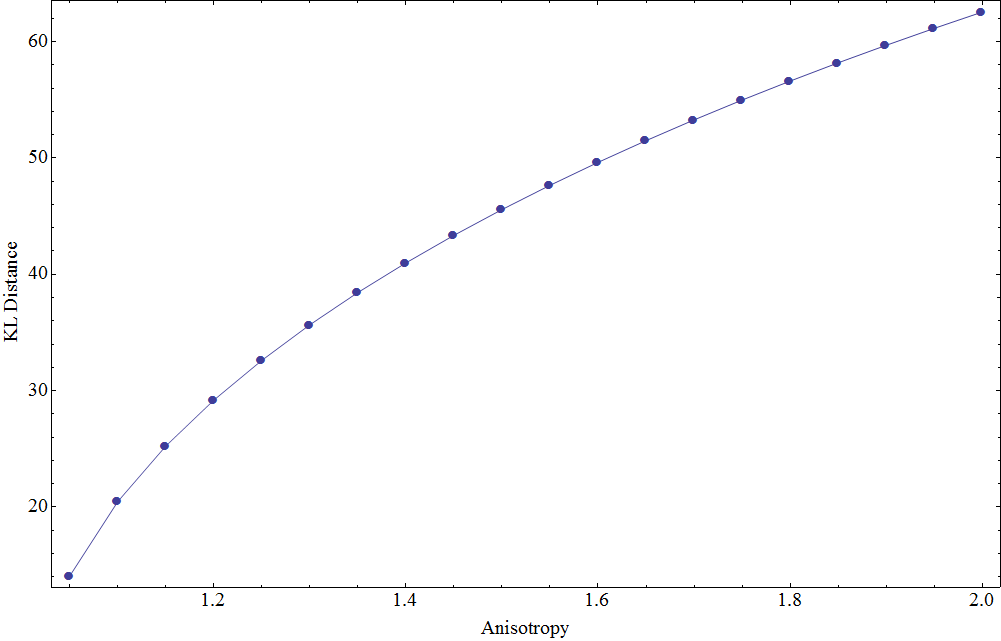}
            \caption{\label{fig:anisinf}Average Kullback-Leibler distances between infinite Heisenberg models with a certain anisotropy and the isotropic one. We show only results for the branch on the right of the critical point $\Delta=1$, as we do not have analytical expressions for the left branch. We see an increase of the distance as we go further from the critical point.
             }
      \end{center}

\end{figure}

We see an increase towards $h = 1$ in the distances between Ising models and a decrease towards $h = 1$ in the distances to the Heisenberg model, see Figs. \ref{fig:isinginf} and \ref{fig:isingheisinf}, situations that match the finite cases. The singularities in $h=1$ arise because our analytical formulas are not well defined there. Anyway, there are strongly marked peaks in that critical region. In Fig. \ref{fig:anisinf} we see also a pattern that matches our finite cases, but only its right wing, as we do not have an analytical expression for the left wing $(\Delta < 1)$.

\section{2D Systems: A superconductor-insulator phase transition}\label{sect:2D}
%%%%%%%%%%%%%%%%%%%%%%%%%%%%

Strictly speaking, the discussion of Section \ref{sect:RG} is only valid for one-dimensional systems whose wavefunction can be described by an MPS. It is however tempting to extend the analysis beyond this framework. We have considered the two-dimensional free fermion Hamiltonian
\beq\label{eq:2D-FF}
H=\sum_{\langle i,j \rangle} \big( c^\dagger_i c_j+ c^\dagger_j c_i- \gamma \; (c^\dagger_i c^\dagger_j+c_j c_i)\big)
-2 \lambda \sum_i c_i^\dagger c_i.
\eeq

This system has a rich phase diagram \cite{Li06}. There is a region of the parameter space with non-zero measure for which the entropy of a part of size $L \times L$ (cut off from an infinite lattice) scales as
\beq\label{eq:2d-scaling}
S_L \sim L \log L.
\eeq 

\begin{figure}[h!]
\begin{center}
\includegraphics[width=80mm,height=80mm]{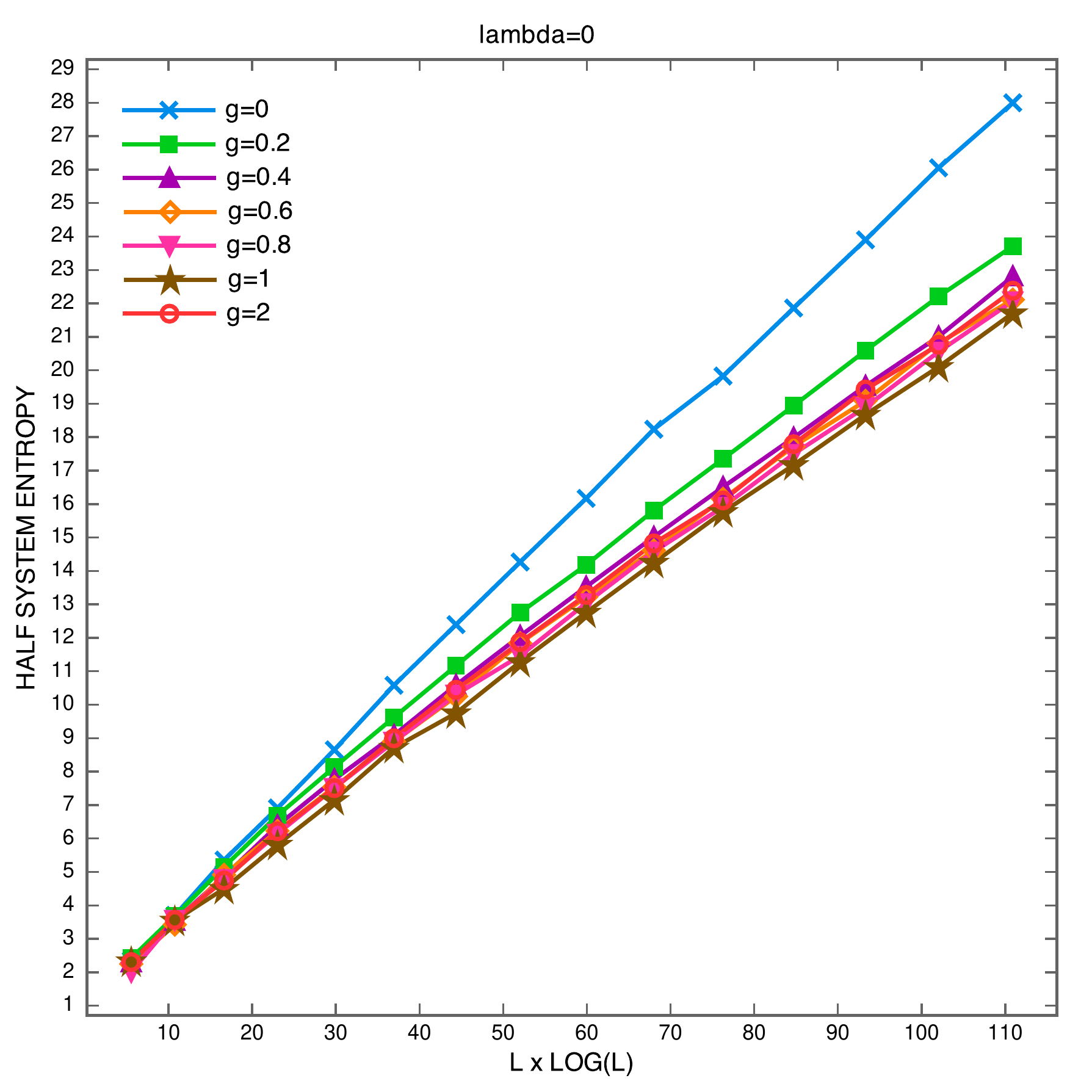} 
\caption{Entanglement entropy for half of a system described by the Hamiltonian (\ref{eq:2D-FF}) in its ground state. Finite size and open boundary conditions are assumed. $\lambda=0$, each curve corresponds to a a different value of $\gamma$.}\label{FIG:2D-Entropy-1}
\end{center}
\end{figure}

\begin{figure}[h!]
\begin{center}
\includegraphics[width=80mm,height=80mm]{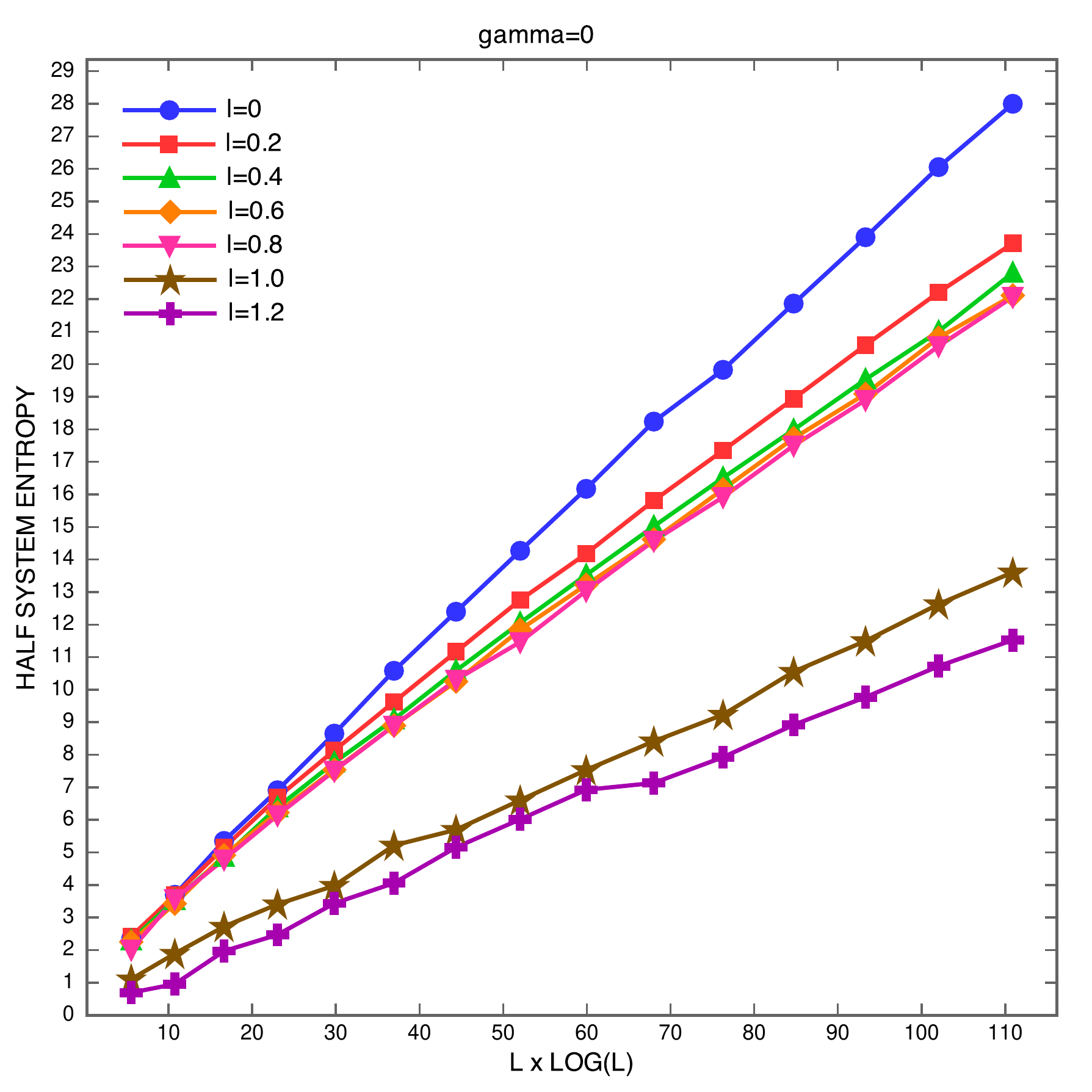} 
\caption{Entanglement entropy for half of a system described by the Hamiltonian (\ref{eq:2D-FF}) in its ground state. Finite size and open boundary conditions are assumed. We have set $\gamma=0$, and each curve corresponds to a different value of $\lambda$.}\label{FIG:2D-Entropy-2}
\end{center}
\end{figure}

First, we have plotted the entropy of half of the system for various sizes. See Fig. \ref{FIG:2D-Entropy-1}, \ref{FIG:2D-Entropy-2}. These plots cannot be compared directly with the results of Ref. \cite{Li06}: we consider half of a finite size system of size $L \times L$, whereas their authors consider a region of size $L \times L$ embedded in an infinite lattice. However, we reproduce two results that are consistent with the analysis of the authors: (i) the scaling law (\ref{eq:2d-scaling}) seems to be approximately obeyed, (ii) the order of our curves match that of the authors. Moreover the scaling behaviour (\ref{eq:2d-scaling}) should actually be completed with subleading terms \cite{Barthel06} which possibly play an important role in our case.

\begin{figure}[h!]
\begin{center}
\includegraphics[width=120mm,height=120mm]{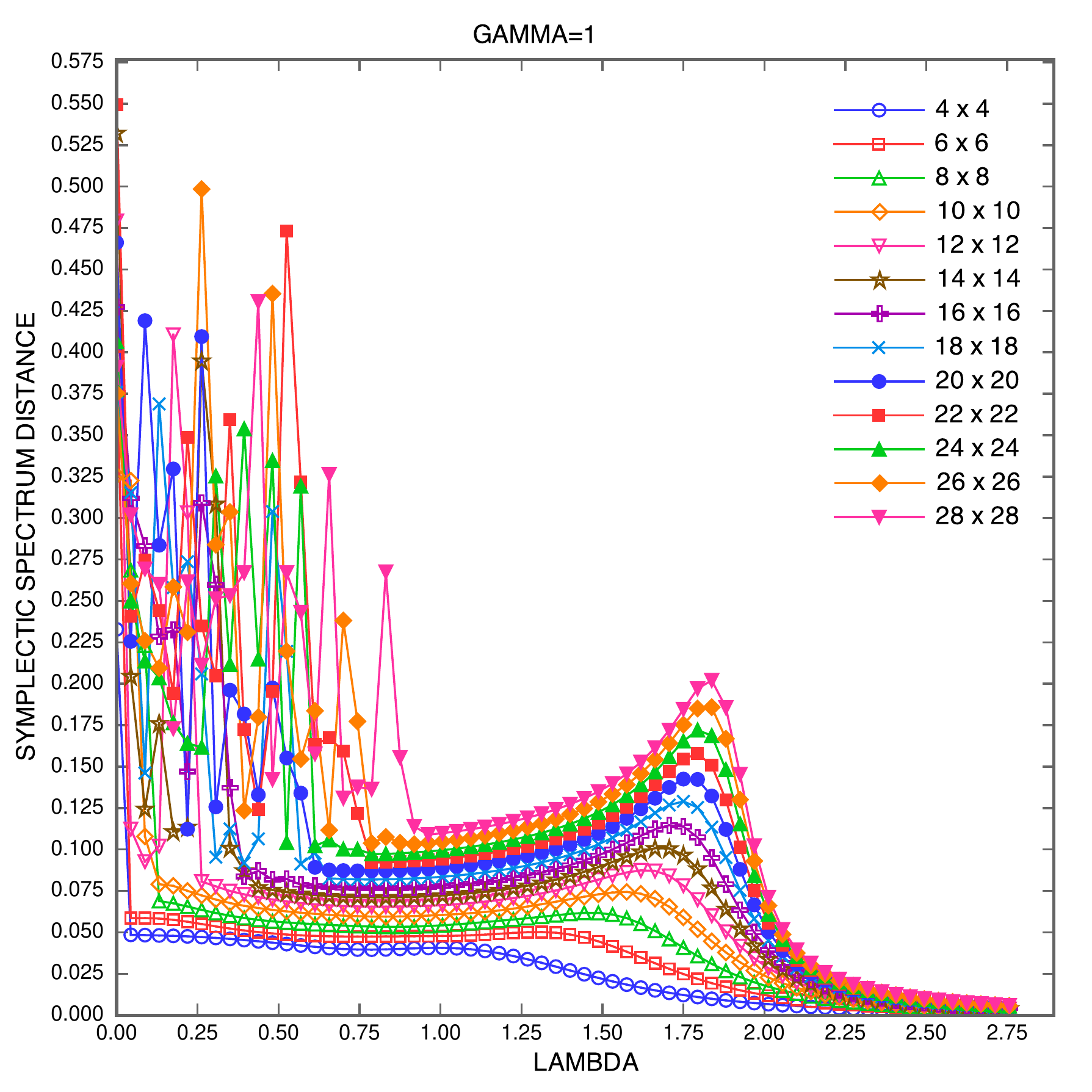} 
\caption{Entanglement spectrum susceptibility as a function of $\lambda$ for different system sizes ($\gamma=1$). The system has been cut in two equal parts.}\label{FIG:2D-SES-SUSCEP}
\end{center}
\end{figure}

\begin{figure}[h!]
\begin{center}
\includegraphics[width=120mm,height=120mm]{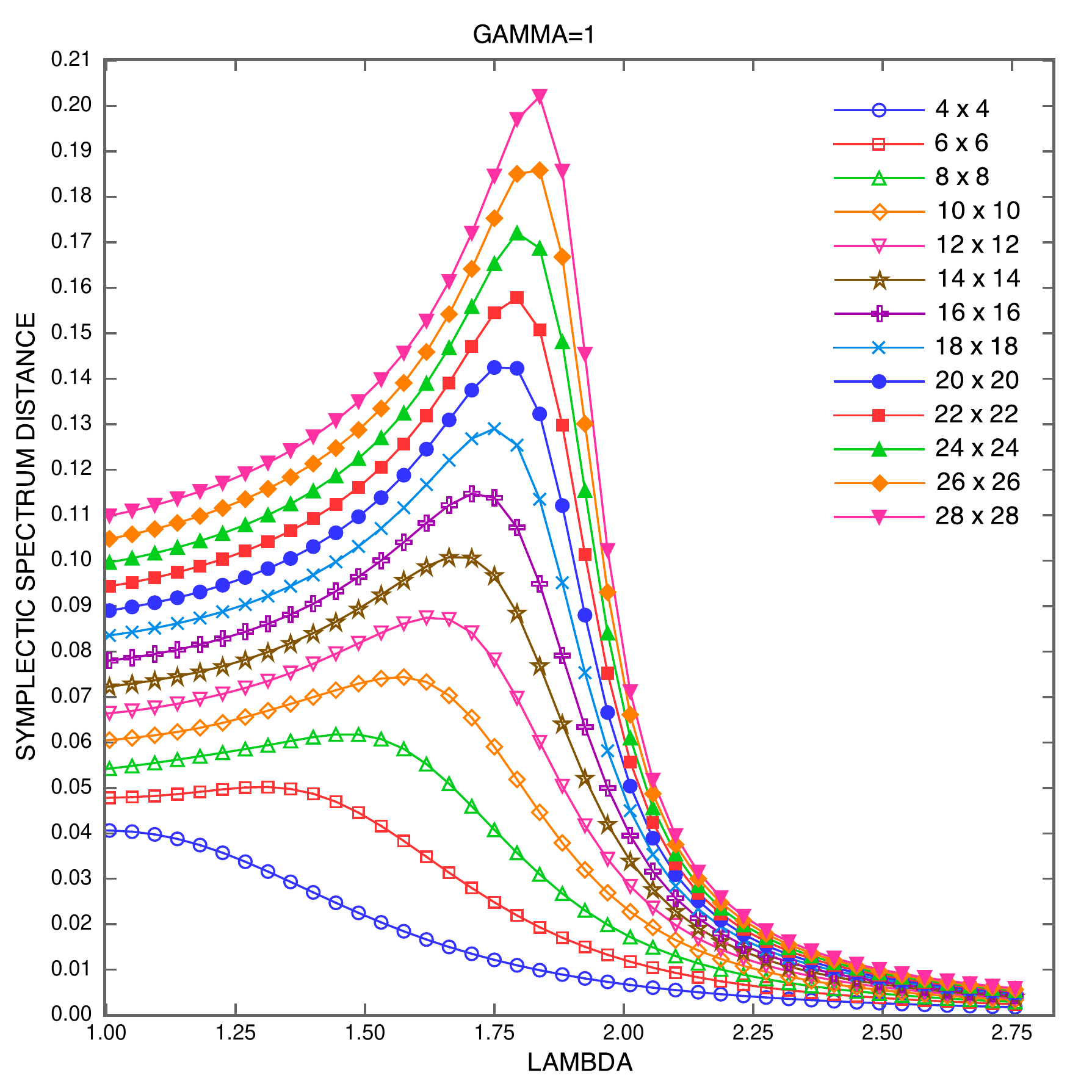} 
\caption{Entanglement spectrum susceptibility as a function of $\lambda$ for different system sizes  ($\gamma=1$). The system has been cut in two equal parts. Here only the region $\lambda>1$ has been plotted compared to previous figure, in order to improve clarity.}\label{FIG:2D-SES-SUSCEP-PART}
\end{center}
\end{figure}

The line $\lambda=2$ is critical, and separates two distinct phases. As in a previous section, we made a numerical experiment to see whether the quantity 
\beq
E(\lambda)=\lim_{\delta \lambda \to 0} \frac{\sqrt{\sum_i (\nu_i( \lambda)-\nu_i( \lambda+d \lambda))^2}}{\delta \lambda}.
\eeq
allows to detect the phase transition. Fig. \ref{FIG:2D-SES-SUSCEP} seems to indicate it does. We are not able to explain the wide oscillations at the left of the diagram. We suspect they can be related to two properties of the Hamiltonian: (i) although both the phase $\{0 < \lambda < 2\}$ ('phase II') and the phase $\{\lambda \geq 2\}$ ('phase III') satisfy an area law $S_L \sim L$, the two-point correlators exhibit exponential decay in phase III, and power law decay in phase II. (ii) the lines $\{\lambda=0 \}$ and $\{\gamma=0 \}$ are critical; they violate the area law and follow Eq. \ref{eq:2d-scaling}. Possibly these lines exert an influence when they are approached. Of course, we could choose to just elude this issue, and present \emph{part} of our numerics. This option would result in Fig.\ref{FIG:2D-SES-SUSCEP-PART}.

%\section{Comparison of conformal field theories}
%%%%%%%%%%%%%%%%%%%%%%%%

%\section{Classical statistical physics}
%%%%%%%%%%%%%%%%%%%

%\section{Ising model vs Heisenberg model}
%%%%%%%%%%%%%%%%%%%%%%%%%%%%%%%%%%

\section{Summary}

We have introduced the entanglement spectrum as a possible distance measure between quantum states. To show that it fulfills the required conditions, we have proved physical equivalence between quantum states that have the same Schmidt spectrum, by considering their representatives in their equivalence classes and showing that there is a fixed local unitary correspondence between all Wightman functions for these states. We have also seen that the difference between two entanglement spectra sets bounds on the difference between the same observables of different states. This suggests that entanglement spectrum can be used as a distance between states, and also as a distance between models when considering the distance between their ground states. We have seen that the distance between close models peaks around criticality, and have checked it for free fermions and Ising and Heisenberg spin chains in 1D and for free fermions in 2D.

\chapter{Tensor networks for frustrated systems}

\pagestyle{fancy}
\fancyhf{}
\fancyhead[LE]{\thepage}
\fancyhead[RE]{MULTIPARTITE ENTANGLEMENT}
\fancyhead[LO]{CH.5 Tensor networks for frustrated systems}
\fancyhead[RO]{\thepage}

In this last chapter of the first part of the thesis we are going to analyse a concept that arises very often in the ground states of certain hamiltonians: geometrical frustration. This phenomenon takes its name from the fact that the geometry of a system sometimes makes it impossible for it to minimize simultaneously the energy of all of its local interactions. As a consequence, the system arranges itself in complex configurations that minimize the global energy but do not minimize it locally, and very often comes with a high degeneration. This has the implication that computations become very expensive and therefore the search for strategies to deal efficiently with these systems it is an important field of study. 

This chapter is divided in two main sections: the first one devoted to a general analysis of geometrical frustration with some simple computations with exact diagonalisations, and a second one dedicated to a method to efficiently compute properties of the ground state for some frustrated systems.

\section{Geometrical frustration}

An early work on frustration is a study of the classical Ising model on a triangular lattice with nearest-neighbour spins 
coupled antiferromagnetically by G. H. Wannier, published in 1950 \cite{Wannier50}, which we will review later on. Since then, frustration has been studied thoroughly both in classical \cite{Toulouse77} and quantum systems \cite{Dawson04}. Here
we will focus on different variations of the quantum Ising model: a 1D chain with second-neighbours coupling, where frustration arises because a spin cannot be coupled antiferromagnetically both
with his first and second neighbours, and the quantum version of the triangular lattice from Wannier's paper. We compute Von Neumann entropies and find the parameters that characterize its maximum and therefore its quantum phase transition, and also find how the maximum scales with the size of the system, which will allow us to identify the universality class of these frustrated systems.

\subsection{Quantum Ising chain with second neighbour couplings}

\begin{figure}[h!]
      \begin{center}
      \includegraphics[scale=0.75]{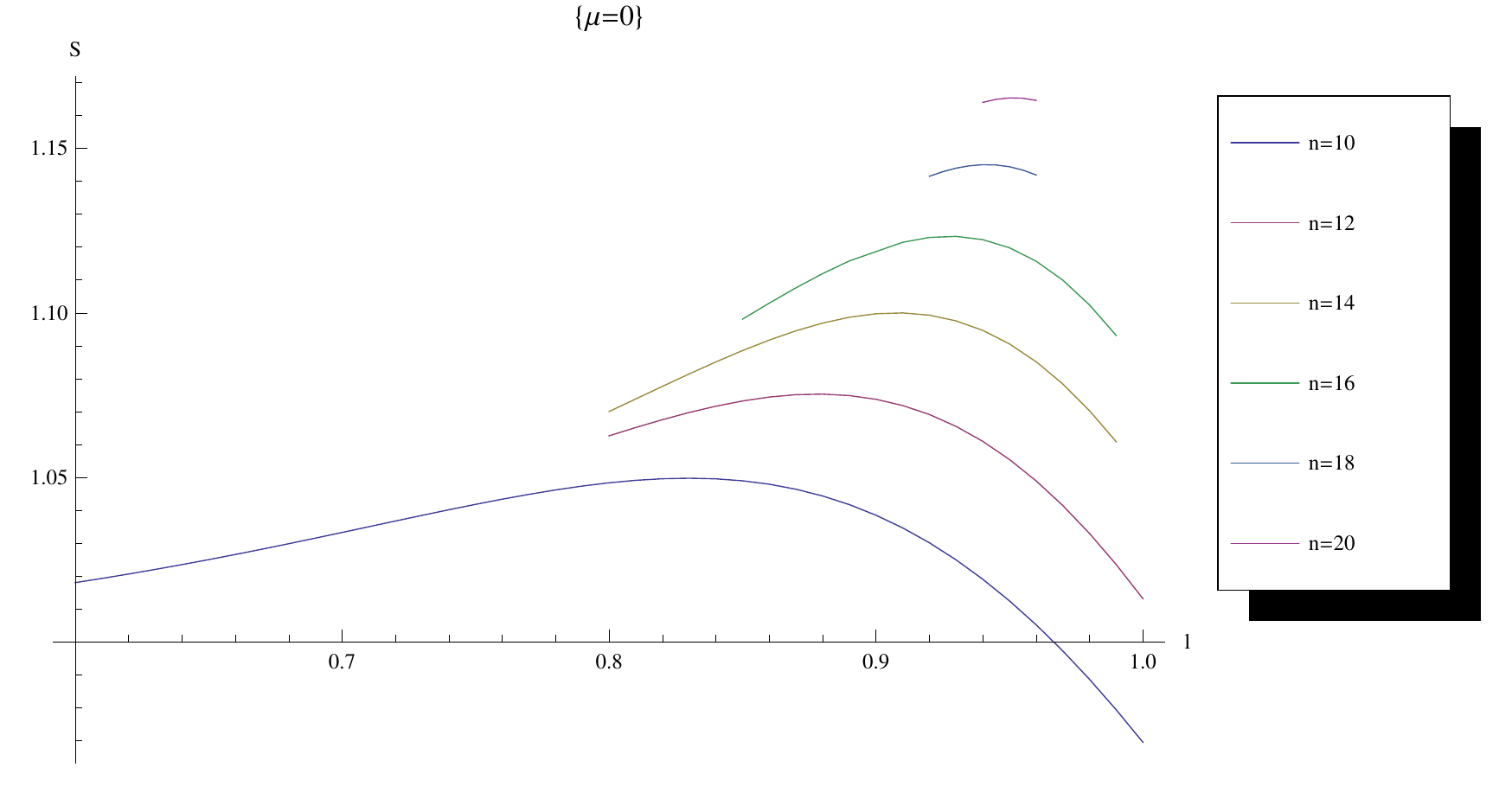}
      \caption{Entropy for varying transverse magnetic field, for four different chains with $N=10-20$ spins, 
without second neighbour coupling.}
\label{fig:smu0}
 \end{center}
      
\end{figure}

\begin{figure}[h!]
      \begin{center}
      \includegraphics[scale=0.75]{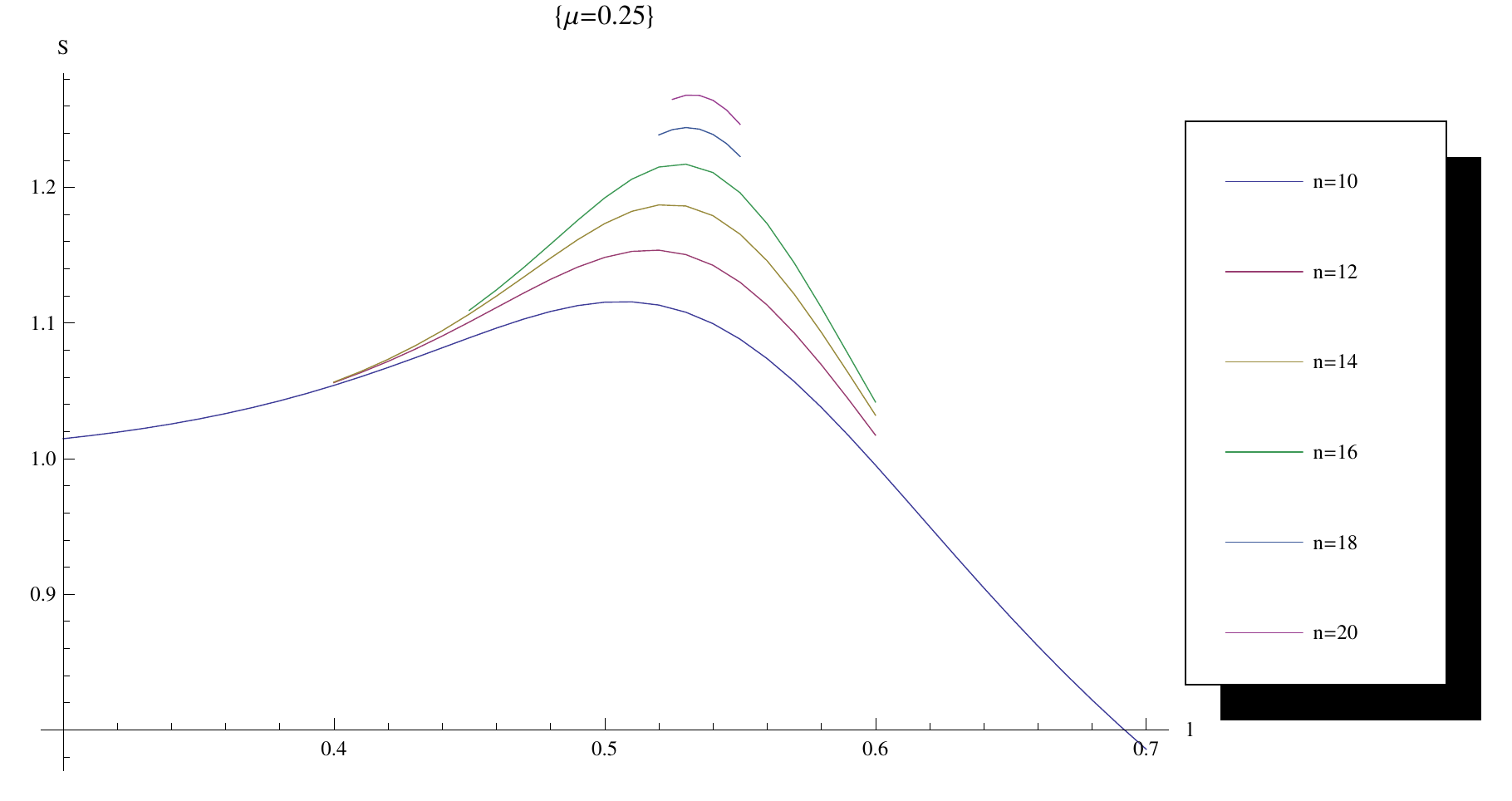}
      \caption{Entropy for varying transverse magnetic field, for four different chains with $N=10-20$ spins 
with $\mu$ = 0.25 second neighbour coupling.}
\label{fig:smu025}
 \end{center}      
\end{figure}

\begin{figure}[h!]
      \begin{center}
      \includegraphics[scale=0.62]{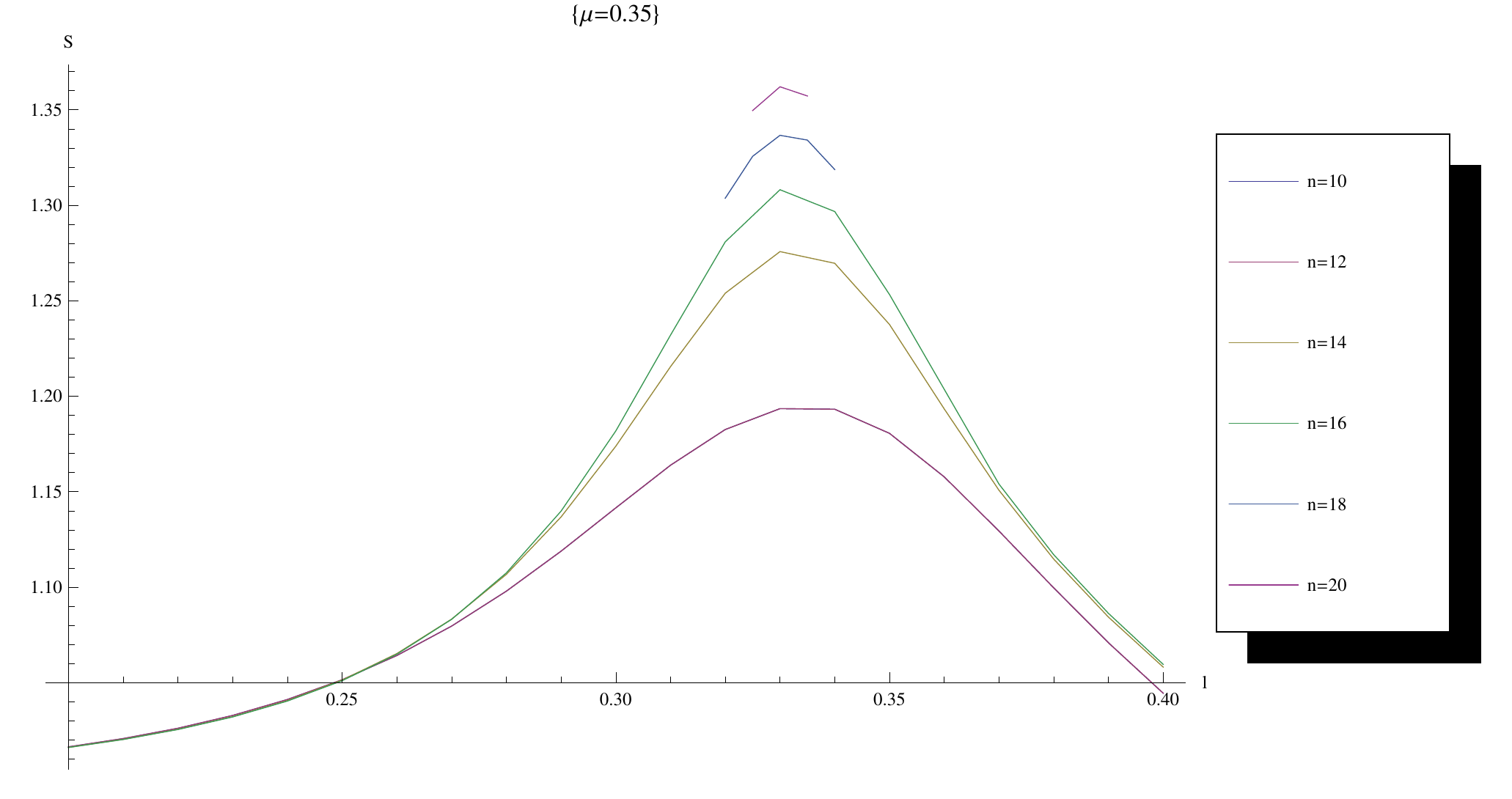}
      \caption{Entropy for varying transverse magnetic field, for four different chains with 10,12,14 and 16 spins,
 with $\mu$ = 0.35 second neighbour coupling.}
       \label{fig:smu035}
 \end{center}
\end{figure}

In this section we show a simple quantum model that exhibits frustration: an Ising chain of N spins with a transverse magnetic field and second neighbour couplings of strength $\mu$
\begin{equation}
{\cal{H}} = \frac{1}{2} \left( \sum_{i=1}^N \sigma_{i}^{x} \sigma_{i+1}^{x} +  \mu\sum_{i=1}^N \sigma_{i}^{x} \sigma_{i+2}^{x} + \lambda \sum\limits_{i} \sigma_{i}^{z} \right)
\end{equation}
In this system there arises frustration because one cannot have antiferromagnetic behaviour at the same time with the nearest neighbour and the second one.
We have diagonalized the system and computed the Von Neumann entropy of half a chain for $N$ from 10 to 20 and different $\mu$ (Fig. \ref{fig:smu0}, \ref{fig:smu025}, \ref{fig:smu035}).

We can see different interesting things with these pictures: the entropy maximums are higher with increasing $N$ and with 
increasing $\mu$. The position of those maximums gets higher with increasing $\mu$, but with increasing $N$ it depends
 on the value of $\mu$: for small $\mu$, the position of the maximum goes to higher fields with increasing $N$, but for big
 $\mu$, the opposite happens.

\begin{figure}[h!]
      \begin{center}
      \includegraphics[scale=0.80]{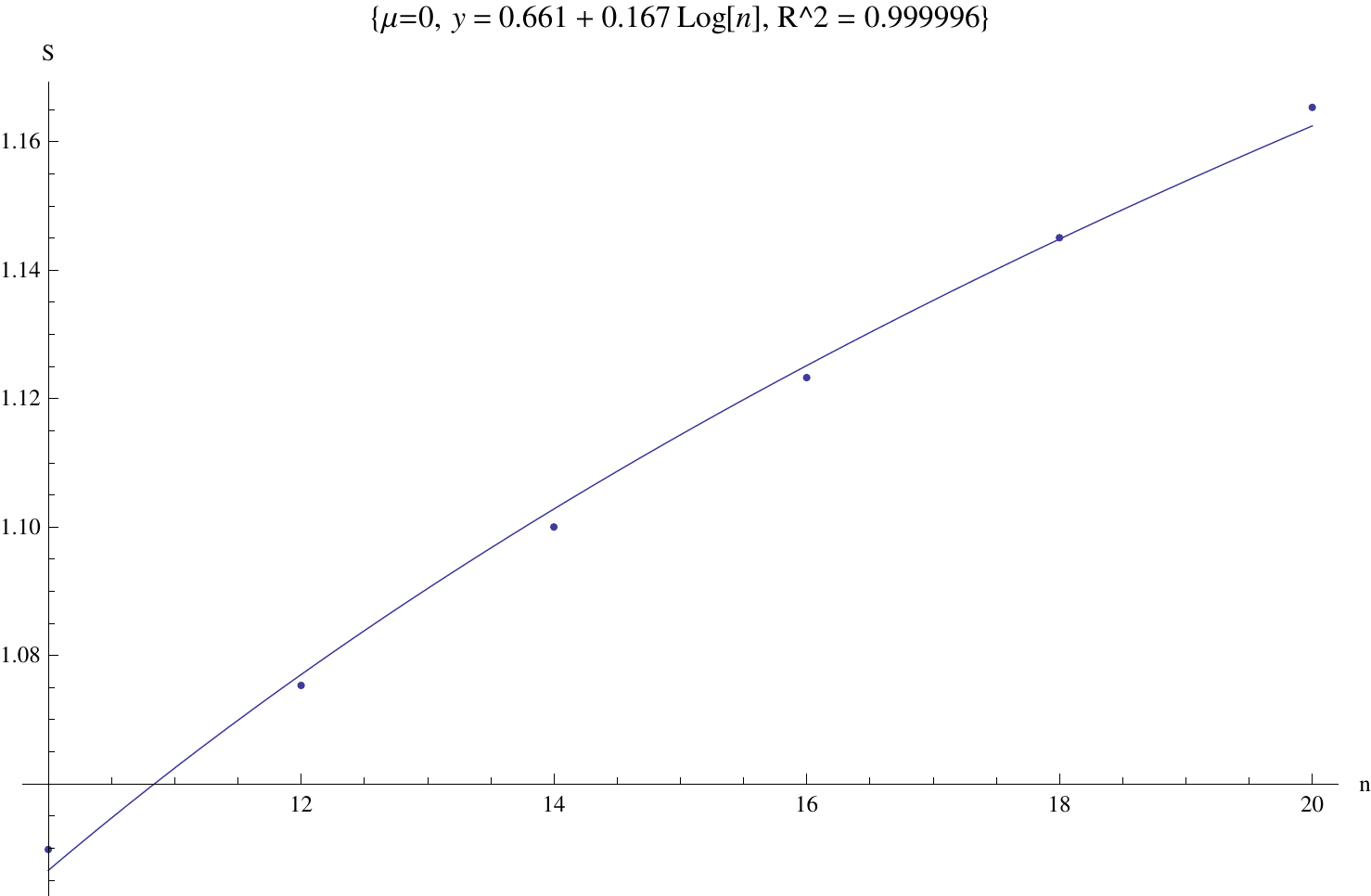}
      \caption{Logarithmic scaling of entropy for $\mu$=0 .}
      \label{fig:regressiomu0}
 \end{center}      
\end{figure}

\begin{figure}[h!]
      \begin{center}
      \includegraphics[scale=0.80]{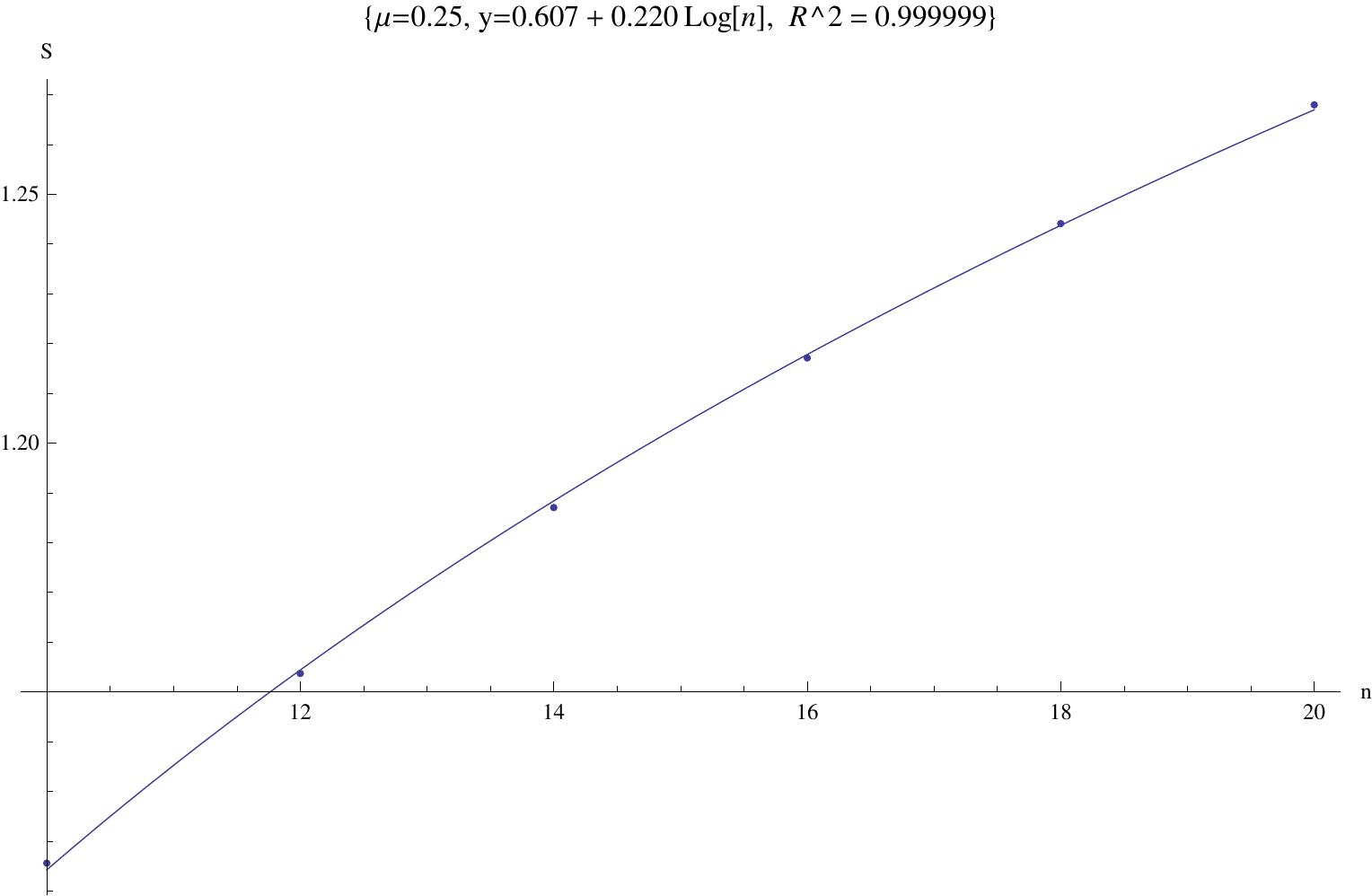}
      \caption{Logarithmic scaling of entropy for $\mu$=0.25}
      \label{fig:regressiomu025}
 \end{center}      
\end{figure}

\begin{figure}[h!]
      \begin{center}
      \includegraphics[scale=0.80]{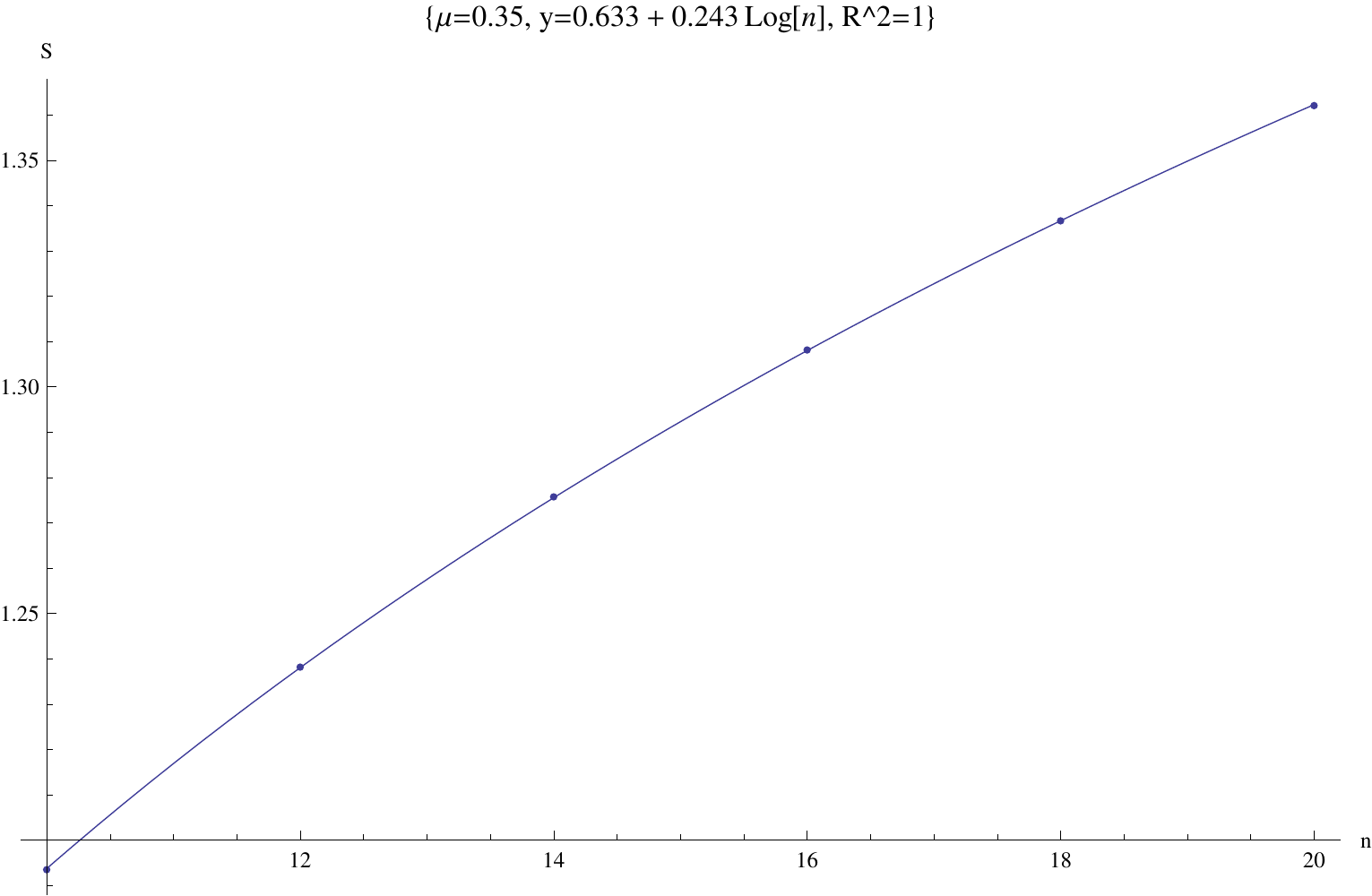}
      \caption{Logarithmic scaling of entropy for $\mu$=0.35}
      \label{fig:regressiomu035}
 \end{center}      
\end{figure}

We also plot the entropy maximums for each $\mu$ to find the scaling of the entropy in each case (Fig. \ref{fig:regressiomu0}, \ref{fig:regressiomu025}, \ref{fig:regressiomu035}).

We can see a logarithmic scaling in the three figures, which is not perfect because of the small size of our chains. This fits with the well-known logarithmic scaling expression for the entropy of a part of length l of an infinite chain \cite{Holzhey94}
\begin{equation}
S = \frac{c}{3} \log l,
\label{eq:infinitelogscaling}
\end{equation}
where c is the \emph{central charge} of a conformal field theory (CFT), which is known to describe the quantum Ising chain in its critical point, and to describe phase transitions in general.
What is interesting is that the coefficient of the linear slope, which depends on the central charge, gets higher with increasing $\mu$, something that indicates that different second-neighbour couplings imply different associated CFTs and therefore a change in the \emph{universality class}.

It must be noted that in our case it doesn't matter whether we try to fit our results to the infinite-size system scaling (Eq. \ref{eq:infinitelogscaling}) or to the finite-size scaling
\begin{equation}
S = \frac{c}{3} \log \left(\frac{L}{\pi} \sin \frac{\pi l}{L}\right),
\end{equation}
because we are keeping $\frac{l}{L}$ constant $(=\frac{1}{2})$, so the fit will be the same.

\subsection{Classical Triangular spin systems}

\subsubsection{Description of the model}

 In this section we consider a triangular lattice of spins, where they can take values +1 or -1 and interact by the
 antiferromagnetic Ising hamiltonian without any external field:

\begin{equation}
{\cal{H}} = \sum\limits_{\lbrace n,n \rbrace} S_{i} \cdot S_{j}
\end{equation}

Where we consider only interaction between first neighbours. This kind of systems with triangular 
structure and antiferromagnetic interaction shows geometrical frustration, and different configurations with the same minimum energy appear, so that the ground state will be highly degenerated.\\

We are going to analyse how this degeneration grows with the size of the system, by looking for all the possible configurations of a specific 
system and counting how many of them show minimum energy. We consider a finite triangular lattice, with $l$ rows, so that the jth-row has $j$ spins in it. The simplest approach is just to find the energy of each 
possible configuration given a lattice of $N$ spins, (where $N=1+2+...+l$) and sum the number of all possible configurations 
with the lowest energy. The number of total possible configurations is $2^{N}$ , so the amount of memory and time needed 
will increase exponentially with the number of spins. We can't get further from 12 rows with this method.
In a second approach, we use a different strategy which lets us get the same results and go further away with much better performance. Given
 a triangular lattice with k rows, if we know all the possible degenerate configurations in the ground state, we can use 
this information to get the degenerate configurations for a lattice with $k+1$ rows without having to explore the whole space again. The reason for this is that the spins in the new row $k+1$ only interact with the ones in row $k$, so in order to find the number of degenerate states we 
only need the configurations of row $k$. The important difference between this approach and the previous one is that the total number of configurations that it must explore before finishing is approximately $2^{k+k+1}$ and not $2^{N}$. For a small 
number of rows this isn't really significant, but we may notice that adding an extra row of k spins means and extra effort
 of $2^{k}$ times the previous number of configurations in the first case, while in the second the extra effort for any extra
 row is always $2^2=4$ times. With this second method we have been able to compute the degeneration for a system of up to 17 rows, the results are shown in Table \ref{table:nonlin}.

 \vspace{3 mm}

\begin{table}[h!]

\centering
\begin{tabular}{c c c}
\hline\hline
Rows & n & Degeneration \\ [0.5ex] % inserts table %heading
\hline
2 & 3 & 6 \\
3 & 6 & 26\\
4 & 10 & 160 \\
5 & 15 & 1386 \\
6 & 21 & 16814 \\
7 & 28 & 284724 \\
8 & 36 & 6715224 \\
9 & 45 & 220240306 \\
10 & 55 & 10032960146 \\
11 & 66 & 634271091558 \\
12 & 78 & 55607968072800 \\
13 & 91 & 6757401238296442 \\
14 & 105 & 1137661035904122264 \\
15 & 120 & 265265658215457903864 \\
16 & 136 & 85635780217381861437248 \\
17 & 153 & 38267278120418832223426206 \\ 
[1ex]
\hline
\end{tabular}
\caption{Degeneration of the number of configurations of minimum energy for a triangular antiferromagnetic lattice of up to 17 rows. $n$ is the total number of spins.}
\label{table:nonlin}
\end{table}

\begin{figure}
\begin{center}
 
\includegraphics[scale=0.68]{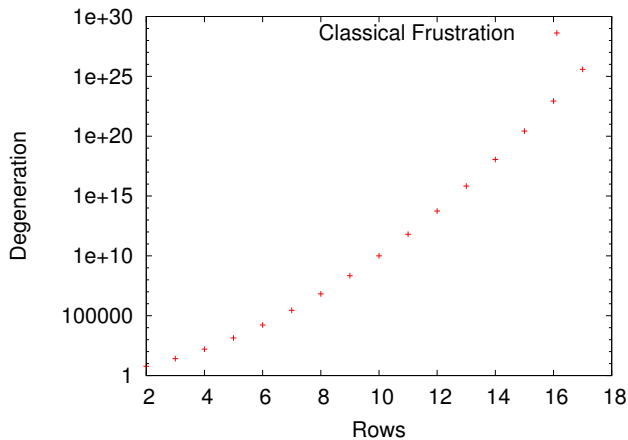}
\caption{Ground state degeneration in spin triangular
 antiferromagnetic lattices.}
 \end{center}
 \end{figure}

\subsubsection{Review of the Wannier paper}

Based on the previous results, we take a look at a paper by Wannier \cite{Wannier50}, written in 1950, which analyses the 
properties of an antiferromagnetic triangular Ising net. It applies statistical mechanics, through a dual relationship
 between the triangular and honeycomb nets, to compare results like the energy and entropy with the ferromagnetic case. We will focus on the antiferromagnetic results.

Concerning the energy, it says that the internal energy is one third of the value obtained in a ferromagnetic triangular lattice. This fits with the results obtained with our method, as the minimum energy for each number of rows is exactly one third of 
the one that we would expect in the ferromagnetic case, which exhibits no frustration at all. An easy way to see this is that if we divide the complete lattice in small triangles pointing up of 3 spins each, each triangle would contribute with -3 to the total energy with a ferromagnetic interaction (one 
for each bound) while with antiferromagnetic interaction it would contribute with -1 (as two of the bounds would contribute 
with the same energy but different signs).\\

\begin{figure}
\begin{center}
\includegraphics[scale=0.68]{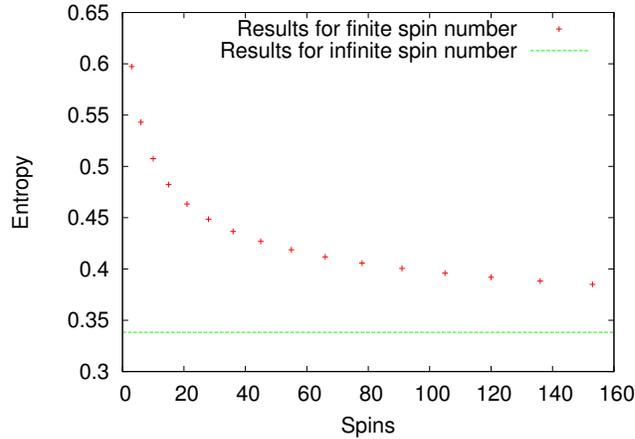}
\caption{Classical entropy of the system, where we compare the results obtained for finite lattices with the result 
analytically obtained by Wannier for an infinite lattice. We see that the finite results tend to the infinite result for increasing number of spins.}
\label{fig:entroclass}
\end{center}
 \end{figure}

The entropy is calculated in the paper taking an infinite lattice, while our program takes a finite number of rows. We can then take a look at how the entropy per spin changes as a function of the number of spins in the lattice. We represent the entropy as the logarithm of the degeneration, without physical
 constants, so that it fits the units given by the paper. Figure \ref{fig:entroclass} shows how the finite-spin entropy approaches the calculated value as the number of spins increases.

% GNUPLOT: LaTeX picture

\subsection{Quantum triangular spin systems}

\subsubsection{Description of the model}
In this section we consider the same geometry as in the previous section, but using as interaction the quantum Ising model of the first section, except for two differences: this time there will be no second-neighbour coupling,
but we will allow different interaction strengths for the vertical and horizontal couplings
\begin{equation}
{\cal{H}} = J_{h}\sum\limits_{\lbrace h,n \rbrace} \sigma_{i}^{z} \sigma_{j}^{z} +  J_{v}\sum\limits_{\lbrace v,n \rbrace}
\sigma_{i}^{z} \sigma_{j}^{z} + \lambda \sum\limits_{i} \sigma_{i}^{x}
\end{equation}
Where $J_v$ and $J_h$ account for the interaction between spins of neighbouring rows and spins of the same row respectively (vertical and 
horizontal interaction).\\

Using the previous hamiltonian for different number of rows and different interaction strengths, we diagonalize the hamiltonian matrix 
and study the energy and Von Neumann entropy of the ground state, looking for its maximums and therefore the phase transitions.

\subsubsection{Results and scaling}

\begin{figure}[h!]
\begin{center}
\includegraphics[scale=0.68]{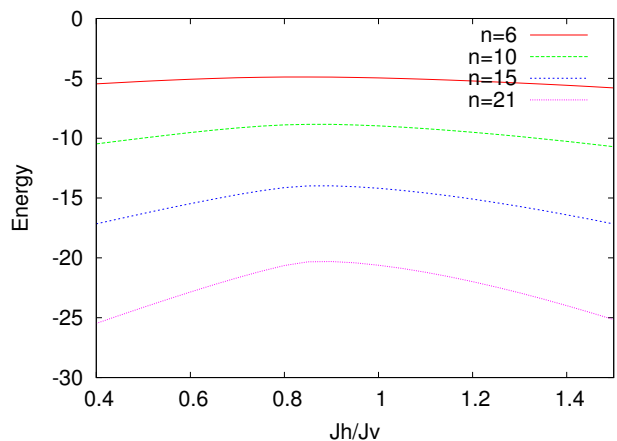}
\caption{Energy dependence on the ratio Jh/Jv for different number of spins. We see that for bigger systems the ground state energy gets lower and that the energy peak gets more pronounced, therefore making the phase transition easier to identify.}
\label{fig:energiawl}
\end{center}
 \end{figure}

In figures \ref{fig:energiawl}-\ref{fig:coefscale} we have shown results with different fields and different sizes of the system. All have been 
represented keeping the vertical interaction with a constant value of 1 and varying the horizontal interaction. We have computed the entropy between the last row and the rest of the system. Other entropies have been considered, as the case of the single spin on the top of the system in relation to the rest, but they weren't so useful to study the phase transition.

After getting results up to 6 rows (21 spins) we have represented the area scaling of the maximum of entropy for different
magnetic fields, plotting linear regressions which show a possible area law. But it should be taken into
account that when taking the entropy from the last row, we are not able to distinguish scaling caused by area from 
scaling by volume. This may be solved by studying the entropy between other parts of the system, but in general we would need bigger systems to distinguish area from volume well. Next section presents a new method to deal more efficiently with systems with frustration.

\begin{figure}[h!]
\begin{center}
\includegraphics[scale=0.68]{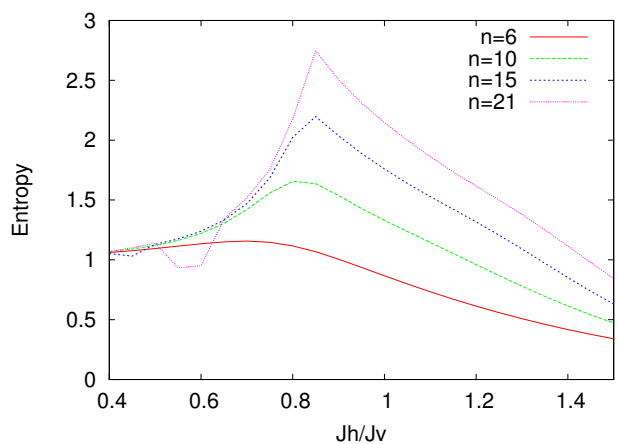}
\caption{Entropy dependence on the ratio Jh/Jv for different number of spins. Here the maximum of entropy in the
 phase transition is really clear, specially for the biggest systems. Irregular behaviour shown for small fields is 
only caused by a lack of precision of the program.}
\end{center}
 \end{figure}
 
 \begin{figure}[h!]
 \begin{center}
\includegraphics[scale=0.68]{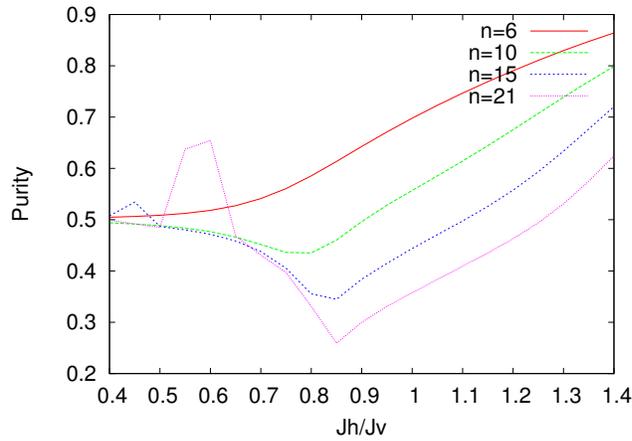}
\caption{Purity dependence on the ratio Jh/Jv for systems with different number of spins. The minimum of purity corresponds to the phase transition and to maximum entanglement.}
\end{center}
 \end{figure}
 
 \begin{figure}[h!]
 \begin{center}
\includegraphics[scale=0.68]{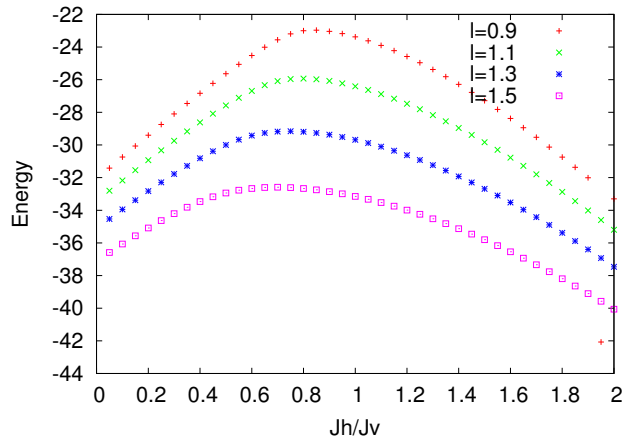}
\caption{Energy dependence on the ratio Jh/Jv for the case of 21 spins. Results are shown for
 four different external magnetic field values $l$.}
 \end{center}
 \end{figure}
 
 \begin{figure}[h!]
 \begin{center}
\includegraphics[scale=0.68]{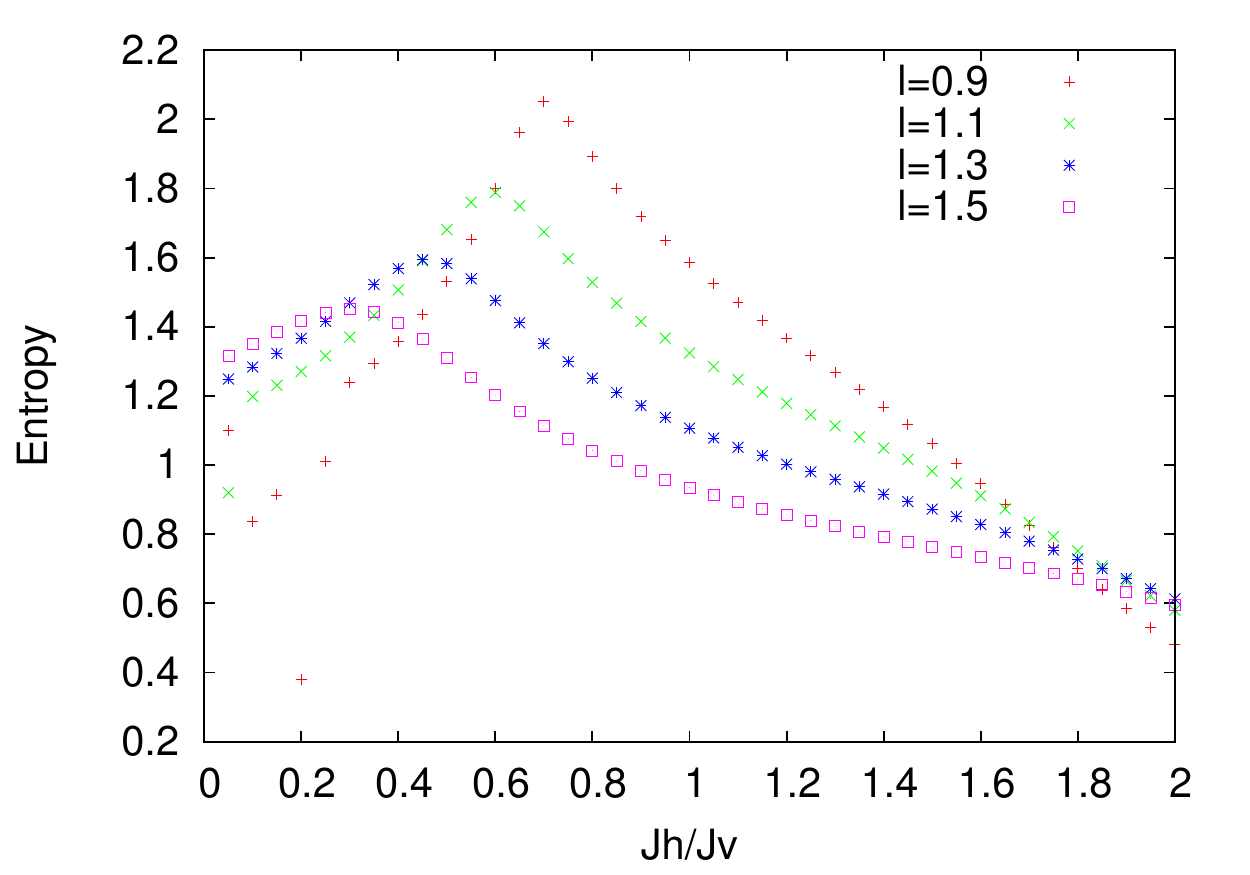}
\caption{Entanglement entropy dependence on the ratio Jh/Jv for the case of 21 spins. As can be seen, changing the external field produces changes in the value of entropy and its position, with the peak getting more pronounced for stronger fields. As mentioned before, irregular behaviour is a consequence of lack of precision of the program.}
\end{center}
 \end{figure}
 
 \begin{figure}[h!]
 \begin{center} 
\includegraphics[scale=0.68]{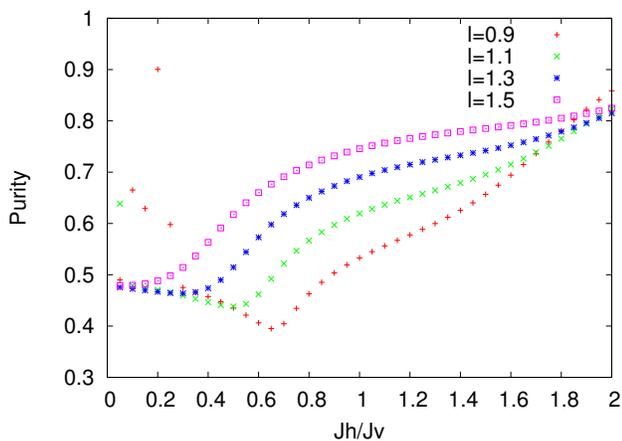}
\caption{Purity dependence on the ratio Jh/Jv for the case of 21 spins.}
 \end{center}
 \end{figure}

 \begin{figure}[h!]
 \begin{center}
\includegraphics[scale=0.3]{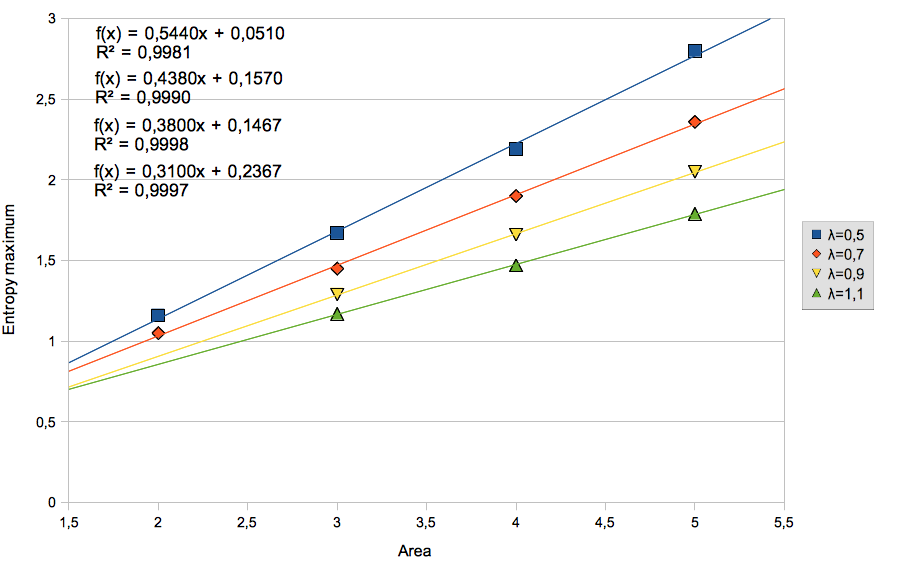}
\caption{Scaling with area of the entropy maximum for different external fields. As we already said, with this example we are not able to distinguish between scaling with area and scaling with volume.  }
 \end{center}
 \end{figure}
 
 \begin{figure}[h!]
 \begin{center}
\includegraphics[scale=0.48]{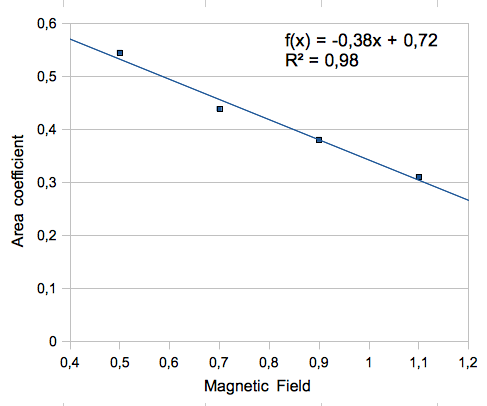}
\caption{Slope coefficient for different external field values. The 4 slope values correspond to the previous figure linear regressions.}
\label{fig:coefscale}
 \end{center}
 \end{figure}

\section{Triangular simplices as tensor networks}

We are going to study a method to better deal with frustrated systems than exact diagonalisation, which we have seen to fail for moderately large lattices. We consider a frustrated anti-ferromagnetic triangular lattice Hamiltonian and show that the properties of the manifold of its degenerated ground state are 
represented by a novel type of tensor networks. These tensor networks
are not based on ancillary maximally entangled pairs, but rather on triangular W-like simplices. 
Anti-ferromagnetic triangular frustration
is then  related to ancillary W-states in contrast to ferromagnetic order
which emerges from the contraction of GHZ-like triangular simplices.
We further discuss the outwards entangling power of various simplices. 
This analysis suggests the emergence
of distinct macroscopic types of order from the classification of entanglement residing on the simplices that define a tensor network. 

\subsection{Introduction to tensor networks}

Tensor networks stand as a powerful variational approach to quantum mechanical systems that can compute relevant quantities much more efficiently than exact diagonalisation. In general they represent very well states that have relatively low long-range entanglement, which are in fact the vast majority of states that arise in nature.
They have a great advantatge in that they escape the sign problem that hampers Monte Carlo simulation \cite{Loh90}. Popular 
tensor networks are those that adapt their connectivity to the natural setting of a
particular system. Relevant instances of tensor networks include Matrix Product States (MPS) \cite{White92} introduced in Sect. \ref{sect:RG} and their complement Matrix Product Operators (MPO), for translational invariant
systems in one dimension, Projected Entangled Pairs States 
(PEPS) \cite{Verstraete04} for translational invariant
systems in two dimensions, and MERA structures \cite{Vidal07} for 
scale invariant dynamics. All these approximation techniques are based on the use of ancillary
maximally entangled states that are projected in different manners along the network.
Yet, this strategy seems to capture the physics of frustration in a poor way. We here shall 
propose the construction of a novel type of tensor networks based on triangular
ancillary states which are related to W-type entanglement (see Sect. \ref{sec:3q}) and, as a consequence,
are suited to describe geometric frustration.

\subsection{Tensor network representation}

\begin{figure}[h!]
      \begin{center}
            \includegraphics[scale=0.27]{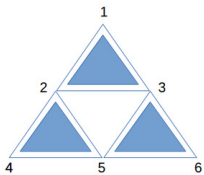}
            \caption{\label{fig:triangle}Example of triangular lattice geometry. Only up triangles have to be
considered so as not to double count each link.
             }
      \end{center}
      
\end{figure}

We shall take the quantum anti-ferromagnetic 
triangular lattice model from previous section, this time with $Jh=Jv=J$
\begin{equation}
H= J \sum_{\{ i,j\} \in T} \sigma^z_i \sigma^z_j  + \lambda \sum_i \sigma^x_i ,
\label{eq:antifmham}
\end{equation}
where $\{ i,j \} \in T$ represents nearest neighbor
interaction on the triangular geometry shown
in Fig. \ref{fig:triangle}. It is easy to see that even in the case of $\lambda=0$ and $J>0$  
there is no arrangement of spins that minimize every term in the
Hamiltonian.

We may then
consider the equal superpositions of all valid states on the computational
basis that carry the same minimum energy $E_0=-n/3$,
\begin{equation}
  \vert \psi_0\rangle =\frac{1}{\sqrt M}\sum_i \vert \psi_i\rangle \qquad H |\psi_i\rangle=E_0 
|\psi_i\rangle , \quad i=1\ldots,M
\end{equation}
where $M$ corresponds to the degeneracy of the ground state manifold.
The degeneracy is lifted
when an external transverse field is applied. 
The special
case where ${\lambda\to 0}$ produces a particular combination of all $|\psi_i\rangle$.
In both cases, namely in the equal superposition or in the limit to zero of the 
external field, the resulting states carry a large entropy, as we shall discuss later.

The question we shall address here is whether a frustrated state such as $|\psi_0\rangle$ accepts a natural representation in terms of tensor networks.
Let us start by observing that in the case of 3 spins forming a triangle, the
equal superposition of valid ground states is
\begin{eqnarray}
\nonumber
  | \psi_0 \rangle &&= \frac{1}{\sqrt 6} \left(|001\rangle+|010\rangle+|100\rangle \right. \\
  && \left. +|011\rangle+|101\rangle+|110\rangle \right) \nonumber \\  && = \frac{1}{\sqrt 2}
 \left( | W \rangle + | \bar W \rangle \right) ,
\end{eqnarray}
where $| W\rangle = \frac{1}{\sqrt 3}\left(|001\rangle+|010\rangle+|100\rangle\right)$ and $| \bar W\rangle = \frac{1}{\sqrt 3}\left(|011\rangle+|101\rangle+|110\rangle\right)$. This introduces in a natural way $W$-states and its special tripartite entanglement described in Sect. \ref{sec:3q}. Frustration in this state amounts
to equally superpose all the possibilities of assigning frustrated triangles, 
hence the emergence of $W$-type entanglement. Note that if we relax
the requirement of equal superposition of possible ground states but retain
isotropy, a freedom on the relative weight of $| W\rangle$ {\sl vs.} $| \bar W \rangle$
states appears. 

We next consider a larger structure, as the one shown in Fig. \ref{fig:triangle}, and 
compute the ground state in the limit of zero external transverse field.
The result reads
\begin{eqnarray}
\nonumber
 | \psi_0 \rangle &&= \alpha \left(|001100\rangle+|010001\rangle+|011101\rangle\right)
\\ &&+ \beta \left(|001101\rangle+|010011\rangle+|001110\rangle \right.+
\nonumber
\\ &&\left.|010101\rangle+|011001\rangle+|011100\rangle\right)
\nonumber
\\ &&+ \gamma \left(|001010\rangle+|010010\rangle+|011000\rangle\right) \nonumber
\\ &&+ \delta |011010\rangle \ .
\label{eq:ground6}
\end{eqnarray}
with $\alpha\sim -.24$, $\beta\sim .19$, $\gamma\sim -.16$ and $\delta\sim .15$.
It is convenient to analyse this state by looking at the distribution of spins
on all triangles pointing up, since the triangles pointing down only provide
a redundant description of the system. There are three up-triangles to be considered,
respectively formed by the qubits \{1,2,3\}, \{3,4,5\} and \{3,5,6\}.
The relevant observation is that each state forming the superposition of
the ground state in Eq. \ref{eq:ground6} is formed by a member of either a $W$ state or a $\bar W$ state.
It is furthermore possible to verify that all the
states in the equal superposition of valid ground states are made of elements of $W$- and
$\bar W$-like states in the up-triangles.
Though we may find a down-triangle with the configuration
111, this does not invalidate the fact that all up-triangles remain a member of genuine $W$ tripartite entanglement. 

Let us now prove that this is the general case, namely that
the equal superposition of valid ground states $| \psi_0\rangle$
is made of all possible combinations of $W$ and $\bar W$ configurations
on all ancillary simplices. To prove this result we first 
consider the Hamiltonian
\begin{equation}
  H_W=\sum_{i,j,k\in T^*} (z_i+z_j+z_k-1)^2 ,
\end{equation}
where $z_i=\frac{1+\sigma^z_i}{2}$  and $T^*$ spans the set
of up-triangles. Expanding this Hamiltonian we find
\begin{equation}
  H_W=\frac{1}{2}\sum_{i,j\in T} \sigma^z_i \sigma^z_j + \sum_i \sigma^z_i + 1 ,
  \label{eq:wham}
\end{equation}
This construction shows that the ground state of the Hamiltonian $H_W$ belongs to
the manifold spanned by $W$ states and that it carries $E_0=0$ energy. This intuitive technique to construct frustrated Hamiltonians indicates
that the way to represent the frustrated triangular dynamics 
needs to cancel the linear terms in $\sigma^z$.
This can be done as follows
\begin{eqnarray}
  \nonumber
  H&&=\sum_{i,j\in T} \sigma^z_i \sigma^z_j \\
  \nonumber
&&=
  \sum_{i,j,k\in T^*} \left( (z_i+z_j+z_k-1)^2 \right.\\
&&\left. + (z_i+z_j+z_k-2)^2 -2 \right) \ ,
\end{eqnarray}
where now ${i,j,k}$ are indexes of the sites that form up-triangles $T^ *$.
The original anti-ferromagnetic triangular Hamiltonian is recovered as a
sum of two conditions, one trying to produce a superposition of $W$ states and another
of $\bar W$ states. Both conditions can not be fulfilled simultaneously,
so that the energy picks a penalization of one unit for each up-triangle coming
from one of the two terms, 
hence $E_0=-n/3$, which is the number of up-triangles.

\begin{figure}[h!]
      \begin{center}
            \includegraphics[scale=0.27]{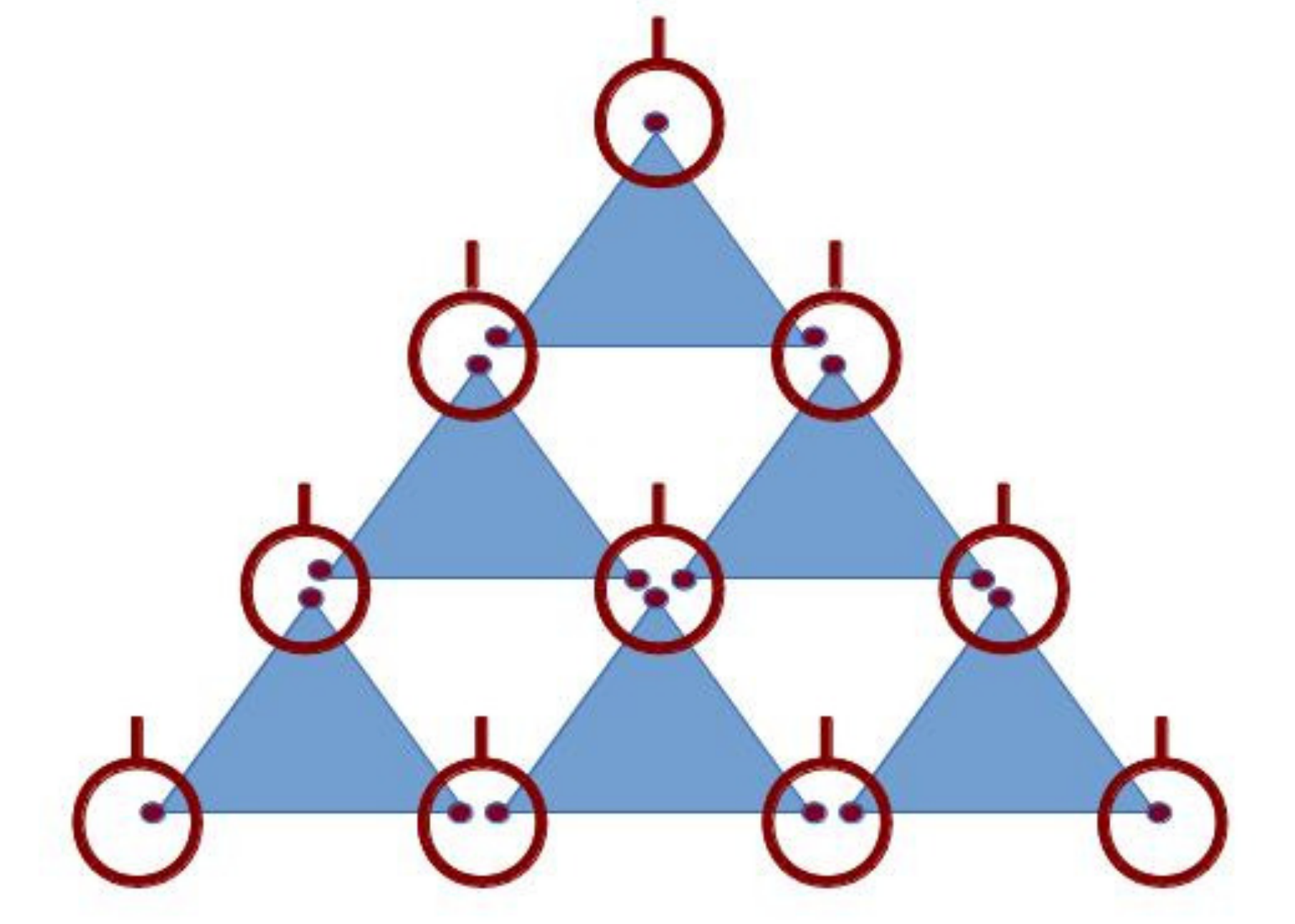}
            \caption{\label{fig:tensornetwork}Tensor network based on a triangular simplex described 
by the tensor $A^ i_{\alpha\beta\gamma}$,  which projects the underlying ancillary indexes onto
a physical one.
             }
      \end{center}
      
\end{figure}

The above arguments allow for the construction of a tensor network that
represents the state $| \psi_0\rangle$ in an exact way. We first consider
the filing of up-triangles all across the triangular lattice with $W$ and
$\bar W$ ancillary states, that is ancillary quantum degrees of freedom
of dimension $\chi=2$,
 as shown in Fig. \ref{fig:tensornetwork}.

\begin{figure}[h!]
      \begin{center}
            \includegraphics[scale=0.22]{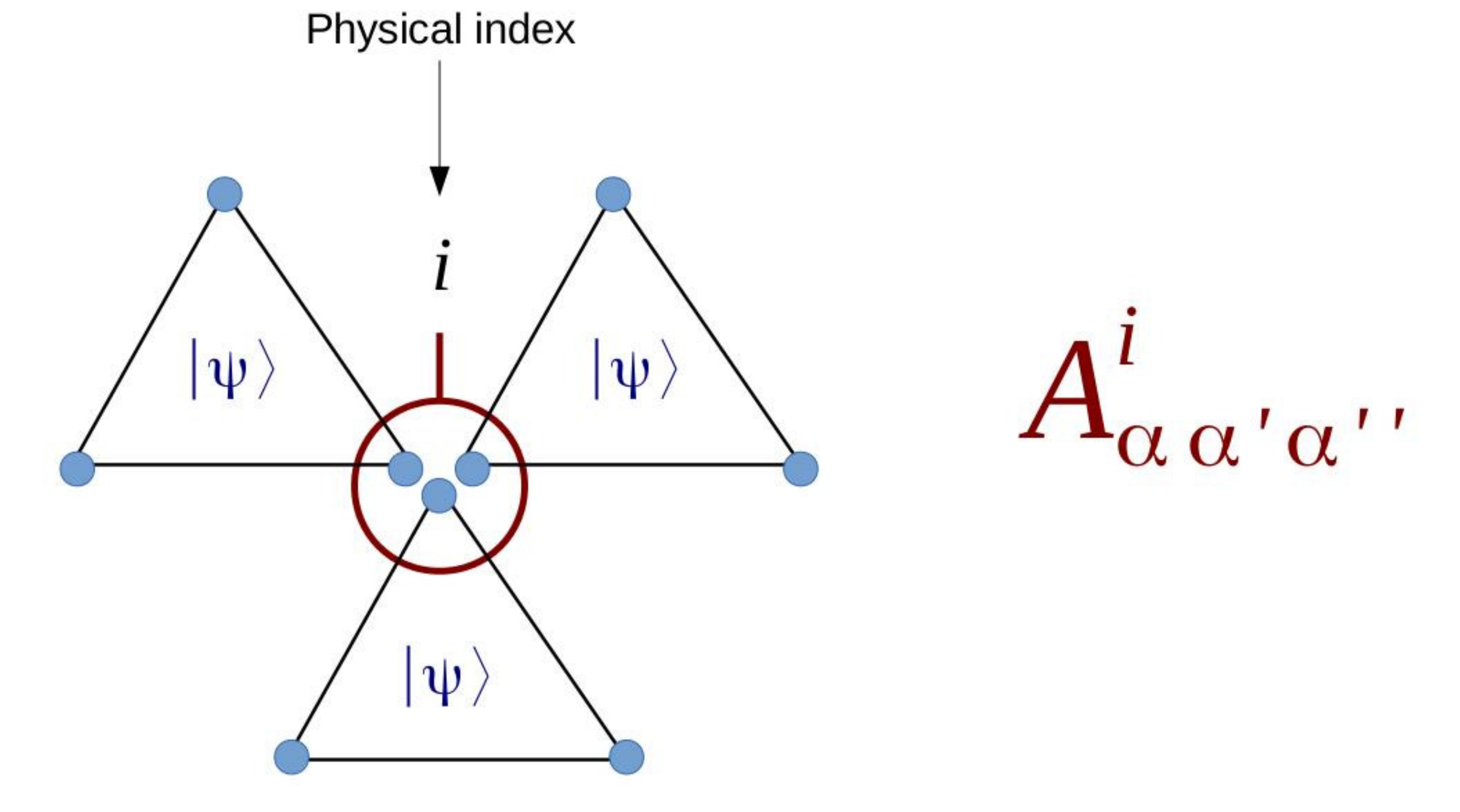}
            \caption{\label{fig:tensor}Detail of the way three concurrent qubits with
            indexes $\alpha$, $\beta$ and $\gamma$, coming from
 entangled triangles are projected into a physical index $i$ defining the
tensor $A^i_{\alpha\beta\gamma}$. The triangular anti-ferromagnetic 
equal superposition of ground states is obtained if $|\psi\rangle=
\frac{1}{\sqrt 2}\left( | W \rangle + | \bar W \rangle \right)$ and $A^i_{\alpha\beta\gamma}=\delta^i_\alpha \delta^i_\beta \delta^i_\gamma$.
             }
      \end{center}
      
\end{figure}
%\end{comment}

We then consider the projection
\begin{equation}
A^i_{\alpha\beta\gamma}=\delta^i_\alpha \delta^i_\beta \delta^i_\gamma .
\end{equation}
That is, the tensor only takes non-vanishing +1 value if all ancillary indexes 
$\alpha, \beta,\gamma$ agree at
that point, and pass their value to the physical index $i$.
It is easy to check that the contraction of ancillary indexes for this tensor network 
reproduces the state $| \psi_0\rangle$. Notably, a similar construction
based on the use of $GHZ$ states at each simplex does describe global ferromagnetic
order. The interplay between $GHZ$ and $W$ entanglement at the level of 
ancillary simplices is the key to distinct types of order at large scales. These triangular simplices are a particular case of
the PESS from Ref. \cite{Arovas08}.

A first consequence from the above construction is to observe that
the entanglement of state $|\psi_0\rangle$ is bounded to obey
area law scaling at most. 
As mentioned previously,  $|\psi_0\rangle$
is made of an exponential superposition of states. Thus, in principle, this
state could display a volume law scaling of entanglement.
Yet, $|\psi_0\rangle$ is
described by a local tensor network that only links each spin to its nearest neighbor.
That immediately sets a bound on the entropy of the state. Indeed, the entanglement
entropy of the state $| \psi_0\rangle$ will scale as the area law at most. 
Moreover the ancillary states carry dimension $\chi=2$, so the bound for
the entanglement entropy of a region A with boundary $\partial A$ is just $S(A)\le
\log 3 \ \partial A$, being 3 the options that emerge outwards from each qubit. 
This is fully consistent with the idea that local interactions of translational
invariant systems produce ground states that obey area law scaling for the entanglement
entropy \cite{Eisert10}.
 
Let us now show that the triangular frustrated system is a particular case of 
the Exact Cover NP-complete problem, closely related to 3-SAT problem. Exact Cover
is a decision problem based on the fullfilment of 3-bit clauses.
To be precise, we are asked to decide whether a set of $n$ bits
accept an assignment such that a set of clauses involving three bits are all satisfied.
Each clause is obeyed if the three bits involved in it take values 001, 010 or 100.
It was proven in Ref. \cite{Garcia12} that there is an exact tensor network that describes the 
possible solutions of this problem and that the hard part
of deciding the instance is found in the contraction of the tensor network. 

In our case, the construction we have
proposed previously can be seen as the particularization of the 
Exact Cover to the problem of a regular triangular lattice.
This implies that triangular frustration is a sort of simple and regular version of Exact Cover.
Indeed, Exact Cover clauses involve bits that have no geometrical proximity relation. 

This observation can be translated to a statement about frustration cycles.
The triangular model produces frustration at the level of single triangles.
Instead, Exact Cover produces frustration over a non-local 
and non-homogeneous triangular lattice. Typical cycles of frustration in
Exact Cover are of $\log n$ size: therefore,  the NP  Exact Cover 
problem is much harder than regular triangular
lattices models because of the long scale cycles for frustration. 

\subsection{Entangling power}

\begin{figure}[h!]
      \begin{center}
            \includegraphics[scale=0.27]{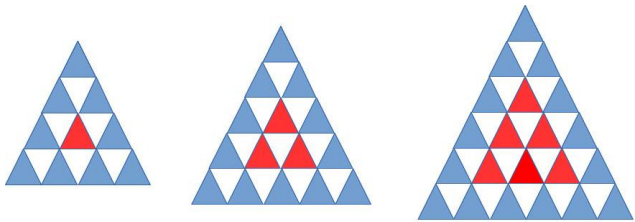}
            \caption{\label{fig:outwards}Outwards entangling power of simplices can be
assessed by computing the entropy of the reduced density matrix corresponding to the
red area made of $n_A$ spins, with $n_A = 3, 6, 10$ in this figure.
             }
      \end{center}
     
\end{figure}

We now turn to the issue of how ancillary entanglement develops into large scale correlations.
We shall refer to this property as {\sl outwards entangling power} of an ancillary simplex.
We may visualize this process by first focusing on a single ancillary triangle.
The different superpositions which are accepted on this ancillary state 
propagate outwards distinct configurations. For instance, the trivial case where
the internal state corresponds to a 000 configuration can only propagate a global
product state made of zeros. In the case where a $W$ ancillary state is used,
the global state retains only a small amount of entanglement due to the dominant
role of 0 versus 1 ancillary states. A remarkable 
jump in entanglement entropy is obtained when $W$ and $\bar W$ are accepted at
the ancillary level. Then the global state gains a complex structure, and 
entanglement appears to scale. The outwards growth of entanglement from
an ancillary triangle can be systematically analysed. In Fig. \ref{fig:outwards} and Table \ref{table:outwards} we 
present how much entanglement is observed as the size of the system increases.

%\begin{comment}
%\end{comment}

\begin{table}[h!]
\centering
  \begin{tabular}{ | c | c  c  c  |}
    \hline
      Simplex & $n_A=3$ & $n_A=6$ & $n_A=10$ \\ \hline
    $|GHZ\rangle$& 1 & 1 & 1\\
    $|W\rangle$ & $\log_2 3$ & $\log_2 3$ & $\log_2 3$ \\
    $|W\rangle$,$|111\rangle$ & 2 & 3&4 \\
    $|W\rangle$,$|\bar W\rangle$ & 2.183 & 3.126 & 5.053 \\
    	$|W\rangle$,$|\bar W\rangle$, $|111\rangle$ & 1.815 & 2.756 & 4.314 \\
   \hline 
  \end{tabular}
  \caption{Outwards entangling power of different triangular simplices. At each simplex,
an equal superposition of its allowed states (on the left) propagates the entanglement entropy (on the right, in ebits)
to the rest of the system. $n_A$ is the number of spins of one of the subsystems (see Fig. \ref{fig:outwards}).}
  \label{table:outwards}
\end{table}

\begin{figure}[h!]
      \begin{center}
            \includegraphics[scale=0.18]{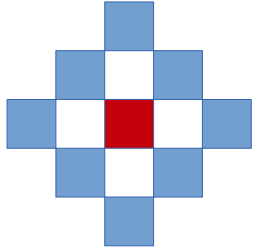}
            \caption{\label{fig:squares}Square network of 24 spins, with the red (or dark) area showing the 4 spins from which we are computing the entanglement entropy.
             }
      \end{center}  
\end{figure}

The relation between $GHZ$ and $W$ entanglement on triangular ancillary states and
the emergence of distinct types of long distance order suggest interesting
generalizations for larger simplices. Let us here analyze the case of a tensor network created from a square four-qubit simplex.
We shall constrain our analysis to simplices that are symmetric under the exchange of
ancillary particles, namely
\begin{eqnarray}
  |\psi\rangle &&= \alpha_{0} |0000\rangle
  \nonumber
\\
&&
+ \alpha_1 \left(|0001\rangle +|0010\rangle +|0100\rangle +|1000\rangle \right) 
  \nonumber
\\
&&
+ \alpha_2 \left(|0011\rangle +|0101\rangle +|0110\rangle + \right. \nonumber 
\\
&& \left.|1001\rangle +|1010\rangle +|1100\rangle \right)
  \nonumber
\\
&&
+\alpha_3 \left(|0111\rangle +|1011\rangle +|1101\rangle +|1110\rangle  \right) \nonumber
\\
&& + \alpha_4 |1111\rangle \ .
\end{eqnarray}
We then consider a tensor network as shown in Fig. \ref{fig:squares} , where every checked square
contains an ancillary state and all links are counted just once. The tensor defined
at every site is the product of delta functions of the physical index with each of the
two ancillary qubits meeting there. 
For the case of 24 qubits shown in Fig. \ref{fig:squares} , we have scanned the entropy of the inner square as a function of the coefficients of this ancillary state.  Maximum
entanglement is obtained for the state corresponding to $\alpha_0=\alpha_2=\alpha_3=\alpha_4= - \alpha_1$.

%\begin{comment}
%\end{comment} 

The above results suggest a connection between microscopic entanglement at
the level of a simplex and the emergence of long-range correlations in the system.
It is tempting to argue that distinct classes of entanglement might
be responsible for different types of long-range order, as we found in the case of the
triangular lattice.
For four qubits, different types of maximally entangled states can be tried (see Chapter 1). States with maximal hyperdeterminant may play a special role in 3D networks based on tetrahedrons.

Similarly, the spatial symmetries which are found on a simplex are related to symmetries 
at large scales. It is easy to see that if the triangular couplings are chosen as
positive in the diagonal directions and negative in the horizontal direction, then frustration
disappears and the exact simplex describing the model is a superposition of 
$|001\rangle$ and $|110\rangle$ states.

\subsection{Practical validation}

Here we take the concrete case of a 3x3 periodic triangular network. We have computed various properties of the ground state of the Ising hamiltonian with external transverse field: energy, transverse magnetisation and entropy (of two rows against the other one). We have computed the exact values with Mathematica, and approximate values with triangular simplices (TS) and PEPS (we have used the gradient descent optimisation method with backtracking line search to find the ground state in these last two cases). The magnetization and entropy with $\lambda=0$ are somewhat arbitrary as we have there a degeneration of 42, so we have many ground states with different magnetisation and entropy while for $\lambda\neq 0$ there is no degeneration.  We take as the "right" ground state the evolution of the ground state with $\lambda\neq 0$ as we make $\lambda \rightarrow 0$.

\begin{table}[h!]
\centering
\begin{tabular}{ | c | c  c  c |}
    \hline
      Model & Exact Energy & TS Energy & PEPS Energy(*)  \\ \hline
    $\lambda=0$ & -9 & -8.99
& -8.97 \\
    $\lambda=0.5$ & -11.13 & -11.02
& -10.75\\
    $\lambda=1$  &  -13.82 & -13.63
& -13.00\\
    $\lambda=2$
& -20.72 & -20.56 & -20.02  \\
   \hline

\end{tabular}
   \caption{Energies for the ground states of Ising Hamiltonians with different external transverse field computed with different methods. (*) PEPS Energy is the best estimation}
   \label{tableenergies}

\end{table}

\begin{table}[h!]
\centering
\begin{tabular}{ | c | c  c  c |}
    \hline
      Model & Exact Mag & TS Mag & PEPS Mag \\ \hline
    $\lambda=0$ & -3.65 & -3.43 & -0.08
\\
    $\lambda=0.5$ & -4.83 & -4.62 & -3.96
\\
    $\lambda=1$ & -5.92 & -5.83 & -4.73
\\
    $\lambda=2$  &  -7.69 & -7.79 & -7.64
\\

   \hline

\end{tabular}
   \caption{Magnetisations for the ground states of Ising Hamiltonians with different external transverse field computed with different methods.}
   \label{tablemag}

\end{table}

\begin{table}[h!]
\centering
\begin{tabular}{ | c | c  c  c |}
    \hline
      Model & Exact $S$ & TS $S$ & PEPS $S$  \\ \hline
    $\lambda=0$ & 1.96 & 2.06 & 0.01
\\
    $\lambda=0.5$ & 1.73 & 1.81 & 0.07
\\
    $\lambda=1$ & 1.48 & 1.53 & 0.13
\\
    $\lambda=2$  &  0.86 & 0.80 & 0.51
\\

   \hline

\end{tabular}
   \caption{Entropies for the ground states of Ising Hamiltonians with different external transverse field computed with different methods.}
   \label{tableentropies}

\end{table}
   
The TS ansatz has beaten PEPS in two aspects: as is evident from tables \ref{tableenergies}, \ref{tablemag} and \ref{tableentropies}, the numerical results are much better for TS than for PEPS, and the amount of computational resources (time and space) is much higher in PEPS than TS (typical convergence time is around 30 seconds for TS and several hours for PEPS). This is directly related to the fact that the TS ansatz is only considering 4 different coefficients for the whole state ($|000\rangle,|W\rangle,|\bar W\rangle $ and $ |111\rangle$), while the PEPS is using $2^7 = 128$, as each site has 6 links with the environment and one physical index.

Therefore, we conclude that tensor networks constructed from non-trivial simplices are an interesting tool and could be the natural way to encode 
different levels of entanglement and symmetries at large order.

\part{Quantum computation}

\chapter{Cloud quantum computation} \label{ch:ibm}

\pagestyle{fancy}
\fancyhf{}
\fancyhead[LE]{\thepage}
\fancyhead[RE]{QUANTUM COMPUTATION}
\fancyhead[LO]{CH.6 Cloud quantum computation}
\fancyhead[RO]{\thepage}

\section{Introduction to quantum computation}

This second part of the thesis is devoted to the most promising future, and partly present, technological advance that quantum information theory is going to bring: quantum computers and their new algorithms that should provide a computational advantage in many difficult problems for classical computers.

Richard Feynman introduced the idea of quantum computing as the correct way to simulate quantum physics in a talk in 1981 \cite{Feynman82}. The appearance of some quantum algorithms providing an important speedup on important problems represented a breakthrough, like Shor's algorithm for factorization \cite{Shor94} and Grover's algorithm for searching in unordered databases \cite{Grover96}.

Many different quantum computing models have been proposed. Cirac and Zoller introduced the ion trap quantum computer \cite{Cirac95} and other models have appeared, one of the most quickly developing right now being the superconducting circuit quantum computer \cite{Steffen11}. 

The practical implementation includes an ion trap computer of 14 qubits in Innsbruck \cite{Monz11} and superconducting circuit computers by IBM and Google, among many others. D-Wave claims to have a quantum computer of 2000 qubits, but it is not yet clear if it really performs quantum computations, and it is certainly not a universal quantum computer.

\section{The IBM Quantum Experience}

The quantum computer by IBM has been the first to be available to the general public, who can write programs and send them to the computer through an online application called the IBM Quantum Experience \cite{IBM}, creating the new field of cloud quantum computation.

The computer is a superconducting circuit quantum computer of 5 qubits. Initially the architecture was such that qubit number 2 (Q2) was at the center and the other qubits were at the four edges, all connected to each other through Q2. Therefore, the only 2-qubit gate available, the CNOT gate, involved necessarily Q2. This has recently changed, but our work was done with the initial architecture.

The gates available were initially the Pauli gates $X$,$Y$ and $Z$, the Hadamard gate $H$, the single-qubit phase gates $S$ and $T$ adding phases of $\pi/2$ and $\pi/4$ respectively and their conjugates $S^{\dagger}$ and $T^{\dagger}$, and the entangling 2-qubit CNOT gate. All these gates except for the $T$ and $T^\dagger$ gates are called Clifford gates and are simple to simulate classically, a result called the Gottesman-Knill theorem \cite{Gottesman99}, and it is the $T$ gate that makes all the difference and allows quantum speedup. Recently arbitrary single-qubit phase gates were added to the gate set.

\begin{figure}[h!]
      \begin{center}
            \includegraphics[scale=0.21]{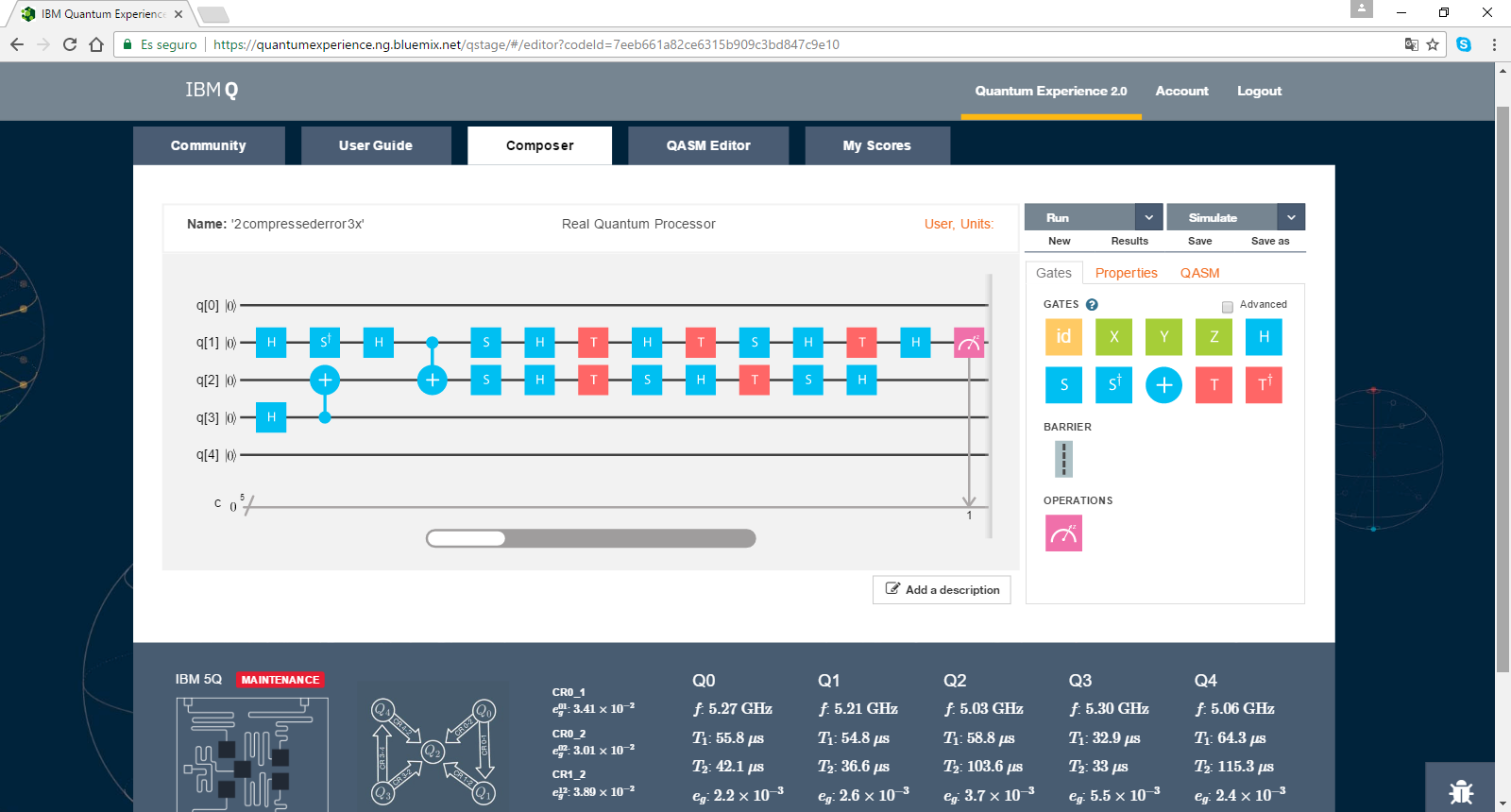}
            \caption{\label{fig:ibmq} Screenshot of the graphical interface of the IBM Quantum Experience. On the right there is the available gate set, and at the bottom information about the computer is shown, like the architecture, the qubits connectivity and several parameters of each qubit, like the energy relaxation time (T1) and the dephasing time (T2).
             }
      \end{center}  
\end{figure}

The programming application is a graphical interface very similar to the canonical way of representing quantum circuits: the 5 qubits are represented as lines where the gates can be inserted. Up to 39 gates can be applied to each qubit. It is also possible to draw the circuit using command lines of a language specially designed to do that, called IBMQASM. Figure \ref{fig:ibmq} shows a screenshot of the graphical interface.

Once the circuit is drawn, running it and receiving the results is done in a couple of seconds, provided the computer is active at the moment and that there is not a very big queue, which so far has never happened to us. As computations are naturally subject to both systematic and statistical errors, IBM provides access to a classical simulator that implements an error model of the quantum computing hardware and therefore allows simulation of a circuit before actually performing the computation.

\section{Mermin inequalities in a quantum computer}

In this section, violation of Mermin inequalities is tested on the five qubit IBM quantum computer. The theoretical aspects of Mermin inequalities have been explained thouroughly in Sect. \ref{sec:mermin} and we shall not repeat them here, but we write the explicit expressions for the Mermin inequalities of 3, 4 and 5 qubits, the ones that will be tested experimentally. The Mermin polynomial for 3-qubits is
\begin{equation}
M_3 = (a_1 a_2 a'_3 + a_1 a'_2 a_3 + a'_1 a_2 a_3) - (a'_1 a'_2 a'_3) \, ,
\label{m3}
\end{equation}
with a classical bound of $\langle M_3 \rangle ^{LR}\leq 2$ and a quantum bound of $ \langle M_3 \rangle ^{QM} \leq 4$.
The  Mermin polynomial that will be experimentally checked for 4-qubits is
\begin{eqnarray}
\label{m4}
M_4 =&& -(a_1 a_2 a_3 a_4) \\ \nonumber 
&&+ (a_1 a_2 a_3 a'_4 + a_1 a_2 a'_3 a_4 + a_1 a'_2 a_3 a_4 + a'_1 a_2 a_3 a_4) \\ \nonumber
&&+ (a_1 a_2 a'_3 a'_4 + a_1 a'_2 a_3 a'_4 + a_1 a'_2 a'_3 a_4  \\ \nonumber 
&&+ a'_1 a_2 a_3 a'_4 + a'_1 a_2 a'_3 a_4 + a'_1 a'_2 a_3 a_4) \\ \nonumber 
&&- (a_1 a'_2 a'_3 a'_4 + a'_1 a_2 a'_3 a'_4 + a'_1 a'_2 a_3 a'_4 + a'_1 a'_2 a'_3 a_4) \\ \nonumber 
&& -(a'_1 a'_2 a'_3 a'_4)\nonumber, ,
\end{eqnarray}
with a classical bound of $\langle M_4 \rangle ^{LR}\leq 4$ and a quantum bound of $ \langle M_4 \rangle ^{QM} \leq 8\sqrt{2}$\hspace{1mm}.

In the 5-qubit case, the Mermin polynomial reads
\begin{eqnarray}
\label{m5}
M_5 =&& -(a_1 a_2 a_3 a_4 a_5) \\ \nonumber 
&&+ (a_1 a_2 a_3 a'_4 a'_5 + a_1 a_2 a'_3 a_4 a'_5 + a_1 a'_2 a_3 a_4 a'_5 \\ \nonumber 
&&+ a'_1 a_2 a_3 a_4 a'_5 + a_1 a_2 a'_3 a'_4 a_5 + a_1 a'_2 a_3 a'_4 a_5 \\ \nonumber
&&+ a'_1 a_2 a_3 a'_4 a_5 + a_1 a'_2 a'_3 a_4 a_5 + a'_1 a_2 a'_3 a_4 a_5 \\ \nonumber 
&&+ a'_1 a'_2 a_3 a_4 a_5) \\ \nonumber
&&-(a_1 a'_2 a'_3 a'_4 a'_5 + a'_1 a_2 a'_3 a'_4 a'_5 + a'_1 a'_2 a_3 a'_4 a'_5
\\ \nonumber
&& + a'_1 a'_2 a'_3 a_4 a'_5 + a'_1 a'_2 a'_3 a'_4 a_5)\, ,
\end{eqnarray}
with a classical bound of $\langle M_5 \rangle ^{LR}\leq 4$ and a quantum bound of $\langle M_5 \rangle ^{QM} \leq 16$.

The experimental verification of multipartite Mermin inequalities faces
the problem of a good control of three or more qubits, including the generation of entangled
states, and the possibility of performing different measurements on each one. Violation
of Mermin inequalities has been reported for three qubits \cite{Pan00} and four qubits
\cite{Zhao03}, where all qubits are made out of photons, and for up to 14 qubits with a quantum computer based on ion traps \cite{Lanyon14}.

In the case of superconducting qubits,
violation of the CHSH inequality was
achieved in Ref. \cite{Ansmann09}, whereas the GHZ construction and the 3-qubit Mermin inequality violation was demonstrated by Ref. \cite{Dicarlo10}.
For a general review of theoretical and experimental progress in Bell inequalities, see Ref. \cite{Brunner14}.

\subsection{Circuit implementation}

Let's talk about the implementation in the IBM quantum computer. In the test of Mermin inequalitites, only  GHZ-like states have to be created. This requires the use of a Hadamard gate on a control
qubit followed by $CNOT$ gates targeted to the rest. In order to implement this kind of action we shall need to operate
$CNOT$ gates targeted to different qubits, which cannot be done directly because of the restriction that all $CNOT$ gates have to be targeted to Q2. This can be solved using the relation
$CNOT_{1\to 2} =(H_1 \otimes H_2) CNOT_{2\to 1} (H_1 \otimes H2)$, where $H_1$ and $H_2$ are Hadamard
gates on qubits 1 and 2, whereas $CNOT_{1\to 2}$ is the control-NOT gate which is controlled by qubit 1.\\

In our choice of settings, the needed GHZ-like states have relative phases, as in the case of 3-qubits, where $|\phi\rangle=1/\sqrt{2}(|000\rangle+i|111\rangle)$. These phases are implemented using $S$ and $T$ gates.
Measurements can only be done on the $\sigma_z$ basis, but they can be simulated in another basis with the help of additional gates, namely an $H$ gate for $\sigma_x$ and an $S^\dagger$ gate followed by an $H$ gate for $\sigma_y$. \\

Another relevant issue to be considered is that not all of the qubits are equally robust in the present
quantum computer, some have relaxation and decoherence times larger than others, although all of them are of the order of $T = \mathcal{O} (100 \mu s)$. We shall adapt our circuits to minimize the  number of gates on
the qubits that behave more poorly. For example gates that implement phases that can be put freely in any qubit are allocated to the most robust ones. \\

Figures 1 and 2 represent the three circuits for the 3, 4 and 5 qubit Mermin inequalities. In principle we need to perform as many experiments as the number of terms in the Mermin inequalities \eqref{m3}, \eqref{m4} and \eqref{m5}. However due to our limited access to the computer and the symmetry of particle exchange of the states and the inequalities, only one experiment for a term representative of each number of primes $(a'_i)$ is run. In our choice of settings, the number of primes amounts to the number of $\sigma_y$ measurements, whereas the non primes ($a_i$) correspond to $\sigma_x$ measurements.
 We thus have 2 experiments for 3-qubits, 5 experiments for 4-qubits and 3 experiments for 5-qubits. Each experiment is run 8192 times, the maximum available, except for the 3-qubit experiments, which have been run only 1024 times. When computing the expected value of the whole polynomial, each experiment is given the corresponding weight. In the errors discussion we compare results obtained when using the symmetry with results obtained without using it, computing all the terms, for the three-qubit case.

\begin{figure}[h!]
\centering
\includegraphics[scale=0.35]{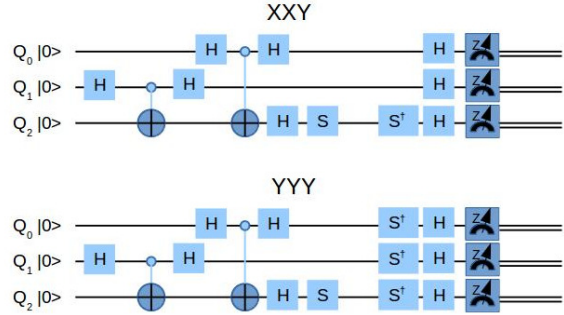}
\label{circuit3}
\caption{The two circuits used for the three-qubit Mermin inequality. The first circuit corresponds to $\sigma_x\sigma_x\sigma_y$ experiment, and the second circuit to $\sigma_y\sigma_y\sigma_y $ experiment. The $S^\dagger$ gates make the difference between a $\sigma_x$ and a $\sigma_y$ measurement.}
\end{figure}

\begin{figure}[h!]
\centering
\includegraphics[scale=0.35]{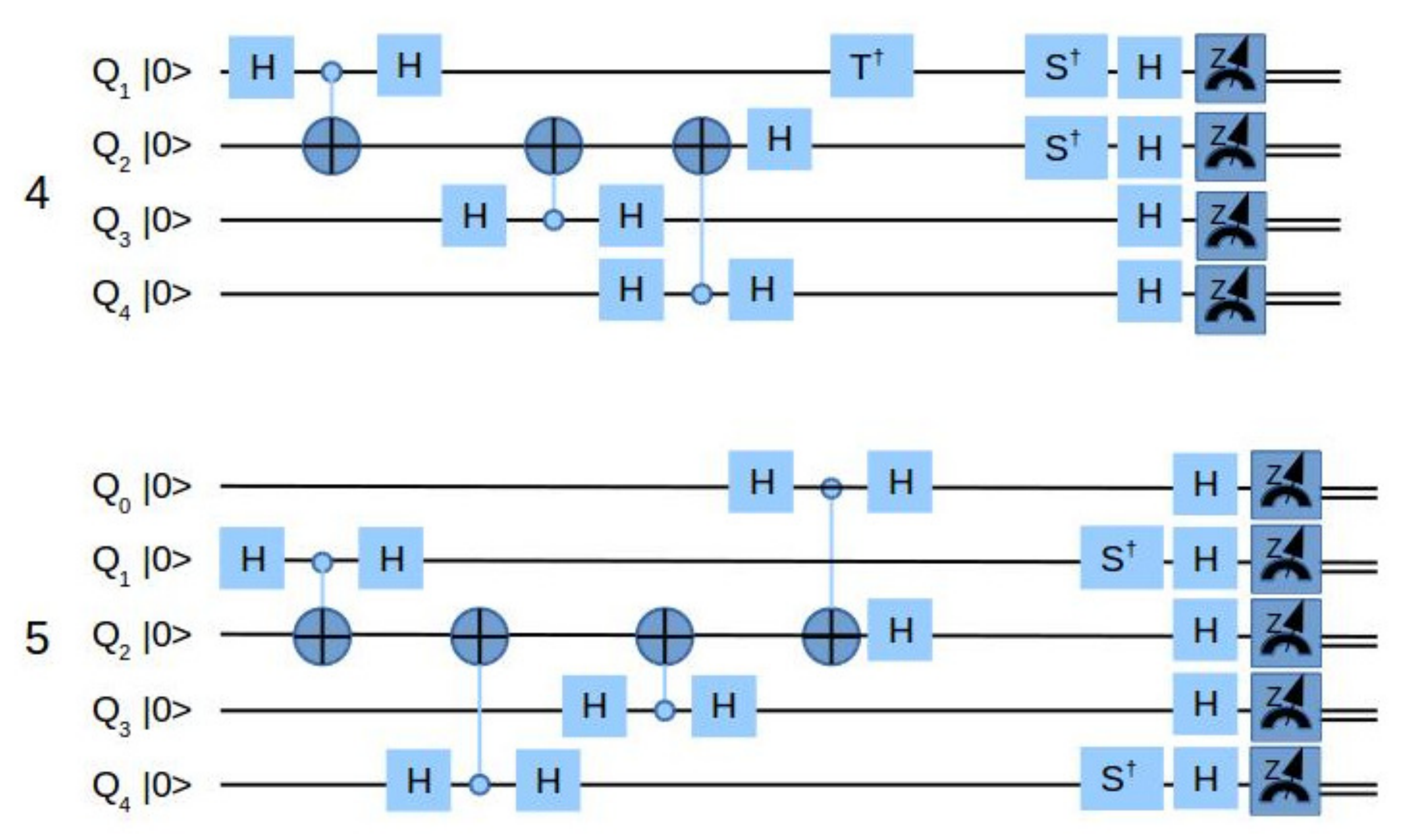}
\label{circuit45}
\caption{Two of the circuits used for the four-qubit and five-qubit Mermin inequalities. The first circuit corresponds to $\sigma_y\sigma_y\sigma_x\sigma_x$ experiment, whereas the second corresponds to $\sigma_x\sigma_y\sigma_x\sigma_x\sigma_y $ experiment. The $S^\dagger$ gates make the difference between a $\sigma_x$ and a $\sigma_y$ measurement. In order to change from $\sigma_x$ to $\sigma_y$, one has to add an $S^\dagger$ gate, or remove it to do the opposite. With this technique one can obtain all circuits needed to test the inequalities. }
\end{figure}

\subsection{Results}

We shall now give a detailed discussion of the results for the 3-qubits case and an abridged one for the 4 and 5-qubit cases, as much of it is basically the same.

In order to check the violation of the inequality, one has to choose the settings and the corresponding state that maximally violate it. One possibility is to choose settings $a_i=\sigma_x$ and $a'_i = \sigma_y$ for all the qubits. The state that maximizes the quantum violation in this case is $|\phi\rangle=1/\sqrt{2}(|000\rangle+i|111\rangle)$.

The 3-qubit Mermin inequality has 4 terms as shown in Eq. \ref{m3}. In principle,  four different circuits are needed, one for each term. The state will be the same for all of them, but the settings change. However, one can use the symmetry of the state and the inequality to reduce the number of measurements needed if there is limited access to the experimental setting as in our case. All the terms that have the same number of primes $(a'_i)$ are represented by the same circuit by symmetry.   We then considered only two different experiments, with 1024 runs each, the $\sigma_x \sigma_x \sigma_y$ experiment and the $\sigma_y \sigma_y \sigma_y$ experiment. The results are shown in table \ref{results3} .

\begin{table}[h!]
\centering
\resizebox{\linewidth}{!}{%
\begin{tabular}{c | c | c | c | c | c | c | c | c}
\hline
Result XXY & \textbf{000} & \textit{001} & \textit{010} & \textbf{011} & \textit{100} & \textbf{101} & \textbf{110} & \textit{111} \\
Probability & 0.229 & 0.042 & 0.024 & 0.194 & 0.043 & 0.203 & 0.231 & 0.033 \\
\hline
Result YYY & \textbf{000} & \textit{001} & \textit{010} & \textbf{011} & \textit{100} & \textbf{101} & \textbf{110} & \textit{111}   \\
Probability & 0.050 & 0.188 & 0.188 & 0.028 & 0.258 & 0.026 & 0.041 & 0.221 \\
\end{tabular}%
}
\caption{Table of detailed results for the two 3-qubit experiments. In bold are results of even parity, in italic results of odd parity. Counts for each result are expressed in probabilities computed out of 1024 runs. Computation of the expected value of XXY gives $\langle XXY \rangle = 0.715$ and for YYY gives $\langle YYY \rangle = -0.710.$ The combination 3  $\langle XXY \rangle - \langle YYY \rangle$ gives $\langle M_3 \rangle_{exp} = 2.85\pm 0.02$. }
\label{results3}
\end{table}

Eight probabilities for each term are obtained. In order to translate these probabilities to the expected values that appear in the inequality, one has to arrange the results in two groups according to the parity of the number of 1 (which represent the value -1.) The expected value of the term is obtained by summing all the probabilities of the results of even parity and subtracting the results of odd parity. The correctly weighted sum of the expected values of each term gives the final result $\langle M_3 \rangle_{exp} = 2.85\pm 0.02$.

In the case of 4-qubits, the use of settings $a_i=\sigma_x$ and $a'_i = \sigma_y$ implies that the state that maximizes the quantum violation is $|\phi\rangle=1/\sqrt{2}(e^{ i \pi/4}|0000\rangle+|1111\rangle)$. With these settings and this state, 5 experiments are performed, one for each term with different number of primes \eqref{m4}, with 8192 runs for each experiment. A result of $\langle M_4 \rangle_{exp} = 4.81\pm 0.06$ was obtained.

In the case of 5-qubits, the use of settings $a_i=\sigma_x$ and $a'_i = \sigma_y$ implies that the state that maximizes the quantum violation is $|\phi\rangle=1/\sqrt{2}(|00000\rangle+|11111\rangle)$. With these settings and this state, 3 experiments are performed, one for each term with different number of primes \eqref{m5}, with 8192 runs for each experiment. A result of $\langle M_5 \rangle_{exp} = 4.05\pm 0.06$ was obtained. This is clearly a poor violation, which is still compatible
with local realism. Improvement of the quantum computer is needed to obtain more accurate results.

\begin{table}[h!]
\centering
\begin{tabular}{c | c | c | c | c | c |}

     & LR & QM & EXP  \\
\hline
3 qubits & 2 & 4 & \textbf{2.85$\pm$ 0.02}   \\
4 qubits & 4 & 8 ${\sqrt 2}$ & \textbf{4.81$\pm$ 0.06}   \\
5 qubits & 4 & 16 & \textbf{4.05$\pm$ 0.06}   \\

\end{tabular}
\caption{Table of results. LR corresponds to the local realism bound for each Mermin  inequality, QM to the quantum bound and EXP is the experimental result. }
\label{tabresults}
\end{table}

The results obtained from the IBM quantum computer are subject to
different kind of errors. The stability of the quantum computer is still poor
and the same experiments run at different times provide results that differ more than the expected
behaviour of statistical fluctuations. As an example, one month after the original runs, the 3-qubit experiment was run again to compare results. That time, a result of $\langle M_3 \rangle_{exp} = 2.57\pm 0.02$ was obtained, clearly showing the previous point. An additional run was done computing separately the four terms of \eqref{m3}, without assuming any symmetry, and a similar result was obtained, $\langle M_3 \rangle_{exp} = 2.57\pm 0.02$, showing that it is safe to assume the symmetry of party exchange.

We may get an estimation of the statistical error as a dispersion
around the mean. We may, as well, treat the results as a multinomial
distribution, using the expression $\delta p = \sqrt{p(1-p)/N}$, which for N=8192 gives $\delta p = \mathcal{O} (10^{-2})$. The different Mermin inequalities for 3, 4 and 5 qubits
require a different number of experiments to be done, which are
considered as independent. We may then add in quadrature its errors,
which is the figure we associate in the explicit results. In this sense, the 5-qubit
result obtained in the present quantum computer does not have sufficient statistical
significance to discard local realism.

Furthermore, some of the issues related to the elimination of loopholes can
not be addressed. Experiments suffer from errors
related to stability,
loss of coherence and lack of full fidelity of the quantum gates. This is
clearly seen as the
violation of Mermin inequalities will deteriorate progressively as
the numbers of qubits,
and gates used in the experiment, increase. We may think of the
experimental verification
of Mermin inequalitites as a test of the overall fidelity of the whole Mermin circuits. We will give a more detailed error discussion about this computer in the next section.

\section{Compressed quantum computation} \label{sec:compressed}

We have also tested the IBM quantum computer with a different problem. The notion of compressed quantum computation \cite{Josza09} refers to the possibility of simulating certain circuits with exponentially less qubits than those theoretically needed. In the present section we test the performance of the IBM quantum computer with a compressed simulation of the transverse field 1D--Ising interaction. The quantum Ising model is an integrable model and an exact circuit can construct its ground state \cite{Verstraete09}. Compressed quantum computation has also been applied to the XY-model and compressed quantum metrology \cite{Boyajian13,Boyajian15,Boyajian16}. Moreover, the compressed simulation of the Ising spin chain (consisting of $2^5=32$ qubits) has been realized in an experiment using NMR quantum computing \cite{Li14}.

On the IBM computer it is possible to simulate a four qubit Ising chain using only $\log_2(4) = 2$ qubits. In principle we could simulate a chain of up to $2^5 = 32$ qubits with the 5 qubits available, but the number of gates needed for the simulating circuit to have an acceptable precision would be far too large, and we are for now constrained to this small case.  In order to realize this computation, we decompose the circuits for the compressed simulation into the available gate set. We run these circuits on the quantum computer and measure the order parameter that displays the quantum phase transition. Given that the size of the system is finite, we do observe smoothed changes of the order parameter that agree with the theoretical predictions within errors.

We make in this section a more detailed treatment of the errors. There are two sources of errors that have to be considered separately. First, it is necessary to run an experiment often enough so that statistical errors are reduced. This is an easy task since it only implies repetition of experiments. Second, systematic errors must be estimated. The situation here is particularly subtle, as a cloud computer is run by teams unrelated to its users. The problem of how to estimate a systematic error without knowing the detail of the computer is non-trivial. Nevertheless, an approach to the correct assessment of systematic errors can be done, using independent controlled circuits of similar complexity to the one of interest. This idea of estimating systematic errors produced by a black box might be of relevance for all future cloud quantum computation.

\subsection{Review of compressed quantum computation}

Let us now briefly review the notion of compressed quantum computation \cite{Josza09}. It has been shown that matchgate circuits running on $n$-qubits can be compressed into circuits using exponentially less qubits. Matchgates are two--qubit gates of the form $A\oplus B$, where the unitary $A$ ($B$) is acting on span$\{\ket{00},\ket{11}\}$ (span$\{\ket{01},\ket{10}\}$) respectively, and the determinants of $A$ and $B$ coincide. The compression is possible if the circuit consists of matchgates acting only on neighbouring qubits, the input state is a computational basis state, and the output is the expectation value of $Z$ of a single qubit. It has been shown that the computational power of a $n$-qubit matchgate circuit is equivalent to that of a universal quantum computer running on only $\lceil \operatorname{log}(n) + 3\rceil$ qubits. That is, the output, which is also in the compressed computation obtained by measuring a single qubit, coincides. Moreover, the circuit size of the compressed computation coincides with the original size up to a factor $\log(n)$. An important fact to note here is that the computation is indeed performed by the quantum computer, as the allowed classical side computation is restricted to ${\cal O}(\log(n))$ space. Note that any polynomial--sized circuit that can be compressed can also be efficiently simulated classically (as a function of $n$) as the dimension of the Hilbert space corresponding to the compressed circuit is linear in $n$.

Compressed quantum simulation of the transverse field Ising model has already been realized in an experiment using NMR quantum computing \cite{Li14}.
Here, we also simulate this model with open boundary conditions, whose evolution is governed by the Hamiltonian
\begin{align}
	H(J) = \sum_{k=1}^{n} Z_k + J \sum_{k=1}^{n-1} X_k X_{k+1},
\end{align}
where $X_k$ ($Z_k$) denote $X$ ($Z$) acting on qubit $k$, respectively.
In the limit $n \rightarrow \infty$, the system undergoes a quantum phase transition at $J=1$ that is reflected in the discontinuity of the second derivative of the transverse magnetization.

The magnetization, $M(J)$, can be measured as follows \cite{Sadchev11,Verstraete09,Kraus11}. The system is initially prepared in the ground state of $H(0)$ and adiabatically evolved to the ground state of $H(J)$ by changing the parameter $J$ adiabatically. In order to perform digital adiabatic evolution over a time period $T$, the Hamiltonian $H(J)$ is discretized into $L+1$ steps. The evolution is then governed by a product of $L$ unitaries which are then approximated up to second order in $\Delta t = \frac{T}{L+1}$ using Suzuki-Trotter expansion. The evolution is indeed adiabatic and the approximation is valid if $T,\text{ }L \rightarrow \infty$ and $\Delta t \rightarrow 0$.
The transverse magnetization, $M(J)$, is obtained by measuring $Z$ on a single qubit.
As this adiabatic evolution together with the measurement of the magnetization is a matchgate circuit, the whole computation can be compressed into a universal quantum computation running on only $m=\operatorname{log}(n)$ qubits. We assume here that $n$ is a power of two. Due to the symmetry of the Ising model the compression to even $\operatorname{log}(n)$ qubits, instead of $\operatorname{log}(n)+3$ qubits, which are required for an arbitrary matchgate circuit, is possible. This exact simulation of the circuit has been shown to be as follows \cite{Kraus11}.
\begin{enumerate}
\item Prepare the input state $\rho_{in} = \frac{1}{2^{m-1}} \identity^{\otimes m-1} \otimes \ket{+_y}\bra{+_y}$, where $Y \ket{+_y}=\ket{+_y}$,
\item evolve the system up to the desired value of J by applying $W(J) = \prod_{l=1}^{L(J)} U_d R_l^T R_0^T$,
\item measure $Y$ on qubit $m$ to obtain the magnetization $M(J) = \\ -\trace{W(J) \rho_{in} W(J)^\dagger \; \identity \otimes Y_m}$.
\end{enumerate}
Here, the $m$--qubit unitary operators $R_0 = \identity \otimes e^{2 \Delta t Y_m}$, $R_l = [1 - \cos(\phi_l)] (\ketbra{1}{1} + \ketbra{2n}{2n}) + \cos(\phi_l) \identity + \sin(\phi_l) \sum_{k=1}^{n-1} \ketbra{2k+1}{2k} - h.c.$, and $U_d = \identity + (e^{i \phi_l} - 1)\ketbra{2n}{2n}$, where $\ket{k} = \bigotimes_{i=1}^{m} \ket{k_i}$ with $k_i$ such that $k = 1 + \sum_{i=1}^{m} k_i 2^{m-i}$, $\phi_l = 2 J_l \Delta t$, and $J_l = \frac{l}{L} J_{max}$ stem from the compression of the adiabatic evolution. 

\subsection{Implementation and results}

In order to perform this computation with the IBM quantum computer, we have to decompose the unitaries, which are required for the state preparation and the evolution into the Clifford+T gate set. In the following, we will outline the steps for the case of two qubits, which simulate a four-qubit spin chain. In Appendix \ref{app:extension}, we explain how the computation can be performed for more qubits once some improvements of the quantum computer are available.

\begin{figure}[ht]
	\centering
	\resizebox{1.0\linewidth}{!}{\includegraphics{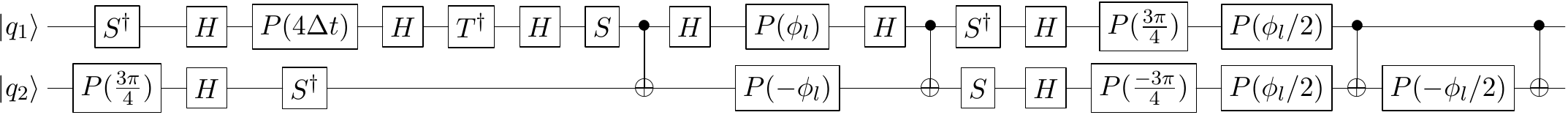}}
	\caption{Decomposition of one adiabatic step of the 2-qubit circuit into $CNOT$ and single qubit gates.}
	\label{fig:step}
\end{figure}

We exchange qubits 1 and 2 in the following due to the special role of qubit 2 in the IBM computer. The input state $\rho_{in} = \frac{1}{2} \ket{+_y}\bra{+_y} \otimes \identity$ is prepared by applying $S H$ to qubit 1 and $CNOT(3,2) H_3$ to qubit 2 and an auxiliary qubit, qubit 3, which is discarded afterwards. This procedure is uneconomical, however it is necessary as implementing gates probabilistically is currently not possible. 
To simulate the adiabatic evolution, products of the gates $U_d, R_l^T$, and $R_0^T$ have to be applied. $R_0$ is a single qubit gate and, in the case of a two qubit circuit, $U_d=\proj{0}_1\otimes \one_2+ \proj{1}_1\otimes P_2(\phi_l)$, where $P(\phi_l)$ denotes a $\phi_l$-phase gate. The circuit depicted in Figure \ref{fig:step} implements one step in the adiabatic evolution, namely $U_d R_l^T R_0^T$, in terms of $CNOT$ and single qubit gates. Note that only the gates depending on $\phi_l$ change from step to step as $l$ is incremented in each step. The decomposition into the gate set is performed using results on decomposing arbitrary two-qubit gates into Bell diagonal gates and decomposing Bell diagonal gates into single qubit unitaries and $CNOT$ gates \cite{Vidal04, Kraus01}. All single qubit gates but phase gates depending on $\phi_l$ can be easily implemented in the Clifford+T gate set. For decomposing arbitrary phase gates we use the algorithm described in \cite{Kliuchnikov16}, where phase gates are approximated using Clifford+T gates. As there is a trade-off between the circuit depth, which is restricted here, and the quality of the approximation, we are forced to introduce a noticeable error (see Fig. 2).

The circuit depicted in Fig. \ref{fig:step} has a circuit depth of 18. Hence, the total circuit, where many of these adiabatic steps have to be used before $Y_1$ is measured, exceeds the current circuit depth limit if we choose a total step number $L$ such that the evolution is indeed adiabatic. Thus, in order to keep the circuit depth feasible, we calculate the two-qubit unitary, $W(J)$, and decompose this unitary into Clifford+T gates. We approximate the single qubit unitaries as well as possible respecting the limit of the circuit depth. We provide the realized circuits in Appendix \ref{app:circuits}.

In Fig. \ref{fig:results} we present the results for the two-qubit circuit described above, that simulates the magnetization of a four-qubit spin chain. We measured, as in the NMR experiment \cite{Li14}, the magnetization for 12 values of $J$, $J=\left\{\frac{1}{6}, \frac{2}{6}, \ldots, 2 \right\}$. We also use the same parameters for the digital adiabatic evolution, $L=2400$, $\Delta t=0.1$. The solid line represents the real magnetization of the four-qubit spin chain. The black circular symbols show the theoretically obtained magnetization using digital adiabatic evolution. However, due to the restricted circuit depth, the circuit has to be approximated by a feasibly sized Clifford+T gate circuit, as described above. The dark gray, diamond shaped symbols depict the magnetization after this step, assuming that the quantum computer works perfectly. Hence, the difference between the diamond shaped and the circular symbols reflects the error made in using a feasible circuit size. Finally, the orange, filled, triangular-shaped symbols denote the actual measurement outcomes obtained using the IBM quantum computer on Sept. 9th 2016. We also provide the measurement outcomes obtained using the IBM simulator, that implements an error model of the hardware. Remarkably, there is a huge discrepancy between the output of the simulator and the actual measurement outcomes, indicating that the simulator provides pessimistic predictions here. In the figure we also illustrate the error we estimated with the validating sets that we describe in next subsection. As can be seen, the results we obtain lie, on average, within the corresponding error bars. Moreover, we also reprint here the results obtained for the same simulation with a NMR quantum computer. There, however, a rescaling, which accounts for some of the errors, was performed. Because there the experimental data (without any rescaling) is given only for the simulation of a $2^5 = 32-$qubit spin chain, a fair comparison between these results seems to be unfeasible.
    	\begin{figure}[ht]
			\centering
			\resizebox{1.0\linewidth}{!}{\includegraphics{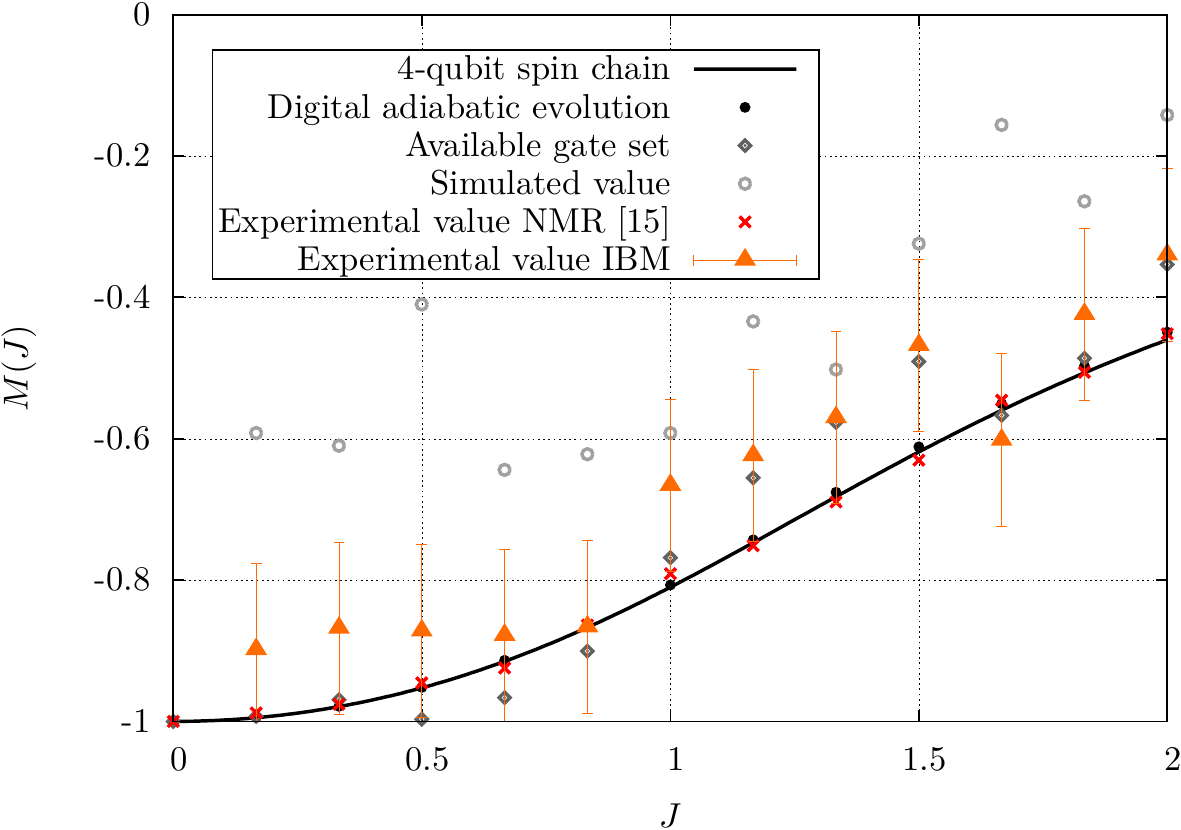}}
			\caption{The magnetization of the two-qubit circuit simulating the four-qubit spin chain (for details see main text).}
			\label{fig:results}
	\end{figure}

\subsection{Error discussion}

Let us now investigate the errors which occur in the computation. Quantum process tomography \cite{Chuang97}, which would completely characterize the performance of the computation, is very demanding even for quantum devices to which the user has access to, as the number of runs would scale exponentially with the system size. The fact that complete knowledge of the computation might not be required has been used to develop several different schemes to benchmark errors of particular gates, such as in randomized benchmarking \cite{Knill08}, and twirling protocols \cite{Lu15}. Here, we propose a method, which is particularly suited for the situation where the user of a quantum computer does not have direct access to the quantum device and where the number of runs is limited. We will first analyze the errors that occur after applying a single gate and then consider those which occur in an actual quantum computation, involving many gates. Note that the maximum allowed number of runs of one computation is limited to 8192, which allows to estimate the statistical error.

To estimate the error that occurs after applying a single gate from the gate set, we perform the following procedure. We apply the single gate $A$ to the initial state $\rho(0)$. Ideally this would yield the state $A \ket{0}$. However, due to systematic errors, in the preparation as well as in the application of the gate, a state $\rho_A(0)$ is obtained. For state tomography, we perform three experiments measuring $\expect{X}$, $\expect{Y}$, and $\expect{Z}$ with 8192 runs each. In order to measure $X$ and $Y$ the gates $H$ and $H S^\dagger$ are applied respectively before the $Z$ measurement. An estimate, $\widehat{\rho_A}(0)$, of the state $\rho_A(0)$ is then determined using the direct inversion method, i.e., $\widehat{\rho_A}(0) = 1/2 \identity + 1/2 \sum_i \expect{\sigma_i} \sigma_i$, where $\{\sigma_i\}_i = \{X, Y, Z\}$. The fidelity, $F= \sqrt{\bra{0} A^\dagger  \widehat{\rho_A}(0) A \ket{0}}$, is presented in Table \ref{tab:fidelities} for different choices of $A$.

\begin{table}[h!]
\centering
\footnotesize
  \begin{tabular}{ | c | c | c | c | c | c | c | c |}
    \hline
    Gate $A$ & $\identity$ & $H$ & $T$ & $S$ & S$^\dagger$ & $X$ & $CNOT$ \\ \hline
    Fidelity $F$ & 0.9813 & 0.9963 & 0.9961 & 0.9964 & 0.9920 & 0.9665  & 0.9794 \\ \hline
  \end{tabular}

  \caption{Fidelities of the estimate of the real state, $\widehat{\rho_A}(0)$, with respect to the ideal state $A \ket{0}$.}
  \label{tab:fidelities}
\end{table}

Note that $\widehat{\rho_A}(0)$ might not correspond to a physical state, as the length of the corresponding Bloch vector might be larger than $1$. Note further that IBM provided a Bloch measurement, which outputs a Bloch vector which is constructed in a similar way as described above. However, the Bloch vector is rescaled with the factor $1/\eta$ to take systematic errors into account. Here, $\eta$ is given by the difference of the probabilities of measuring the state $\ket{0}$ when $\ket{1}$ ($\ket{0}$) was prepared respectively, i.e., $\eta = p(0,\rho(0)) - p(0,\rho(1))$. A typical value for $1/\eta$ would be $1.05$. IBM Bloch measurement gave results that are much more precise than those that we can produce, but they have removed this measurement in one of the last updates, saying that it was confusing the users.

Knowing the errors of a single gate is of course not sufficient to gain an estimate of the error obtained in an actual quantum computation, as it does not give any information about the error which accumulates during the computation due to e.g. a drift in the quantum computation. However, without knowing all the details of the experimental setup the derivation of a suitable error model is unfeasible. Due to that we propose here a different method to estimate the error, which is suitable in case the user of the quantum computer does not have direct access to it. The idea is to use a set of circuits which are approximately of the same length and complexity as the circuits of interest and whose output can be determined classically. We will call these circuits \textsl{validating circuits} in the following. They are chosen of the same length and complexity to ensure that they give rise to similar errors as the circuits of interest. Moreover, they are chosen to be classically simulatable such that the error can in fact be determined. As an example, consider a circuit of length $N$ containing $n$ $A$ gates, which are supposed to be the most erroneous ones. Then, a set of validating circuits is a set of circuits, $\{U_i\}_i$ where, for each $i$, $U_i$ contains $N$ gates in total and $n$ $A$ gates, while the other gates as well as the order in which the gates are applied may differ from the ones used in the original circuit.
Given that the outcome of these circuits can be computed classically, the error of the quantum computer running these circuits can be determined. One can then use
this error in order to estimate the error occurring in the circuit of interest, whose output cannot be computed easily.

As any computation performed on a few qubits can be simulated classically, the error can be determined directly without the use of a validating circuit set. However, once larger quantum computers become available such an approach might be very useful to estimate the expected error. Note that in order to derive the validating circuits, which have to be classically simulatable, one might use the results presented in Ref. \cite{Nest11}. There, it was shown that if in two classically efficiently simulatable gate sets (strong simulation) the Clifford gates and the matchgates are grouped in a particular way, then the output of the computation can also be simulated efficiently (weak simulation).

Here, the circuits of interest perform the compressed simulation of the Ising model, which will be derived below. Let us, for the sake of genuine error analysis, assume that the output of these circuits is unknown to us. In contrast, we assume that the output of the validating circuits is known. In order to construct them we consider two of the circuits performing the compressed simulation of the Ising model (for details see Appendix \ref{app:validating}). We keep the number of $CNOT$ and $T$ gates constant in order to keep the same complexity level, but exchange the other gates with random Clifford gates and perform the measurement. We repeat the procedure ten times obtaining 20 validating circuits in total. In Appendix \ref{app:validating} (Table \ref{tab:validation}), we present the error of the 20 validating circuits. The average error is $0.122$, which is in good agreement with the experimental and theoretical results (see Fig. \ref{fig:results}).

\chapter{Gapless adiabatic quantum computation}

\pagestyle{fancy}
\fancyhf{}
\fancyhead[LE]{\thepage}
\fancyhead[RE]{QUANTUM COMPUTATION}
\fancyhead[LO]{CH.7 Gapless adiabatic quantum computation}
\fancyhead[RO]{\thepage}

\section{Introduction to adiabatic quantum computation}

In the previous chapter we have studied and presented some examples of the conventional quantum computation paradigm: the so-called \emph{circuit model}. Here we are going to talk about a different computational model called \emph{adiabatic quantum computation}, introduced in Ref. \cite{Farhi00}. The basic idea is to identify a potentionally complicated hamiltonian whose ground state is the solution of our problem of interest. Then, another system with a simple hamiltonian is prepared and initialized in its ground state. The computation consists on making the system evolve slowly from its initial state to the problem final state so that the system is kept all the time on the ground state of the instantaneous hamiltonian, in order to ensure convergence to the desired state. Such evolution is guaranteed by the \emph{adiabatic theorem}, an old theorem from 1928 which says that ``A physical system remains in its instantaneous eigenstate if a given perturbation is acting on it slowly enough and if there is a gap between the eigenvalue and the rest of the Hamiltonian's spectrum'' \cite{Born28}.

The basic procedure is described in Ref. \cite{Farhi00}. A quantum state evolves with the Schr\"odinger equation
\begin{equation}
i \frac{d}{dt} \ket{\psi(t)} = H(t) \ket{\psi(t)}.
\end{equation}
The adiabatic theorem tells us how to follow this evolution when $H(t)$ is slowly varying. Consider a family of hamiltonians $\tilde{H}(s)$ with s a continuous parameter going from 0 to 1, $0 \leq s \leq 1$, and take $H(t) = \tilde{H}(t/T)$, so that $T$ controls the rate at which $H(t)$ varies. We define the instantaneous eigenvectors and eigenstates as
\begin{equation}
H(s) \ket{l;s} = E_l(s) \ket{l;s},
\end{equation}
where l orders the states according to their energy, $l=0$ being the ground state. Suppose $\ket{\psi(0)}$ is the ground state of $H(0)$ , that is $H(0) = \ket{l=0;s=0}$. Then according to the adiabatic theorem, if the gap between the two lowest levels, $E_1(s)-E_0(s)$, is strictly greater than 0 for all $s$, then
\begin{equation}
\lim_{T \rightarrow \infty} |\braket{l=0;s=1}{\psi(T)}| = 1,
\end{equation}
which means that the evolved state will be arbitrarily close to the ground state of the final hamiltonian if we make T big enough. Let us define the minimum gap by
\begin{equation}
g_{min} = \min_{0 \leq s \leq 1} (E_1(s) - E_0(s)).
\end{equation}
A closer look to the adiabatic theorem tells us that ``T big enough'' means taking
\begin{equation}
T >> \frac{\xi}{g^2_{min}},
\label{eq:tmin}
\end{equation}
where
\begin{equation}
\xi = \max_{0\leq s \leq 1} | \bra{l=1; s} \frac{d\tilde{H}}{ds}  \ket{l=0;s} |
\label{eq:overlap}
\end{equation}
is the overlap between the two lowest levels of two infinitely close hamiltonians, which dictates how easily the ground state will jump to the first excited state during the evolution.

Thus Eq. \ref{eq:tmin} represents a necessary condition on the minimum  time to keep the evolution adiabatic. It is not a sufficient condition, but we will not deal with this issue here, treatment of sufficient conditions can be found in Refs. \cite{Marzlin04,Tong07}. 

It may seem more natural to define the minimum computation time as the maximum of the quotient of the overlap and the gap, thus forcing them to be calculated at the same $s$, instead of calculating the minimum and maximum separately, but it might be more difficult to compute for some models, and Eq. \ref{eq:tmin} will always work as an upper bound of that quantity \cite{Farhi17}. For the concrete computation on this thesis it doesn't make a difference.

As a quantum computing model, adiabatic quantum computation has been shown to be polinomially equivalent to conventional quantum computing in the circuit model \cite{Aharonov07}. It has a distinct advantage in the possibility of getting around the problem of energy relaxation, one of the sources of errors for circuits that we talked about in the previous chapter, as the system is kept all the time in the ground state. In practice, there are problems during the computation. Quantum effects as opposed to classical tend to happen close to phase transition points, and it is exactly there that the gap gets close to zero and so the required computation time starts to become very large. We try to deal precisely with this problem in the next section, trying to find if the numerator in Eq. \ref{eq:tmin} could go to zero at the same rate as the gap, thus enabling computation in finite time, but we finally show that this is not the case, at least for the important Ising with transverse field model.

\section{Analysis of the computation time}

We want to compute the time required for an adiabatic evolution of the Ising model with transverse field between the 0 and
critical values of the field, and are especially interested in finding the result for an infinite chain, which means a zero
energy gap. We want to compute the details and see if the evolution time goes to infinity or not in that limit.

If the gap goes to zero, we need the overlap between the first excited state and the ground state affected by the Hamiltonian derivative to go also to zero and cancel it out for the time to be finite, according to Eq. \ref{eq:tmin}. We have to find the gap and the overlap separately. We need to diagonalise the hamiltonian first.

\subsection{Diagonalisation of the hamiltonian}

Let us begin with the expression for the Ising Hamiltonian with transverse field for an N-spin 1D chain with periodic boundary conditions
\begin{equation}
H=-\frac{1}{2}\sum _{j=1}^N \left( \sigma_j^x \sigma _{j+1}^x + s\sigma _j^z \right),
\label{eq:isingh}
\end{equation}
where we use $s$ for the transverse field having in mind that we will use it as the adiabatic parameter.
The spectrum of this hamiltonian can be found analitically with a series of transformations. Several sources describe the procedure, here we write a resumed version following mainly Ref. \cite{Depasquale08}. We begin by performing a Jordan-Wigner transformation which maps spin operators onto fermionic creation and annihilation operators
\begin{equation}
\begin{split}
& a^{\dagger}_j = \left( \prod_{m<j} \sigma_m^z \right) \frac{\sigma_j^x+i\sigma_l^y}{2}, \\
& a_j = \left( \prod_{m<j} \sigma_m^z \right) \frac{\sigma_j^x-i\sigma_l^y}{2}.
\end{split}
\end{equation}
\label{eq:jordanwignerh}
The resulting transformed hamiltonian can be written as a sum of two terms, a generic and a parity term, 
\begin{equation}
H=-\frac{1}{2} (H_G+H_P) 
\end{equation}
with
\begin{equation}
  \begin{split}
    & H_G=\sum _{j=1}^N \left( s-2s a_j^\dagger a_j + a_j^\dagger a_{j+1}^\dagger + a_j^\dagger a_{j+1} - a_j a_{j+1} -
      a_j a_{j+1}^\dagger \right), \\
    & H_P= - \left( a_N^\dagger a_1^\dagger + a_N^\dagger a_1 - a_N a_1 -
      a_N a_1^\dagger \right) \left( \exp \left(i \pi \sum _{j=1}^N a_j^\dagger a_j \right) + 1 \right).
  \end{split}
\label{eq:hghp}
\end{equation}
The parity term $H_P$ appears because of the finitude of the chain, and depends on the parity of the whole system. Now we can
perform a deformed Fourier transformation imposing that these two terms end up having the same form, at the price of having two different sets of momenta ($k$ and $k+\alpha$) corresponding to the two parity sectors. The transformation reads
\begin{equation}
c_k = \frac{1}{N} \sum_j a_j e^{-i\frac{2\pi}{N}(kj+\alpha_j)} \hspace {15mm} 
\end{equation}
where $\alpha_{j+1} = \alpha_j+\alpha$, and the value of $\alpha$ is determined by the parity
\begin{equation}
\begin{split}
& \exp \left(i \pi \sum _{j=1}^N a_j^\dagger a_j \right) = +1 \rightarrow \alpha = 0. \\
    & \exp \left(i \pi \sum _{j=1}^N a_j^\dagger a_j \right) = -1 \rightarrow \alpha = \frac{1}{2}.
\end{split}
\end{equation}
The hamiltonian after the deformed Fourier transformation gets the following form
\begin{equation}
\begin{split}
H=\frac{1}{2}\sum _{k=-(\frac{N-1}{2})}^\frac{N-1}{2} &\left( 2 \left( s- \cos \left( 2 \pi \frac {k+ \alpha}{N}  \right) 
\right) c_k^\dagger c_k \right.\\
&\left. + i \sin \left( 2 \pi \frac {k+ \alpha}{N}  \right) \left( c_{-k}^\dagger c_k^\dagger 
+ c_{-k} c_k \right) -s \right).
\end{split}
\end{equation}
Finally we perform a Bogoliubov transformation
\begin{equation}
    b_k = u_k c_k -i v_k c_{-k}^\dagger, 
\end{equation}
with
\begin{equation}
   \begin{split}
    & u_k = \cos \frac{\theta_k}{2}, \hspace {5mm} v_k = \sin \frac{\theta_k}{2}, \\
    & \cos\theta_k=\frac{s-\cos \frac{2\pi k}{N}}{\epsilon_k}, \hspace {5mm} \sin\theta_k=\frac{\sin\frac{2\pi k}{N}}
{\epsilon_k}, \\
    & \epsilon _{k} = +\sqrt{1+s^2-2s\cos \left( 2 \pi \frac {k+ \alpha}{N} \right). } 
  \end{split}
\end{equation}
The hamiltonian takes with this last transformation a diagonal form
\begin{equation}
H = \sum _{k=-(\frac{N-1}{2})}^\frac{N-1}{2} \epsilon _{k} \left( b_k^\dagger b_k - \frac{1}{2} \right). \\
\end{equation}
analogous to the harmonic oscillator. Note that $\epsilon _{k}$ is symmetric on $k$ and defined positive. The $\alpha$ dependence of the Hamiltonian tends to disappear as 
$ N \rightarrow \infty $, because for large N it is the same to compute $\epsilon_k$ for $\alpha=1,2,3$ as 
for $\alpha=1/2, 3/2, 5/2...).$ 

\subsection{Computation of the gap}

\begin{figure}[h!]
\includegraphics[scale=0.45]{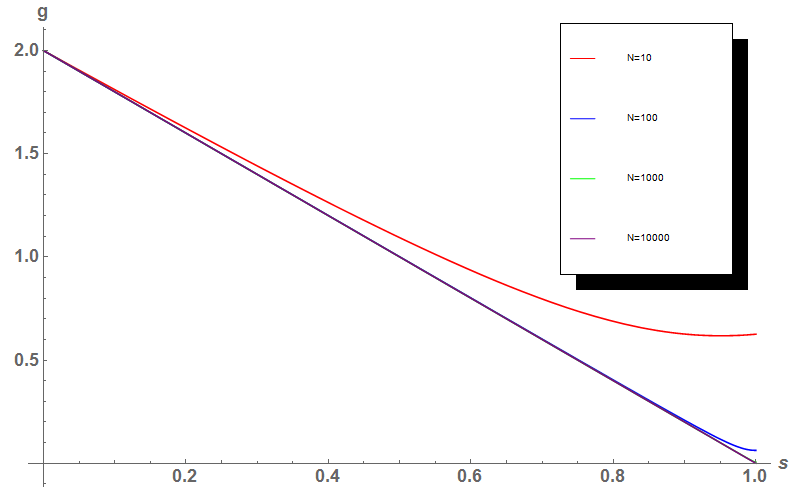}
\caption{Gap evolution between $s=0$ and $s=1$ for different sizes of the chain. For larger chains the gap tends to 0 at $s=1$.}
\label{fig:gapevolution}
\end{figure}

For $|s|<1$, our range of interest, only states with an even number of fermions
(in pairs of momenta $k+\alpha$ and $-k+\alpha$) are physical, the others having the wrong parity. It is possible to visualize this easily by noting that for this region the number of fermions can be mapped in the spin space to the number of domain walls between spins with different orientations. Because of the periodic boundary conditions, the number of domain walls must be even. Moreover, the state with 0 fermions
of even parity has always less energy than the state with 0 fermions of odd parity, something we checked numerically. The energy of the ground state is therefore
\begin{equation}
    E_0 = - \frac{1}{2} \sum _{k=-(\frac{N-1}{2})}^\frac{N-1}{2} \epsilon _{k}
\end{equation}
with $\alpha=0$. The relevant gap corresponds to the difference in energies between the ground state and the first excited 
state in this parity sector, which is obtained by creating a pair of fermions with $k=1/2$ and $k=-1/2$
\begin{equation}
    \begin{split}
    & E_1 = - \frac{1}{2} \left( \sum _{k=-(\frac{N-1}{2})}^\frac{N-1}{2} \epsilon _{k} \right) +2 \epsilon _{1/2},  \\
    & g = E_1-E_0 = 2 \epsilon _{1/2} = 2 \sqrt{1+s^2-2s\cos \left( \frac {\pi}{N} \right)}.
  \end{split}
\end{equation}
In order to find the minimum value for the gap with respect to $s$, we just derive and find
\begin{equation}
 g_{min} = 2 \sin \left(\frac{\pi}{N}\right),
\end{equation}
for $s_{g_{min}} = \cos \left(\frac{\pi}{N}\right)$. Therefore for an infinite chain we get $g_{min} = 0$ and $s_{g_{min}} = 1$, as expected. The evolution of the gap between $s=0$ and $s=1$ for different sizes of the chain is shown in Fig. \ref{fig:gapevolution}.

\subsection{Computation of the overlap}

\begin{figure}[h!]
\includegraphics[scale=0.45]{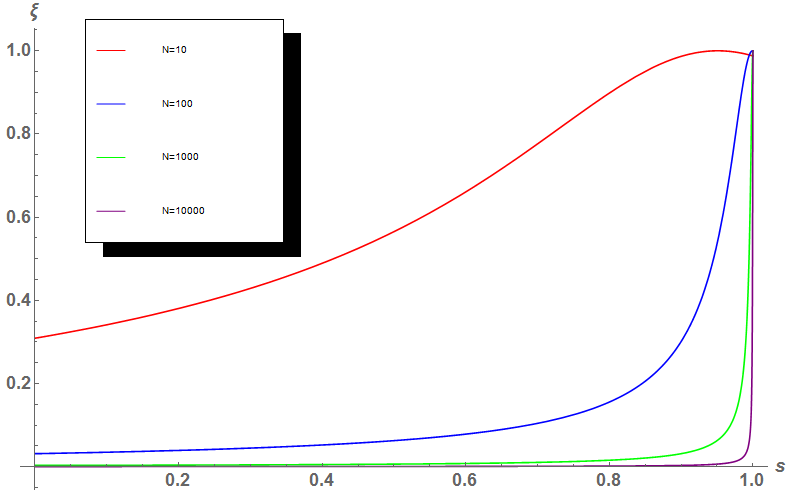}
\caption{Overlap evolution between $s=0$ and $s=1$ for different sizes of the chain. For larger chains the overlap tends to 1 at $s=1$.}
\label{fig:overlapevolution}
\end{figure}

We now evaluate $\xi$ from Eqs. \ref{eq:tmin} and \ref{eq:overlap} to see if it can cancel the zero gap and allow a finite time for the adiabatic computation. We need to express the derivative of the hamiltonian with respect to the adiabatic parameter $s$ after the three transformations. Below is the sequence of the expression of the derivative after each transformation 
\begin{equation}
\begin{split}
&\frac{dH}{ds}=-\frac{1}{2}\sum _{j=1}^N \sigma _j^z \hspace{35 mm} \rm{Initial\hspace{1 mm}derivative}.\\
&\frac{dH}{ds}=-\frac{1}{2}\sum _{j=1}^N 1-2 a_j^\dagger a_j \hspace{24 mm} \rm{After \hspace{1 mm} Jordan-Wigner\hspace{1 mm}  transformation}.\\
&\frac{dH}{ds}=-\frac{1}{2}\sum  _{k=-(\frac{N-1}{2})}^\frac{N-1}{2} 1-2 c_k^\dagger c_k \hspace{15 mm} \rm{After \hspace{1 mm} Fourier\hspace{1 mm}  transformation}.\\
&\frac{dH}{ds}=-\frac{1}{2}\sum  _{k=-(\frac{N-1}{2})}^\frac{N-1}{2} 1- (1+\cos\theta_k) b_k^\dagger b_k - 
(1-\cos\theta_k) b_{-k} b_{-k}^\dagger \\
& \hspace{30mm} - i \sin\theta_k (b_k^\dagger b_{-k}^\dagger - b_{-k} b_k). \\
\end{split}
\end{equation}
The last expression is the final derivative after Bogoliubov transformation. We have said that the first excited state is obtained by creating two fermions with $k=
\pm 1/2$, so we can write
\begin{equation}
| l=1 \rangle = b_{1/2}^\dagger b_{-1/2}^\dagger | l=0 \rangle.
\end{equation}
Introducing this in the expression for the overlap, taking the appropiate complex conjugate
\begin{equation}
\left| \left\langle l=1 \left| \frac {dH}{ds} \right| l=0 \right\rangle  \right| = 
\left| \left\langle l=0 \left| b_{-1/2} b_{1/2} \frac {dH}{ds} \right| l=0 \right\rangle  \right|,
\end{equation}
only the term with two creation operators from $dH/ds$ will contribute
\begin{equation}
\frac{1}{2} \left| \sum  _{k=-(\frac{N-1}{2})}^\frac{N-1}{2} \sin\theta_k  
 \left\langle \left| l=0 \hspace{1 mm} b_{-1/2} b_{1/2} b_k^\dagger b_{-k}^\dagger \right| l=0 \right\rangle  \right|,
\end{equation}
and now only the terms with $k=-1/2$ and $k=1/2$ will contribute.
\begin{equation}
  \begin{split}
    & \frac{1}{2} \left| \sin\theta_{1/2}  
    \left\langle l=0 \left| b_{-1/2} b_{1/2} b_{1/2}^\dagger b_{-1/2}^\dagger \right| l=0 \right\rangle \right.\\  
    & \hspace{3mm} \left. +  \sin\theta_{-1/2}  
     \left\langle l=0 \left| b_{-1/2} b_{1/2} b_{-1/2}^\dagger b_{1/2}^\dagger \right| l=0 \right\rangle \right| = \\ 
    & \frac{1}{2} \left| \sin\theta_{1/2}  
    \left\langle l=0 \left| b_{-1/2} b_{1/2} b_{1/2}^\dagger b_{-1/2}^\dagger \right| l=0 \right\rangle \right.\\
     & \hspace{3mm} \left. -\sin\theta_{1/2}  
    \left\langle l=0 \left| b_{-1/2} b_{1/2} ( - b_{1/2}^\dagger b_{-1/2}^\dagger ) \right| l=0 \right\rangle  \right| = \\
    & \sin\theta_{1/2} = \frac{\sin\frac{\pi}{N}}{\epsilon_{1/2}} = \frac{\sin\frac{\pi}{N}}{\sqrt{1+s^2-2s\cos \left( \frac {\pi}{N} \right)}}.
  \end{split}
\end{equation}
Now we have to find the maximum of this expression with respect to s, which corresponds to the minimum denominator, already
found when we were searching for the minimum gap
\begin{equation}
\max_{0\leq s \leq 1} \frac{\sin\frac{\pi}{N}}{\sqrt{1+s^2-2s\cos \left( \frac {\pi}{N} \right)}} =
\frac{\sin\frac{\pi}{N}}{\sin\frac{\pi}{N}} = 1.
\end{equation}
The evolution of the overlap between s=0 and s=1 is shown in Fig. \ref{fig:overlapevolution}.

\subsection{Conclusion}

\begin{figure}[h!]
\includegraphics[scale=0.38]{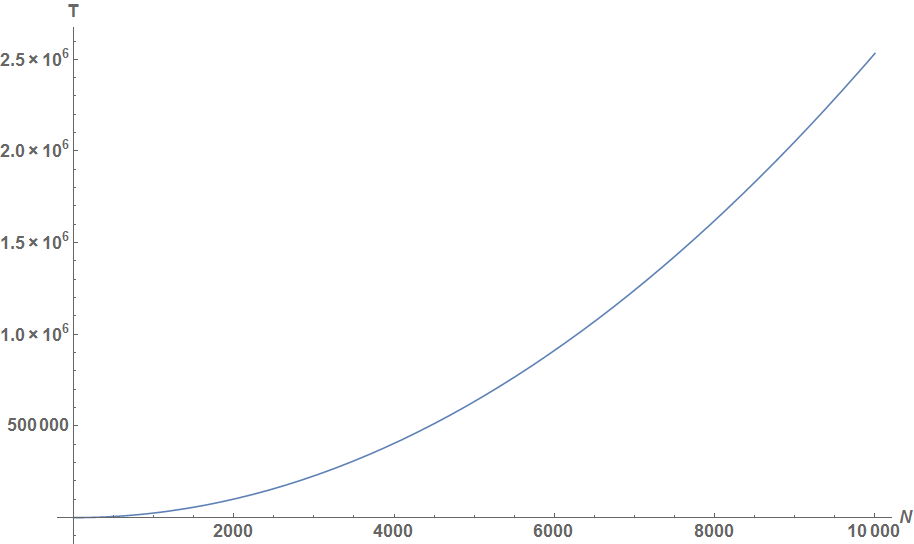}
\caption{Scaling of the total computation time with the size of the chain, which goes as $N^2$ for large $N$.}
\label{fig:timescaling}
\end{figure}

As the maximum overlap for the infinite size limit (and for all sizes) is 1 and not 0 as desired, the time for an adiabatic evolution in an infinite Ising chain with transverse field from $s=0$ to $s=1$ remains infinite because of the gap. In particular,
\begin{equation}
T\textgreater \textgreater \frac{ \max_{0\leq s \leq 1} \left| \left\langle l=1 \left| \frac {dH}{ds} \right|
 l=0 \right\rangle  \right| }{g^2_{min}} 
= \frac {1}{4 \sin^2 \frac{\pi}{N}} \simeq N^2
\end{equation}
where the last approximation is taken for $N \rightarrow \infty$. So the time for an adiabatic quantum computation involving this system would grow as $N^2$ for large $N$. Fig. \ref{fig:timescaling} shows the scaling of the total computation time with N.

\chapter{Algorithm for computation of the diagonal ensemble}

\pagestyle{fancy}
\fancyhf{}
\fancyhead[LE]{\thepage}
\fancyhead[RE]{QUANTUM COMPUTATION}
\fancyhead[LO]{CH.8 Algorithm for computation of the diagonal ensemble}
\fancyhead[RO]{\thepage}

In this last chapter we give and test a new algorithm for the computation of the \emph{diagonal ensemble} of a hamiltonian, a concept we will soon define. We start by giving a motivation for this in the context of quantum thermodynamics, a link between quantum information and condensed matter physics that is developing quickly recently \cite{Gogolin15}. 

\section{Motivation: The problem of thermalisation}

One of the fundamental postulates of quantum mechanics says that states evolve unitarily according to the Schr\"odinger equation. This implies determinism and, in case of a finite size system, recurrence of the wave function, and at first sight is apparently in some tension with the concepts of statistical mechanics and the second law of thermodynamics. As soon as 1929 Von Neumann already tried to prove ergodicity and a tendency to evolve into states of maximum entropy to develop the H-theorem in the quantum context \cite{Vonneumann29}, and the field of quantum statistical mechanics emerged.

But some basic questions remain, in particular how quantum states taking extremal values for the entropy arise from microscopic dynamics, and how the concepts of \emph{equilibration} and \emph{thermalisation} appear at this fundamental level. We first define these two concepts.

\emph{Equilibration} of a certain property means that the corresponding value 
evolves over time, after starting from some non-equilibrium state, to reach and remain close to a
certain value for an extended period of time.

If equilibration happens, then it happens towards the time-averaged state. Let us consider an initial state of the system, $\ket{\psi(0)} = \sum_n c_n \ket{E_n}$, written here in its energy eigenbasis, which evolves unitarily under its hamiltonian $H$ as
\begin{equation}
\ket{\psi(t)} = e^{-iHt} \ket{\psi(0)}.
\label{eq:schrodinger}
\end{equation}
If we take its corresponding density operator,
\begin{equation}
\rho(t)=\ket{\psi(t)}\bra{\psi(t)},
\end{equation}
its time average is a mixed state given by
\begin{equation}
\begin{split}
&\lim_{T \rightarrow \infty} \frac{1}{T} \int_0^T \rho(t) dt = \lim_{T \rightarrow \infty} \frac{1}{T} \int_0^T \sum_{nm} c_n c_m^* e^{-it(E_n-E_m)} \ket{E_n} \bra{E_m} dt = \\
& \lim_{T \rightarrow \infty} \frac{1}{T} \sum_{nm,E_n=E_m} c_n c_m^* \ket{E_n}  \bra{E_m} \hspace{1mm} T + \\
& \lim_{T \rightarrow \infty} \frac{1}{T} \sum_{nm,E_n \neq E_m} c_n c_m^* \ket{E_n}\bra{E_m} \frac{1}{-i (E_n-E_m)} \left(e^{-iT(E_n-E_m)} - 1 \right)
\end{split}
\end{equation}
Each of the terms in the second sum will become smaller as T grows, so in the infinite limit it will go to zero. Assuming there is no degeneracy in the system, we find that the long-time average of the expectation value is given by
\begin{equation}
\lim_{T \rightarrow \infty} \frac{1}{T} \int_0^T \rho(t) dt = \sum_{n} |c_n|^2 \ket{E_n} \bra{E_n}.
\label{eq:diagens}
\end{equation}
The final state is diagonal in the energy eigenbasis and is thus called \emph{diagonal ensemble} \cite{Srednicki99}.

\emph{Thermalisation} means that some local property of the system  equilibrates towards the value of the thermal state, which is
\begin{equation}
\rho_{th}(\beta) = \frac{e^{-\beta H}}{\tr ({e^{-\beta H}})}
\end{equation}
where $\beta$ is the inverse temperature. Since the whole system begins as a pure state and evolves unitarily according to the Schr\"odinger equation,
the idea of thermalisation applies to a subsystem. In this sense, although the whole system remains in a pure state, 
a small part is seen to thermalise as a result of the interaction with the rest of the system, which acts as a bath.

Two different types of thermalisation can exist. Let us consider an initial state of the whole system, $\ket{\psi(0)}$, which evolves unitarily with the hamiltonian $H = H_S+H_B+H_I$, the hamiltonian of the subsystem plus the hamiltonian of the bath plus the interaction, according to Eq. \ref{eq:schrodinger}. The state of the subsystem at a time $t$ will be
\begin{equation}
\rho_{S}(t) = \textrm{tr}_{B} \ket{\psi(t)} \bra{\psi(t)}.
\end{equation}
Then the subsystem is said to experiment \emph{strong thermalisation} if the instantaneous expectation values converge to the thermal ones at large times, i.e
\begin{equation}
\|\rho_{S}(t)-\textrm{tr}_B(\rho_{\mathrm{th}}(\beta))\| \xrightarrow[{t\to\infty}]{} 0
\end{equation}
whereas the subsystem is said to experiment \emph{weak thermalisation} if it only converges to the thermal state after time averaging, i.e. if the subsytem of the diagonal ensemble converges to the subsystem of the thermal ensemble
\begin{equation}
\lim_{T \rightarrow \infty} \frac{1}{T} \int_0^T \rho_S(t) dt = \textrm{tr}_B \left(\sum_{n} |c_n|^2 \ket{E_n} \bra{E_n}\right) = \textrm{tr}_B(\rho_{\mathrm{th}}(\beta)).
\end{equation}
The following step is to ask if and which systems do thermalise. Many systems have been analyzed in this respect, for example Ref. \cite{Banuls11} studies the Ising model with transverse and parallel fields and finds that depending on the initial state, it can show strong thermalisation, weak thermalisation or no thermalisation at all, at least for the studied time scales. The Eigenstate Thermalisation Hypothesis (ETH) gives some theoretical substance to the question \cite{Srednicki94,Rigol09}. In its simplest formulation, it states that the expectation value $\bra{E_k} O \ket{E_k}$ of a few-body observable $O$ in an individual hamiltonian eigenstate $\ket{E_k}$ equals the thermal average $O$ at the mean energy $E_k$, which means that the knowledge of a single many-body eigenstate suffices to compute thermal averages.

Many systems have been shown to fulfil ETH, although it is not true in general. It is known that integrable systems do not usually thermalise, as they have some restrictions in their evolution because of its conserved quantities and thus some dependencies on the initial states. Systems that exhibit a \emph{many-body localisation} (MBL) phase are also expected not to thermalise.

\section{Computation of the diagonal ensemble}

Putting some elements together from the previous section, it is possible to see if a system exhibits weak thermalisation by computing the diagonal ensemble and comparing observables of a subsystem with those of a subsystem of the thermal ensemble. The problem in practice is that due to the exponential grow of the Hilbert space, exact computation of the diagonal ensemble is impossible for moderately large systems, and one does not expect to observe thermalisation in very small systems. Even with MPS/MPO techniques, the diagonal ensemble continues to prove a formidable challenge because of its highly non-local character. Direct approximation of the diagonal ensemble via constraint overlap maximisation was addressed in Ref. \cite{Nebendahl15}. Here we are going to test a new evolution algorithm which hopefully can work efficiently with an MPO implementation.

We are going to use the same system from Ref. \cite{Banuls11}, the Ising model in 1D with transverse and parallel fields
\begin{equation}
H=J\sum_i^{N-1} \sigma_z^i \sigma_z^{i+1}+g \sum_i^N \sigma_x^i+h \sum_i^N \sigma_z^i,
\end{equation}
with J=1, g=1.05, h=0.5 and open boundary conditions.

Our test observable is $\langle \sigma_x \rangle _{N/2} $, i. e. $\sigma_x$ in the middle of the chain, which will be our subsystem for study. The results are given for the initial state $\ket{\psi}=\frac{1}{\sqrt{2}} (|0\rangle + |1\rangle)^{\otimes N}$, although we also checked other initial states finding no difference in the results regarding the convergence to the diagonal ensemble.

Instead of using standard Schr\"odinger evolution and then computing the time average of the results, it is better to evolve the states with the following operator
\begin{equation}
\ket{\rho(t)} = \exp(-M^2 t) |\rho(0)\rangle
\label{eq:diagevolution}
\end{equation}
where $\ket{\rho(t)}$ is the vectorized form of the density matrix $\rho(t)$, and
\begin{equation}
M= \sqrt{\frac{\Gamma}{2}} \left( H \otimes \mathbb{I} - \mathbb{I} \otimes H^T\right).
\end{equation}
The master equation \ref{eq:diagevolution} makes the state converge directly to the diagonal ensemble. It is not a physical evolution, but for our purposes works more efficiently. The $\Gamma$ factor controls the velocity of the convergence. We will use always $\Gamma=0.7$ and control the later approximations with the size of the time step. It is an operator with a size double of the original hamiltonian, which can be interpreted as the union of the physical system and an ancilla system. It applies to the vectorized form of the density matrix of the system, of a size that also doubles the one from the original ket. The vectorization is done because it is most convenient for its MPO implementation, and it is of common use in practical algorithms.

We show now why the operator brings the state to the diagonal ensemble

\begin{equation}
\begin{split}
&\ket{\rho(t \rightarrow \infty )} = \lim_{t\rightarrow \infty} \exp(-M^2 t) \ket{\rho(0)} = \\
& \lim_{t\rightarrow \infty} \sum_{nm} e^{-t \frac{\Gamma}{2}(E_n-E_m)^2} \ket{E_n E_m} \bra{E_n E_m} \hspace{2mm} c_n c_m^* \ket{E_n E_m} = \\
&\sum_{n} |c_n|^2 \ket{E_n E_n}.
\end{split}
\end{equation}

which is the vectorized version of the expression from Eq. \ref{eq:diagens}.
\subsection{Results with Taylor approximation}

\begin{figure}[h!]
\centering
\includegraphics[scale=0.85]{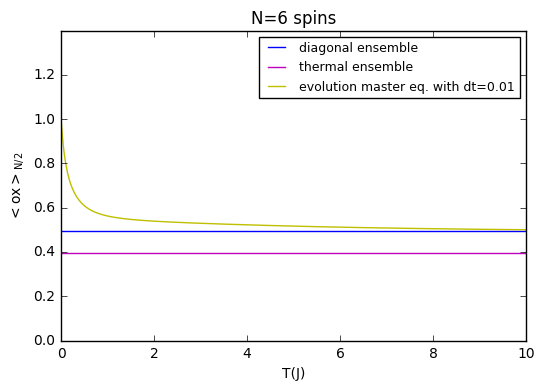}
\caption{Taylor results for 6 spins}
\label{fig:tay6}
\end{figure}

\begin{figure}[h!]
\centering
\includegraphics[scale=0.85]{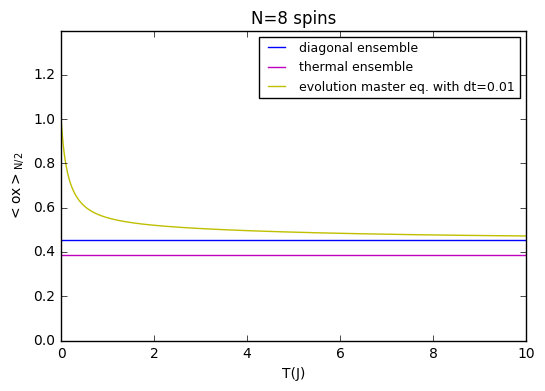}
\caption{Taylor results for 8 spins}
\label{fig:tay8}
\end{figure}

\begin{figure}[h!]
\centering
\includegraphics[scale=0.85]{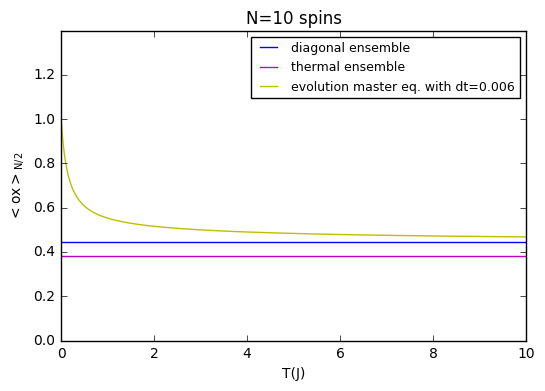}
\caption{Taylor results for 10 spins}
\label{fig:tay10}
\end{figure}

Before presenting the new algorithm, we give results obtained using the first-order Taylor approximation
\begin{equation}
e^{-M^2 \delta t} \approx \mathbb{I} + M^2 \delta t.
\label{eq:taylor}
\end{equation}
Evolution is computed in time steps of $\delta t$ until T=10J. %We used sparse matrices and computed in the energy eigenbasis.
We give in Tab. \ref{tab:taylor} and Figs. \ref{fig:tay6}, \ref{fig:tay8} and \ref{fig:tay10} the results of $\langle \sigma_x \rangle _{N/2}$ for the state evolving with operator from Eq. \ref{eq:taylor} together with the values for the diagonal ensemble and the thermal ensemble for $N=6,8,10$. We find convergence to the diagonal ensemble if we make $\delta t$ small enough; the maximum $\delta t$ that gives correct convergence decreases approximately as $\delta t_{max} \sim 1/N^2$. For $ \delta t > \delta t_{max}$ the evolution starts to suffer from large oscillations and does not converge to any value at all. We also observe the diagonal ensemble approaching the thermal ensemble as we increase N, which would imply thermalisation if there was convergence, but it is not conclusive at all with these small sizes.

\vspace{20mm}

\begin{table}[h]
\centering
  \scalebox{1.1}{
  \begin{tabular}{ c | c | c | c | c | c }
    \hline
    N &  $\langle \sigma_x \rangle _{diag}$ & $\epsilon(\langle \sigma_x \rangle_{10J}/\langle \sigma_x \rangle_{diag})$ & $\delta t_{max}$ & $\langle \sigma_x \rangle _{th}$ & $\epsilon(\langle \sigma_x \rangle_{diag}/\langle \sigma_x \rangle_{th})$ \\ \hline
    6 & 0.4944 & 0.012 & 0.019 & 0.3946 & 0.25 \\ \hline
    8 & 0.4525 & 0.042 & 0.010 & 0.3861 & 0.18 \\ \hline
    10 & 0.4428 & 0.056 & 0.006 & 0.3817 & 0.16 \\ \hline
  \end{tabular}
  }

   \caption{Table of results for the evolution with the Taylor approximation. N is the number of spins. $\langle \sigma_x \rangle _{diag}$ is the value of the observable in the diagonal ensemble and $\epsilon(\langle \sigma_x \rangle_{10J}/\langle \sigma_x \rangle_{diag})$ is the relative difference between the state after evolution at T=10J and the diagonal ensemble, which increases with N (it takes longer to converge when N grows).  $\delta t_{max}$ is the maximum time step size such that the 1st order Taylor approximation behaves properly. It decreases with N approximately as $1/N^2$. $\langle \sigma_x \rangle _{th}$ is the value of the observable in the thermal ensemble, and $\epsilon(\langle \sigma_x \rangle_{diag}/\langle \sigma_x \rangle_{th})$ is the relative difference between the diagonal ensemble and the thermal ensemble, which decreases with N as we go closer to the thermodynamic limit, thus being compatible with thermalisation, although not conclusive at all with these small sizes.  }
   \label{tab:taylor}
  
\end{table}

\subsection{Order reduction Trotter algorithm}

The Taylor expansion is not useful for larger N, as we can see by the increasing differences between evolved state and diagonal ensemble, and by the decreasing $\delta t$ needed to reproduce the full exponential faithfully, which means that we will need more and more time steps to obtain the results. 

We can try to do a trotterisation of the exponential \cite{Lie88}, having in mind an implementation with MPO/MPS. But the standard Trotter decomposition is also not useful because of the $\exp(H^2)$ terms in the evolution operator from Eq. \ref{eq:diagevolution}, which mix all terms of the hamiltonian even in this case of nearest-neighbour interactions, and thus it doesn't lead to local enough pieces that we can simulate with a standard TEBD-like algorithm \cite{Vidal03}. But using an extra ancilla qubit we can first decompose this term using the relationship:
\begin{eqnarray}
&&\exp(\delta \sigma_x \otimes H)\exp(\delta \sigma_y \otimes H)\exp(-\delta \sigma_x \otimes H)\exp(-\delta \sigma_y \otimes H) \nonumber \\ 
= && (\mathbb{I}+\delta \sigma_x \otimes H + \frac{1}{2}\delta^2 H^2)(\mathbb{I}+\delta \sigma_y \otimes H + \frac{1}{2}\delta^2 H^2)\times \nonumber \\
&&(\mathbb{I}-\delta \sigma_x \otimes H + \frac{1}{2}\delta^2 H^2)(\mathbb{I}-\delta \sigma_y \otimes H + \frac{1}{2}\delta^2 H^2) + \mathcal{O}(\delta^3) \nonumber \\
= && \mathbb{I} + 2\delta^2 (i \sigma_z - \mathbb{I} + \mathbb{I}) \otimes H^2 + \mathcal{O}(\delta^3) \nonumber \\ 
= && \mathbb{I}+2i\delta^2 \sigma_z \otimes H^2 + \mathcal{O}(\delta^3) \nonumber \\ = && \exp(2i\delta^2 \sigma_z \otimes H^2).
\end{eqnarray}
So we can use the approximation
\begin{equation}
\begin{split}
\exp(-(\sigma_z \otimes M^2) (\delta t)^2) \approx & \exp(-( \sigma_x \otimes M) \bar{\delta t})\exp(-( \sigma_y \otimes M)\bar{\delta t}) \\ & \exp(( \sigma_x \otimes M) \bar{\delta t})\exp(( \sigma_y \otimes M) \bar{\delta t}),
\end{split}
\end{equation}
where $\bar{\delta t} = \sqrt{\frac{i}{2}} \delta t$.

Then in principle each of the four resulting exponentials can be implemented efficiently with MPO with a proper Trotter decomposition, as we have eliminated the problem terms $\exp(H^2)$. We can name this approximation an \emph{Order reduction Trotter} approximation because of the reduction of the exponent.

We have computed the evolution with the exact exponential and with the approximated exponentials. In order to get the diagonal ensemble for large times, a projection of the state into the $|0\rangle$ of the ancilla has to be done. Renormalization after each step, which can be applied either to the vectorized state or to the density matrix, is not needed, although it might be convenient to prevent numerical instabilities. Hermitization and positivization of the density matrix after each step are not essential, although they improve slightly the results or allow to slightly increase the time step. 

\begin{figure}[h!]
\centering
\includegraphics[scale=0.60]{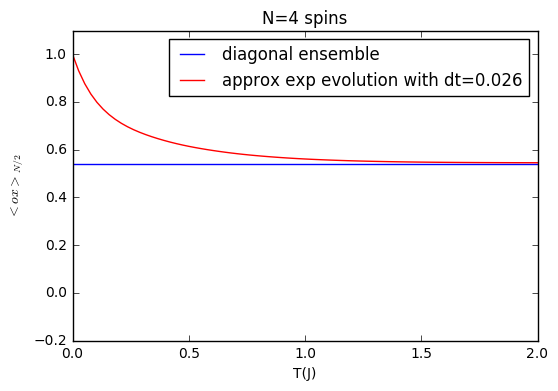}
\caption{Order reduction Trotter evolution with a time step slightly below $\delta t_c$. We can see that convergence to diagonal ensemble is perfect.}
\label{fig:newtrot0026}
\end{figure}

\begin{figure}[h!]
\centering
\includegraphics[scale=0.60]{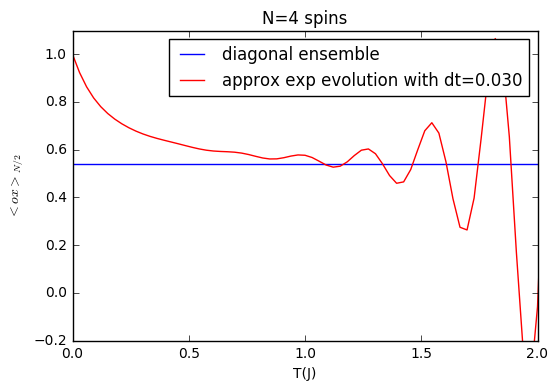}
\caption{Order reduction Trotter evolution with a time step slightly above $\delta t_c$. We can see that it does not converge at all to the diagonal ensemble.}
\label{fig:newtrot0030}
\end{figure}

A remarkable feature of the approximation is that, below a certain critical $\delta t_c(N)$, that depends on the size of the system, the convergence to the diagonal ensemble is exact, and one does not need to further decrease the time step in order to improve accuracy. This is already quite apparent in Figs. \ref{fig:newtrot0026} and \ref{fig:newtrot0030} where results are shown for N=4 and two different time step sizes, and can be perfectly checked numerically. It can be explained by the following argument.

The operator we need to study is 

\begin{equation}
\begin{split}
\lim_{P \rightarrow \infty}  & \left( \Pi_0 \exp(-( \sigma_x \otimes M) \bar{\delta t})\exp(-( \sigma_y \otimes M) \bar{\delta t}) \right. \\
& \left. \exp(( \sigma_x \otimes M) \bar{\delta t})\exp(( \sigma_y \otimes M) \bar{\delta t})\Pi_0 \right)^P,
\end{split}
\end{equation}

where $\Pi_0$ is the projector onto the $|0\rangle$ state of the ancilla. Expanding in the energy eigenstate basis and applying the projector and simplifying one obtains the following operator

\begin{equation}
\begin{split}
& \sum_{nm} \left( \lim_{P \rightarrow \infty} \left( \cosh (2\Delta(E_n-E_m) + \frac{\sqrt{2}}{4}\exp{i\frac{3\pi}{4}}(\cosh(4\Delta(E_n-E_m))-1)  \right)^P \right. \\
& \left. \hspace{5mm} |E_nE_m\rangle \langle E_nE_m| \right)
\end{split}
\label{eq:limitp}
\end{equation}

where $\Delta = \bar{\delta t} \sqrt{\Gamma/2}$ and $E_n$ is the spectrum of the original hamiltonian.
This operator drives any state to its diagonal ensemble if we only consider terms $n=m$. So we have to prove that terms $n \neq  m$ do not contribute to the sum in the limit.

In order to prove this, we study the modulus of the complex number inside brackets that multiplies the matrix. In the case $n=m$ the modulus is 1. So we expect that terms with $\mod<1$ will not contribute in the infinite limit, while terms with $\mod>1$ will instead prevail.

\begin{figure}[h!]
\centering
\includegraphics[scale=0.60]{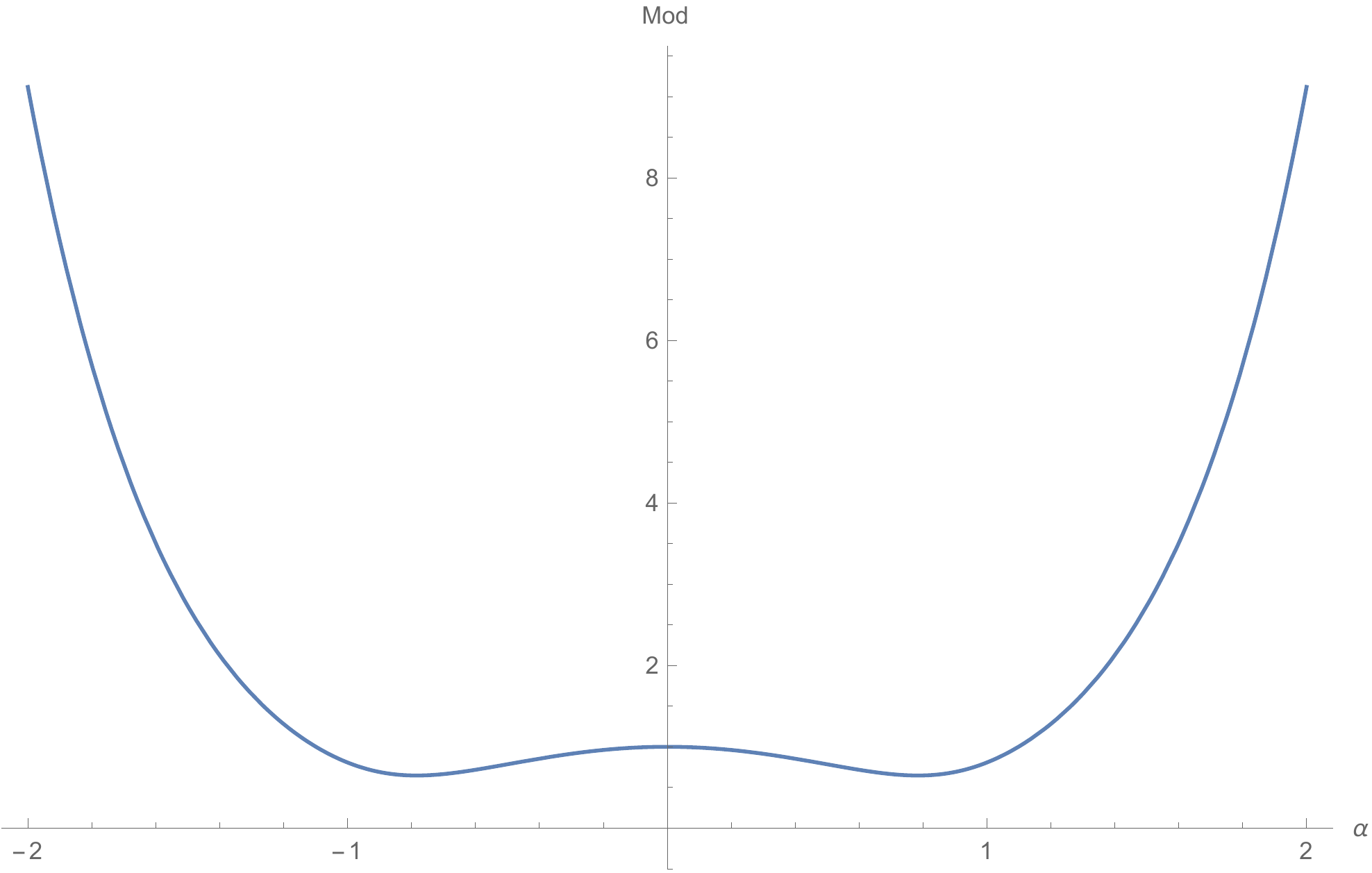}
\caption{Modulus of the coefficient in front of the basis in Eq. \ref{eq:limitp} with $\alpha$. For $\alpha=0$, the diagonal ensemble, the modulus is 1, while for $\alpha$ below a certain value the modulus is always smaller than 1, thus allowing exact convergence to the diagonal ensemble.}
\label{alpha}
\end{figure}

Fig. \ref{alpha} depicts the value of the modulus with $\alpha=\delta t \sqrt{\Gamma/2} (E_n-E_m)$. We can see a whole region of small $\alpha$ where the modulus is smaller than 1. This means that, as expected, there exists a certain $\delta t_c$, dependent on $(E_n-E_m)_{max}$, the highest energy difference in the spectrum, below which all terms with $n\neq m$ will not contribute and the approximate evolution will converge exactly to the diagonal ensemble.

Obtaining the critical value from the graph, we get the following relationship

\begin{equation}
\delta t_c = \frac{2}{\Gamma}\frac{1.2074}{{(E_n-E_m)^2}_{max}}
\end{equation}

This expression matches perfectly with the numerical results. We obtain $\delta t_c(n=2) = 0.129$ and $\delta t_c(n=4) = 0.027$. Assuming that $(E_n-E_m)_{max}$ increases linearly with $N$, $\delta t_c$ will decrease with $1/N^2$. This apparently is not very good news, as it is the same scaling of the Taylor expansion. Nevertheless, we plan to complete the project in the future by implementing the algorithm with MPO and check if there is a significant speedup in the computation of the diagonal ensemble, which would allow us later to check thermalisation for this and many other systems.

\chapter{Conclusions and outlook}

\pagestyle{fancy}
\fancyhf{}
\fancyhead[LE]{\thepage}
\fancyhead[RE]{CONCLUSIONS}
\fancyhead[LO]{}
\fancyhead[RO]{\thepage}

In this thesis we have addressed two of the most studied topics in the modern science of quantum information, multipartite entanglement and quantum computation. 

After beginning with a review of multipartite entanglement, first chapter focuses on the study of systems of four qubits, where we have used the hyperdeterminant as an entanglement measure because of its potential for representing genuine multipartite entanglement. The state with maximum hyperdeterminant has been found, together with some interesting properties. It is an interesting path for the future to find its parent hamiltonian and to understand how could it be more useful than other states as a resource in information processing tasks.

Second chapter turns the attention to absolutely maximally entangled states (AMEs), those that have maximally mixed states in all possible reductions. We have explored their relationships with classical codes and combinatorial designs, and have introduced the concept of multiunitarity of a matrix as a property deeply linked with them. A complete characterization of AMEs for all Hilbert spaces remains to be done, and as in the previous chapter, a better understanding of its aplications in information processing.

In the third chapter we have turned our attention to the field of Bell inequalities, focusing on an operational approach which is very useful to find classical and quantum bounds. We have found new Bell inequalities for qutrits with their bounds, with a table of results that suggest a structure that remains to be completed. Also we have provided a way to build Bell inequalities starting from maximally entangled states, making an identification between the computational basis of the state and the measurement basis of the Bell inequality. One surprising result is that all states maximally violating these inequalities are variants of the GHZ state, and we would like to know which Bell inequalities are maximally violated by AMEs and states with maximum hyperdeterminant.

Going from fundamental to slightly more practical issues, we begin in fourth chapter to deal with the hamiltonians that originate entangled states. We have proposed there the use of the entanglement spectrum as a way to measure the distance between theories, using some theoretical arguments using the concept of the renormalization group and bringing numerical evidence that it behaves as we would expect from a distance measure. And in the last chapter of the first part, we have dealt with the concept of frustration, which appears in certain ground states and makes computations especially difficult, and have proposed a new tensor network model using triangular simplices. Efficient algorithms remain to be created in order to solve practical problems for real-life hamiltonians.

The second part of the thesis deals with the more practical and technological subject of quantum computation. We begin by introducing the field of cloud quantum computation, which could be said to have been created last year with the appearance of the first quantum computer available to the general public, a computer we have tested with two different algorithms. We have found that, while clearly performing quantum computations, the errors are still large and the power still too limited to be useful, but it provides a great proof of concept and an experimentation basis for future improvements. We have also provided a possible error analysis to define the uncertainty of the results of these computers of which we don't have direct access to. In the few months that have passed since its appearance many improvements have already been done and we believe that this and many other computers will cause an explosion of this field in the near future.

Seventh chapter deals with a different model of computation, adiabatic quantum computation, and in particular with the analysis of whether the computation time really grows unboundedly when the gap of the hamiltonian goes to zero. We have found that this is indeed the case for the Ising model, as the overlap between the ground state and the first excited states never goes to zero and is in fact 1 for the critical point.

Finally, in the last chapter we have proposed a new algorithm to compute the diagonal ensemble efficiently, with the motivation to study thermalisation of systems. We have proved that the algorithm does provide a correct result, although it remains to be worked out if it is really more efficient than others with an implementation with MPO that we plan to do in the near future.

Many of the subjects touched in this thesis will continue to be topical in the next years and especially the promised technological advances in quantum computing should allow us to solve some of the unanswered questions of this thesis, and in general of the whole field.

\appendix
\appendixpage
\noappendicestocpagenum
\addappheadtotoc

\chapter{Bell inequalities}

\pagestyle{fancy}
\fancyhf{}
\fancyhead[LE]{\thepage}
\fancyhead[RE]{Appendix A: Bell inequalities}
\fancyhead[LO]{}
\fancyhead[RO]{\thepage}

\section{Maximizing settings: mutually unbiased bases and multiplets of optimal settings} \label{mubs}

In Chapter 3 we have shown that two remarkable sets of measurement settings optimize the violation of Bell inequalities. These are the \emph{mutually unbiased bases} (MUB) and \emph{multiplets of optimal settings} (MOS). Two orthonormal bases $\{|\phi_0\rangle ,\dots ,|\phi_{d-1}\rangle \}$ and $\{|\psi_{0}\rangle ,\dots ,|\psi_{d-1}\rangle \}$ are mutually unbiased if
\begin{equation}
|\langle \phi_{j}|\psi_{k}\rangle |^{2}={\frac {1}{d}},\quad \forall j,k\in \{0,\dots ,d-1\} \, .
\end{equation}
If $d$ is a prime power number, i.e. $d=p^n$ for $p$ prime and $n\in\mathbb{N}$, then there exists a maximal set of $d+1$ MUB. In prime dimensions such set is given by the eigenvectors bases of the $d+1$ generalized Pauli operators defined in Eq.\ref{genpauli}
\begin{equation}
X, Z, XZ, XZ^2, ..., XZ^{d-1}.
\end{equation}
We say that a set of normal operators is MUB if their eigenvectors bases are MUB.
For example, the optimal settings for Mermin inequalities for qubits are MUB. Indeed, if one setting is fixed to $\sigma_x$ then the other setting has to be a linear combination of the form $\alpha \sigma_y + \beta \sigma_z$ in order to maximize the eigenvalue of the commutator. This restriction implies that the settings are MUB.

In the qutrit case, the optimal settings for the CGLMP inequality, $A=\lambda_3$ and $A'=\frac{2}{3} (\lambda_1+\lambda_6) + \frac{1}{6} (\lambda_3+\sqrt{3}\lambda_8)$, are not MUB. However, for three qutrits the optimal settings, $A=\lambda_3$ and $A'=\frac{1}{\sqrt{3}} (\lambda_2+\lambda_4+\lambda_6)$, are MUB.

For Bell operators with complex settings the optimal settings have a more regular structure. The elements of the basis $X^i Z^j, X^k Z^l$ are MUB except for the case where $j=l$ and $i=k$. So it is clear that in 3 and 4-qutrit cases the optimal settings are mutually unbiased ($A=X$ and $A'=Z$) while in the 2 and 6-qutrit cases $A=X$ and $A'$ is a combination that includes $X$ \eqref{complexap}, so it cannot be unbiased with respect to $A$.

We have introduced the notion of {\sl multiplets of optimal settings} (MOS) which
 denotes any set of matrices that maximize the 2-qutrit and 6-qutrit inequalities,
  and all the 2 qudit inequalities. One is obtained from the other by applying a phase matrix and
 then the Fourier transform 
 and they have the property that both the commutator and the anticommutator of any pair of MOS are nilpotent matrices, i.e., matrices $M$ such that $M^k=0$ for some integer $k$. If one of the settings is set to $X$ then the other one has the following form
\[ MOS = e^{i\phi} \left( \begin{array}{ccccccc}
0 & 0 & 0 & ... & ... & ... & 1 \\
-1 & 0 & 0 & ... & ... & ... & 0 \\
0 & -1 & 0 & ... & ... & ... & 0 \\
...... & ... & ... & ... & ... & ... & ...\\
...... & ... & ... & ... & ... & ... & ...\\
...... & ... & ... & ... & ... & ... & ...\\
0 & ... & ... & ... & 0  & -1 & 0
\label{genmap}
\end{array} \right),\] 
where $\phi$ is a global phase that has to be tuned when changing between different forms of equivalent Bell inequalities. So, it is the same as $X$ but with opposite signs in all elements except for the first one, and a global phase.

\section{From two to three qutrits}
\label{2to3}

From the 2-qutrit inequality \eqref{cglmpim} it is possible to derive the 3-qutrit inequality \eqref{3qutrits}, under the assumption of symmetry for an additional third party. Starting from Eq.\ref{cglmpim} it follows the sequence
\begin{eqnarray}
&& \hspace{3.8cm} [w (ab) - (a'b+ab') + w (a'b')]_A \leq \sqrt{3}, \nonumber\\ 
&& \hspace{3.1cm} [-i (w (ab) - (a'b+ab') + w (a'b'))]_H \leq \sqrt{3}, \nonumber\\
&&\hspace{2.3 cm}  [\frac{w^2-w}{\sqrt{3}} (w (ab) - (a'b+ab') + w (a'b'))]_H \leq \sqrt{3}, \nonumber\\
&& \hspace{0.5 cm} [(1-w^2)(ab)+(w-w^2)(a'b+ab')+ (1-w^2)(a'b'))]_H \leq 3, \nonumber\\
&&  [(ab)-w^2(ab+a'b+ab') + w(a'b+ab')+(w+2)(a'b')]_H \leq 3, \nonumber\\ 
&&  [(ab)-w^2(ab+a'b+ab') + w(a'b+ab'+a'b')+ 2(a'b')]_H \leq 3.\nonumber\\
\end{eqnarray}
This form of the 2-qutrit CGLMP inequality suggests an 8-term symmetric inequality for three qutrits, where all terms with the same number of primes should have the same coefficients. By inserting $c$ and $c'$ according to this last requirement we have
\begin{eqnarray}
&[(abc)-w^2(abc'+a'bc+ab'c)+& \nonumber\\
&w(a'bc'+ab'c'+a'b'c)+2(a'b'c')]_H \leq 3.&
\end{eqnarray}
Thus, the symmetric 3-qutrit inequality \eqref{3qutrits} is obtained.

\section{Generalization to $d$ dimensions} \label{ddimensions}

In Ref. \cite{Collins02} the bipartite CGLMP is extended to $d$ outcomes. Its expression in the probability language reads
\begin{eqnarray}
C_{22d} = && \sum_{k=0}^{[d/2]-1} \left( 1-\frac{2k}{d-1} \right)\\
&& \bigl(p(a=b+k)+p(b=a'+k+1)+\nonumber \\
&&p(a'=b'+k)+p(b'=a+k) \nonumber \\
&& -( p(a=b-k-1) + p(b=a'-k)+\nonumber \\
&&p(a'=b'-k-1) + p(b'=a-k-1))\bigr)\le 2. \nonumber
\end{eqnarray} 
Let us write these inequalities in term of operators. In order to do this let us start from a different form for \eqref{cglmpim} presented for example in Ref. \cite{Chen02}
\begin{eqnarray}
C_{223} = &&  [ab+ab'+a'b-a'b']_H \nonumber \\
&& +\frac{1}{\sqrt{3}} [-ab+ab'+a'b-a'b']_A \le 2 \, .
\label{reim}
\end{eqnarray}     
In order to transform from probabilities to operators we have to establish a match between the number of variables and the number of equations. The variables here are the joint probabilities $p (a=b+k)$, with $k$ running from 0 to $d-1$, so there are $d$ unknowns. We need therefore $d$ equations. One equation is given by the normalization condition, i.e., the sum of probabilities is 1. For $d=2$, a second equation is enough, and that is the definition of expectation value of the product
\begin{equation}
ab = p(a=b)-p(a=b+1) \, .
\end{equation}
For $d=3$ there are 3 equations. Apart from the normalization of probabilities, two extra equations are needed, and those can be the hermitian and antihermitian parts of the expected value of the product, as in Eq. \ref{reim}. It appears to be an accident that the CGLMP for $d=3$ can be expressed solely with the antihermitian part by inserting powers of $w$ as in Eq. \ref{cglmpim}.

For $d=4$ we add the hermitian part of the expected values of the squares of products, and for $d=5$ we add their antihermitian part. The concrete expressions read as follows
\begin{eqnarray}\label{C42}
C_{224}&=&\frac{1}{3}\left( 2  [ab+ab'+a'b-a'b']_H \right.\\
      && + 2  [-ab+ab'+a'b-a'b']_A \nonumber \\
      && \left. +  [(ab)^2+(ab')^2+(a'b)^2-(a'b')^2]_H \right)\nonumber
\end{eqnarray}
and
\begin{eqnarray}\label{C52}
C_{225}&\!\!=\!\!&\frac{1}{2}\bigl( [ab+ab'+a'b-a'b']_H+\\
&&  [(ab)^2+(ab')^2+(a'b)^2-(a'b')^2 ]_H\bigr)+ \nonumber \\
&&\frac{2}{5} \bigl((3s_1+s_2)  [-ab+ab'+a'b-a'b']_A+ \nonumber \\
&&(-s_1+3s_2)  [-(ab)^2\!+\!(ab')^2\!+\!(a'b)^2\!-\!(a'b')^2 \bigr) ]_A\, ,\nonumber
\end{eqnarray} 
where the numbers $s_1$ and $s_2$ are the imaginary parts of $e^{2\pi i/5}$ and $e^{4\pi i/5}$, respectively. The classical bounds for these operators are $\langle C_{42}\rangle_{LR}=2$ and $\langle C_{52}\rangle_{LR}=2$.

It is possible to derive the general expression of the Bell operator for any number of levels $d$ as follows
\begin{equation}
C_{22d} = N \left( \sum_{k=1}^{[d/2]} r_{k,d}\mathrm{H}_{(ab)^k} + \sum_{k=1}^{[(d-1)/2]} i_{k,d} \mathrm{A}_{(ab)^k} \right) \le 2,
\end{equation}
where $r_{k,d}$ and $i_{k,d}$ are constants related to real and imaginary parts of $w$ (in general related to both of them), $N$ is a normalization constant such that the maximal classical value of $C_{22d}$ is 2, and also
\begin{eqnarray}
\textrm{H}_{(ab)^k}&\equiv&  [(ab)^k+(ab')^k+(a'b)^k-(a'b')^k ]_H, \nonumber \\
\textrm{A}_{(ab)^k}&\equiv&  [-(ab)^k+(ab')^k+(a'b)^k-(a'b')^k ]_A.\nonumber
\end{eqnarray}
All these inequalities are maximally violated by d-dimensional MOS. The numerical violation ratios increase with $d$, and can be found for example in Ref. \cite{Acin02}.

\chapter{Compressed quantum computation}

\pagestyle{fancy}
\fancyhf{}
\fancyhead[LE]{\thepage}
\fancyhead[RE]{Appendix B: Compressed quantum computation}
\fancyhead[LO]{}
\fancyhead[RO]{\thepage}

\section{Validating circuit sets}
\label{app:validating}

In this section we present some details about the validating circuit sets from Chapter 6. As explained in the main text, we introduce the concept of validating circuits in order to estimate the error that occurs in a cloud quantum computation. To this end, circuits of similar complexity as the circuit of interest, the so-called validating circuits, are considered. Assuming that the outcome of the validating circuits can be computed classically, the error is determined by comparing the real computational outcome to the ideal one. Here, we construct 20 validating circuits for the compressed simulation of the Ising model by randomly exchanging Clifford gates with other Clifford gates in circuits 2 and 3 of Figure \ref{fig:cheatingcircuits}, where the number of $T$-gates and $CNOT$-gates is not changed. We choose circuit 2 and 3 as they are of different complexity, and they together are representative for the kind of circuits that we are dealing with in simulating the Ising spin chain. 

In Table \ref{tab:validation}, we present the error $e$ of the 20 validating circuits. We perform a $Y$ measurement on one of the qubits and calculate the error given by the difference between the measured value and the ideal value, $e = |\expect{Y_{measured}}  - \expect{Y_{ideal}}|$, of the 20 validating circuits. The average error is $0.122$.

\begin{table}[h!]
\centering
\footnotesize
  \begin{tabular}{| c | c | c | c | c | c | c | c | c | c | c |}
    \hline
    $C_2$  & 0.038 & 0.076 & 0.030 & 0.130 & 0.066 & 0.166 & 0.270 & 0.128 & 0.260 & 0.000
                    \\ \hline
    $C_3$ & 0.034 & 0.202 & 0.070 & 0.152 & 0.216 & 0.076 & 0.078     &  0.248 & 0.144 & 0.056
                    \\ \hline
  \end{tabular}
 \caption{Table of the error $e$ in measuring $Y$ on one qubit in the validating circuits, which are constructed by altering two of the circuits of interest, $C_2$ and $C_3$, 10 times each.}
 \label{tab:validation}
 \end{table}

Note that the accuracy of the error analysis can be improved by constructing validating circuits for each different value of $J$ and performing separate error analysis leading to individual error bars for each value of $J$ in Figure \ref{fig:results}. Instead, here we construct validating circuits only for two different values of $J$ and average over them as explained above, as we are limited in the number of computations that we can perform on the IBM quantum computer and we have consumed all our credits. 

\section{Circuits for the simulation of the four--qubit Ising chain}
\label{app:circuits}

\begin{figure}[ht]
	\centering
	\resizebox{1.0\textwidth}{!}{\includegraphics{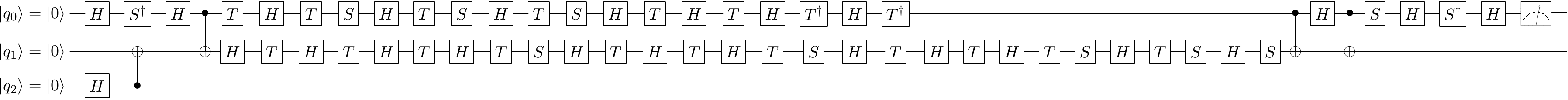}} \newline\vspace{1mm}
	\resizebox{1.0\textwidth}{!}{\includegraphics{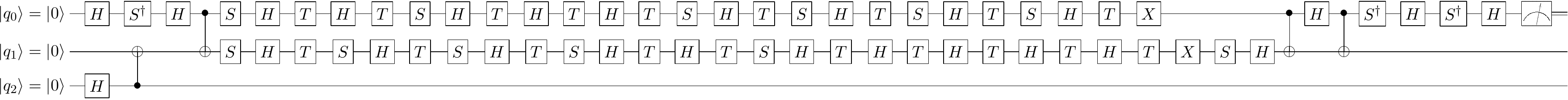}} \newline\vspace{1mm}
	\resizebox{1.0\textwidth}{!}{\includegraphics{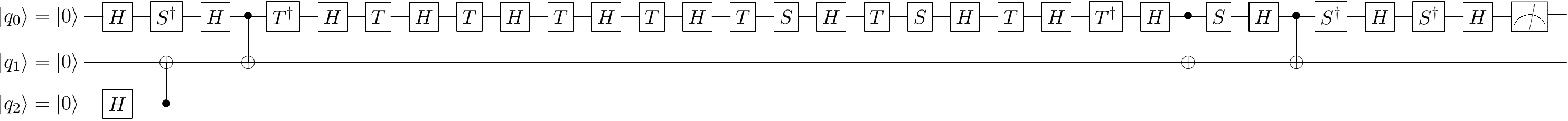}} \newline\vspace{1mm}
	\resizebox{1.0\textwidth}{!}{\includegraphics{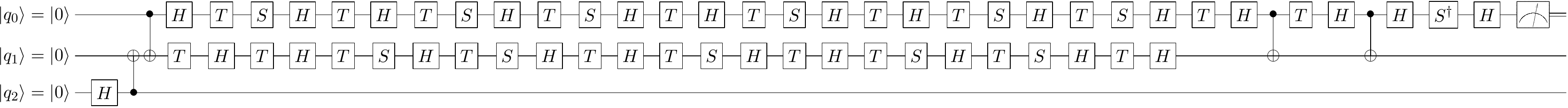}} \newline\vspace{1mm}
	\resizebox{1.0\textwidth}{!}{\includegraphics{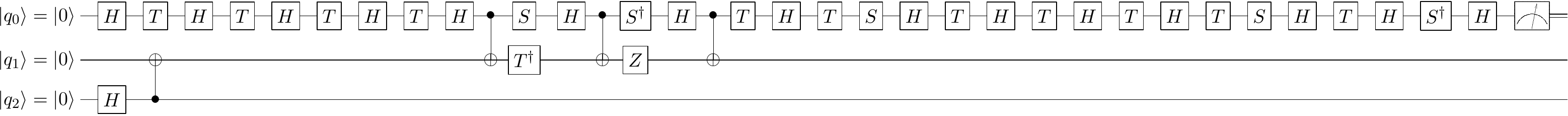}} \newline\vspace{1mm}
	\resizebox{1.0\textwidth}{!}{\includegraphics{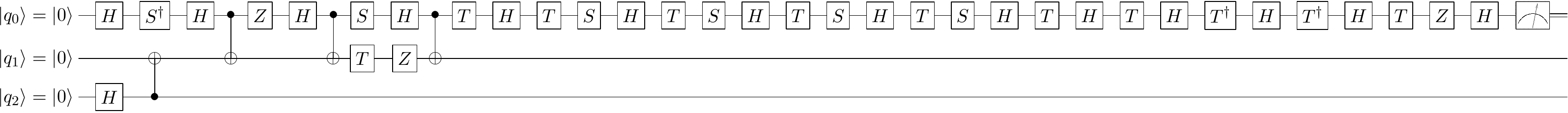}} \newline\vspace{1mm}
	\resizebox{1.0\textwidth}{!}{\includegraphics{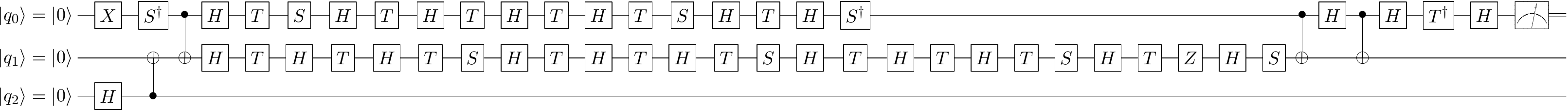}} \newline\vspace{1mm}
	\resizebox{1.0\textwidth}{!}{\includegraphics{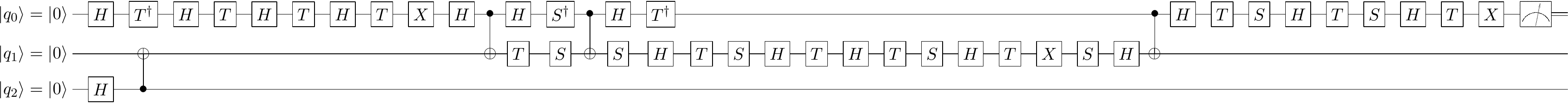}} \newline\vspace{1mm}
	\resizebox{1.0\textwidth}{!}{\includegraphics{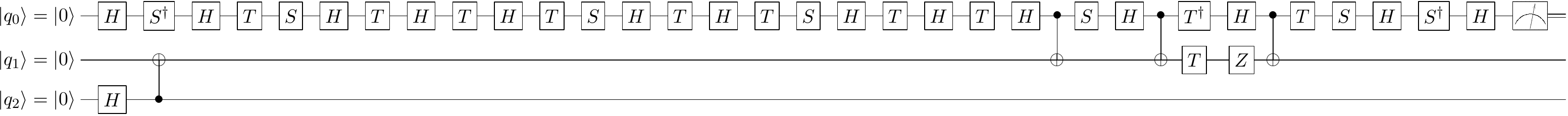}} \newline\vspace{1mm}
	\resizebox{1.0\textwidth}{!}{\includegraphics{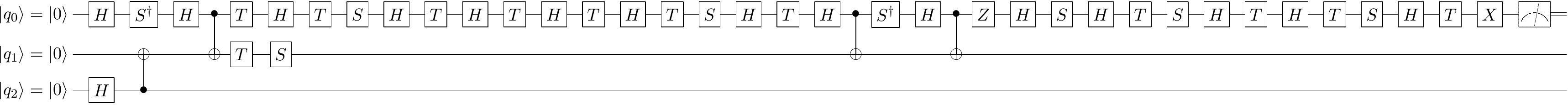}} \newline\vspace{1mm}
	\resizebox{1.0\textwidth}{!}{\includegraphics{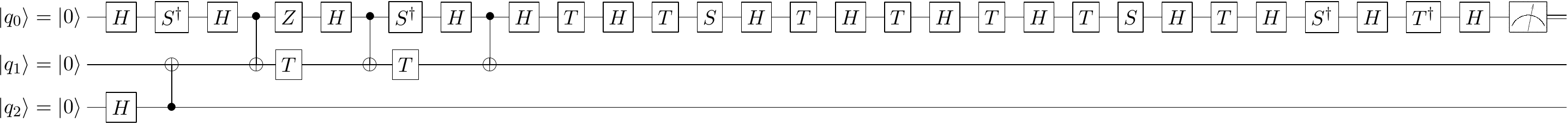}} \newline\vspace{1mm}
	\resizebox{1.0\textwidth}{!}{\includegraphics{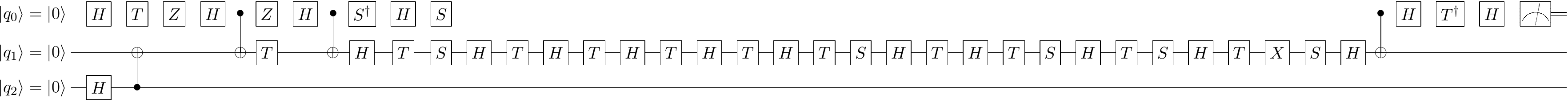}} \newline\vspace{1mm}
	\caption{Circuits implementing digital adiabatic evolution in order to simulate the magnetization of a four-qubit spin chain using two qubits. The twelve circuits correspond to values $J=\left\{\frac{1}{6}, \frac{2}{6}, \ldots, 2 \right\}$ as in Ref. \cite{Li14}. }
	\label{fig:cheatingcircuits}
\end{figure}

In this section we explicitly give the circuits simulating the magnetization of a $4$-qubit spin chain using $2$ qubits. We measure the magnetization of the spin chain at 12 equidistantly distributed values of $J$. In particular, we choose $J=\left\{\frac{1}{6}, \frac{2}{6}, \ldots, 2 \right\}$, as in Ref. \cite{Li14}. We also choose the parameters of the adiabatic evolution, $\Delta t=0.1$ and $L=2400$. See Sect. \ref{sec:compressed}  for an explanation of these parameters. As explained in the main text, we compute the unitary $W(J)$ performing the whole adiabatic evolution and decompose this unitary into the available gates set, as a step-wise implementation of the adiabatic evolution is not possible at the moment due to the current limit in circuit depth. We entangle qubit 2 with an auxiliary qubit, qubit 3, which is discarded afterwards in order to prepare $\identity$ on qubit 2. In each circuit we measure qubit 1 in order to obtain the magnetization $M(J)$. The explicit circuits for each value $J$ are given in Figure \ref{fig:cheatingcircuits}.

% \begin{widetext}
 
 %\end{widetext}

\section{Extension to more qubits}
\label{app:extension}

In the following, we argue that with the current version of the IBM quantum computer it seems unfeasible to run the compressed simulation of the Ising spin chain using three or more qubits and hence, simulating an eight or more-qubit spin chain. Nevertheless, we show that the computation will become possible once several improvements that IBM announced are implemented.

\begin{figure}[ht]
	\centering
	\resizebox{0.6\linewidth}{!}{\includegraphics{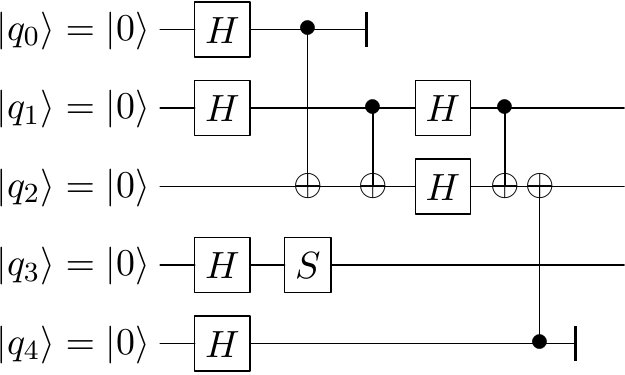}}
	\caption{Circuit for preparation of the three-qubit state $\rho_{in} = \frac{1}{4} \identity \otimes \ket{+_y}\bra{+_y}$.}
	\label{fig:preparation3}
\end{figure}

At the moment, performing the computation using three or more qubits seems not possible due to the restriction in circuit depth, the limited gate set, and the fact that gates cannot be implemented probabilistically. We show as an example that even preparing the initial state $\rho_{in}$ is a difficult task. To obtain the initial state, two of the qubits have to be prepared in a completely mixed state, while one qubit is prepared in $\ket{+_y}$. See Figure \ref{fig:preparation3} for a possible, but very uneconomical way to do so using a circuit of depth six and consuming two auxiliary qubits that are discarded in the process. Note that there seems to be no less wasteful way to prepare $\rho_{in}$ as applying gates probabilistically is not possible at the moment.

Nevertheless, once the improvements that IBM announced are available, implementing the circuit for more qubits will become possible. Here, we show as an example how to implement the circuit for three qubits, though the method can be generalized to more qubits. To this end, we assume that the following improvements are available. We assume that advanced classical processing is available. In particular, we assume that it is possible to apply gates probabilistically. Furthermore, we assume that subroutines are available, i.e., user-defined gates can be declared and used.

In this case the circuit can be implemented as follows. The initial state $\rho_{in} = \frac{1}{4} \identity \otimes \ket{+_y}\bra{+_y}$ is prepared by performing either a Pauli $X$ or $\identity$ with probability $1/2$ on both of the qubits for which we want to prepare $\frac{1}{2}\identity$ individually (which we will denote as qubits 1 and 2 in the following), and furthermore performing a single qubit unitary that rotates $\ket{0}$ to $\ket{+_y}$ for the remaining qubit, which we will denote as qubit 3 in the following.

\begin{figure}[ht]
	\centering
	\resizebox{0.6\linewidth}{!}{\includegraphics{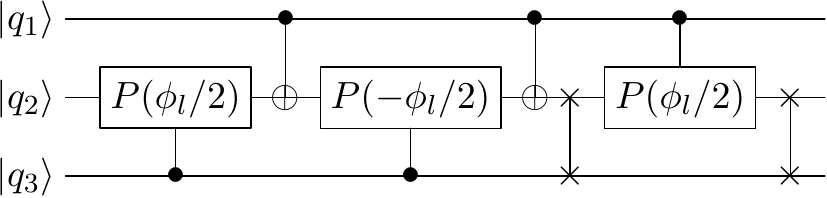}}
	\caption{Circuit implementing a $\Lambda_{1, 2} P_3(\phi_l)$ gate.}
	\label{fig:ccphase}
\end{figure}

\begin{figure}[ht]
	\centering
	\resizebox{0.8\linewidth}{!}{\includegraphics{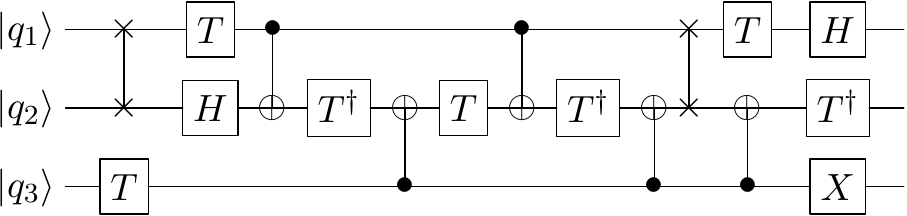}}
	\caption{Circuit implementing the operator $A$.}
	\label{fig:agate}
\end{figure}

After the initial state is prepared, the system is evolved adiabatically. In each step of this adiabatic evolution the unitary $U_d R_l^T R_0^T$ has to be applied. The unitaries $U_d$, $R_l^T$, and $R_0^T$ are given in the main text. The unitary $R_0$ is a single qubit unitary and hence can be implemented easily. We have $U_d = \Lambda_{1,2}P_3(\phi_l)$, where $\Lambda_{i_1, \ldots, i_n} G$ denotes a gate $G$ controlled by qubits $i_1, \ldots, i_n$. A possible implementation of this controlled phase gate is depicted in Figure \ref{fig:ccphase} \cite{Nielsen00}. Recall that the two swaps can be implemented using three $CNOT$ gates, while phase gates that are controlled by one qubit may be decomposed into two controlled not gates and three single qubit unitaries \cite{Nielsen00}.
Implementing $R_l^T$ is more tricky. First, one performs a basis transformation by applying $A^\dagger$, where $A = \ketbra{8}{1} + \sum_{k=1}^{7} \ketbra{k}{k+1}$. In the new basis the unitary $R_l^T$ is given by the unitary $\Lambda_{1,2} O^T(\phi_l)$, where $O(\phi_l) = e^{i \phi_l Y_3}$ followed by a single qubit unitary $O(\phi_l)$ \cite{Boyajian13}. Finally, the basis change has to be undone, i.e., $A$ is applied. The controlled rotation can be implemented in a similar way as shown above for $U_d$. In order to implement $A$, we use that this unitary can be decomposed into a Toffoli gate $\Lambda_{2, 3} X_1$ followed by a $CNOT(2,3)$, and a Pauli $X_3$ \cite{Boyajian16}. This circuit can be further decomposed using the decomposition of the Toffoli gate suitable for the IBM quantum computer \cite{IBM} and some simplifications, yielding the circuit depicted in Figure \ref{fig:agate}.

\begin{figure}[!ht]
	\centering
	\resizebox{0.8\linewidth}{!}{\includegraphics{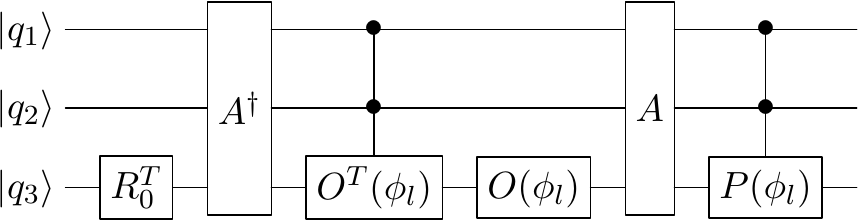}}
	\caption{Circuit implementing one step of the adiabatic evolution using three qubits.}
	\label{fig:3step}
\end{figure}

Altogether we obtain a circuit that implements one step of the adiabatic evolution depicted in Figure \ref{fig:3step}.
This circuit can be packed into a user-defined three-qubit gate depending on the free parameter $\phi_l$ and the adiabatic evolution is performed by applying these gates with increasing $l$ successively. Finally, measuring $Z$ on qubit 3 yields the magnetization of the eight-qubit spin chain.

As IBM announced, that advanced classical processing, and user-defined gates will become available in future, implementing the circuit for three or more qubits will become feasible, as long as the number of steps (recall that we used L=2400 steps before) is not an issue. Otherwise, similar methods as those used in the two qubit circuit will have to be applied.

%\addcontentsline{toc}{chapter}{Bibliography}

\end{document}